\documentclass[12pt, a4paper]{report}
\usepackage{pdfpages}
\usepackage{amsmath}
\usepackage{mathtools}
\usepackage{multirow}
\usepackage{rotating}

\usepackage{cleveref}
\usepackage{fyp}
\usepackage{xspace}
\usepackage{amsthm}
\usepackage{mathpartir}
\usepackage{amssymb}
\usepackage{amsfonts}
\usepackage{semantic}
\usepackage{centernot}
\usepackage{enumerate}
\usepackage{graphics}
\usepackage{amsmath}
\usepackage{xcolor}
\usepackage{relsize}
\usepackage{enumitem}
\usepackage{tikz}
\usepackage{breqn}
\usepackage{multicol}
\usepackage{shorthand}
\usepackage{actor}
\usepackage{lang}
\usepackage{detectEr}

\usepackage{pgfplotstable}
\usepackage{pgfplots}

\makeatletter
\renewcommand{\@chapapp}{}
\newenvironment{chapquote}[2][2em]
  {\setlength{\@tempdima}{#1}%
   \def\chapquote@author{#2}%
   \parshape 1 \@tempdima \dimexpr\textwidth-2\@tempdima\relax%
   \itshape}
  {\par\normalfont\hfill--\ \chapquote@author\hspace*{\@tempdima}\par\bigskip}
\makeatother

\begin{document}

\title{Towards Runtime Adaptation of Actor Systems}
\author{Ian Cassar}
\date{29$^{th}$ May 2015}
\supervisor{Dr. Adrian Francalanza}
\department{Faculty of ICT}
\universitycrestpath{crest}
\submitdate{29$^{th}$ May 2015} 

\include{}

\begin{acknowledgements}
I would honestly like to express my gratitude towards my supervisor Dr. Adrian Francalanza, whom despite being abroad and busy with his own research he always found enough time and patience to provide me with essential guidance and support. His excellent guidance, patience and tutoring made it possible for me to understand a lot of complex material that helped me make further progress in my work. I have learnt a lot from him and I must say that it has been quite a privilege working alongside him for the second time.

I would also like to thank my parents for always offering me their full support during my tough times. Even when I started to have doubts in my abilities, they never stopped believing in me, and always provided me with essential moral support. Moreover I would like to thank my loving girlfriend Melanie whom despite being busy with her final year in medical school, she was always by my side when I needed her most. The support of my loved ones kept me persevering and believing in myself even when I was facing tough situations. I therefore dedicate this dissertation to all of them.

Finally I would like to acknowledge that the research work disclosed in this dissertation was partially funded by the \textit{Master It!} scholarship scheme (Malta).
\end{acknowledgements}
       
\pagebreak 
\parbox{14cm}{
\vspace{6cm}
\begin{large}
	\begin{chapquote}{An Old English Proverb}
	\begin{huge}
		``When there's a will, there's a way.''	
	\end{huge}\medskip
	\end{chapquote}
\end{large}
}

\includepdf[pagecommand={\thispagestyle{plain}},pages={1}]{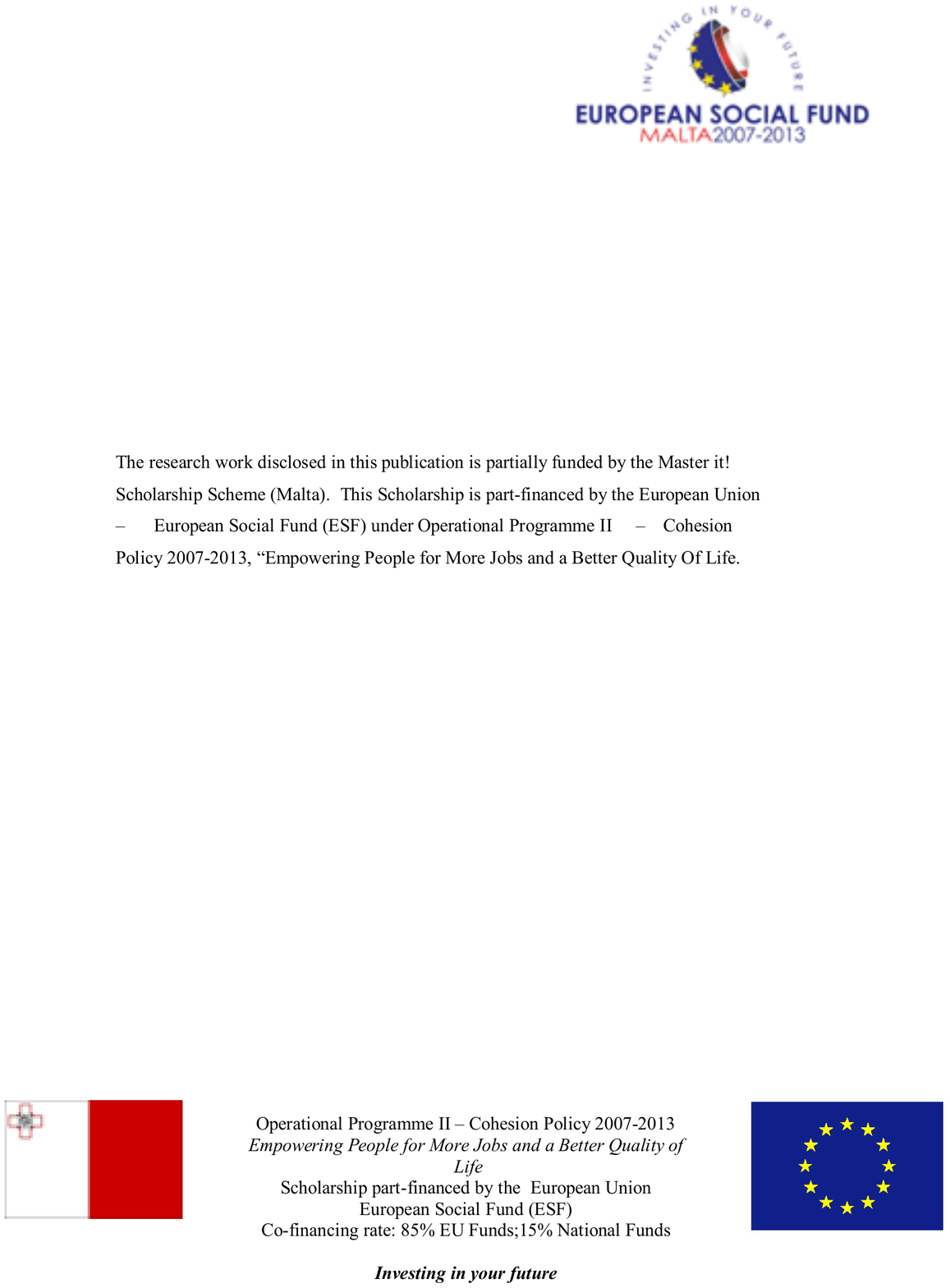}
      
\begin{abstract}
Runtime Adaptation (RA) is a technique prevalent to long-running, highly available software systems, whereby system characteristics are altered \emph{dynamically} in response to runtime events, while causing \emph{limited disruption} to the execution of the system. Actor-based, concurrent systems are often used to build such long-running, reactive systems which require suffering from \emph{limited downtime}. This makes runtime adaptation an appealing technique for mitigating erroneous behaviour in actor-systems, since mitigation is carried out while the system executes.

In this dissertation we focus on providing effective adaptations that can be \emph{localised} and applied to specific concurrent actors, thereby only causing a temporary disruption to the parts of the system requiring mitigation, while leaving the rest of the system intact. We make the application of localised adaptations efficient through \emph{incremental synchronisation}, whereby the specifier can strategically suspend specific parts of the system, whenever this is strictly required for ensuring that adaptations are effectively applied. We also study static analysis techniques to determine whether the specified incremental synchronisation is in some sense \emph{adequate} for local adaptations to be carried out.

We thus identify a number of generic adaptations that can be applied to \emph{any} actor system, regardless of its design and the code that it executes. We implement the identified adaptations as an \emph{extension} of an \emph{existing} Runtime Verification tool for actor-systems, thereby creating a RA framework for monitoring and mitigating actor systems. In parallel to our implementation we also develop a \emph{formal model} of our RA framework that further serves to guide our implementation. This model also enables us to better understand the subtle errors that erroneously specified adaptation scripts may introduce. We thus develop a \emph{static type system} for detecting and rejecting erroneous adaptation scripts prior to deployment, thereby providing the specifier with assistance for writing valid scripts. Although the static typesystem analyses scripts with respect to certain assumptions, we do not assume that the monitored system abides by these assumptions. We therefore augment our RA framework with \emph{dynamic checks} for halting monitoring whenever the system deviates from our assumption. Based on this dynamically checked model of our RA framework, we prove \emph{type soundness} for our static type system. 

As a result of this dissertation we thus implement and formalise a \emph{novel} a Runtime Adaptation framework for actor systems as extension to an existing Runtime verification tool. Furthermore, exploring the mixture of static and dynamic typechecking, in the context of runtime verification, for the purpose of adaptation is also quite novel. This exploration lead to the developing a \emph{novel} type analysis technique for detecting erroneously specified runtime adaptation scripts.  

\end{abstract}

\tableofcontents

\listoffigures


\mainmatter

\chapter{Introduction}
\label{sec:intro}
Runtime Adaptation (RA) \cite{Kell08survey,Kalareh:phd,JacquesSilva12} is a technique prevalent to long-running, highly available software systems, whereby system characteristics  (\eg its structure, locality \etc) are altered \emph{dynamically} in response to runtime events (\eg detected hardware faults or software bugs, changes in system loads), while causing \emph{limited disruption} to the execution of the system.   Numerous examples can be found in service-oriented architectures \cite{Oreizy08SOA,RA-SOA:2008}  (\eg cloud-services, web-services, \etc) for self-configuring, self-optimising and self-healing purposes \cite{ibm2005architectural}; the inherent component-based, decoupled organisation of such systems facilitates the implementation of adaptive actions \emph{affecting a subset} of the system while allowing other parts to continue executing normally. 

Actor systems \cite{Agha:1986,actorsinscala,Cesarini:2009} consist of \emph{independently}-executing  components called \emph{actors}. Every actor is \emph{uniquely-identifiable} with a unique process id, has its own \emph{local memory}, and can also \emph{spawn} other actors and interact with them through \emph{asynchronous messaging}. 
Actors systems are often 
used to build the aforementioned service-oriented systems with limited downtime \cite{Armstrong07, yaws:12, actorsinscala}.  Actor-oriented coding practices such as \emph{fail-fast} design-patterns \cite{Cesarini:2009,actorsinscala} already advocate for a degree of RA that contribute towards building robust, fault-tolerant systems. Such systems are developed using mechanisms such as \emph{process linking} and \emph{supervision trees}, in which supervisor actors can detect crashed actors and 
respond through adaptations such as  \emph{restarting} the actors, \emph{reverting} them through a previous (or different) implementation version, or else \emph{killing} further actors 
that may potentially be affected by the 
crash.

This dissertation studies ways how runtime adaptation for actor systems can be \emph{extended} to be able to respond to sets of runtime events that go beyond actor crashes. More specifically, we would like to observe \emph{sequences of events}  that allow us to invoke an \emph{adaptation} action to mitigate the effects of the detected misbehaviour. This technique, however, can also be used to observe \emph{positive} (liveness) events that allow us to adapt the system to execute more \emph{efficiently} (\eg by switching off unused parts). More generally, we intend to develop a framework for extending actor-system functionality through RA, so as to improve aspects such as resilience and resource management. 

Due to the inherent concurrency of actor-systems, adaptation actions can be achieved using generic, actor-level manipulations (\eg restart misbehaving actors) to mitigate only the concurrent components (actors) that contributed to the detected misbehaviour, thus leaving the other actors unaffected. Although similar adaptations are already being employed by using certain system design conventions (\eg fail-fast, supervision trees, \etc), it can be quite challenging to implement effective adaptations that can be applied on any system without requiring it to be developed in a particular manner. This is because effective adaptations require the monitor to temporarily synchronise with the components in need of adaptation, so as to achieve a tighter level of control upon them. This allows for the adaptations to apply the required mitigation in a \emph{timely manner} thereby increasing the effectiveness of the the applied mitigation. However, actor systems \emph{do not} natively support synchronisation. Implementing such an RA framework is even more challenging as to our knowledge, runtime adaptation \wrt actor systems has not yet been explored, and is therefore quite novel.   

\section{Aims and Objectives} \label{sec:intro:obj}
We aim to introduce Runtime Adaptation for actor systems by \emph{extending existing} Runtime Verification (RV) tools, such as \cite{diana04,FJN+11bip,elarva:2012,FraSey14}, with RA functionality. The appeal of such an approach is that RV tools \emph{already} provide mechanisms for \emph{specifying} the system behaviour needed to be observed, together with \emph{instrumentation mechanisms} for observing such behaviour. We also aim to introduce adaptations that can be \emph{localised} to individual components, thereby allowing for specific actors to be mitigated while leaving the other actors unaffected. 

As a proof-of-concept, we aim to focus on one of these actor-based RV tools called \detecterGen \cite{FraSey14} --- an RV tool for monitoring invariant (safety) properties about long-running (reactive) actor-based systems written in Erlang \cite{Armstrong07}; presently this tool is only able to perform detections that are exclusively asynchronous. Using this tool we want to investigate ways how to insert a degree of synchrony so as to allow for detections to be \emph{replaced} by effective adaptation actions that respond to behaviours detected, while reusing as many elements as possible  from the existing technology.


\begin{figure}[t]
 \centering
 \begin{tikzpicture}[>=latex,auto,thick]
    \begin{scope}[draw=blue!50,fill=blue!20,minimum size=0.60cm, align=left]
	    \node (Incrementor) at (-2.1,0) [shape=rectangle,draw,fill]  {\;\,\quad\textbf{Incrementor}\quad};
		\node (Decrementor) at (2.1,0) [shape=rectangle,draw,fill]  {\;\quad\textbf{Decrementor}\quad};
		\node (Interface) at (0,-1.6) [shape=rectangle,draw,fill]  {\;\quad\textbf{Common-Interface}\quad};	 
    \end{scope}
    
    \begin{scope}[draw=black,fill=black, align=right, font=\small]
	    \node at ([xshift=0.23cm,yshift=-0.25cm]Incrementor.north west) [shape=rectangle,draw,fill,font=\small]  {${\color{white}j}$ };
		\node at ([xshift=0.20cm,yshift=-0.23cm]Decrementor.north west) [shape=rectangle,draw,fill,font=\small]  {${\color{white}k}$\! };
		\node  at ([xshift=0.21cm,yshift=-0.22cm]Interface.north west) [shape=rectangle,draw,fill,font=\small]  {${\color{white}i}$ };	 
    \end{scope}
    
    \begin{scope}[align=center]
    	\node (in) at (0,-2.9)  {(1) $\etuple{\eatom{inc},3,cli}$};  
    	\node (outGood) at (-3.5,-2.9)  {$(3)\;\etuple{\eatom{res},4}$};  
    	\node (outErr) at (3.5,-2.9)  {$\color{red}(3)\;\eatom{err}$};  
    \end{scope}    
    \begin{scope}[draw=black]
	    \draw[
            ->] (in) [align=center] to node [left]{} (Interface); 	  	    
	    \draw[->] (Interface) [align=center,bend right=25] to node [left]{\\[1mm] $(2)\;\etuple{\eatom{inc},3,cli}$} (Incrementor); 	  	    
	    \draw[->] (Incrementor) [align=center,bend right=25] to node [left]{} (outGood); 
	 \end{scope}
     \begin{scope}[draw=blue]
        \draw[dashed,-] (4,-2.2) to (-4,-2.2);
        \node (ext)  at (-5,-2.5) {\textcolor{blue}{External View}};  
     \end{scope}
    \begin{scope}[draw=red, fill=red]
	    \draw[->] (Interface) [align=center,bend left=25] to node [right]{\\[1mm] $\color{red}(2)\;\etuple{\eatom{inc},3,cli}$} (Decrementor); 	
	    \draw[->] (Decrementor) [align=center,bend left=25] to node [right]{} (outErr); 	  	    	    
    \end{scope}   
  \end{tikzpicture}
	\caption[An example server actor implementation]{An example server actor implementation offering integer increment and decrement services.}
	\label{fig:sys}
\vspace{-2mm}
\end{figure}
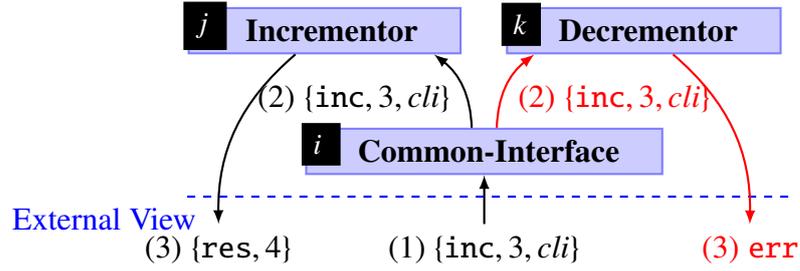
 
%
\begin{example} 
   \label{ex:intro}
In \figref{fig:sys} we present a simple running example system to which we refer in different parts throughout this dissertation. This figure depicts an actor-based server consisting of: 
\begin{itemize}[leftmargin=5mm] \setstretch{0.5}
  \item a front-end \emph{common-interface} actor with identifier $i$, receiving client requests; 
  \item a back-end \emph{incrementor} actor with identifier $j$, handling integer increment requests; 
  \item and a back-end \emph{decrementor} actor  $k$, handling decrement requests.  
\end{itemize} 
  A client sends service requests to actor $i$ of the form $\tup{\textsl{tag},\textsl{arg},\textsl{ret}}$ where: \textsl{tag} selects the type of service, \textsl{arg} carries the service arguments and \textsl{ret} specifies the return address for the result 
  (typically the client actor ID). The interface actor forwards the request to one of its back-end servers (depending on the \textsl{tag}) whereas the back-end servers process the requests, sending results (or error messages) to \textsl{ret}. With respect to this simple system we can define several safety (invariant) RV properties including \Propref{prop:intro:1}. \bigskip
  
  \begin{property} \label{prop:intro:1}
  	\emph{For every increment request, $\{\eatom{inc},n,cli\}$, that is received by the common-interface $i$, the requesting client, $cli$, should \emph{never} be sent an error message \eatom{err}.}  
  \end{property}\bigskip
  
  \noindent With an RA framework developed as \emph{extension} of an RV setup, RV properties can easily be augmented with adaptation actions that the monitor can apply upon detecting a violation. For instance, RV \propref{prop:intro:1} can thus be extended into RA \propref{prop:intro:2} below.\bigskip
  
  \begin{property} \label{prop:intro:2}
  	\emph{For every increment request, $\{\eatom{inc},n,cli\}$, that is received by the common-interface $i$, should the requesting client, $cli$, be sent an error message \eatom{err},  this should be mitigated by: restarting the common-interface $i$, and purging the mailbox of the system actor that sent the error message to the client.}  \exqed
  \end{property} \bigskip

\end{example}

\noindent To fulfill our aim of developing such a runtime adaptation framework for actor systems, we divide our work into the following three objectives:
\begin{enumerate}[label=(\roman*),leftmargin=7mm]
	\item As a minor objective, we first want to carry out a preliminary study \wrt the different techniques by which synchronous monitoring can be introduced in asynchronous actor systems, and thus integrate them within the \detecterGen\ RV tool. Although it is generally accepted that synchrony increases monitoring overheads, we are not aware of any studies that attempt to quantify and assess by how much, especially \wrt actor systems. Hence we also want to carry out an impact assessment based on the studied synchronous monitoring techniques (\ie the ones we integrate in \detecterGen). This would aid us in identifying the ideal synchronisation mechanism for introducing effective adaptation actions within \detecterGen\ in an \emph{efficient} way. 
	\item More importantly, we want to identify and implement \emph{generic adaptation actions} that are relevant in the context actor systems. To be able to effectively apply these adaptations we want to build upon the efficient synchronous monitoring candidate technique identified in (i), so as to permit the monitor to temporarily achieve tighter control over the system (or parts of it) thereby being able to effectively execute the required mitigation.	Using this synchronisation technique and the identified adaptations, we want to extend \detecterGen\ along with its specification language resulting in an effective RA framework for actor systems. This framework should allow the specifier to apply the necessary adaptations on \emph{any} actor-based system (written in Erlang) regardless of the code it executes or the way it is designed and implemented. Providing the monitor with more control can however lead to introducing errors in the system. To be able to better identify and study these errors we want to develop a \emph{formal model} \wrt the implementation constraints of our RA framework.
	\item  Formalising our RA framework puts us in a position to identify and study in more depth the errors that our monitors may introduce, without having to deal with the complexities of the implementation. We conjecture that some of the identified errors can be statically detected by thoroughly inspecting the adaptation scripts prior to deployment. We therefore want to look into static analysis techniques that can assist the specifier into writing error-free adaptation scripts. 
\end{enumerate}

\noindent Each objective contributes in its own way towards developing a \emph{more refined} Runtime Adaptation framework for actor systems. For instance in (i) we address performance aspects related to developing an \emph{efficient} (yet effective) RA framework; while in (ii) we identify the relevant adaptation actions and address their implementability issues. This contributes to having a more effective RA framework that provides the specifier with an adequate number of \emph{generic adaptations} that can be applied to any actor-based (Erlang) system. Furthermore, formalisation also contributes towards developing a more understandable RA framework. In fact in (iii) we plan to exploit the understandability of the formal model to address certain incorrectness that the RA monitors may now introduce. This contributes towards a more conducive RA framework which provides guidance to the specifier for writing \emph{error-free} adaptation scripts.
 
\section{Document Outline} 
In this thesis we address our three objective in Chapter \ref{chp:syn-asyn}, \ref{chp:runtime-adaptation} and \ref{chp:typ-sys}. Each of these chapters \emph{presents} and \emph{evaluates} the contributions required for fulfilling the respective objective. More specifically, we structure our document in the following manner:
\begin{itemize}[leftmargin=5mm]
	\item  In \Cref{chp:background} we provide the reader with the necessary background material for better understanding our work. We define and differentiate between Runtime Verification, Adaptation and Enforcement. 
	We also 
	identify a \emph{spectrum} of online monitoring approaches ranging from complete synchronous and complete asynchronous monitoring on opposite ends of the spectrum. Finally we present the actor model as implemented by a host language called Erlang. This is followed by an overview of the specification language used by our target RV tool, \detecterGen, for specifying RV properties for Erlang systems. 
	\item  In \Cref{chp:syn-asyn} we explore the design space for synchronisation mechanisms and assess the respective overheads induced as a result of increasing synchrony when monitoring actor systems. Based on the results of this impact assessment, we identify an efficient synchronisation technique that guides the implementation of our RA adaptation framework presented in \Cref{chp:runtime-adaptation}. 
	\item  In \Cref{chp:runtime-adaptation} we identify relevant adaptations and build on the efficient synchronisation technique identified in \Cref{chp:syn-asyn}. This allows for the implementation of more effective adaptations which require the monitor to achieve tighter control over the system. We then formalise the behaviour of our runtime adaptation scripts as operational semantics. This allows for better understanding the core concepts of our runtime adaptation framework without having to deal with the complexities of the implementation. 
	\item  As erroneous adaptation scripts can introduce errors, in \Cref{chp:typ-sys} we present a static type system 
	for identifying potential potential problems, that may arise from erroneous scripts, prior to deployment. Finally we evaluate the type system by proving type soundness for typed monitors.
	\item Finally, in \Cref{chp:related-work} we present other work related to different aspects of our research. We then conclude this document and provide future work suggestions in \Cref{chp:conc}. 
	\item The appendix chapters \ref{chp:app-mon-opt}, \ref{chp:app-syn-asyn} and \ref{chp:aux-lemmas} respectively provide: information about \detecterGen's monitor optimisations; additional RV properties used in our impact assessment along with detailed result tables; and the proofs for auxiliary lemmas required when proving theorems introduced in \Cref{chp:runtime-adaptation,chp:typ-sys}.
\end{itemize}

\chapter{Background}
\label{chp:background}
In this chapter we provide the reader with the necessary background information for better understanding our work. More importantly we aim to disambiguate terms and techniques that may sometimes be confused for one another, while we also want to establish what we truly mean when we refer to certain terms during the course of this dissertation.

For instance, in \secref{sec:rv-ra-re} we define and disambiguate between three monitoring techniques, namely Runtime Verification, Runtime Adaptation and Runtime Enforcement. This is intended to clarify what we mean by these three terms, 
as the distinction between them 
is not always made clear in the current literature.

Similarly, in \secref{sec:syn-asyn-survey} we disambiguate between two different classes of monitoring techniques found in the current literature, namely Offline and Online monitors; whereby we expand on the latter by looking into different definitions of online monitoring. 
Based on these definitions we devise a spectrum of online monitoring techniques \wrt component-based systems. Furthermore, we also define our own terms to identify and distinguish between the techniques in the spectrum. This is intended to help the reader understand the design space that we need to explore for augmenting the \detecterGen\ RV tool with synchrony in \Cref{chp:syn-asyn}.

Following this, in \secref{sec:erlang} we present an overview of the actor model with respect to Erlang $-$ a programming language that implements this model. This is intended to aid the reader in understanding the core concepts of actor-systems, along with technical (implementation-oriented) issues regarding the Erlang programming language. 

Finally, in \secref{sec:detecter-primer} we present a detailed overview of the \detecterGen\ RV tool, particularly about its specification logic. This is aimed to aid the reader in understanding the syntax and semantics of the logic used by this tool, upon which we build and extend in this dissertation.

\section{Runtime Verification, Adaptation and Enforcement}
\label{sec:rv-ra-re}
Ensuring formal correctness for actor-based, concurrent systems is a difficult task, primarily because exhaustive, static analysis verification techniques such as model-checking quickly run into state-explosion problems. This is typically caused by multiple thread interleavings of the system being analysed, and by the range of data the system can input and react to. 

We therefore look into three different (yet related) dynamic monitoring techniques that overcome the issues of static analysis techniques by verifying only a single execution, normally the currently executing one. In literature the boundaries between these techniques are often blurred; we therefore want to disambiguate between these monitoring techniques and establish upfront what we mean by the terms Runtime Verification (RV), Runtime Adaptation (RA) and Runtime Enforcement (RE).

The first technique that we consider is Runtime verification (RV) \cite{Bauer:ltl,Leu:RV:Overv}. This is a lightweight dynamic verification technique that provides an appealing compromise towards ensuring a degree of formal correctness without running into such scalability issues. These issues are avoid by only verifying that a specific execution of a \emph{system} (usually the currently executing one) satisfies or violates a specific correctness property. 

\begin{figure}[ht!]
	\centering
	\includegraphics[width=0.65\textwidth]{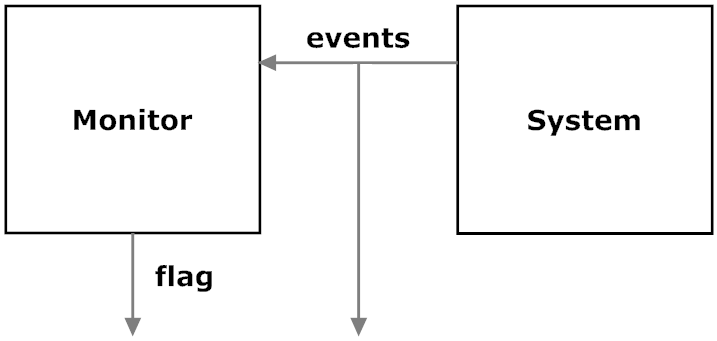}
	\caption{Runtime Verification}
	\label{fig:back:rv}
\end{figure}

In most runtime verification settings \cite{Bauer:ltl,FGP12DistribRV,FraSey14,taxonomy:Delgado:2004} a correctness property $\hV$ is generally specified in terms of a logic with precise formal semantics, and then automatically synthesised into an executable \emph{monitor} $\mondef{\hV}$. This monitor is essentially the executable implementation of the abstract property, which actually checks whether a given execution trace of a specific \emph{system}, violates (or satisfies) the property it was derived from. As illustrated in \figref{fig:back:rv}, the monitors in runtime verification usually have a \emph{passive} role \cite{Bauer:ltl,FraSey14}. In fact, they are generally concerned exclusively with receiving system events, analysing them and \emph{detecting} (flagging) violations (or satisfactions) of their respective correctness properties. Hence RV monitors refrain from directly modifying the system's behaviour in any way.

\begin{figure}[ht!]
	\centering
	\includegraphics[width=0.7\textwidth]{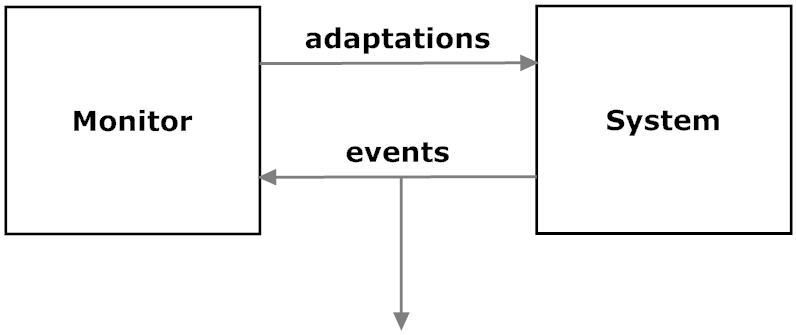}
	\caption{Runtime Adaptation}
	\label{fig:back:ra}
\end{figure}

Monitors in Runtime Adaptation \cite{Oreizy08SOA,Kell08survey,JacquesSilva12,saudrais09} break the passivity of RV monitors. In fact, RA monitors are generally provided with more control over the system, enabling them to automatically execute adaptation actions after analysing a particular sequence of system events. For instance as shown in \figref{fig:back:ra}, rather than flagging violations, RA monitors can therefore execute adaptation actions upon detecting an event sequence that denotes incorrect behaviour. In this way they aim to mitigate and reduce the effects of the detected violation. The adaptation actions executed by the monitor, do not necessarily correct or revert the detected misbehaviour \cite{Rinard12,Kell08survey}; instead they change certain aspects of the system as it executes, with the aim of preventing either future occurrences of the same error, or of other errors that may potentially occur as a side-effect of the detected violation. 

However, the use of runtime adaptation goes beyond self-healing \cite{ibm2005architectural} \ie detecting and mitigating misbehaviour. In fact RA can also be used to optimise \cite{ibm2005architectural,Kell08survey} the system's behaviour based on the information collected by the monitor, \eg switch off redundant processes when under a small load, or increase processes and load balancing when under a heavy load. These type of adaptations contribute towards making the monitored system more efficient. 

\begin{figure}[ht!]
	\centering
	\includegraphics[width=0.8\textwidth]{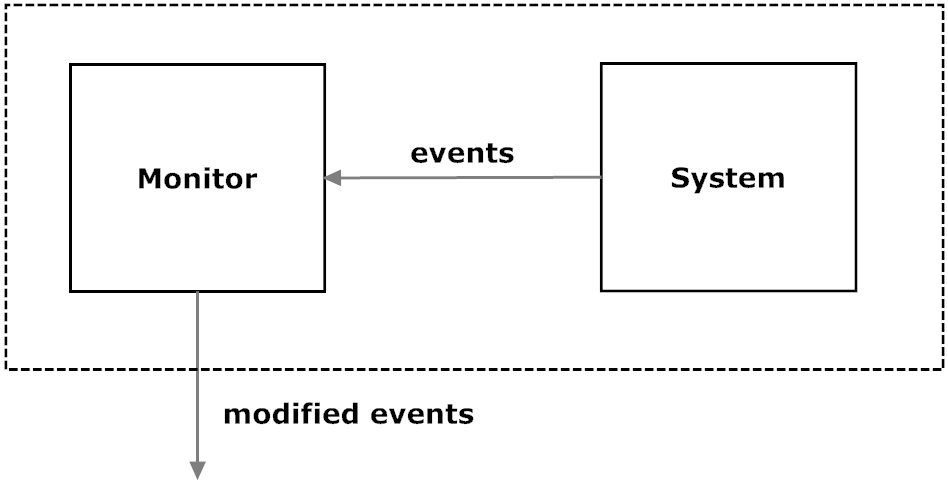}
	\caption{Runtime Enforcement}
	\label{fig:back:re}
\end{figure}

Runtime Enforcement (RE) \cite{FalconeFM12,ligatti10,ligatti05} is the strictest monitoring technique, whereby correct behaviour is kept in line by anticipating incorrect behaviour and countering it before it actually happens. In fact as depicted in \figref{fig:back:re}, in runtime enforcement techniques the monitor and the system are so \emph{tightly coupled} to the extent that the monitor acts as a \emph{proxy} which wraps around the system and analyses its external interactions (see the dotted-line in \figref{fig:back:re}). In this way the monitor is thus able to either drop incorrect events by preventing certain system actions from occurring; or else add system events by executing actions on behalf of the system, that should have been executed by the system but where not \cite{ligatti10,ligatti05}. 

The distinction between RV, RA and RE is however often blurred in the current literature. For instance, we find a number of RV tools such as \cite{larva-CPS09,polyLarva-CFMP12,elarva:2012,chen-rosu-2005-tacas} that allow for user defined actions to be executed whenever certain conditions are met. This distinction is further blurred as Runtime Adaptation may also be used to achieve Runtime Enforcement by anticipating system misbehaviour and executing stricter and more corrective adaptations. This may require adaptations to be tailor-made for the respective system, to totally prevent incorrect behaviour from manifesting. However, unlike RE, RA typically allows for violations to occur, but then executes remedial actions to \emph{mitigate the effects} \cite{Kell08survey} of the committed violations. During the course of this thesis we focus on Runtime Adaptation as defined in \figref{fig:back:ra}.

\section{Synchronous and Asynchronous monitoring}
\label{sec:syn-asyn-survey}
As developing a runtime adaptation framework on top of a runtime verification setup requires introducing a degree of synchrony in our monitoring, we look into the different types of monitoring techniques found in the current literature. Furthermore, as monitors are usually derived from high-level properties, the same property may be converted into \emph{different types of monitors} which may apply different instrumentation techniques for checking 
whether the property was satisfied or not.

In \secref{sec:off_on} we therefore conjecture that there are two main classes of monitoring approaches, namely, Online and Offline. Particularly, we look into the latter class (\ie offline) as this is often ambiguated with asynchronous monitoring, whereas for the former class (\ie online) in \secref{sec:syn_asyn} we discuss the different definitions that one can find in the current RV literature. Based on these definitions, in \secref{sec:spectrum} we present a \emph{spectrum} of \emph{online monitoring approaches} \wrt to component-based systems, that vary in the level of coupling and control that they posses over the system's components. 

\subsection{Offline and Online monitoring}
\label{sec:off_on}
We divide runtime monitoring into the following two main classes: $(i)$ Online monitoring and $(ii)$ Offline monitoring. Although our work deals exclusively with online monitoring, we opt to define offline monitoring such that we are able 
to distinguish it 
from other monitoring techniques.

\subsubsection{Online Monitoring}
In an online monitoring setup \cite{Leu:RV:Overv,Bauer:ltl,lola:runtime,scala}, an executing system is dynamically monitored for violations (or satisfactions) \emph{during the course of its execution}. Online monitors are therefore developed in such a way that they execute alongside the system by verifying its execution in an incremental fashion \cite{Leu:RV:Overv,Bauer:ltl}. This means that the monitor must be able to receive notifications about relevant events occurring in the executing system and make a decision based on the current information collected so far. 

As online monitors are developed to verify currently executing systems, they are also capable of making early detections. This is sometimes exploited by RV tools (such as in \cite{larva-CPS09,polyLarva-CFMP12}) to allow for the execution of user defined actions that may potentially be used to mitigate the corruption suffered by the system as a result of a property violation. The primary disadvantage of online monitoring is that it imposes an inevitable runtime overhead on the system 
given that additional monitoring code is added to the system. As runtime overheads are a very undesirable side effect of online monitors, a good deal of effort is usually devoted to create highly efficient monitors that keep this overhead as low as possible.

\subsubsection{Offline Monitoring}
Unlike online monitoring, in offline monitoring \cite{Leu:RV:Overv,Bauer:ltl,chen-rosu-2007-oopsla,Exago,MeredithR10} the system is not directly monitored as it executes, instead the relevant system events are recorded as an execution trace inside a data store. Once the monitored system terminates (or whenever a satisfactory number of events have been recorded) the collected execution trace is forwarded to the offline monitor. The offline monitor, which is entirely independent from the system, then proceeds by inspecting the system events recorded in the trace, and given that the trace provides enough information, it deduces whether the property it represents was satisfied or violated.

Offline monitors are particularly suitable for properties that can only be verified by \emph{globally analysing} the complete execution trace which is generated once the system stops executing. In fact some properties may require to be globally analysed using a \emph{backward traversal} \cite{MeredithR10,SyncVSAsync:Rosu:2005} of the trace. This monitoring mode is less intrusive than online monitoring, as it does not interfere with the system except for the logging of events, thus imposing much less overheads. This, however, comes at the cost of \emph{late detections}, given that violations are detected either after the system stops or else after the system is done executing a considerably large amount of actions which are recorded and then inspected by the monitor. 

\subsection{Varying definitions of Online Monitoring approaches}
\label{sec:syn_asyn}
In the current literature we often encounter different definitions of online monitoring which usually vary in two aspects: $(i)$ the level of coupling between the monitor and the system, and $(ii)$ the level of control that the monitor has over the system. 

The definition of synchronous online monitoring presented in \cite{taxonomy:Delgado:2004,lola:runtime,BauFal12}, requires the system to work in lockstep with the monitoring code, whereby the \emph{entire} monitored system is bound to block and wait until the necessary monitoring code is done handling an event generated by a system component. These type of online monitors are therefore very tightly coupled to the system and have a very tight level of control over the system to the extent that they cause \emph{all} system components to block even when only a single component generates an event. 
As this extreme synchronous monitoring technique uses a high amount of synchronisation is generally appealing for inherently synchronous systems \cite{lola:runtime,BauFal12} such as circuits and embedded systems.

The work presented in \cite{chen-rosu-2007-oopsla,CCMSC2008}, 
implement a less stringent definition of synchronous online monitoring in which the monitors are only allowed to block the concurrent component which generates the event. In this way other concurrent components may carry on with their normal execution, even while the component which generated the event is 
interrupted until the required monitoring is carried out. Although these monitors still posses a relatively tight level of coupling and control over the system, they are able to provide timely detections 
without resorting to a high use of synchronisation as opposed to the definitions presented in \cite{taxonomy:Delgado:2004,lola:runtime,BauFal12}.

Yet another online monitoring approach is proposed in \cite{cc-saferAsync,CCPHD2013}, allowing the user to manually specify synchronisation checkpoints where the system (or certain parts of it) should temporarily stop and wait for the monitor to finish handling monitoring requests. In this way the system is able to asynchronously send monitoring events to the monitor, 
without waiting for the monitoring code to complete. Explicit checkpoints can then be employed to temporarily block certain parts of the system thereby increasing the monitor's level of control over the system which allows the lagging monitor to catch up with the system execution. The use of checkpoints are especially useful when monitoring for the occurrence of certain safety-critical events, in which late detection is not feasible. Checkpoints therefore permit the user to optimize the synthesised monitors by manually determining the level coupling and control that the monitor possesses over the system.

In \cite{SyncVSAsync:Rosu:2005} synchronous monitoring is defined as a monitoring approach which provides timely detections. This definition describes an even less stringent monitoring approach in which the monitors are loosely-coupled from the system yet retain enough control over the system so that they can provide timely detections. We conjecture that implementing synchronous monitoring \wrt this definition requires monitoring most of the system events asynchronously, yet utilise synchronisation at strategic points which might lead to a violation. In this way violations are still detected in a timely manner.

Finally, in this thesis we follow the definitions for asynchronous online monitoring given in \cite{cc-saferAsync,SyncVSAsync:Rosu:2005,CCPHD2013,elarva:2012}. These definitions state that although the synthesised monitors still execute alongside the system, they are however \emph{loosely-coupled} from the system to the extent that they have barely any control over the monitored system and may suffer from \emph{late detections}. In fact whenever a specified event is performed by the system, an event notification is asynchronously forwarded to the monitor without suspending any part of the system. The monitor is then able to verify the received event notifications at its own pace independently from the system. Importantly note that unlike offline monitors (defined in \secref{sec:off_on}), asynchronous monitors still execute alongside the system and still have to analyse the systems trace of events in an incremental fashion. 
Due to this lack of synchrony and control, asynchronous monitors are presumed to impose considerably lower overheads compared to other approaches. However, most of the time an asynchronous monitor is unable to detect violations immediately and carry out effective mitigation actions.  

\subsection{The Spectrum of Online Monitoring Approaches}
\label{sec:spectrum}
\begin{figure}[htbp]
\centering
	\begin{tikzpicture}
	\draw (-5,0) -- (9.5,0);
	
	\foreach \x in {-4.5,-1.125,2.25,5.625,9}
	   \draw (\x cm,3pt) -- (\x cm,-3pt);
	
	\draw (-3.65,0) node[below=3pt] {Tightly Coupled};	
	\draw (-4.5,0) node[above=3pt] {CS}; 
	\draw (-1.125,0) node[above=3pt] {SMSI}; 
	\draw (2.25,0) node[above=3pt] {AMC}; 
	\draw (5.625,0) node[above=3pt] {AMSD}; 
	\draw (9,0) node[above=3pt] {CA}; 
	\draw (8.2,0) node[below=3pt] {Loosely Coupled};	
	\end{tikzpicture}\vspace{-5mm}
	\caption{The Online monitoring Spectrum.\label{fig:pic}}
\end{figure}
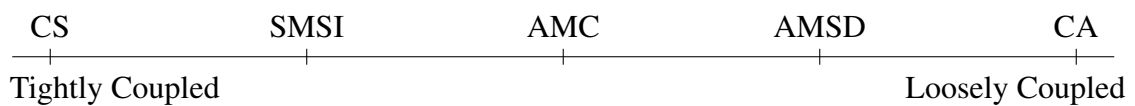

Based on the range of definitions presented in \secref{sec:syn_asyn}, we devise a \emph{spectrum} of online monitoring approaches which we explain \wrt \emph{component-based systems}, whereby components represent concurrent entities (\eg node, thread, actor, \ldots) that can be in either two states, namely \emph{blocked} or \emph{running}. As depicted in Figure~\ref{fig:pic}, this spectrum ranges from a tightly coupled \emph{completely-synchronous} monitoring approach on one end, to a loosely-coupled \emph{completely-asynchronous} monitoring approach on the other end. 

The spectrum also presents currently known online monitoring approaches that lie in between these two extremes. These middle way approaches provide a tradeoff between the level of coupling and control that the monitor has upon the system. It is generally assumed that monitoring approaches that are closer to the synchronous end of the spectrum tend to posses a higher level of control over the system, which comes at the expense of higher the overheads. Conversely, approaches which are closer to the opposite end are generally presumed to be more efficient. 
However, so far we are not aware of any publications that conducted a thorough investigation to confirm (or reject) this general assumption. \bigskip

\subsubsection{Completely-synchronous Monitoring (CS)}
\begin{figure}[htbp]
	\centering
	\includegraphics[width=0.62\textwidth]{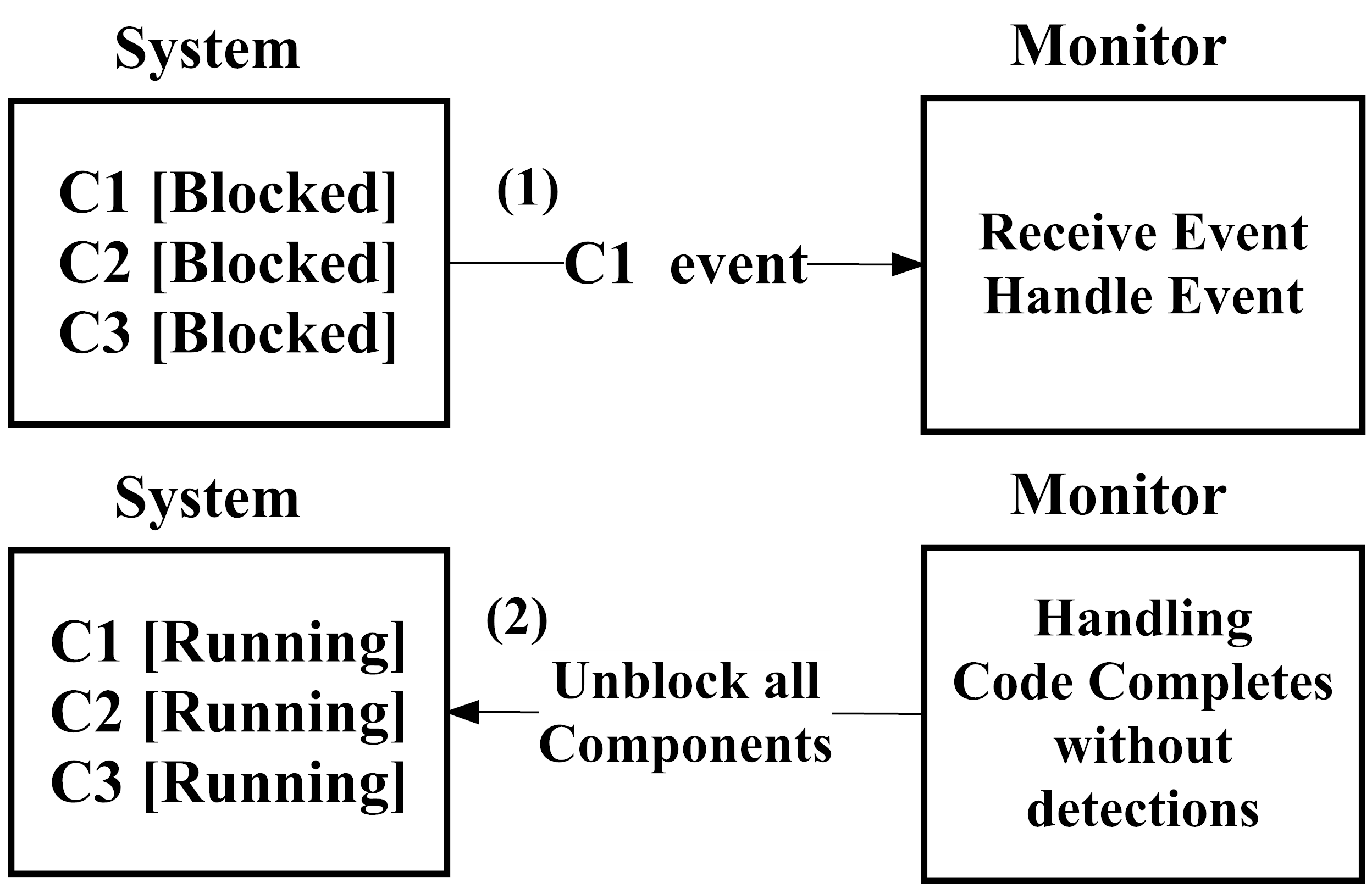}
	\caption{Completely-synchronous Monitoring (CS)}
	\label{CS}
	\vspace{-1mm}
\end{figure}

As depicted in \figref{CS}, \emph{Completely-synchronous monitoring} \cite{taxonomy:Delgado:2004,lola:runtime,BauFal12} refers to the synchronous extreme of the spectrum. This implies that the system and the monitors are extremely \emph{tightly coupled} together to the extent that whenever an event occurs in \emph{one} component of the system, such as $C1$, the \emph{entire} system 
execution is \emph{interrupted} (as shown by $(1)$ in \figref{CS}). The system components remain blocked until the necessary monitoring checks are performed, in which case the monitor unblocks the components (as shown by $(2)$). 

Since completely-synchronous monitoring generally provides immediate detections, it can therefore be used to effectively mitigate incorrect behaviour. This approach is however highly intrusive as it introduces an unnecessarily high level of synchronisation that might lead to an infeasible performance degradation on the monitored system. Completely-synchronous monitoring is generally applied on small synchronous systems \cite{lola:runtime,BauFal12} such as embedded systems and circuits.

\subsubsection{Synchronous Monitoring with Synchronous Instrumentation (SMSI)}
\begin{figure}[htbp]
	\centering
	\includegraphics[width=0.62\textwidth]{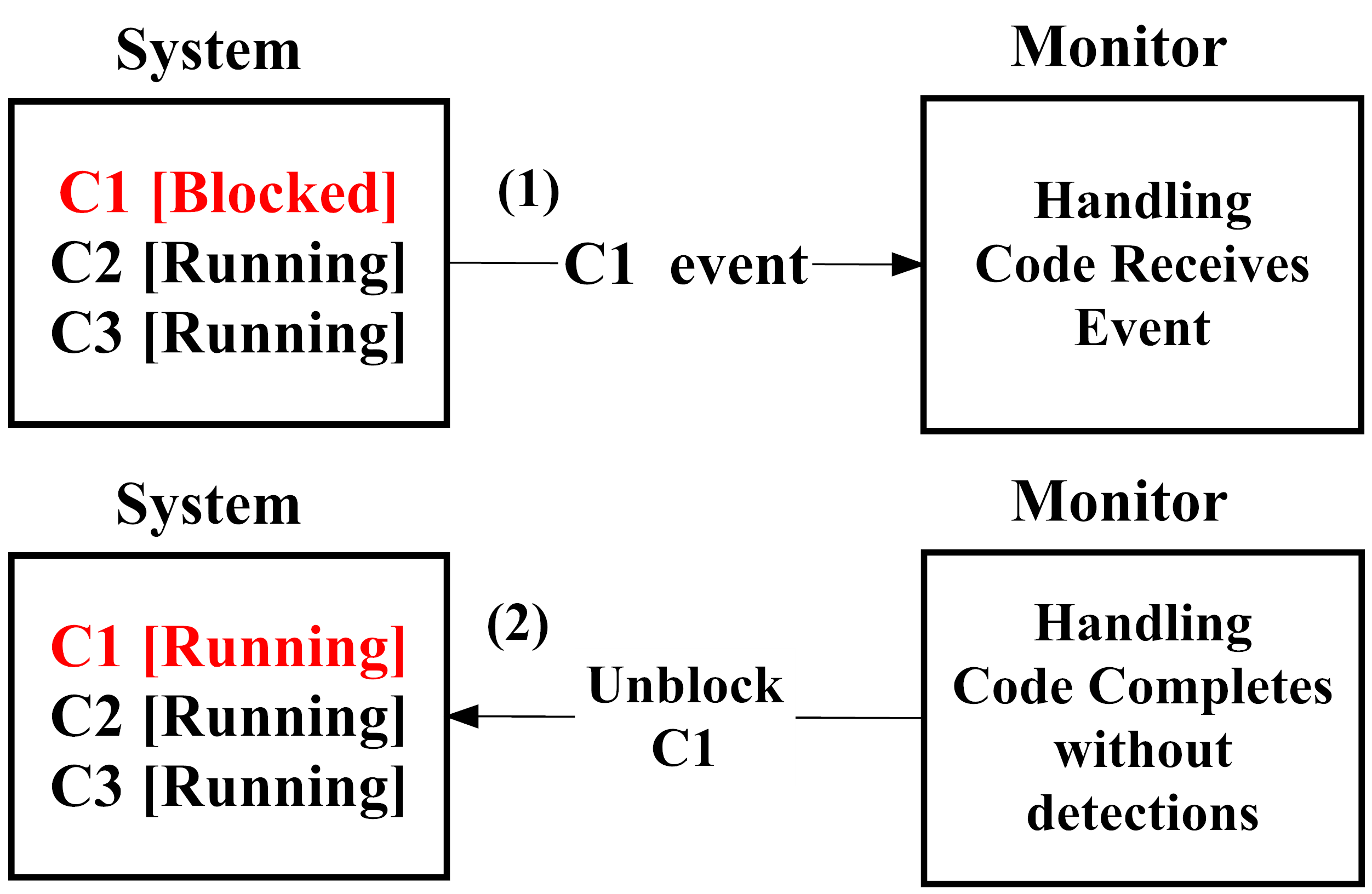}
	\caption{Synchronous Monitoring with Synchronous Instrumentation (SMSI)}
	\label{SMSI}	\vspace{-1mm}
\end{figure}

\emph{Synchronous Monitoring with Synchronous Instrumentation} \cite{CCMSC2008,chen-rosu-2007-oopsla} is a monitoring approach which is closer  to a completely-synchronous approach yet is less intrusive. In fact as shown in \figref{SMSI}, this approach assumes that whenever a \emph{single component} of the monitored system, such as $C1$, 
executes a specified event, then \emph{only the execution of this component} is interrupted (as shown by $(1)$ in \figref{SMSI}). This component remains blocked until monitoring completes, in which case the component is reset to a running state (as shown by $(2)$). 

Hence synchronous instrumentation monitoring reduces the level of overall synchronisation by relinquishing control over the other components, thereby allowing them to keep executing whenever a certain component generates an event. However, this monitoring technique still posses a relatively high control over the system as it still needs to synchronously inspect each and every event, even though it only interrupts the normal execution of the component which generated the event until monitoring completes. 

\subsubsection{Asynchronous Monitoring with Checkpoints (AMC)}
\begin{figure}[htbp]
	\centering
	\includegraphics[width=0.65\textwidth]{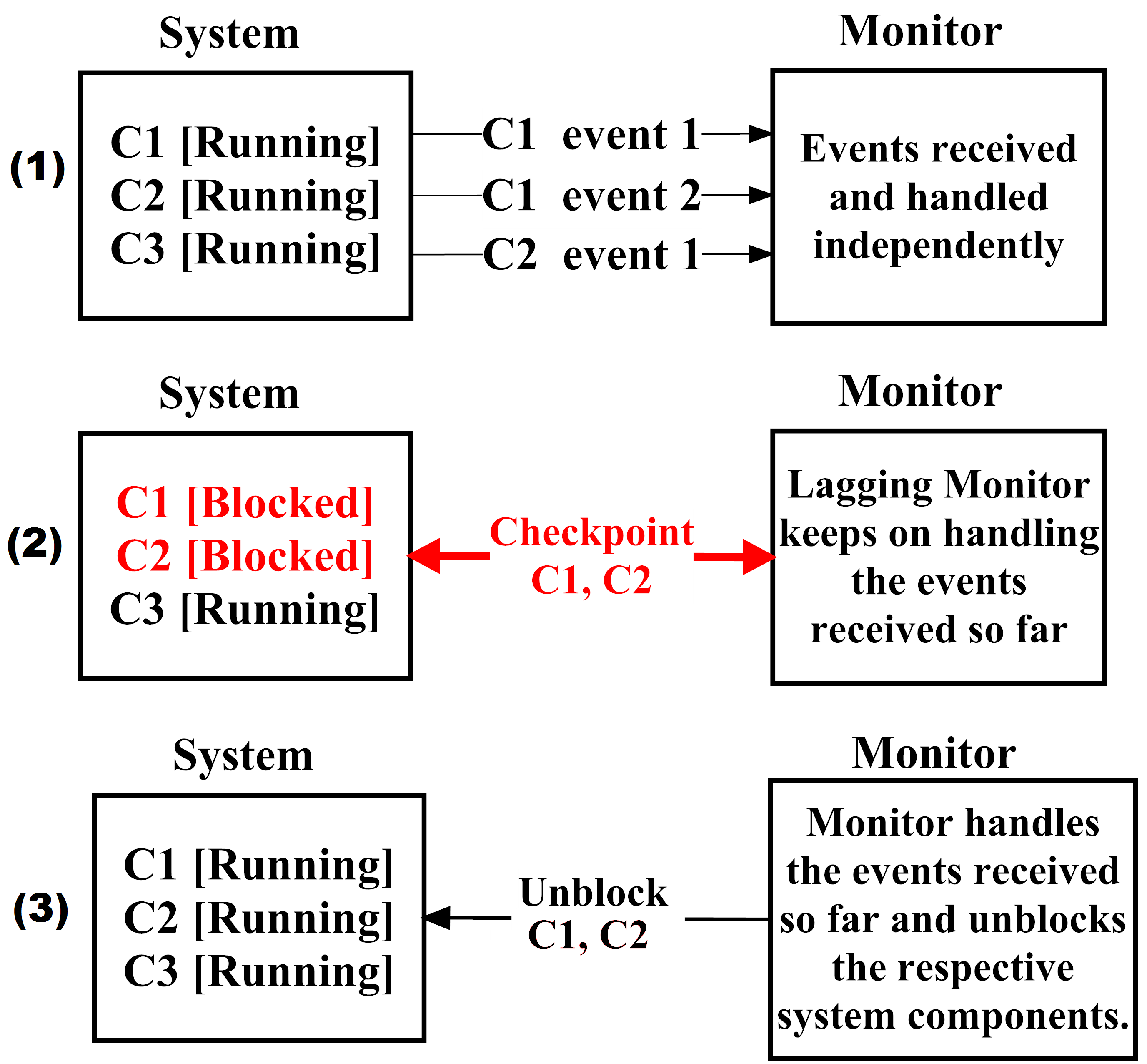}
	\caption{Asynchronous Monitoring with Checkpoints (AMC)}
	\label{AMC}	\vspace{-1mm}
\end{figure}

This approach \cite{cc-saferAsync,CCPHD2013,taxonomy:Delgado:2004} allows for an asynchronous decoupling between the system (or parts of it) and monitor executions yet also provides the user with the ability to specify checkpoints where the decoupled system and monitor executions should synchronise. We conjecture that in terms of component-based systems, checkpoints may also be associated with specific system components. For instance as shown by $(1)$ and $(2)$ in \figref{AMC}, an RV user may use a checkpoint specify that certain system components such as $C1$ and $C2$ should temporarily block their execution and synchronise. Such synchronisation can be done during periods where the system is not required to be responsive, and therefore affords to wait for the monitor to catch-up with the system execution. 

As shown by $(3)$, once the monitor synchronises with the system, it unblocks the respective system components and allows them to continue generating further event notifications. An appealing feature of this monitoring approach is that it enables the user to manually determine the level of synchrony and control that the monitor can posses over the system. The user may also use this mechanism to specify that the system should stop and wait for the monitor's analysis to complete when certain \emph{safety-critical events} occur during execution \cite{cc-saferAsync}. This ensures that safety-critical violations are detected in a timely fashion thus permitting the monitor to effectively react to the violation and possibly invoke mitigating actions. To further reduce the level of monitor intrusion, 
the user may specify that the system should keep on executing whenever it generates non-safety critical events.

\subsubsection{Asynchronous Monitoring with Synchronous Detection (AMSD)}
\begin{figure}[t!]
	\centering
	\includegraphics[width=0.75\textwidth]{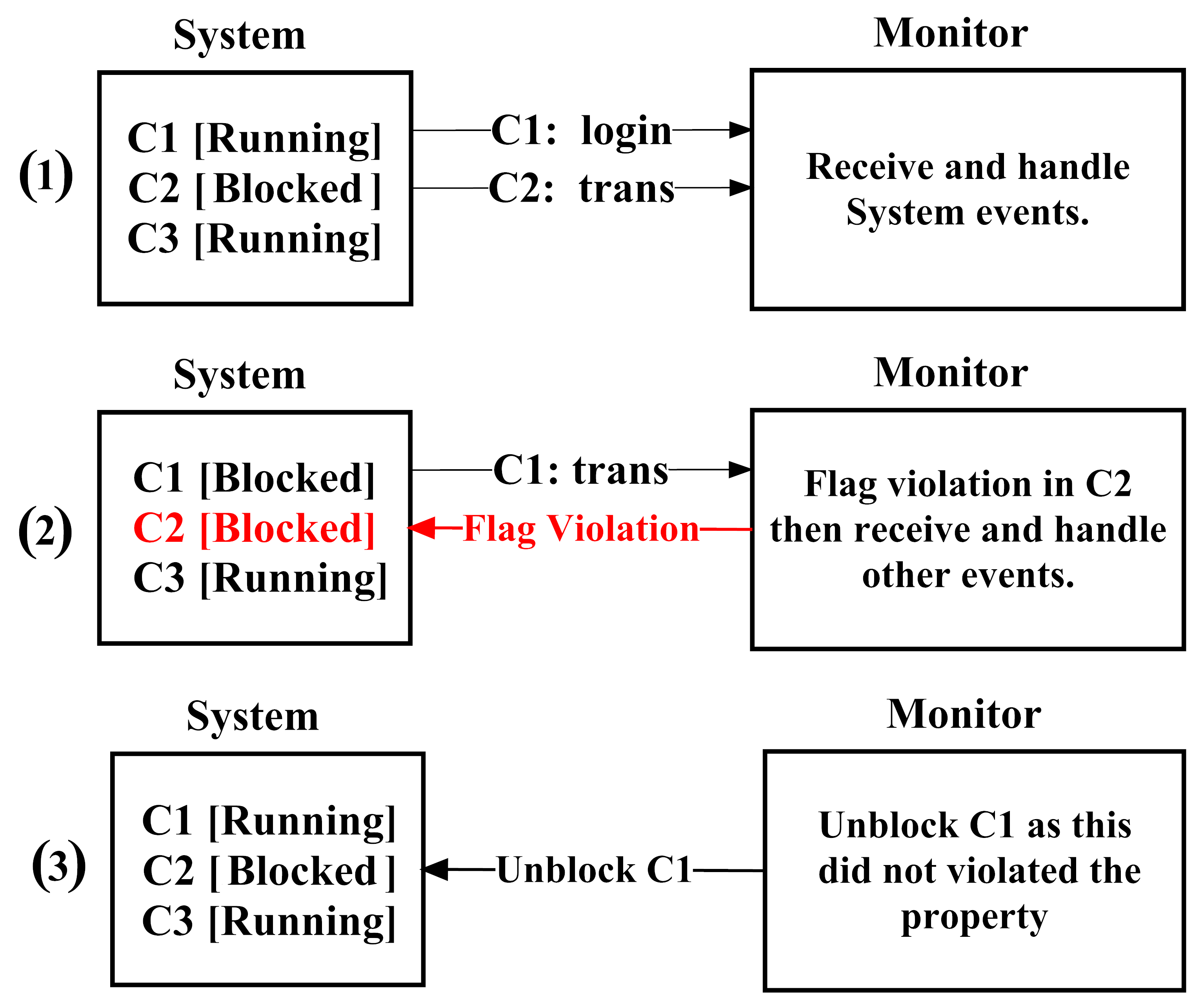}
	\caption{Asynchronous Monitoring with Synchronous Detection (AMSD)}
	\label{AMSD}	
\end{figure}

\emph{Asynchronous Monitoring with synchronous detection} (based on the definition given in \cite{SyncVSAsync:Rosu:2005}) is yet another monitoring approach that lies closely to completely-asynchronous monitoring in our spectrum. 
This approach uses minimal synchronisation to ensure synchronous (timely) detections, by synchronising only for system events that might directly contribute to a violation. In \figref{AMSD} we illustrate this concept \wrt an example that assumes a simple transaction system which must satisfy the following invariant property:
\begin{property} \label{prop:trans-sys}
	\emph{A user cannot make a transaction before a login.}
\end{property}

\noindent\Propref{prop:trans-sys} is thus only violated whenever a transaction event (\textsf{trans}) is perceived without a preceeding login event. This means that to timely detect violations of \propref{prop:trans-sys} it suffices that the monitor synchronises only with a system component whenever this performs a transaction event. As shown by $(1)$ in \figref{AMSD}, only component $C2$ is blocked, as this produced a transaction event that may lead to a property violation. By contrast $C1$ kept on executing even though it produced a login event. In $(2)$ we see that the monitor managed to detect that $C2$ violated \propref{prop:trans-sys}, as it did not login prior to performing the transaction. The detection was made in a timely fashion as $C2$ was not allowed (in $(1)$) to execute any further after committing the transaction event that lead to the violation. Furthermore in $(2)$ one can notice that $C1$ blocked waiting for the monitor's verdict as a result of producing a transaction event, in which case the monitor detects that $C1$ had already performed a login event meaning that it did not violate \propref{prop:trans-sys}. Hence in $(3)$ the monitor allows $C1$ to proceed. 

\subsubsection{Completely-asynchronous Monitoring (CA)}
\begin{figure}[htbp]
	\centering
	\includegraphics[width=0.75\textwidth]{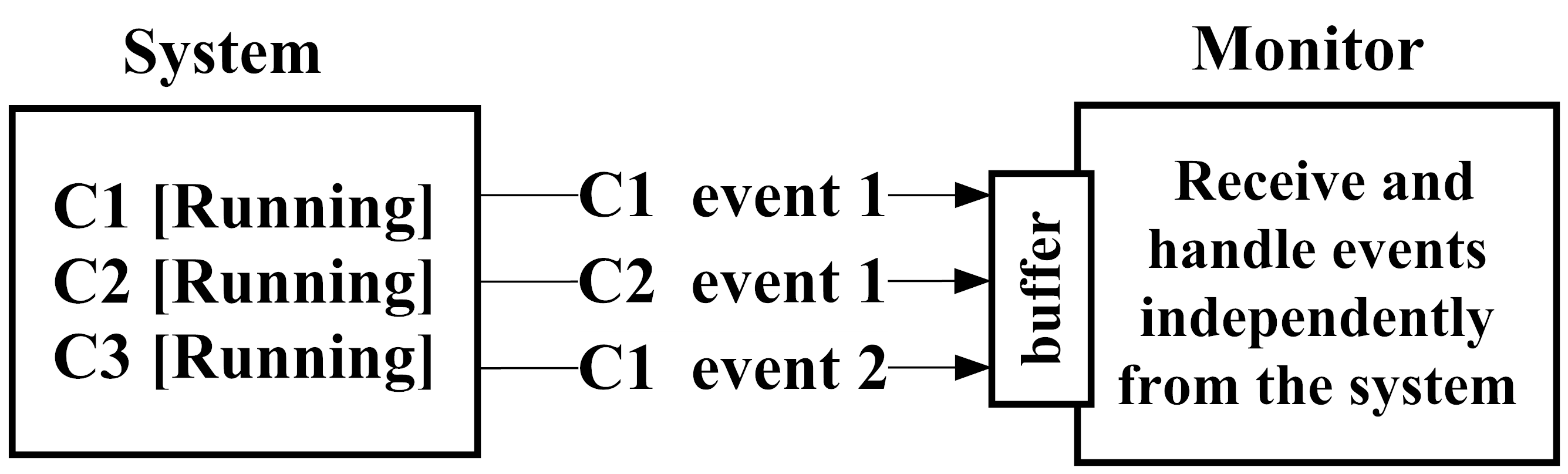}
	\caption{Completely-asynchronous Monitoring (CA)}
	\label{CA}	
\end{figure}

Finally, in \emph{Completely-asynchronous monitoring} \cite{cc-saferAsync,FraSey14,CCPHD2013,elarva:2012} the monitors are designed to be as \emph{loosely coupled} as possible from the system they are monitoring. In fact completely-asynchronous monitors are designed to listen for system events and handle them in the ``background'' without interfering in any way with the system execution. Hence, as shown in \figref{CA}, in completely-asynchronous monitoring, the system is allowed to proceed immediately after placing an event notification (\eg $C1 \textsl{ event 1}$) in the monitor's buffer. The monitor can then independently read the event notifications from its buffer and carry out the necessary checks at its own pace. 

Although this approach imposes minimal intrusion over the monitored system, it suffers from late detection which makes it highly-unlikely for an asynchronous monitor to be able to effectively mitigate the detected misbehaviour. As offline monitoring \cite{taxonomy:Delgado:2004, SyncVSAsync:Rosu:2005}, is inherently asynchronous, it is sometimes classified as being an asynchronous monitoring approach. However, we conjecture that offline monitoring generally requires prerecorded (generally complete) traces, while CA monitoring can incrementally analyse partial traces that can be extended as the system proceeds with its execution.

\subsubsection{Conclusion}
We have introduced the concept of a \emph{spectrum of online monitoring approaches}, and explained each approach \wrt component based systems. The spectrum consists in:
\begin{itemize}
	\item \emph{Completely-synchronous monitoring} on one end --- this achieves an extremely high level of control over the system that could potentially deter monitoring efficiency; and
	\item \emph{Completely-asynchronous monitoring} on the opposite end --- this is presumed to provide efficient monitors with very low intrusion over the monitored system, by sacrificing the level of control over the monitored system, including timely detections.
\end{itemize}
We have also identified a number of middle-way approaches that lie in between these two online monitoring extremes in our spectrum. We also conjecture that there may be other middle-way approaches apart from the ones that we identified which still need to be explored. The identified approaches however allow us to better understand the design space that we need to explore and assess in \Cref{chp:syn-asyn} in order to achieve our first objective (see objective $(i)$ in \secref{sec:intro:obj}), \ie that of increasing the level of synchrony in the \detecterGen\ RV tool while preserving monitoring efficiency. 

\section{The Actor-Model and Erlang}
\label{sec:erlang}
The component based systems that we consider in this thesis are actor-based \cite{actorsPaper} systems written in a programming language called Erlang \cite{Armstrong07,ErlangOTP}. In Erlang \emph{processes} (\ie actors) are threads of execution that are uniquely identified by a \emph{process identifier} and posses their own \emph{local} memory.  Erlang processes execute \emph{asynchronously} to one another, interacting through \emph{asynchronous messages} instead of sharing data. 

In fact, processes can only communicate by explicitly sending a \emph{copy} of their data to the destination process (using the unique identifier as address). These messages are received at a process \emph{mailbox} (a form of message buffer) and can be exclusively read at a later stage by the process owning the mailbox. Since asynchronous messages  may  reach a mailbox in a different order than the one intended, the (mailbox) read construct uses \emph{pattern matching} to allow a process to retrieve the first message in the mailbox matching a pattern, possibly reading messages out of order.  Whenever a retrieved pattern is not matched, the executing process \emph{blocks}, waiting for a message matching this pattern to reach its mailbox. Apart from the mailbox, actors possess their own private state known as the \emph{process dictionary}, in which they can associate data entries with a unique key, which can then be used in different points during the actor's execution to retrieve or modify the stored data entry. 

Processes may also be \emph{registered} under a specific name. This allows other actors to send messages to a registered process by using its name rather then its process id. This is useful to avoid having to communicate the process id of an (important) actor, to the other actors. Furthermore, processes may \emph{spawn} other processes at runtime, and communicate locally-known (\ie private) process identifiers as messages. A Process may not only use the known process identifiers to communicate messages, but also to forcibly \emph{terminate} or \emph{link} processes \cite{ErlangOTP}. Implementation wise, Erlang processes are relatively lightweight \cite{Armstrong07}, and language coding practices \cite{ErlangOTP} recommend the use of concurrent processes wherever possible.  These factors ultimately lead to systems made up of independently-executing components (\ie processes) that are more scalable, maintainable, and use the underlying parallel architectures more efficiently \cite{Cesarini:2009}.   

At its core, Erlang is \emph{dynamically-typed}, and function calls with mismatching parameters are typically detected at runtime. This implies that Erlang provides very few static guarantees thereby making it prone to more dynamic errors; this makes Runtime Adaptation an appealing approach for detecting and mitigating dynamic errors. Function definitions are named and organised in uniquely-named modules. Within a particular module, there can be more than one function with the same name, as long as these names have different arities (\ie number of parameters).  Thus, in Erlang, every function is uniquely identified by the triple \emph{module\_name:function\_name/arity} \eg \textsf{math:mul/2} identifies a function named \textsf{mul} which takes two arguments and is defined in the \textsf{math} module. 

Erlang also offers a number of fault-tolerance mechanisms including \emph{process linking} \cite{Armstrong07,ErlangOTP}, whereby whenever a process fails \ie terminates abnormally, all the other actors linked to it are terminated as well. However, Erlang also provides a \emph{trapping} mechanism whereby rather then failing, the linked processes are sent a message in their mailbox notifying them that a linked process has failed. Furthermore, linking may be either \emph{bi-directional} or \emph{uni-directional} \cite{Armstrong07}. This means that if an actor $A$ links with $B$ using a bi-directional link, then should either one of the linked actors crash, then the other actor crashes as well (or is notified in case of trapping). Whereas if $A$ links with $B$ using a uni-directional link, then only $A$ gets terminated (or notified) in case the linked actor $B$ crashes, \ie if $A$ crashes, $B$ remains unaffected.

Erlang systems are often structured by the \emph{supervisor} pattern which is built on linking and trapping, in which processes at the system fringes are encouraged to \emph{fail-fast} when an error occurs (as opposed to handling the error locally). In this way the linked supervisor processes \cite{Cesarini:2009} is kept responsible to carry out the necessary adaptations for detecting the crash and handling the error. However, misbehaviour does not necessarily lead to crashing an actor, for instance if an invalid computation is saved to a database, then none of the actors are likely to crash. Therefore, a more refined adaptation mechanism built over a runtime verification setup provides a good alternative for detecting and mitigating such misbehaviour.

As actors in Erlang are inherently processes that interact using asynchronous message passing, they can be formally represented as \emph{Labelled Transition Systems (LTS)} \cite{fredlund:unifiedsemantics,fredlund:phd,Carlsson-Core-Erl:01,FraSey14} where at an abstract level, computation can be viewed as a graph of states connected by labelled arcs denoting the communicated messages.     

\section{\detecterGen $-$ A concurrent RV tool}
\label{sec:detecter-primer}
\label{sec:detecter-primer-2}
In \cite{FraSey14}, the authors present \detecterGen\ $-$ a provably correct, actor-based runtime verification tool written in Erlang. 
This tool uses a syntactic subset of the modal $\mu$-calculus called Safe Hennessy-Milner logic (sHML) \cite{aceto:SHML} to specify safety properties over Labelled Transition System (LTS). As Erlang programs can be formalised in terms of an LTS, \SHML can therefore be used to specify properties for Erlang systems as well.   

Moreover, the monitors synthesised by this tool, employ a \emph{completely asynchronous} (CA) monitoring approach for detecting the runtime violations of the respective formulas. This implies that the synthesised monitors barely have any control over the monitored system. Asynchronous monitoring is in fact achieved using the tracing mechanism offered by the EVM OTP platform libraries \cite{ErlangOTP} which do not require instrumentation at source-code level. Instead VM directives generate trace messages for specified execution events that are sent to a specially-designated \emph{tracer} actor \ie to the synthesised monitor.

Hence to introduce synchrony in \detecterGen\ it is likely that we need to \emph{completely redesign} the mechanism for forwarding trace messages to the monitor, such that the system and the monitor are allowed to synchronise whenever a trace event is exchanged.


\subsection{The Syntax}\label{sec:language}
The tool's \SHML syntax \cite{FraSey14,aceto:SHML} is defined inductively using the BNF in \figref{fig:logic}. It is parametrised by a set of boolean expressions, $\bV,\bVV\in\Bool$, equipped with a \emph{decidable} evaluation function, $\bV\Downarrow\vV$ where $\vV \in \sset{\btt,\bff}$, and a set of system actions $\actE,\actEE\in\Act$. 
It assumes two distinct denumerable sets of variables: 
\begin{itemize} \itemsep0em
	\item  {\setstretch{1.1}\emph{Term variables}, $\xV,\xVV,\ldots\in\Vars$ --- used in action patterns to quantify over data values, and to evaluate boolean conditions defined in if-statements; and }
	\item  \emph{Formula variables}, $\hVarX,\hVarY,\ldots\in\LVars$ ---  used to define \emph{recursive} (logical) formulas.
\end{itemize}
Note that even though we work up-to $\alpha$-conversion (for both variable sets), \detecterGen\ accordingly renames duplicate variables during a pre-processing phase. 

Formulas $\hV,\hVV\in\FRM$ include: truth \mtru, and falsity \mfls, conjunctions,  $\hV\!\mand\!\hVV$, modal necessities, \mnec{\patE}{\hV} for specifying particular actions that the system may perform, maximal fixpoints to describe recursive properties, $\mmax{\hVarX}{\!\hV}$, and if-statements to analyse system data, \mboolE{\bV}{\hV}{\hVV}. A necessity formula, \mnec{\patE}{\hV} specifies a patten $\patE$ denoting the following system actions:
\begin{itemize} \itemsep0em 
	\item  {\bfseries\emph{message input}}: \eg $\mrecv{i}{3}$ $-$ process $i$ must receive the value 3;
	\item  {\bfseries\emph{message output}}: \eg $\msendD{i}{j}{3}$ $-$ process $i$ must send $j$ the value 3;
	\item  {\setstretch{0.8}{\bfseries\emph{function call}}: \eg $\mcall{i}{\eatom{mod},\eatom{fun},[1,\eatom{test}]}$ $-$ process $i$ must call a function \eatom{fun}, which is defined in module \eatom{mod}, using arguments $[1,\eatom{test}]$; and}
	\item 	{\setstretch{0.8}{\bfseries\emph{function return}}: \eg $\mret{i}{\eatom{mod},\eatom{fun},2,\etuple{res,5}}$ $-$ process $i$ must return from function \eatom{fun} (which takes 2 arguments) with return value $\etuple{res,5}$.}
\end{itemize}

\begin{remark} Although from a theoretical standpoint \SHML specifications should limit themselves to input and output actions so as to treat individual components as \emph{black-boxes}, monitoring for function calls and returns is essential. This is because certain output and input actions in actual Erlang implementations may be abstracted away inside the function calls of system libraries, making them (directly) unobservable to the instrumentation mechanism. However, these actions (and the data associated with them) can still be observed indirectly, through the calls and returns of the functions that abstract them. \exqed
\end{remark}

A pattern $\patE$ in a necessity $\mnec{\patE}{\hV}$ may also contain term variables that \emph{pattern-match} with actual (closed) system actions, thus acting as a \emph{binder} for these variables in the continuation subformula \hV; to improve readability, we sometimes denote term variables that are not used in the continuation subformula as underscores, $\_$. Similarly a recursive definition \mmax{\hVarX}{\hV}, serves as a binder for recursive variable \hVarX\ in \hV. \medskip

\begin{example}   \label{ex:background}
Recall \exref{ex:intro} in which we define safety \Propref{prop:intro:1} for monitoring invariant behaviour in the actor-based server illustrated in \figref{fig:sys}. Using \detecterGen's syntax we can now formally represent this property as the following formula:
  \begin{align} \vspace{-5mm}
  \setlength{\abovedisplayskip}{-5pt}%
\setlength{\belowdisplayskip}{-5pt}%
\setlength{\abovedisplayshortskip}{-5pt}%
\setlength{\belowdisplayshortskip}{-5pt}%
    \label{intro:1}
    \hV \;\deftxt\;\;& \mmax{\,\hVarY}{\mnec{\mrecv{i}{\tup{\eatom{inc},x,y}}}
        \begin{mlbrace}(\,\mnec{\msendD{j}{y}{\tup{\eatom{res},\smash{x+1}}}}\,\hVarY)   \; \mand \; (\,\mnec{\msendD{\_\,}{y}{\eatom{err}}}\,\mfls) \end{mlbrace}} \vspace{-5mm}
  \end{align}
The tool \detecterGen allows us to specify the recursive safety property \eqref{intro:1} (above) as follows: it requires that, from an \emph{external} viewpoint, \emph{every} increment request received by $i$, \ie action $\mrecv{i}{\tup{\eatom{inc},x,y}}$, is followed by a service answer from  $j$ to the client address bound to variable $y$, with value $x+1$ as specified by action $\msendD{j}{y}{\tup{\eatom{res},\smash{x+1}}}$ (and then recurs through variable \hVarY).  However, increment requests followed by an error message sent from \emph{any} actor back to $y$ (as defined by action $\msendD{\_\,}{y}{\eatom{err}}$), represent a violation (\ie \mfls). The \detecterGen compiler is then used to automatically synthesise a concurrent actor-based monitor corresponding to property \eqref{intro:1}. It also adds directives to Erlang VM to forward system execution events to the synthesised monitor \cite{FraSey14}. \exqed
\end{example}

\exref{ex:background} is just a pathological example showing that \SHML is ideal for specifying \emph{infinite computation} (using maximal fixpoints) in the form of \emph{invariant}, safety properties such as formula \eqref{intro:1}. In the case of actor-based systems, this essential as the actor model is usually used to develop \emph{non-terminating, reactive} systems \cite{Agha:1986,actorsinscala,Armstrong07}.

\begin{figure}[ht!]
 \begin{small}
	 \noindent\textbf{BNF Syntax.}\vspace{-3mm}
	  \begin{align*}
	    \hV,\hVV \in \FRM & \;\bnfdef\; \mtru \;\bnfsep\; \mfls \;\bnfsep\; {\hV}\mand{\hVV} \;\bnfsep\; \mnec{\actE}{\hV}  \;\bnfsep\; \hVarX \;\bnfsep\; \mmax{\hVarX}{\hV} \;\bnfsep\;  \mboolE{\,\bV\,}{\,\hV\,}{\,\hVV} 
	  \end{align*}
	\noindent\textbf{Satisfaction Semantics.}\vspace{-3mm}
	\begin{align*}
	 (\actV, \mtru) \in \relR &\quad \text{always} \\
	 (\actV, \mfls) \in \relR &\quad \text{never} \\
	 (\actV, \mboolE{\bV}{\hV}{\hVV}) \in \relR &\quad \text{if }  (\mcondEval{\bV}{\btt} \text{ and } (\actV,{\hV})\in \relR)\text{ or }(\mcondEval{\bV}{\bff} \text{ and } (\actV,{\hV})\in \relR)\\  
	 (\actV, {\hV}\mand{\hVV}) \in \relR &\quad \text{if } (\actV, \hV) \in \relR \text{ and }(\actV, \hVV) \in \relR\\
	 (\actV, \mnec{\patE}{\hV}) \in \relR &\quad \text{if }  \actV \wtraS{\actE\,} \actVV \text{ and } \match{\patE}{\actE}{\sigma} \text{ then } (\actVV, \hV\sigma) \in \relR \\
	 (\actV, \mmax{\hVarX}{\hV}) \in \relR  &\quad \text{if } (\actV, \hV\sub{(\mmax{\hVarX}{\hV})}{\hVarX} \in \relR) 
	\end{align*}
	\noindent\textbf{Violation Semantics.}  \vspace{-5mm}
	\begin{align*}
			 \vsat{\actV}{\sV}{\mtru}  &\quad \text{never} \\
			 \vsat{\actV}{\sV}{\mfls}  &\quad \text{always} \\
			 \vsat{\actV}{\sV}{\mboolE{\bV}{\hV}{\hVV}} &\quad \text{if }  (\mcondEval{\bV}{\btt} \text{ and } \vsat{\actV}{\sV}{\hV}) \text{ or }(\mcondEval{\bV}{\bff} \text{ and } \vsat{\actV}{\sV}{\hVV})\\   
			 \vsat{\actV}{\sV}{\hV\mand\hVV} &\quad \text{if } \vsat{\actV}{\sV}{\hV} \text{ or } \vsat{\actV}{\sV}{\hVV}\\
			 \vsat{\actV}{\actE\sV}{\mnec{\patE}{\hV}} &\quad \text{if } \actV \wtraS{\actE\,} \actVV  \text{ and } \match{\patE}{\actE}{\sigma} \text{ and } \vsat{\actVV}{\sV}{\hV\sigma}    \\
			 \vsat{\actV}{\sV}{\mmax{\hVarX}{\hV}}  &\quad \text{if } \vsat{\actV}{\sV}{\hV\sub{(\mmax{\hVarX}{\hV})}{\hVarX}}  
	\end{align*}\vspace{-8mm}
\end{small}
  \caption{The Logic and its Semantics}
  \label{fig:logic}
\end{figure}

\subsection{The Model} \label{sec:detecter:model}
The semantics \cite{FraSey14} for \SHML, given in \figref{fig:logic}, is defined over a formal \emph{Labelled Transition System} (LTS) representation of an arbitrary Erlang program in the style of the operational models (for Erlang programs) presented in \cite{fredlund:unifiedsemantics,fredlund:phd,Carlsson-Core-Erl:01,FraSey14}. In these operational models, the runtime semantics for Erlang programs is formally represented in the form of a set of actors $\actV,\actVV\in\Actr$ composed in parallel. These actors can reduce either by using internal (unobservable) $\tau$-actions, $\actV\!\traS{\tau}\!\actVV$, or else by using external observable $\actE$ actions, $\actV\!\traS{\actE}\!\actVV$, \ie actions that an external actor such as the monitor can perceive. Weak transitions \cite{FraSey14} for modeled Erlang systems \ie $\actV\,(\traS{\tau})^{*}\!\!\traS{\actE}(\traS{\tau})^{*}\!\,\actVV$, are often represented as $\actV\wtraS{\actE}\actVV$.

\subsection{The Semantics} \label{sec:detecter:semantics}
In \cite{FraSey14} the authors define satisfaction semantics for the \SHML\ logic over arbitrary Erlang programs modelled as an LTS (as discussed in \secref{sec:detecter:model}), to denote how an \SHML formula $\hV$ is satisfied by an Erlang program $\actV$.

The satisfaction semantics state that a \emph{satisfaction relation} $\relR\in(\Actr\times\FRM)$ is \emph{always} satisfied by $\mtru$ but \emph{never} satisfied by $\mfls$, and in the case of conjunction formulas, $(\hV\mand\hVV)$, both branches must be satisfied. Conditionals, $\mboolE{\bV}{\hV}{\hVV}$, are only satisfied if the taken branch is satisfied. Modal necessities, $\mnec{\patE}{\hV}$, are satisfied in two ways: either \emph{trivially} \ie when necessity pattern $\patE$ does \emph{not} pattern match action $\actE$; or else since the resulting actors $\actVV$ satisfy $\hV\sigma$ after performing a pattern-matchable event action $\actE$ that yields substitution environment $\sigma:: \Vars\rightharpoonup \Val$. 

Finally recursive definitions, \mmax{\hVarX}{\hV}, are satisfied by a set of actors $\actV$, if the unfolded formula $\hV\sub{(\mmax{\hVarX}{\hV})}{\hVarX}$ (where every occurrence of formula variable \hVarX\ in \hV\ is substituted by \mmax{\hVarX}{\hV}) is also satisfied by $\actV$. This implies that recursive formulas may potentially be satisfied by \emph{infinite computation}. 

In \cite{FraSey14} the authors also argue that even though \SHML\ can specify \emph{infinite behaviour}, the \emph{execution traces} that \emph{violate} safety properties (expressible through \SHML) are always \emph{finite}. This implies that since runtime monitors can only work with finite traces \cite{Bauer:ltl,Leu:RV:Overv}, they are therefore able to detect violations from finite violating traces. Based on this fact the authors formulate the violation semantics (presented in \figref{fig:logic}) and prove that these correspond to their satisfaction semantics.

The violation relation, denoted as $\vsatL$, is the least relation of the form $(\Actr\times\Act*\times\FRM)$ satisfying the rules defined in \figref{fig:logic} (follows from the semantics given in \cite{FraSey14}). Regardless of the contents of a trace $\sV$, all actor systems $\actV$ can \emph{never} violate \mtru, yet \emph{always} violate \mfls. Conjunction formulas, $\hV\mand\hVV$, cause a violation if either $\hV$ or $\hVV$ lead to a violation. Conditionals, \mboolE{\bV}{\hV}{\hVV}, only cause a violation in case the taken branch (\hV\ or \hVV) causes a violation. All actor systems \actV violate a necessity formulas $\mnec{\patE}{\hV}$ whenever a pattern-matchable action  $\actE$ is performed (yielding substitution $\smash{\sigma::\Vars\rightharpoonup\Val}$), and the resulting actors \actVV that are transitioned violate $\hV\sigma$. Note that actors that \emph{do not} perform any pattern-matchable actions trivially satisfy $\mnec{\patE}{\hV}$, and hence cannot lead to a violation. A recursive formula \mmax{\hVarX}{\hV} is violated whenever the unfolded formula $\hV\sub{(\mmax{\hVarX}{\hV})}{\hVarX}$ is also violated. 

The violation relation is only satisfied by \emph{finite} traces, as opposed to the satisfaction relation which can also be satisfied by \emph{infinite} traces. Hence the violation semantics makes it easier for us to understand how the synthesised runtime monitors flag a violation \wrt a finite trace. In this dissertation we therefore focus on the violation semantics.

\section{Conclusion}
We have provided the reader with the necessary background information for better understanding our work. We have also established what the terms Runtime Verification, Runtime Adaptation and Runtime Enforcement refer to in the context of this dissertation, given that the distinction between these techniques is not always made clear in the current literature.

We have also clarified how we distinguish between Offline and Online monitors, and expanded on the latter by looking into different definitions of online monitoring, 
upon which we presented a spectrum of online monitoring techniques \wrt component-based systems. Furthermore, we also define our own terms to identify and distinguish between the techniques in the spectrum. By identifying this spectrum we are now in a better position to understand the design space that we need to explore for introducing synchrony within the \detecterGen\ RV tool in \Cref{chp:syn-asyn}.

Finally we provided an overview of the actor model as implemented by a host language Erlang, followed by a detailed overview of our target RV tool, \detecterGen, on which we build our runtime adaptation framework for actor systems in \Cref{chp:runtime-adaptation}.
	

\chapter{Assessing the Impact of Increasing Monitor Synchronisation} 
\label{chp:syn-asyn}
\newcommand{\madv}{\texttt{advices:}\xspace}
\newcommand{\aadv}{\texttt{after\_advice}\xspace}
\newcommand{\badv}{\texttt{before\_advice}\xspace}
\newcommand{\uadv}{\texttt{upon\_advice}\xspace}

Effective runtime adaptation requires achieving a tighter level of control over the system so as to provide timely detections of violations, thereby allowing the monitor to mitigate erroneous actors before these are allowed to proceed. In this chapter we therefore conduct preliminary investigations for implementing synchronous monitoring mechanisms over an \emph{inherently asynchronous architecture} (\ie the actor model). 

We conduct this investigation by exploring the design space of monitoring mechanisms identified in \secref{sec:spectrum}, and assess the repercussions that a monitored actor system may suffer due to synchrony. Although one might easily infer that synchronous monitoring incurs higher overheads as opposed to its asynchronous counterpart, we set out to investigate and quantify the overhead increase in an \emph{inherently asynchronous setting} such as ours. This is important as in the case of asynchronous platforms where synchrony is not natively supported (such as the actor-model), introducing synchronous monitoring in a naive way might highly affect the performance of the monitored asynchronous system. This opposes inherently synchronous settings \cite{lola:runtime,BauFal12} in which the impact of using synchronous monitoring might not be as significant. Quantifying the impact of synchronous monitoring on asynchronous actor-systems will then guide our design decisions for implementing adequate adaptation mechanisms in \Cref{chp:runtime-adaptation}.

We start our investigations by first introducing tight, coarse-grained synchronisation in \secref{sec:synchr-instr}, as this is more straightforward to understand and implement. Following this in \secref{sec:hybr-instr}, we explore techniques for reducing the level of synchrony to a more fine-grained level, 
while still being able to achieve timely detections. Since synchronous communication is not native to asynchronous platforms (such as the actor model), introducing synchrony requires instrumenting the system to encode a synchronisation protocol that forces the system to wait until the monitor is done verifying a reported event. Encoding such a protocol may therefore introduce overheads that must be minimised as much as possible. 

Hence in \secref{sec:evaluation}, we finally assess and quantify the monitoring overheads when using 
coarse-grained synchrony, fine-grained synchrony and asynchronous monitoring. We conduct our experiments \wrt Yaws $-$ a third-party, actor-based webserver written in Erlang, by monitoring it 
for several safety properties expressed in \detecterGen's logic, using these three different monitoring techniques.

\section{Introducing Synchronous 
Monitoring}
\label{sec:synchr-instr}
\begin{figure}[ht!]
	\centering
	\includegraphics[width=0.65\textwidth]{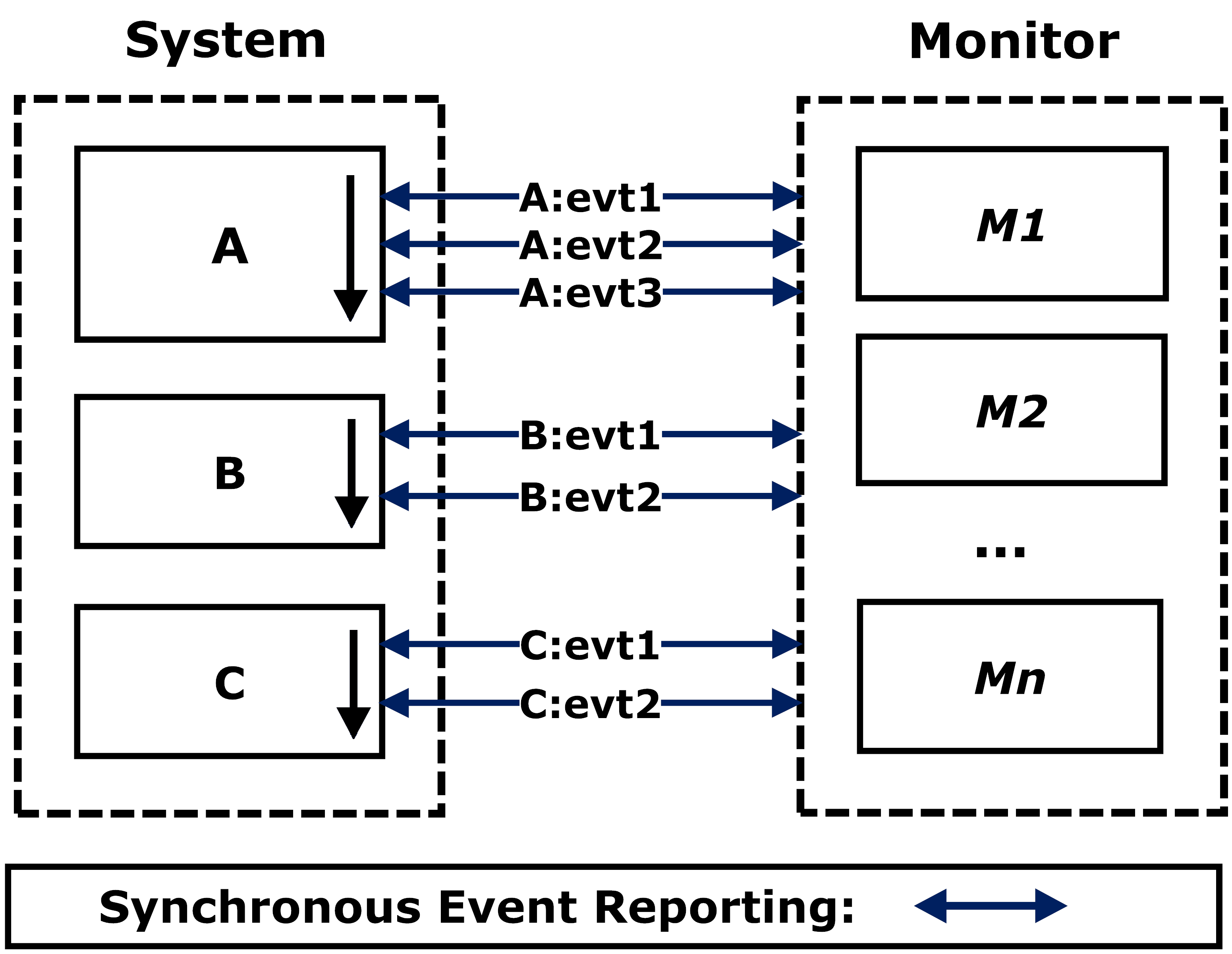}
	\caption{A High-level illustration of Synchronous Instrumentation Monitoring.}
	\label{fig:synch-abs}
\end{figure}

There are a number of ways how one can layer synchronous monitoring atop of an inherently asynchronous computational model, such as the actor model. For instance, one can insert actual monitoring functionality \emph{within} the sequential thread of execution of each actor (in the style of \cite{diana04,chen-rosu-2007-oopsla}) and then have the monitoring code (scattered across independently executing actors) synchronise, as required, in a \emph{choreographed} setup \cite{diana04,FGP12DistribRV}.   Instead, we opt for an \emph{orchestrated} solution as shown in \figref{fig:synch-abs} (below). 

In this setup, monitored actors are instrumented to synchronously report monitored actions as event notifications to a (conceptually) centralised monitoring setup that receives all reported events and performs the necessary checking. Note that although conceptually centralised, the orchestrator monitor is not monolithic \cite{FraSey14} as it consists of independent, concurrent sub-monitors that execute independent of one another. 

Furthermore, it is imperative that such a synchronous event reporting mechanism must prevent a system actor from progressing until the monitor is done checking its synchronously reported event. For instance in \figref{fig:synch-abs}, system actor $A$ must only be allowed to progress up to event \textsf{A:evt2} after the monitor completes checking the previously reported event, \ie \textsf{A:evt1}. However, while $A$ is waiting for the monitor to finish verifying its reported event (\ie \textsf{A:evt1}), the other concurrent actors $B$ and $C$ are still allowed to progress independently until they synchronously report their own events. 

There are various reasons for opting for such an orchestrated setup. The main reason being that it allows us to perform a \emph{like-with-like} comparison with the existing asynchronous setup present in \detecterGen, thus obtaining a more precise comparison between the relative overheads of synchronous and asynchronous monitoring. Furthermore, this setup conforms to the ideology of the actor oriented paradigm \cite{Armstrong07,Coulouris:2011}, which strongly encourages the developer to divide large tasks into smaller subtasks that can execute concurrently whenever this is possible. This is therefore reflected in this setup, as the monitor and the system are kept as two completely separate entities. Also, by using this approach the instrumentation code at the system side is kept minimal, thus leaving the instrumented code close to the original. Moreover, monitoring is \emph{consolidated} into a group of concurrent actors that are separate from the monitored system, thereby improving manageability (\eg parts of the system may crash leaving the monitor unaffected).  

\subsection{Instrumentation through Aspect-Oriented Weaving}
\label{sec:instr-thro-aspect}

We consider Aspect Oriented Programming (AOP) as an ideal candidate for implementing synchronous event reporting. The instrumentation is carried out using an AOP Framework for Erlang \cite{AOPErlang,CgvFypAOP} (in the style of standard AOP tools such as \cite{aspectJ}) to add the necessary instrumentation  in the system code. 

As \detecterGen\ supports the monitoring of output and input actions (\ie message sends and receives) along with function calls and returns, we needed an AOP framework that supports instrumenting all of these actions. The AOP presented in \cite{AOPErlang} was not ideal as this only supports function calls and returns; however this was later augmented with aspects for instrumenting the required input and output actions as a preliminary investigation in \cite{CgvFypAOP}. However this AOP framework \cite{CgvFypAOP} required further refinements, which we conducted during the course of this research, as it was still unstable and too coarse-grained\footnote{The aspects for instrumenting input and output actions did not support for selectively instrumenting specific send and receive operations which match a specific pattern. This was therefore adding unnecessary instrumentation.}.

The extended aspect-based instrumentation requires an \emph{aspect file} that specifies the actions requiring instrumentation, along with a purpose built 
module called \texttt{advices.erl} containing three types of advices used by the AOP injections, namely \badv, \aadv and \uadv advices.  For the function call events specified in the aspect file, the AOP weaves \badv advices reporting the event data to the monitor \emph{before} the function is invoked. For message sends and function returns, the AOP weaves \aadv advices to report the committed action to the monitor once the action is performed. In the case of message sends, this allows for the system to actually send the required message prior to reporting the event to the monitor. In the case of function returns, \aadv advices are required since return values are only known \emph{after} the function actually returns. The 
weaving concerning mailbox reading events (performed using the \texttt{recieve} construct) was not as straightforward. As the Erlang \texttt{recieve} construct may contain multiple pattern-matching guarded clauses \ie \\[1mm]\indent$\quad\texttt{recieve } \textsl{guard}_1 \texttt{ -> } \textsl{expression}_1 \texttt{ ; } \ldots\texttt{ ; } \textsl{guard}_n \texttt{ -> } \textsl{expression}_n \texttt{ end}.\\[1mm]$ 
The AOP weaves \uadv advice for each guarded expression that matches the message patten defined by the receive aspects, as specified in the aspect file. For instance if we assume that only $\textsl{guard}_1$ and $\textsl{guard}_2$ match the receive patterns specified in the aspect file, we get \\{\indent$\quad\texttt{recieve } \textsl{guard}_1 \texttt{ -> } \uadv(..),\textsl{expression}_1 \texttt{ ; } $\\[-2mm] \indent$\quad\qquad\qquad\, \textsl{guard}_2 \texttt{ -> } \uadv(..),\textsl{expression}_2 \texttt{ ; } $\\[-2mm] \indent$\quad\qquad\qquad\, \textsl{guard}_3 \texttt{ -> } \textsl{expression}_3 \texttt{ ; } \ldots\texttt{ ; } $\\[-2mm] \indent$\quad\qquad\qquad \textsl{guard}_n \texttt{ -> } \textsl{expression}_n $\\[-2mm] \indent$\quad\texttt{ end}.$}\\ 
as opposed to the coarse-grained instrumentation in \cite{CgvFypAOP} which adds \uadv to \emph{every} guarded expression thereby introducing unnecessary instrumentation. Also note that at runtime only \emph{one} \texttt{recieve} guarded expression is triggered. In the case if the triggered expression is an instrumented one, the necessary pattern-matched data of the event is extracted and thus can be reported to the monitor by the advice as required.

\subsection{The Synchronous Instrumentation Protocol}

\begin{figure}[ht!]
  \centering
  \begin{tikzpicture}[>=latex,auto,thick]
    \begin{scope}[draw=blue!50,fill=blue!20,minimum size=0.8cm]
      \node (sys) at ( -0.5,1) [shape=rectangle,draw,fill,text width=4.8cm, minimum height = 3cm, text centered]      
          {...\\ Actor $i$ commits event e1;\\ $i$ gets data d1 from e1;\\ $i$ blocks on nonce(e1);\\ $i$ unblocks with nonce1;\\...\\ Actor $j$ commits event e2;\\ $j$ gets data d2 from e2;\\ $j$ blocks on nonce(e2);\\ $j$ unblocks with nonce2;\\...};
       \node (mon) at ( 8,1) [shape=rectangle,draw,fill,text width=5.6cm, minimum height = 3cm]
       {\\[1.9mm]$\;$loop(Nonce)\;$\rightarrow$\\ $\quad$send\_ack(Nonce), \\ $\quad$\{Evt,Nonce2\} = recv\_event(), \\ $\quad$PtrnMatch = handle(Evt), \\ $\quad$if(PtrnMatch) $\rightarrow$ \\ $\qquad$loop(Nonce2); \\ $\quad$else $\rightarrow$\\ $\qquad$ send\_ack(Nonce2)\\ $\quad$end.\\[2mm]\;};  
        \node (sysN) at (-0.5,4) {\textbf{System}}; 
        \node (sysN) at (8,4) {\textbf{Monitor}}; 
    \end{scope}
     \node (sys0) at (1.85,3.4){};
       \node (mon0) at (5.2,3.4){};
       \node (sys1) at (1.85,2.2){};
       \node (mon1) at (5.2,2.2){};
       \node (sys2) at (1.85,1.9){};
       \node (mon2) at (5.2,1.9){};
       \node (sys3) at (1.85,0.1){};
       \node (mon3) at (5.2,0.1){};
       \node (sys4) at (1.85,-0.2){};
       \node (mon4) at (5.2,-0.2){};
    \begin{scope}[draw=red,fill=red, dashed]
    	\draw[<-] (sys0)  to node{\small\textbf{ack}(init\_nonce)} (mon0); 
      \draw[->] (sys1)  to node{\small\textbf{evt}(d1,\{$i$,nonce1\})} (mon1); 
      \draw[->] (mon2)  to node{\small\textbf{ack}(nonce1)} (sys2); 
      \draw[->] (sys3) to node{\small\textbf{evt}(d2,\{$j$,nonce2\})} (mon3); 
      \draw[->] (mon4) to node{\small\textbf{ack}(nonce2)} (sys4); 
    \end{scope}
  \end{tikzpicture}
  \caption{A High-level depiction of the Synchronous Event Monitoring protocol.}
       \label{fig:sync_protocol}
\end{figure}
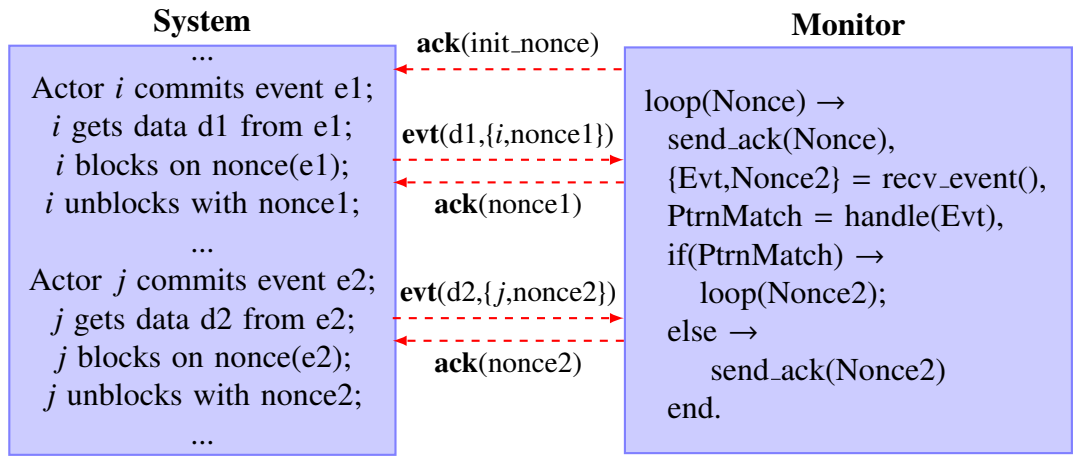

\figref{fig:sync_protocol} depicts a global implementation overview 
of the synchronous instrumentation monitoring (SMSI) protocol for two monitored events, \ie events \textsf{e1} and \textsf{e2}. This protocol denotes the messages exchanged between the instrumentation code injected in the asynchronous actor system and the monitor in order to achieve a synchronous event reporting mechanism. In fact this mechanism is achieved through handshakes over asynchronous messages between the two parties; this technique is a common practice for encoding synchronous message passing on asynchronous models such as the actor model \cite{Armstrong07}. 

This protocol starts with monitoring loop sending an initial acknowledgement message to the system, signalling it to execute until it produces the \emph{next monitored event}. To synchronously report a monitored event, the instrumentation code at the system side extracts the necessary data associated with the event, sends it as a trace message to the monitor, and then pauses by issuing a receive statement that waits for an \emph{unpause-acknowledgement} message from the monitor.  Since, the (acknowledgement) asynchronous messages may get reordered in transit (potentially interfering with this protocol) the instrumentation code at the system side generates a unique \emph{nonce} for every monitored event and sends it along with the event data. In return, when the monitor sends back the acknowledgement message, it includes this unique nonce. This allows the instrumentation code to pattern-match for (and possibly read out-of-order) mailbox inputs containing this nonce, and unblock based on this corresponding acknowledgement. 

The monitoring loop outlined in \figref{fig:sync_protocol} corresponds to the pattern-matching functionality required by a necessity formula $\mnec{\patE}{\hV}$. For instance, if the pattern is not matched (\ie the necessity is trivially satisfied), it acknowledges immediately to the system to continue executing and terminates monitoring by issuing a \texttt{send\_ack(Nonce2)} command to \emph{release} the blocked system actor. However, if the pattern matches there is still a chance that a violation can be detected and therefore proceeds by handling the event. Event handling involves executing internal (non-necessity) actions specified after the modal necessity (\ie corresponding to $\hV$ in $\mnec{\patE}{\hV}$) such as: unfolding recursive definitions, evaluating if-statement conditions, spawning conjunction submonitors, \etc Crucially, if $\hV$ \emph{reduces} into $\mfls$ (\eg after evaluating an if-statement), the monitor \emph{refrains} from sending back the acknowledgement, thus blocking the offending system indefinitely while flagging the violation. In a runtime adaptation setup, this allows for effectively executing adaptation actions instead of flagging the violation. Otherwise, the monitor issues command \texttt{loop(Nonce2)} to reiterate and release the blocked actor using the delegated nonce (\ie \texttt{Nonce2}).

\section{A Hybrid Instrumentation for Timely Detections}
\label{sec:hybr-instr}
Since synchronous instrumentation monitoring (SMSI) is too coarse-grained (as it synchronises for every system event), we therefore devise an alternative fine-grained synchronisation strategy which achieves timely detections with less synchronisation. The main idea is to move away from synchronous instrumentation monitoring, towards using \emph{incremental synchronisation} to obtain \emph{synchronous detection monitoring} (AMSD) for actor systems. This is possible as in order to attain timely detections, the instrumentation need not require certain system components to execute in lockstep with the monitor for \emph{every} monitored event leading to the violation. Instead, (expensive) synchronous event monitoring can be limited to \emph{critical actions}, (\ie system actions contributing \emph{directly} towards a violation), thereby letting the system execute in a decoupled fashion otherwise. 

\begin{figure}[ht!]
	\centering
	\includegraphics[width=0.65\textwidth]{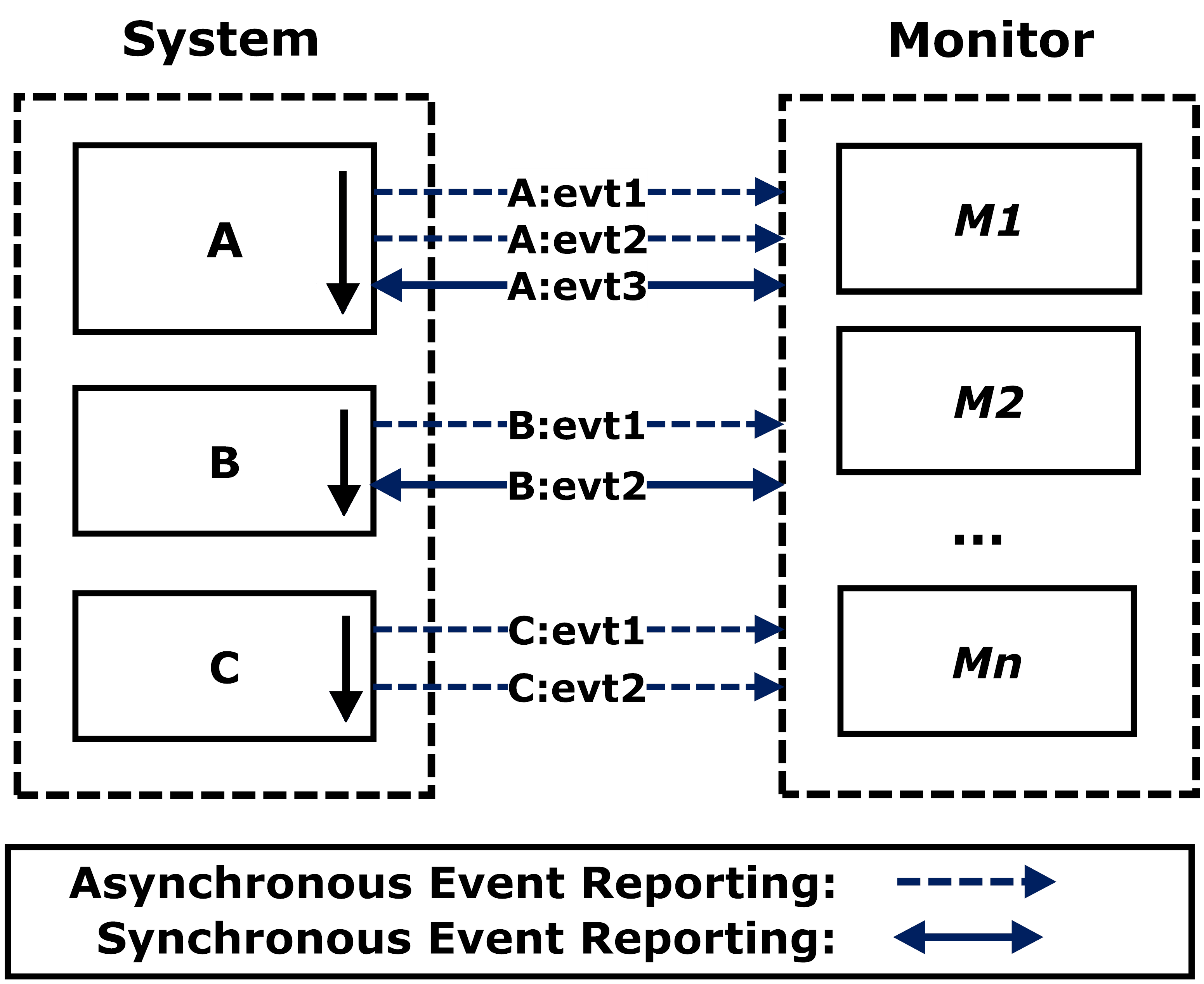}
	\caption{A High-level illustration of Synchronous Detection (Hybrid) monitoring.}
	\label{fig:hybrid-abs}
\end{figure}

For instance in \figref{fig:hybrid-abs} if we assume that events \textsf{A:evt3} and \textsf{B:evt2} are the only system events that may lead to a violation, we can limit synchronous event monitoring to these two events only. Note that for now we make a \emph{simplifying assumption} that for the logic used by \detecterGen\ (given in \figref{fig:logic}) these critical actions are necessity actions preceeding (directly or indirectly) a \mfls\ forumla as explained in \exref{ex:hybrid-instrumentation} below.  

\begin{example}\label{ex:hybrid-instrumentation}
  Recall \Propref{prop:intro:1} which we formalised as formula \eqref{intro:1} in \exref{ex:background} (restated below):  
  {
\setlength{\belowdisplayskip}{8pt}%
\setlength{\abovedisplayskip}{5pt}%
\begin{align*}     
    \hV \;\deftxt\;\;& \mmax{\,\hVarY}{\mnec{\mrecv{i}{\tup{\eatom{inc},x,y}}}
        \begin{mlbrace}(\,\mnec{\msendD{j}{y}{\tup{\eatom{res},\smash{x+1}}}}\,\hVarY)   \; \mand \; (\,\mnec{\msendD{\_\,}{y}{\eatom{err}}}\,\mfls) \end{mlbrace}} \qquad \eqref{intro:1}
  \end{align*} 
  }
  Based on our simplifying assumption, to synchronously detect a violation in this property, only action $\mnec{\msendD{\_}{y}{\eatom{err}}}{}$ needs to be synchronously monitored, as it precedes a \mfls\ forumla. The other actions (\ie $\mnec{\mrecv{i}{\tup{\eatom{inc},x,y}}}$ and $\mnec{\msendD{j}{y}{\tup{\eatom{res},\smash{x+1}}}}$) can however be monitored for asynchronously, without affecting the timeliness of detections. \bqed
\end{example}

\begin{remark} With this simplifying assumption we assume that the actor that commits the last action preceeding a violation (\mfls) is solely responsible for the detected violation, however this is not always the case. 

For instance in the example system given in \figref{fig:sys}, despite the incrementor being the one to send the error message \textsf{err} to the connected client thus violating \eqref{intro:1}, this error was caused by an invalid interaction made by the common-interface in some previous step. Hence strictly speaking to truly achieve a timely detection, the common-interface should also be forbidden from progressing after performing an interaction that potentially contributes to the violation, until the monitor performs the necessary checks. 

Our assumption however permits us to get a clear indication of the associated performance impacts without introducing too much complexities at this stage. Remember that for now 
our main objective is to identify an efficient synchronous monitoring mechanism upon which we can build in order to introduce runtime adaptation in \Cref{chp:runtime-adaptation}.	\bqed
\end{remark}

\subsection{Logic Extensions}
\label{sec:hybrid-logic-extensions}
To achieve synchronous detection monitoring we explore a \emph{hybrid} technique that is capable of \emph{switching} between synchronous and asynchronous event monitoring, and which limits synchronous monitoring to critical system actions. To achieve this we extend the syntax of \detecterGen's logic \cite{FraSey14} (outlined in \secref{sec:detecter-primer-2}) by two constructs: a \emph{synchronous false} formula and a \emph{synchronous necessity} formula, with violation semantics analogous to  that of \mfls\ and \mnec{\patE}{\hV} respectively.
{
\setlength{\belowdisplayskip}{6pt}%
\setlength{\abovedisplayskip}{8pt}%
\begin{align*}
    \hV,\hVV \in \FRM & \;\bnfdef\; \ldots \;\bnfsep\; \msfls  \quad \emph{(synchronous false)} \;\bnfsep\;  \msnec{\patE}{\hV} \quad \emph{(synchronous necessity)} 
\end{align*}
}
In the extended logic, formulas carry additional  instrumentation information relating to how they need to be runtime-monitored. By default, all the monitoring is asynchronous, unless one specifies that  a violation is to be synchronously detected,  \msfls, or that a particular system event needs to be synchronously monitored,  \msnec{\patE}{\hV}. Although synchronous necessities can be used in specification scripts to engineer synchronisation checkpoints (\ie to encode AMC; see \secref{sec:spectrum}) similar to what is presented in \cite{CP12FF}, in this study we limit their use to achieving timely-detections. In fact, rather then using synchronous necessities explicitly we \emph{encode} synchronous falsities, \msfls, in terms of synchronous necessities as explained in \exref{ex:synch-asynch}.

\begin{example} \label{ex:synch-asynch}  Recall the system presented in \figref{fig:sys} for which we extend and refine formula \eqref{intro:1} (restated in \exref{ex:hybrid-instrumentation}) as \eqref{hybrid:3} (shown below). We now distinguish between two kinds of violations in which: result values $w$ that are less than $x\!+\!1$ are defined to be \emph{asynchronously} detected (\ie $\mbool{\smash{w<x+1}}{\mfls}$), while values greater than $x\!+\!1$ are specified to be \emph{synchronously} detected (\ie $\mbool{\smash{w>x+1}}{\msfls}$). We also specify that cases where client $y$ is sent an error (\ie $\msendD{\_\,}{y}{\eatom{err}}$) are to be synchronously detected as well.\\
  \begin{minipage}[c]{0.90\linewidth}
   \begin{align}\label{hybrid:3}\setstretch{0.8}
    \mmax{\,\hVarY}{\mnec{\mrecv{i}{\tup{\eatom{inc},x,y}}}
        \begin{mlbrace}\!(\,\mnec{\msendD{j}{y}{\tup{\eatom{res},w}}} 
        \begin{mlbrace} \mbool{\smash{w=x+1}}{\hVarY}\\ \,\mand\, \mbool{\smash{w>x+1}}{\msfls} \\ \,\mand\,\mbool{\smash{w<x+1}}{\mfls} \end{mlbrace}\,)   
        \; \mand \; (\,\mnec{\msendD{\_\,}{y}{\eatom{err}}}\,\mathbf{\msfls})\! \end{mlbrace}} 
  \end{align}
\end{minipage}\\

The new monitor synthesis algorithm requires a pre-processing phase to determine which events are to be synchronously monitored in order to implement a synchronous fail. For instance, formulas $\mnec{\msendD{\_\,}{y}{\eatom{err}}}\msfls$ and $\msnec{\msendD{\_\,}{y}{\eatom{err}}}{\mfls}$ are both monitored in the same way, in fact the pre-processing function \textsf{syn\_enc}, presented in Def.~\ref{def:syn-infer}, encodes the former into the latter. As shown by Def.~\ref{def:syn-infer}, generally determining the critical actions to synchronously monitor for implementing a synchronous fail is not as straightforward, since the first necessity formula preceding a $\msfls$ may be interposed by intermediate formulas such as conjunctions and if-statements (as in the case of formula \eqref{hybrid:3}).   \bqed 
\end{example}

\begin{definition}\label{def:syn-infer}
	\begin{small}
		$\textsf{syn\_enc}:\FRM\mapsto(\textsc{Bool}\times\FRM)$
			\begin{equation*}\setstretch{1.1}
				\textsf{syn\_enc}(\hV)\defEquals
						\begin{cases} 
							\langle false,\msnec{\patE}\hVV'\rangle  & \text{if }\hV=\mnec{\patE}\hVV\,\land\,\textsf{syn\_enc}(\hVV) =\langle true,\hVV'\rangle \\
							\langle false,\mnec{\patE}\hVV'\rangle  & \text{if }\hV=\mnec{\patE}\hVV\,\land\,\textsf{syn\_enc}(\hVV) =\langle false,\hVV'\rangle \\
							\langle b_{1}\lor b_{2},{\hVV'_{1}}\mand{\hVV'_{2}}\rangle  & \text{if }\hV={\hVV_{1}}\mand{\hVV_{2}}\,\land\,\textsf{syn\_enc}(\hVV_{1}) =\langle b_{1},\hVV_{1}'\rangle \\[-1.2mm] & \;\land\,\textsf{syn\_enc}(\hVV_{2}) =\langle b_{2},\hVV_{2}'\rangle \\
							\langle b_{1}\lor b_{2},\mboolE{c}{\hVV'_{1}}{\hVV'_{2}}\rangle  & \text{if }\hV=\mboolE{b}{\hVV_{1}}{\hVV_{2}}\\[-1.2mm] & \;\land\,\textsf{syn\_enc}(\hVV_{1}) =\langle b_{1},\hVV_{1}'\rangle \\[-1.2mm] & \;\land\,\textsf{syn\_enc}(\hVV_{2}) =\langle b_{2},\hVV_{2}'\rangle \\
							\langle b,\mmax{\hVarX}{\hVV'}\rangle  & \text{if }\hV=\mmax{\hVarX}{\hVV}\,\land\,\textsf{syn\_enc}(\hVV) =\langle b,\hVV'\rangle \\
							\langle true,\mfls\rangle  & \text{if }\hV=\msfls \\
							\langle false,\hV\rangle  & \text{otherwise}
						\end{cases}\\
			\end{equation*}	
	\end{small}
\end{definition}

	The pre-processing function \textsf{syn\_enc} recursively inspects a formula $\hV$ and returns a more implementable version in which the necessities specifying critical actions are identified and converted into synchronous necessities. For cases where $\hV$ reduces to $\msfls$, this function returns a pair containing boolean value $true$ and $\mfls$. The boolean value $true$ indicates that a synchronous falsity \msfls\ was found and converted to \mfls, which is more implementable. This truth value then allows for the \emph{first necessity} action declared prior to $\msfls$ to be converted into a synchronous necessity. This guarantees that the falsity is detected \emph{on-time}. In fact when inspecting necessity actions, $\mnec{\patE}\hVV$, the function recursively inspects the subsequent formula $\hVV$ (\ie it reapplies itself as $\textsf{syn\_enc}(\hVV)$) and if the recursive application returns the pair $\langle true,\hVV'\rangle$, it converts the necessity $\mnec{\patE}\hVV$ into a synchronous necessity $\msnec{\patE}\hVV'$. This is possible as the truth value $true$ in the returned pair, signifies that at least one \msfls\ was found somewhere along $\hVV$ and converted into \mfls, thus changing $\hVV$ into $\hVV'$ which is more implementable. After encoding a normal necessity into a synchronous necessity, the \textsf{syn\_enc} function then returns a pair containing the encoded formula, \ie $\msnec{\patE}\hVV'$, and truth value $false$, where the latter denotes that the synchronous falsity has already been encoded. This prevents prior asynchronous necessities from being unnecessarily converted into synchronous ones. 
	
	To conduct our experiments we therefore augment the \textsf{syn\_enc} function as a pre-processing phase in our compiler to encode formulas containing synchronous falsities into a more implementable versions as shown in \exref{ex:syn-encode}.

	\begin{example} \label{ex:syn-encode}
	In the case of property \eqref{hybrid:3} the compiler uses the \textsf{syn\_enc} function to check whether there exists at least \emph{one} path that leads directly to a synchronous fail. Since action $\mnec{\msendD{j}{y}{\tup{\eatom{res},w}}}$ is followed by a conjunction of three if-statements in which one of them leads to a synchronous falsity, \ie $\mbool{w>x+1}{\msfls}$, then this necessity is converted into a synchronous necessity. More obviously action $\mnec{\msendD{\_\,}{y}{\eatom{err}}}$ is also converted as it is immediately followed by \msfls. The compiler therefore converts property \eqref{hybrid:3} into \eqref{hybrid:4}:
	
		\noindent\begin{minipage}[c]{0.90\linewidth}
		    \begin{equation}\label{hybrid:4}\setstretch{0.8}
			    \mmax{\,\hVarY}{\mnec{\mrecv{i}{\tup{\eatom{inc},x,y}}}
			        \begin{mlbrace}\!(\,\msnec{\msendD{j}{y}{\tup{\eatom{res},w}}} 
			        \begin{mlbrace} \mbool{w=x+1}{\hVarY}\\ \,\mand\, \mbool{w>x+1}{\mfls} \\ \,\mand\, \mbool{w<x+1}{\mfls} \end{mlbrace}\,)   
			        \; \mand \; (\,\msnec{\msendD{\_\,}{y}{\eatom{err}}}\,\mfls)\! \end{mlbrace}}
			\end{equation}
		\end{minipage}
		\begin{minipage}[l]{0.09\linewidth}
			\vspace{1.7cm}
			  \bqed
		\end{minipage}
	\end{example}

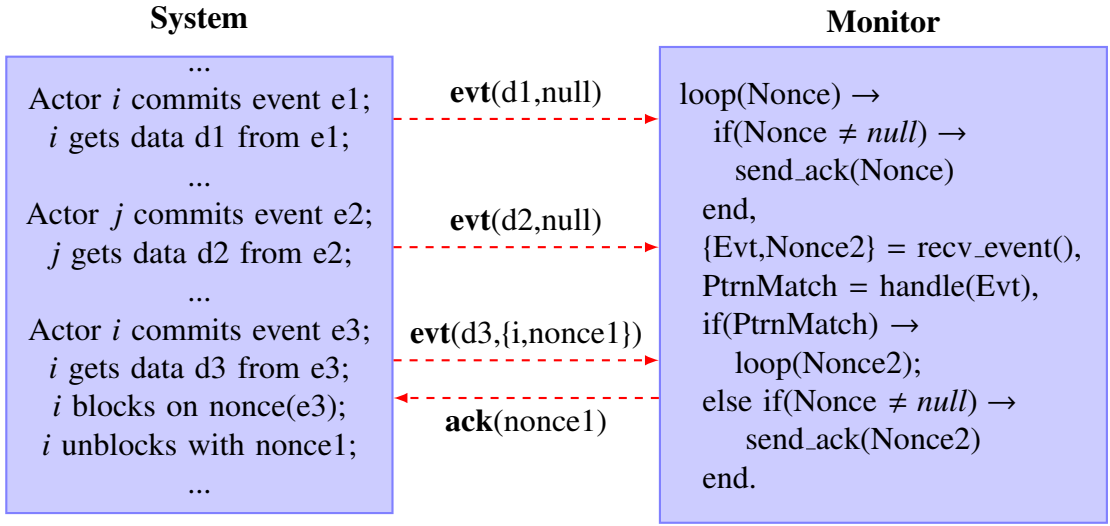
\begin{figure}[t!]
  \centering
  \begin{tikzpicture}[>=latex,auto,thick]
    \begin{scope}[draw=blue!50,fill=blue!20,minimum size=0.8cm]
      \node (sys) at ( -1,1) [shape=rectangle,draw,fill,text width=4.8cm, minimum height = 3cm, text centered] 
          {...\\ Actor $i$ commits event e1;\\ $i$ gets data d1 from e1;\\...\\Actor $j$ commits event e2;\\ $j$ gets data d2 from e2;\\...\\ Actor $i$ commits event e3;\\ $i$ gets data d3 from e3;\\$i$ blocks on nonce(e3);\\ $i$ unblocks with nonce1;\\...\\\; };
       \node (mon) at ( 8,1) [shape=rectangle,draw,fill,text width=5.6cm, minimum height = 3cm] {\\[1mm]$\; $loop(Nonce)\;$\rightarrow$\\$\quad$ if(Nonce $\not{=} null$)\;$\rightarrow$\\ $\qquad$send\_ack(Nonce) \\ $\quad$end, \\ $\quad$\{Evt,Nonce2\} = recv\_event(), \\ $\quad$PtrnMatch = handle(Evt), \\ $\quad$if(PtrnMatch) $\rightarrow$ \\ $\qquad$loop(Nonce2); \\ $\quad$else if(Nonce $\not{=} null$)\;$\rightarrow$\\ $\qquad$ send\_ack(Nonce2)\\ $\quad$end.\\[2mm]\; };  
        \node (sysN) at (-1,4.5) {\textbf{System}}; 
        \node (monN) at (8,4.5) {\textbf{Monitor}}; 
    \end{scope}
       \node (sys1) at (1.4,3.2){};
       \node (mon1) at (5.2,3.2){};
       \node (sys2) at (1.4,1.5){};
       \node (mon2) at (5.2,1.5){};
       \node (sys3) at (1.4,0){};
       \node (mon3) at (5.2,0){};
       \node (sys4) at (1.4,-0.5){};
       \node (mon4) at (5.2,-0.5){};
    \begin{scope}[draw=red,fill=red, dashed]
      \draw[->] (sys1) to node{\textbf{evt}(d1,null)} (mon1); 
      \draw[->] (sys2) to node{\textbf{evt}(d2,null)} (mon2); 
      \draw[->] (sys3) to node{\textbf{evt}(d3,\{i,nonce1\})} (mon3); 
      \draw[->] (mon4) to node{\textbf{ack}(nonce1)} (sys4); 
    \end{scope}
  \end{tikzpicture}
  \caption{A High-level depiction of the Hybrid Monitoring Protocol. 
  }
       \label{fig:hybrid_protocol}
\end{figure}

\subsection{The Hybrid Monitoring Protocol}
In hybrid monitoring, both synchronous and asynchronous event monitoring require code instrumentation, as shown in \figref{fig:hybrid_protocol} (below). \emph{Asynchronous} necessity actions inject advice functions that send a monitoring message, \eg event $e1$, to the monitor containing: the event details, \eg $d1$, and a \emph{null} nonce; this message is reported \emph{without blocking} the system. Upon receiving the null nonce, the monitor determines that it does not need to send an acknowledgment back to the system. Furthermore, \emph{Synchronous} necessity actions are implemented as per synchronous instrumentation (see \secref{fig:sync_protocol}), where the system attaches a \emph{fresh} nonce along with the event data to the monitoring message thereby notifying the monitor that it requires to be acknowledged back.

\section{Evaluation}
\label{sec:evaluation}
By now the extended version of the \detecterGen\ RV tool supports three methods for monitoring actor-based systems, namely:
\begin{itemize}[leftmargin=*,labelindent=5pt,itemindent=-15pt] \itemsep0em
	\item \emph{\textbf{Asynchronous Monitoring (CA)}}, which was the only monitoring mechanism originally supported by \detecterGen;
	\item \emph{\textbf{Synchronous Instrumentation Monitoring (SMSI)}}, introduced in \secref{sec:synchr-instr};
	\item \emph{\textbf{Hybrid Monitoring (AMSD)}}, introduced in \secref{sec:hybr-instr}.
\end{itemize}
Using these three monitoring approaches, we want to carry out experiments in which we monitor for several safety properties written specifically for the Yaws webserver. With these experiments we aim to obtain performance results for each approach in order to be able to compare and contrast the performance overheads imposed by these monitoring techniques. The results of these experiments will then serve as guidance for choosing the appropriate instrumentation mechanism on which we can build our RA framework in \Cref{chp:runtime-adaptation}.

\subsection{The Yaws Webserver} \label{sec:yaws:background}
Yaws \cite{yaws:11,yaws:12} is a high-performance, actor-based HTTP webserver written in Erlang.  For every client connection, this server assigns a dedicated handler actor that listens for and services HTTP client requests.  At its core, its implementation fundamentally relies on the lightweight nature of Erlang processes to be able to efficiently handle a vast amount of 
client connections. 

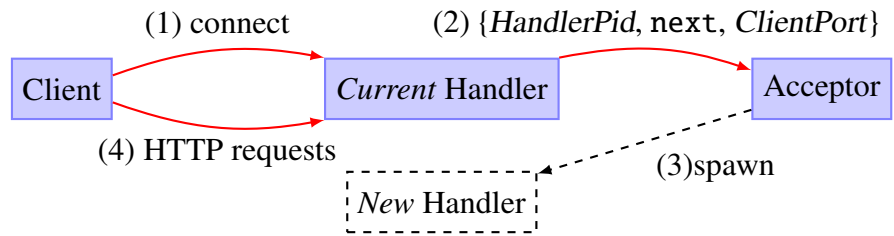
\begin{figure}[t]
  \centering
  \begin{tikzpicture}[>=latex,auto,thick]
    \begin{scope}[draw=blue!50,fill=blue!20,minimum size=0.8cm]
      \node (client) at ( 0,3) [shape=rectangle,draw,fill] {Client};
      \node (curHandler) at ( 5,3) [shape=rectangle,draw,fill] {\emph{Current} Handler};
      \node (acceptor) at ( 10,3) [shape=rectangle,draw,fill] {Acceptor};       
    \end{scope}
   \begin{scope}[draw=black,dashed,minimum size=0.8cm]
    \node (newHandler) at ( 5,1.5) [shape=rectangle,draw] {\emph{New} Handler};
   \end{scope}
    \begin{scope}[draw=red,fill=red]

      \draw[->] (client) [bend left=15] to node [above]{(1) connect} (curHandler); 
      \draw[->] (curHandler) [bend left=15] to node [above]
      {(2) \{\textsl{HandlerPid}, \eatom{next}, \textsl{ClientPort}\}} (acceptor);
      \draw[->] (client) [bend right=15] to node  [below]{(4) HTTP requests} (curHandler); 
     \end{scope}
    \begin{scope}[draw=black, dashed]
      \draw[->] (acceptor) to node{(3)spawn} (newHandler);
    \end{scope}
  \end{tikzpicture}
  \caption{Yaws client connection protocol}
\label{fig:yaws-protocol}
\end{figure}

The Yaws protocol for establishing client connections is depicted in \figref{fig:yaws-protocol}. This protocol relies on an \emph{acceptor} component which upon creation, it spawns a \emph{connection handler} to be assigned to the next client connection. Subsequently, the acceptor blocks waiting for messages in its mailbox, while the unassigned handler waits for the next TCP connection request. Clients send connection requests through standard TCP ports $(1)$, which are received as messages in the \emph{handler}'s mailbox.  The current handler accepts these requests by reading the respective message from its mailbox and $(2)$  sending a message containing its own \emph{pid} and the \emph{port} of the connected client to the \emph{acceptor}. This acts as a notification that the handler is now engaged in handling the connection of a specific client. Upon receiving the message, the \emph{acceptor} unblocks, records the information sent by the handler and $(3)$ spawns a \emph{new} handler listening for future connection requests from other clients. 

Once it is assigned a handler, the connected client then engages \emph{directly} with it by using $(4)$ standard HTTP requests; these normally consist of six (or more) HTTP headers containing information such as the client's User Agent, Accept-Encoding and the Keep-Alive flag status. It is worth highlighting that the HTTP request information is \emph{not} sent in one go, but follows a protocol of messages: it starts by sending the \eatom{http\_req}, followed by six (or more) \eatom{http\_header} messages containing client information, terminated by a final \eatom{http\_eoh} message. The dedicated connection handler inspects the request  and the client information received in the headers, and services the respective HTTP request 
accordingly.

Earlier versions of the Yaws webserver were found to be suffering from a Directory Traversal Vulnerability \cite{Holovaty:2009:DGD:1572516} which was also reported on the reputable exploit-db website \cite{yaws-exploit-misc}. This vulnerability permits for attackers to perform a \emph{Dot-dot slash} attack in which a malicious client provides the webserver with an invalid, malformed URL, containing a sequence of dots and slashes, in an attempt to exploit the vulnerability. This allows external clients to navigate to sensitive directories, pertaining to the server, such as the \emph{System32} directory. 

\subsection{Monitoring for safety properties in Yaws}
\label{sec:monit-safety-prop}
Despite the simplicity and elegance of \detecterGen's \SHML\ specification language, it turns out that it provides enough expressivity to specify numerous safety properties for the Yaws webserver. For brevity in this section we only present two of these properties, and in appendix \secref{sec:app-aux-properties} we present another two properties.
%

\subsubsection{Detecting Malicious Requests}

In the first property we assume the existence of the $\textsl{isMalicious}(..)$ predicate to define \Propref{prop:abs:yaws:1} (below). This (decidable) predicate can determine whether the client will engage in a security-breaching activity by analysing the 6 HTTP headers sent to the handler.\bigskip

\begin{property} \label{prop:abs:yaws:1}
	\emph{Every time a client connection is established (determined from message $(2)$ of \figref{fig:yaws-protocol}), and the assigned handler receives an HTTP-Get request followed by 6 HTTP headers ($h1$ to $h6$) and terminated by the end-of-headers notification, then the communicated headers must not amount to a potentially security-breaching request as determined by the predicate $\textsl{isMalicious}(..)$.}
\end{property}\bigskip

\noindent Using \detecterGen's formal logic presented in \figref{fig:logic} (see \secref{sec:language}), we can now formalise \Propref{prop:abs:yaws:1} as formula (\ref{prop:yaws:1}).  
\bigskip

\noindent\begin{minipage}[c]{0.90\linewidth}
		 {\small\setstretch{1.1}
			\noindent\begin{equation} 
			\begin{array}{l}
			    \mmax{\hVarX}{\big( \\\quad\mnec{\mrecv{\eatom{acceptor}}{\etuple{hPid,\eatom{next},\_}}}\\
			      \quad\mnec{\mret{hPid}{\etuple{\eatom{yaws}, \eatom{do\_recv},\eatom{3},\etuple{\eatom{ok},\etuple{\eatom{http\_req},\eatom{GET},\_ ,\_}}}}}\\
			    \quad\mnec{\mret{hPid}{\etuple{\eatom{yaws}, \eatom{do\_recv},\eatom{3},\etuple{\eatom{ok},h1}}}} \;
			    \mnec{\mret{hPid}{\etuple{\eatom{yaws}, \eatom{do\_recv},\eatom{3},\etuple{\eatom{ok},h2}}}} \\
			     \quad\mnec{\mret{hPid}{\etuple{\eatom{yaws},\eatom{do\_recv},\eatom{3},\etuple{\eatom{ok},h3}}}} \;
			     \mnec{\mret{hPid}{\etuple{\eatom{yaws},\eatom{do\_recv},\eatom{3},\etuple{\eatom{ok},h4}}}}\\
			     \quad\mnec{\mret{hPid}{\etuple{\eatom{yaws},\eatom{do\_recv},\eatom{3},\etuple{\eatom{ok},h5}}}} \;
			     \mnec{\mret{hPid}{\etuple{\eatom{yaws},\eatom{do\_recv},\eatom{3},\etuple{\eatom{ok},h6}}}} \\ \quad
			       \mboolE{(\textsl{isMalicious}(h1,h2,h3,h4,h5,h6))}{\;\mfls\;\\\quad}{\;\mnec{\mret{hPid}{\etuple{\eatom{yaws},\eatom{do\_recv},\eatom{3},\etuple{\eatom{ok},\eatom{http\_eoh}}}}}{\;\hVarX}} \\ \big) }   
			\end{array}\label{prop:yaws:1} \vspace{-3mm}     
			\end{equation}
		}
\end{minipage}	

The logical formula stated in \eqref{prop:yaws:1} specifies this property as a recursive formula which pattern matches the assigned handler to term variable \emph{hPid}. This value is then used to pattern match with the header term variables, \emph{h1} to \emph{h6}, for every iteration of the HTTP request protocol. An if-statement is then used to evaluate the $\textsl{isMalicious}(..)$ predicate to determine whether the received HTTP headers pose a security threat or not. 

We note that, whereas the handler messages to the acceptor  are observed \emph{directly} (\ie by using input action $\mnec{\mrecv{\eatom{acceptor}}{\etuple{hPid,\eatom{next},\_}}}$), the client HTTP messages received by the handler have to be observed \emph{indirectly} through the return values (of the form $\etuple{\eatom{ok},header}$) of the invoked function \eatom{do\_recv}, which is defined in module \eatom{yaws} with arity \eatom{3}.  Instrumentation allowing a direct observation of these actions is complicated by the fact that the client TCP messages are sent through functions from the \eatom{inet} Erlang library, which is part of the Erlang Virtual Machine kernel \cite{ErlangOTP,Armstrong07}.


\subsubsection{Detecting Directory Traversal Exploitation}

In a similar manner we specify yet another property for detecting behaviour that may exploit the Directory Traversal Vulnerability, found in earlier versions of the software, as described in \secref{sec:yaws:background}. For the intents and purposes of this case study, we aim to detect such possible attacks by monitoring for external client requests and compare the requested urls to a \emph{white-list}. We assume that external clients are only allowed to request for two files, namely ``pig.png'' and ``site.html'' as specified in safety \propref{prop:abs:yaws:2}. \bigskip

\begin{property} \label{prop:abs:yaws:2}
	\emph{Every time a client connects, and the assigned handler receives an HTTP-Get request for a specific file stored on our server, followed by 6 HTTP headers ($h1$ to $h6$) and the end-of-headers notification, then the requested file can only refer to either for ``pig.png'' or ``site.html''.}
\end{property} \bigskip

We formalise this property (\ie\, \Propref{prop:abs:yaws:2}) as recursive formula \eqref{prop:yaws:2} which binds the assigned handler with term variable \emph{hPid}, and then uses this value to pattern match with the HTTP-GET request, the 6 headers, and the ending header \eatom{http_eoh}. The file requested in the HTTP-GET request is pattern matched and stored in the $path$ term variable. The matched path value is then inspected by an if-statement that checks whether the requested path points either to ``\eatom{pic.png}'' or ``\eatom{site.html}''. If the path is found to be pointing to any of these two files, the property immediately checks that the server actually sends the requested file to the client and recurs on formula variable $X$; otherwise, a property violation is flagged.

 \vspace{-3mm}
{\small\setstretch{1.1}
\begin{equation} 
\begin{array}{l}
 \mmax{\hVarX}{\Bigl(
 	\\\indent \mnec{\msend{\eatom{acceptor}}{\{hPid,\texttt{next},\_\}}}
	 \\\indent \mnec{\mret{hPid}{\{\eatom{yaws}, \eatom{do\_recv},3,\{\texttt{ok},\{\eatom{http\_req},\eatom{'GET'},\{\eatom{abs\_path},path\},\_\}\}\}}} 
	 \\\indent \mnec{\mret{hPid}{\etuple{\eatom{yaws}, \eatom{do\_recv},\eatom{3},\etuple{\eatom{ok},\_}}}}
	 \\\indent \quad \pmb{\vdots}
	 \\\indent \mnec{\mret{hPid}{\etuple{\eatom{yaws}, \eatom{do\_recv},\eatom{3},\etuple{\eatom{ok},\eatom{http_eoh}}}}}
	 \\\indent \mboolE{path == ``\eatom{/pic.png}'' \texttt{ orelse } path == ``\eatom{/site.html}''}{
 	\\\indent\indent \mnec{\mcall{hPid}{\{\eatom{yaws\_sendfile}, \eatom{send},[\_,path,\_,\_]\}}} \; \hVarX\\\indent }{ \mfls} 
 	\\\indent \; \Bigr)
 \\\Bigr) }\vspace{-7mm}
\end{array}\label{prop:yaws:2}
\end{equation}
}

\subsubsection{Adapting these Properties for Hybrid Monitoring}
\noindent To be able to monitor for these properties using our hybrid approach, we use the extended syntax (introduced in \secref{sec:hybrid-logic-extensions}) to reformulate security properties \eqref{prop:yaws:1} and \eqref{prop:yaws:2} whereby we require violation detections to be synchronous \ie using \msfls\ instead of \mfls.  For instance, from formula \eqref{prop:yaws:2} we obtain the hybrid formula \eqref{prop:yaws:2-hybrid} stated below.

 \vspace{-3mm}
{\small\setstretch{1.1}
\begin{equation} 
\begin{array}{l}
 \mmax{\hVarX}{\Bigl(
 	\\\indent \mnec{\msend{\eatom{acceptor}}{\{hId,\texttt{next},\_\}}}
 	 \\\indent \mnec{\mret{hId}{\{\eatom{yaws}, \eatom{do\_recv},3,\{\texttt{ok},\{\eatom{http\_req},\eatom{'GET'},\{\eatom{abs\_path},path\},\_\}\}\}}} 
	 \\\indent \mnec{\mret{hId}{\etuple{\eatom{yaws}, \eatom{do\_recv},\eatom{3},\etuple{\eatom{ok},h1}}}} 
	 \\\indent \quad \pmb{\vdots}
	 \\\indent \mnec{\mret{hId}{\etuple{\eatom{yaws}, \eatom{do\_recv},\eatom{3},\etuple{\eatom{ok},h6}}}}
	 \\\indent \mnec{\mret{hId}{\etuple{\eatom{yaws}, \eatom{do\_recv},\eatom{3},\etuple{\eatom{ok},\eatom{http_eoh}}}}}
	 \\\indent \mboolE{path == ``\eatom{/pic.png}'' \texttt{ orelse } path == ``\eatom{/site.html}''}{
 	\\\indent\indent \mnec{\mcall{hId}{\{\eatom{yaws\_sendfile}, \eatom{send},[\_,path,\_,\_]\}}} \hVarX\\\indent }{ \pmb{\msfls}} 
 \\\Bigr) }
\end{array}\label{prop:yaws:2-hybrid}
\end{equation} \vspace{-5mm}
} 

All of the other properties used in our experiments where reformulated in the same manner as per hybrid formula \eqref{prop:yaws:2-hybrid}.

\subsection{The Performance Results}
\label{sec:eval-results}

\noindent We measure the respective overheads resulting from the three monitor instrumentation techniques that are now supported by \detecterGen, over Yaws for varying client loads, in terms of: 
\begin{itemize} \itemsep0em \setstretch{1.2}
	\item the average number of \emph{CPU cycles} required per Yaws client request;
	\item the average \emph{memory utilisation} for the Yaws webserver to respond to batches of client requests; and
	\item the average response time of the (monitored) Yaws server, \ie the time taken for the server to respond to a client request.
\end{itemize}

\begin{remark} In our experiments we utilise a version of \detecterGen, presented in \cite{CasFraSaid15}, that provides two monitoring optimisations --- a static optimisation and a dynamic optimisation, along with the original unoptimised monitor synthesis technique given in \cite{FraSey14}. These techniques only optimise the way that the concurrent monitors are synthesised and the way they operate at runtime, but they are independent from the system instrumentation used for event reporting. This means that they can be applied to every monitoring instrumentation technique explored so far (\ie asynchronous, synchronous and hybrid instrumentation). This enables us to ensure that the performance discrepancies obtained in our experiments persist regardless of how the synthesised concurrent monitors are optimised. This allows us to rule out that the discrepancies are caused due to unoptimised monitors. We also provide a brief overview of these optimisations in Appendix \Cref{chp:app-mon-opt}. \bqed
\end{remark}

The experiments were carried out on an Intel Core 2 Duo T6600 processor with 4GB of RAM, running Microsoft Windows 7 and EVM version R16B03. To further confirm the discrepancies between synchronous, hybrid and asynchronous monitoring we run our test-suite and take separate batches of results using $(i)$ unoptimised monitors, $(ii)$ statically optimised monitors and $(iii)$ dynamically (fully) optimised monitors. 

For each property (in our test-suite) and each client load, we take five sets of readings and then average them out. Since results pertaining to a specific monitor optimisation do not show substantial variations when using different properties, we again average the results across all properties and compile them as graphs for each optimisation. We thus obtain the graphs shown in \Cref{fig:unopt_results,fig:sopt_results,fig:s+dopt_results} --- the same results are also compiled as tables in appendix \secref{sec:app-tab-res}.

\begin{figure}
\centering
	\begin{tikzpicture}
	\begin{axis}[width=\textwidth, height=0.4\textwidth,
	xlabel=No. of Client Requests,
	 ylabel=Avg. CPU Cycles (per req),
	symbolic x coords={50,100,200,500,1000,2000},xtick=data, scaled ticks=true, ymode=log]
	\addplot[color=red,mark=diamond*] coordinates {
		(50,27383363)
		(100,32701235)
		(200,46507388)
		(500,82517701)
		(1000,144236086)
		(2000,269164031)
	};
	\addplot[color=blue,mark=*] coordinates {
		(50,28793304)
		(100,34354428)
		(200,48329466)
		(500,84240983)
		(1000,148972852)
		(2000,277172132)
	};
	\addplot[color=black!30!green,mark=square*] coordinates {		
		(50,31496134)
		(100,39037295)
		(200,53796534)
		(500,97003119)
		(1000,175726777)
		(2000,308936637)
	};
	\addplot[color=black,mark=x] coordinates {
		(50,14555590)
		(100,14406795)
		(200,14126670)
		(500,14171124)
		(1000,14365540)
		(2000,14263590)
	};
	\end{axis}
\end{tikzpicture}
	\begin{tikzpicture}
	\vspace{-5mm}
	\begin{axis}[width=\textwidth, height=0.5\textwidth,
	xlabel=No. of Client Requests,
	 ylabel=Avg. Memory Util. (MB),
	 symbolic x coords={50,100,200,500,1000,2000},xtick=data
	]
	\addplot[color=red,mark=diamond*] coordinates {
		(50,30.978)
		(100,32.235)
		(200,32.9475)
		(500,35.825)
		(1000,39.93)
		(2000,45.359)
	};
	\addplot[color=blue,mark=*] coordinates {
		(50,31.408)
		(100,32.39)
		(200,33.375)
		(500,36.26)
		(1000,41.02)
		(2000,49.017)
	};
	\addplot[color=black!30!green,mark=square*] coordinates {
		(50,31.355)
		(100,32.25)
		(200,33.515)
		(500,36.15125)
		(1000,42.24625)
		(2000,51.55416667)
	};
	\addplot[color=black,mark=x] coordinates {
		(50,25.617)
		(100,25.683)
		(200,25.633)
		(500,25.683)
		(1000,26.167)
		(2000,26.25)
	};
	\end{axis}
\end{tikzpicture}
	\indent\begin{tikzpicture}
	\begin{axis}[width=\textwidth, height=0.48\textwidth,
	xlabel=No. of Client Requests,
	 ylabel=Avg. Response Time per req. (ms),
	 symbolic x coords={50,100,200,500,1000,2000},xtick=data,	
	  legend style={
			at={(0.5,-0.25)},
			anchor=north,
			legend columns=-1}
	]
	\addplot[color=red,mark=diamond*] coordinates {
		(50,14.544)
		(100,15.821)
		(200,17.752)
		(500,24.738)
		(1000,40.444)
		(2000,66.023)
	};	
	\addlegendentry{Async} 	
	\addplot[color=blue,mark=*] coordinates {
		(50,17.296)
		(100,17.331)
		(200,19.748)
		(500,27.2999)
		(1000,44.820)
		(2000,75.621)
	};
	\addlegendentry{Hybrid}
	\addplot[color=black!30!green,mark=square*] coordinates {
		(50,17.626)
		(100,19.082)
		(200,22.99009)
		(500,36.27429)
		(1000,60.8191)
		(2000,112.748)
	};
	\addlegendentry{Sync}
	\addplot[color=black,mark=x] coordinates {
		(50,11.3023)
		(100,11.351)
		(200,11.471)
		(500,11.4418)
		(1000,11.454)
		(2000,11.6563)
	};
	\addlegendentry{Baseline}
	\end{axis}
\end{tikzpicture}\vspace{-5mm}
	\caption{Performance impact analysis (using unoptimised monitors)}
	\label{fig:unopt_results}
\end{figure}
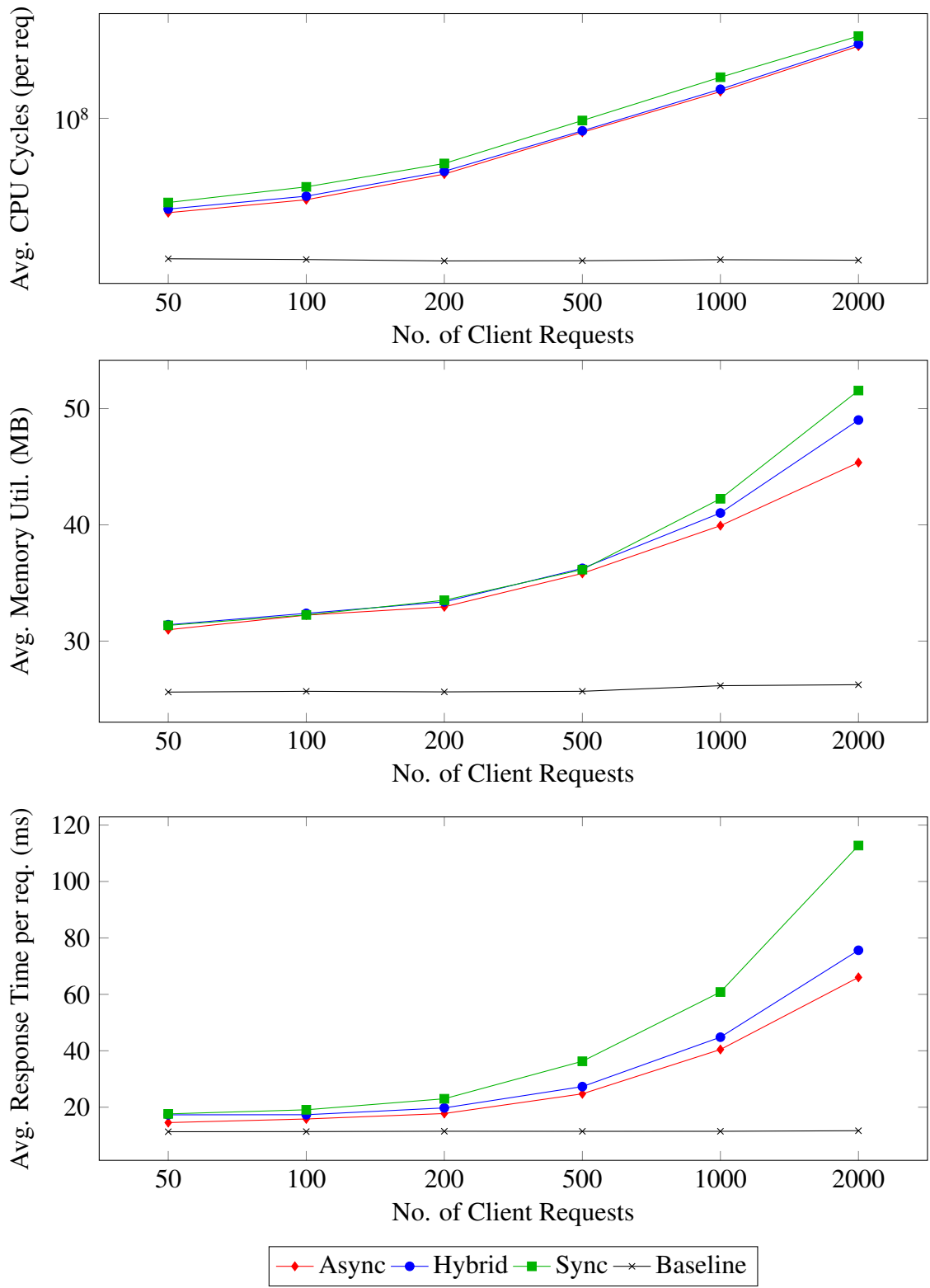

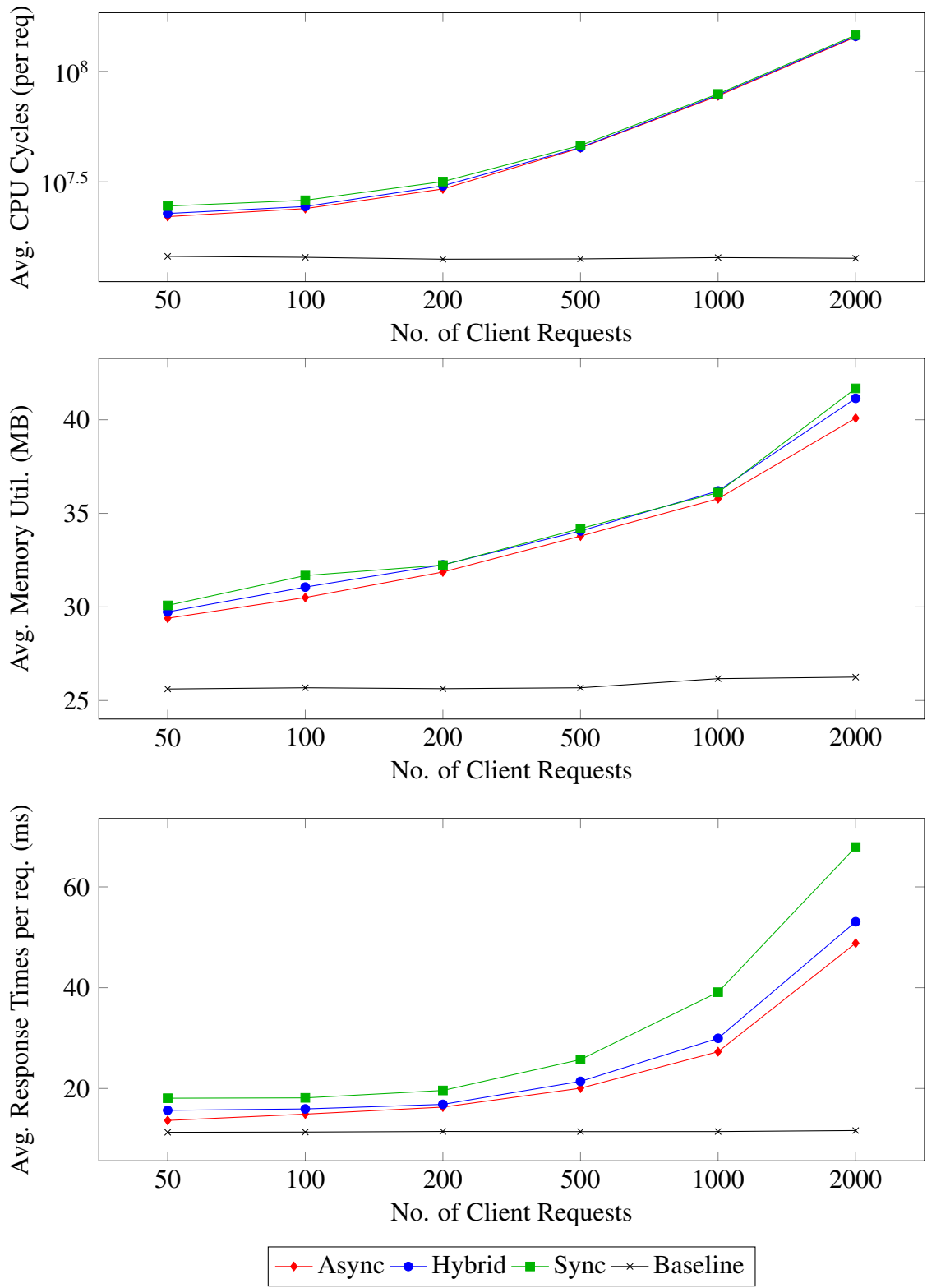
\begin{figure}
\centering
	\begin{tikzpicture}
	\begin{axis}[width=\textwidth, height=0.4\textwidth,
	xlabel=No. of Client Requests,
	 ylabel=Avg. CPU Cycles (per req),
	symbolic x coords={50,100,200,500,1000,2000},xtick=data, scaled ticks=true, ymode=log]
	\addplot[color=red,mark=diamond*] coordinates {
		(50,22022921)
		(100,23939539)
		(200,29374983)
		(500,45044217)
		(1000,77478758)
		(2000,143147839)
	};
	\addplot[color=blue,mark=*] coordinates {
		(50,22750326)
		(100,24468090)
		(200,30349522)
		(500,45283732)
		(1000,78177490)
		(2000,144178965)
	};
	\addplot[color=black!30!green,mark=square*] coordinates {
		(50,24580787)
		(100,26110060)
		(200,31720809)
		(500,46238708)
		(1000,79098419)
		(2000,146011176)
	};
	\addplot[color=black,mark=x] coordinates {
		(50,14555590)
		(100,14406795)
		(200,14126670)
		(500,14171124)
		(1000,14365540)
		(2000,14263590)
	};
	\end{axis}
\end{tikzpicture}
	\begin{tikzpicture}
	\vspace{-5mm}
	\begin{axis}[width=\textwidth, height=0.5\textwidth,
	xlabel=No. of Client Requests,
	 ylabel=Avg. Memory Util. (MB),
	 symbolic x coords={50,100,200,500,1000,2000},xtick=data,
	]
	\addplot[color=red,mark=diamond*] coordinates {
		(50,29.395)
		(100,30.5025)
		(200,31.8675)
		(500,33.785)
		(1000,35.7825)
		(2000,40.0875)
	};
	\addplot[color=blue,mark=*] coordinates {
		(50,29.733)
		(100,31.06)
		(200,32.2525)
		(500,34.06)
		(1000,36.1975)
		(2000,41.1456)
	};
	\addplot[color=black!30!green,mark=square*] coordinates {
		(50,30.081)
		(100,31.677)
		(200,32.23792)
		(500,34.19012)
		(1000,36.10876)
		(2000,41.6771)
	};
	\addplot[color=black,mark=x] coordinates {
		(50,25.617)
		(100,25.683)
		(200,25.633)
		(500,25.683)
		(1000,26.167)
		(2000,26.25)
	};
	\end{axis}
\end{tikzpicture}
	\indent\begin{tikzpicture}
	\begin{axis}[width=\textwidth, height=0.48\textwidth,
	xlabel=No. of Client Requests,
	 ylabel=Avg. Response Times per req. (ms),
	 symbolic x coords={50,100,200,500,1000,2000},xtick=data,
	 legend style={
			at={(0.5,-0.25)},
			anchor=north,
			legend columns=-1}	
	]
	\addplot[color=red,mark=diamond*] coordinates {
		(50,13.654)
		(100,14.909)
		(200,16.304)
		(500,20.055)
		(1000,27.293)
		(2000,48.836605)
	};
	\addlegendentry{Async}
	\addplot[color=blue,mark=*] coordinates {
		(50,15.664)
		(100,15.94043)
		(200,16.84583)
		(500,21.40865)
		(1000,29.9522)
		(2000,53.08243274)
	};
	\addlegendentry{Hybrid}
	\addplot[color=black!30!green,mark=square*] coordinates {
		(50,18.06556)
		(100,18.13647)
		(200,19.60522)
		(500,25.73741)
		(1000,39.10876)
		(2000,67.91071)
	};
	\addlegendentry{Sync}
	\addplot[color=black,mark=x] coordinates {
		(50,11.3023)
		(100,11.351)
		(200,11.471)
		(500,11.4418)
		(1000,11.454)
		(2000,11.6563)
	};
	\addlegendentry{Baseline}
	\end{axis}
\end{tikzpicture}\vspace{-5mm}
	\caption{Performance impact analysis (using statically optimised monitors)}
	\label{fig:sopt_results}
\end{figure}

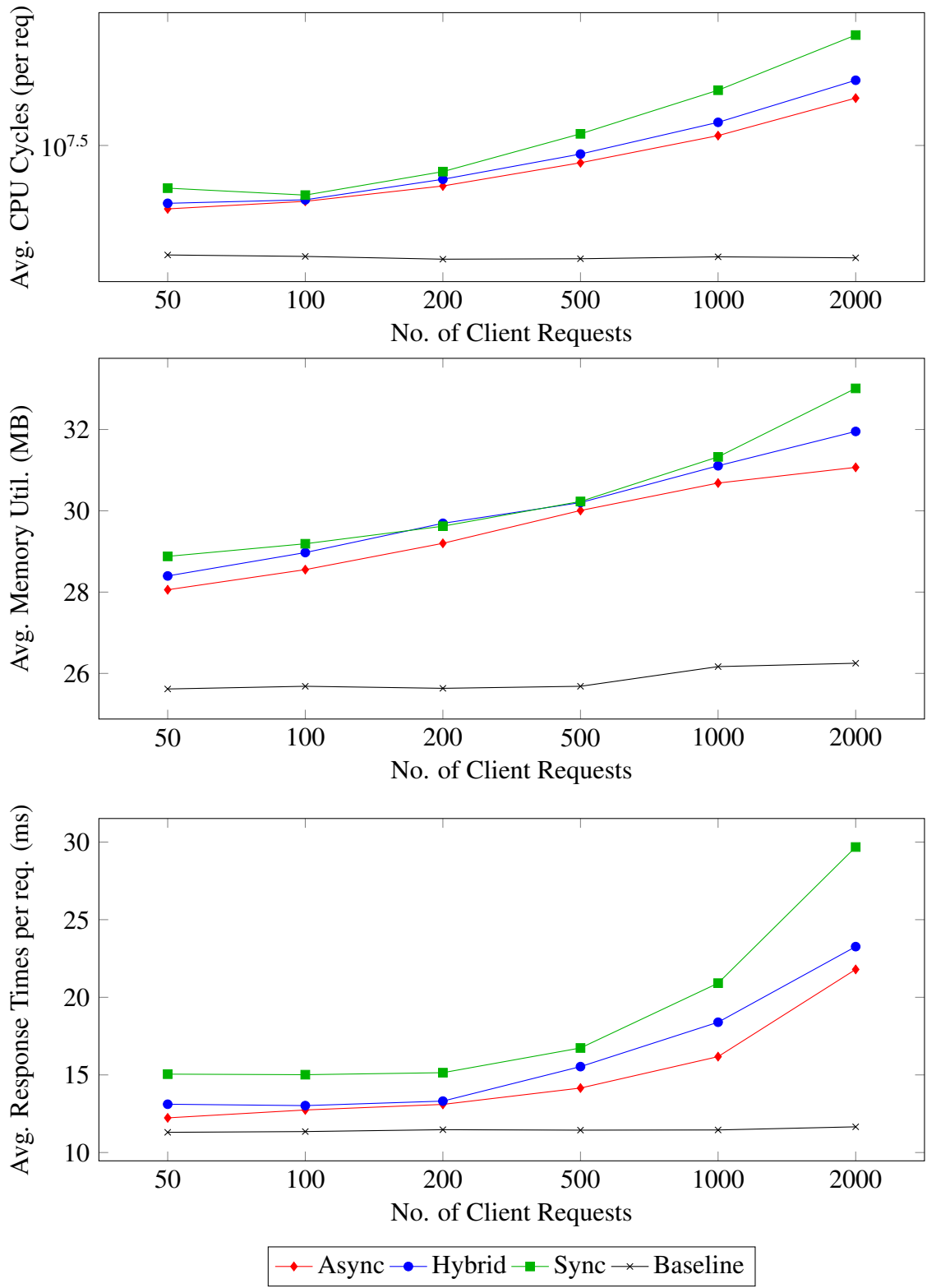
\begin{figure}
\centering
	\begin{tikzpicture}
	\begin{axis}[width=\textwidth, height=0.4\textwidth,
	xlabel=No. of Client Requests,
	 ylabel=Avg. CPU Cycles (per req),
	symbolic x coords={50,100,200,500,1000,2000},xtick=data, scaled ticks=true, ymode=log]
	\addplot[color=red,mark=diamond*] coordinates {
		(50,20174678)
		(100,21285896)
		(200,23732114)
		(500,27972045)
		(1000,33908989)
		(2000,44227348)
	};
	\addplot[color=blue,mark=*] coordinates {
		(50,20988211)
		(100,21515712)
		(200,24884714)
		(500,29754716)
		(1000,37257754)
		(2000,50190536)
	};
	\addplot[color=black!30!green,mark=square*] coordinates {
		(50,23368903)
		(100,22245771)
		(200,26271776)
		(500,34338960)
		(1000,46772984)
		(2000,69139508)
	};
	\addplot[color=black,mark=x] coordinates {
		(50,14555590)
		(100,14406795)
		(200,14126670)
		(500,14171124)
		(1000,14365540)
		(2000,14263590)
	};
	\end{axis}
\end{tikzpicture}
	\begin{tikzpicture}
	\vspace{-5mm}
	\begin{axis}[width=\textwidth, height=0.5\textwidth,
	xlabel=No. of Client Requests,
	 ylabel=Avg. Memory Util. (MB),
	 symbolic x coords={50,100,200,500,1000,2000},xtick=data
	]
	\addplot[color=red,mark=diamond*] coordinates {
		(50,28.0575)
		(100,28.5525)
		(200,29.2)
		(500,30.0075)
		(1000,30.6825)
		(2000,31.0687)
	};
	\addplot[color=blue,mark=*] coordinates {
		(50,28.3975)
		(100,28.9725)
		(200,29.6925)
		(500,30.2025)
		(1000,31.1075)
		(2000,31.9525)
	};
	\addplot[color=black!30!green,mark=square*] coordinates {
		(50,28.8791667)
		(100,29.1891667)
		(200,29.62154762)
		(500,30.23)
		(1000,31.32666667)
		(2000,33.01166667)
	};
	\addplot[color=black,mark=x] coordinates {
		(50,25.617)
		(100,25.683)
		(200,25.633)
		(500,25.683)
		(1000,26.167)
		(2000,26.25)
	};
	\end{axis}
\end{tikzpicture}
	\indent\begin{tikzpicture}
	\begin{axis}[width=\textwidth, height=0.48\textwidth,
	xlabel=No. of Client Requests,
	 ylabel=Avg. Response Times per req. (ms),
	 symbolic x coords={50,100,200,500,1000,2000},xtick=data,
	 legend style={
			at={(0.5,-0.25)},
			anchor=north,
			legend columns=-1}	
	]
	\addplot[color=red,mark=diamond*] coordinates {
		(50,12.2332)
		(100,12.741825)
		(200,13.100325)
		(500,14.153975)
		(1000,16.177675)
		(2000,21.795625)
	};	
	\addlegendentry{Async} 	
	\addplot[color=blue,mark=*] coordinates {
		(50,13.1115)
		(100,13.02285)
		(200,13.3128)
		(500,15.5342)
		(1000,18.39438)
		(2000,23.2638)
	};
	\addlegendentry{Hybrid}
	\addplot[color=black!30!green,mark=square*] coordinates {
		(50,15.04966667)
		(100,15.01426667)
		(200,15.14311667)
		(500,16.73514167)
		(1000,20.91005)
		(2000,29.68005833)
	};
	\addlegendentry{Sync}
	\addplot[color=black,mark=x] coordinates {
		(50,11.3023)
		(100,11.351)
		(200,11.471)
		(500,11.4418)
		(1000,11.454)
		(2000,11.6563)
	};
	\addlegendentry{Baseline}
	\end{axis}
\end{tikzpicture}\vspace{-5mm}
	\caption{Performance impact analysis (using fully optimised monitors)}
	\label{fig:s+dopt_results}
\end{figure}

From the graphs in \Cref{fig:unopt_results,fig:sopt_results,fig:s+dopt_results} we note that \emph{regardless of how the monitors are optimised} \cite{CasFraSaid15}, memory and CPU utilisation suffer from a marginal difference in which the hybrid monitoring overheads generally lie midway between synchronous monitoring $-$ which scored the worst; and asynchronous monitoring $-$ which gave the best performance results. A clearer, and more substantial difference is seen in the average time that the server takes to respond\footnote{Providing low average response times \cite{Holovaty:2009:DGD:1572516} is imperative in reactive systems (such as webservers) to make sure that the client is served as soon as possible.} to each client request when it is under a certain load.  
\figref{fig:unopt_results} shows that unoptimised monitors suffer from a substantial response difference of about 46ms when a synchronously monitored instance of Yaws is placed under the load of 2000 client requests, as compared to its asynchronous counterpart. By contrast hybrid monitoring provided timely detections by suffering only from a 9ms response overhead increase (for the same load) when compared to asynchronous monitoring.  

Furthermore, \Cref{fig:sopt_results,fig:s+dopt_results} demonstrate that even though the monitoring optimisations \cite{CasFraSaid15} help in reducing the overall overheads, synchronous instrumentation monitoring still fared the worst when compared to the rest. In fact this suffers from a 19ms difference (compared to asynchronous) when using statically optimised monitors to synchronously monitor a Yaws instance placed under the load of 2000 client requests, and a 9ms difference with dynamically optimised monitors. Contrasting this, the hybrid approach only suffers from a 4ms and 1.5ms response time difference for statically optimised and dynamically optimised monitors respectively.   

We therefore conclude that regardless of any optimisation used by the synthesised monitors, synchronous instrumentation monitoring incurs substantial overheads when compared to its asynchronous counterpart. By contrast, the hybrid instrumentation results show that timely detections can still be achieved with overheads that are higher yet closer to that of asynchronous monitoring.

The reason for this performance degradation is apparent from \figref{fig:sync_protocol} and \figref{fig:hybrid_protocol}, where one can notice that for every synchronous monitored event, the system execution incurs a responsiveness penalty waiting for the monitor to acknowledge back. A further penalty is also incurred since synchronous event reporting requires executing additional code (\eg to generate \emph{nonces}) when compared to its asynchronous counterpart. Hence hybrid monitoring performs better than synchronous instrumentation monitoring as the former performs less synchronisations. For the same reason, hybrid monitoring still imposes a slightly higher degree of overheads than asynchronous monitoring as it still needs to apply a small level of synchrony in order to achieve timely detections.

\section{Conclusion}
In this chapter we have analysed the design space for introducing synchrony on top of an inherently asynchronous platform. The primary objective of this investigation was to identify a synchronisation technique that allows for a tighter monitoring control 
to be achieved \emph{efficiently}. This contributes towards developing an effective, yet efficient RA framework. 

We started by devising a handshaking protocol for achieving synchronous event monitoring on actor-based Erlang systems which required refining an AOP framework \cite{CgvFypAOP} for Erlang systems to instrument the protocol in existing systems. Using this handshaking protocol we introduced a coarse-grained synchronous instrumentation monitoring approach which synchronised for every reported event. Following this we reduced the level of synchronisation by limiting it to critical actions \ie system actions that can directly contribute to a violation. This lead to creating a hybrid monitoring approach capable of switching between synchronous and asynchronous event reporting, while still ensuring timely detections when required. This however required us to extend the \detecterGen's specification language.

After integrating the synchronous and hybrid monitoring approaches in \detecterGen\footnote{A \emph{stable} release of this version called \detecter\ is also accessible from \texttt{https://bitbucket.org/casian/detecter2.0}}, we carried out a systematic assessment \wrt Yaws, regarding the relative overheads incurred by different instrumentation techniques within an actor setting. The results of this impact assessment show that the hybrid technique yielded response times that are lower than those of synchronous instrumentation approach.  

Therefore, the conducted impact analysis enables us to conclude that since synchrony is not natively supported by asynchronous, actor-based systems, 
this induces inevitable synchronisation overheads. 
However, our analysis also allows us to conclude that although the incremental synchronisation methodology used in our hybrid technique still introduces a certain degree of overheads, these can be minimised and kept to a \emph{feasible} standard if synchronisation is utilised on a \emph{by-need basis}; \ie only when it is strictly required (\eg to timely achieve detections or to effectively apply adaptations).

In the next chapter we therefore build upon the incremental synchronisation methodology that we have identified by this impact assessment, in order to implement an efficient yet effective Runtime Adaptation framework for actor systems.

\chapter{Language Design, Implementation and Formalisation}
\label{chp:runtime-adaptation}
%
%

In this chapter we present the design and implementation details about how we develop our runtime adaptation framework on top of the \detecterGen\ RV tool. Based on the implementation constraints we then develop a formal operational model which provides a higher level description of how our RA scripts behave at runtime. We divide this chapter into two parts: in the first part (\secref{sec:syn-asyn-adaptations} to \ref{sec:yaws-ra-casestudy}) we analyse the implementation challenges, while in the in the second part (\secref{sec:ra-model}) we explain how we developed the formal model for our RA scripts. 

More specifically, in \secref{sec:syn-asyn-adaptations} we identify a number of adaptation actions that should be applicable on \emph{any} arbitrary actor system, regardless of the design and coding practices used in its implementation. We also articulate the constraints that the actor model imposes which makes it more challenging to implement such adaptation actions. We therefore provide a preliminary investigation of how we can overcome these challenges. 

In \secref{sec:lang-design} we present our design decisions for integrating the identified adaptations within the logic used by \detecterGen, thereby enabling users to specify runtime adaptation monitors for mitigating property violations. This is followed by \secref{sec:lang-implementation} in which we expand on the preliminary investigation given in \secref{sec:syn-asyn-adaptations} and provide an overview of how we implement the synchronisation protocol required by our runtime adaptation framework. We conclude the first part by showing how one of the RV scripts written for Yaws (used in the evaluation of \Cref{chp:syn-asyn}), can easily be converted into an RA script for our new framework. 

Finally in \secref{sec:ra-model} we focus on developing a formal model that allows for predicting the runtime behaviour of an RA script without actually running it. This allows us to analyse the behaviour of our scripts without having to deal with the complexities of the implementation, thereby enabling us to further understand the possible errors that erroneous RA scripts may introduce.  

\section{Identifying Adaptation Actions within the Constraints of Actor Systems}
\label{sec:syn-asyn-adaptations}
Actor design practices such as the fail-fast design pattern and systems structured using supervision-trees, permit for the creation of \emph{self-adaptive} actor systems whereby supervisor actors are purposely developed to intervene and mitigate cases when an actor crashes abnormally. As misbehaviour does not necessarily cause an actor to crash, these design strategies are limited in the type of misbehaviour that they can detect (\ie detections only happen when an actor crashes). One way how to to overcome this limitation is to employ an RV monitor which is able to deliberately interfere by crashing certain system actors whenever it detects erroneous behaviour. This would in turn trigger the system's built-in self-adaptive functionality which 
mitigates the crashed actors.

\begin{example} \label{ex:saa:1:fig}
Assume a simple \emph{self-adaptive} actor system illustrated in \figref{fig:sa-sys} (below). This system consists in four components, namely $A$, $B$ and $C$ which are the actual worker components, along with component $S$ which is their \emph{supervisor}. The sole purpose of this supervisor is to apply the necessary mitigation when either one (or more) of the worker components crash.

\begin{figure}[ht!]
	\centering
	\includegraphics[width=0.8\textwidth]{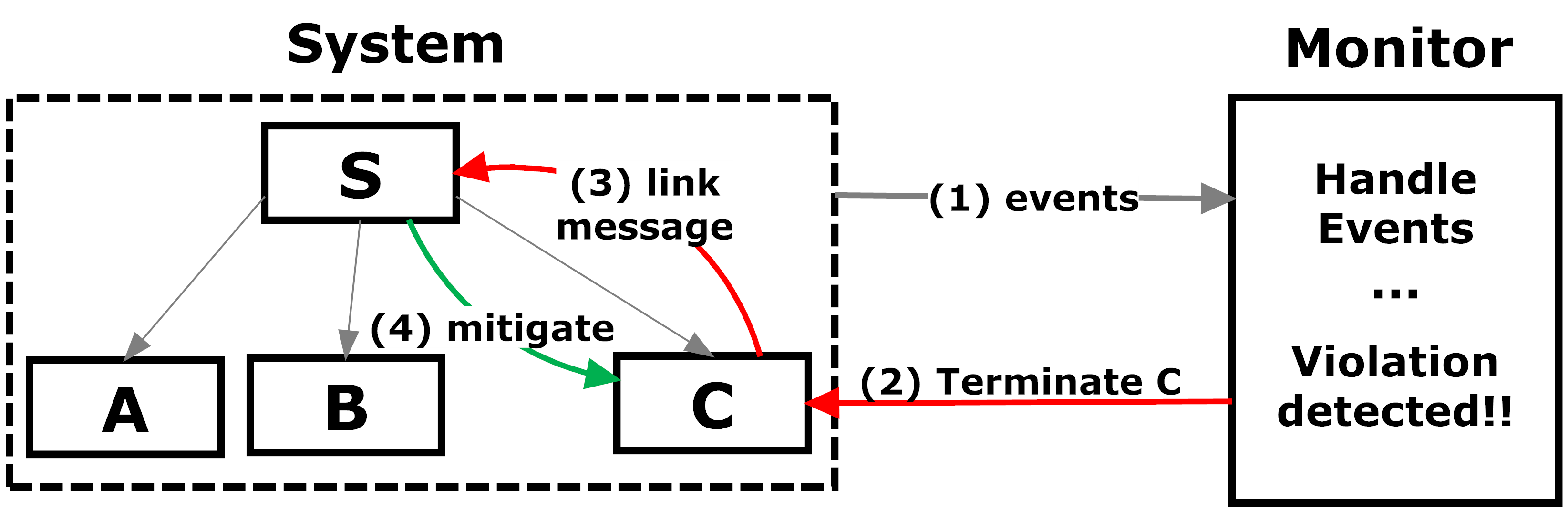}
	\caption[Triggering the self-adaptive functionality of a system.]{Triggering the self-adaptive functionality of a system through Runtime Monitors.}
	\label{fig:sa-sys}
\end{figure}

As explained by \figref{fig:sa-sys}, this simple RA monitor $(1)$ receives and analyses the system events, and upon detecting a violation it $(2)$ deliberately terminates one of the system's components (\ie component $C$) instead of flagging the violation. Upon crashing, $(3)$ this component sends a link message to its supervisor, $S$, which in turn $(4)$ applies the necessary mitigation. \bqed
\end{example}

Although this straightforward RA mechanism is already quite useful and effective, it relies on the assumption that the monitored actor system already contains a certain degree of self-adaptive capabilities. In cases where a monitored actor system lacks (or has limited) self-adaptive functionality, killing an actor would not trigger any type of mitigation mechanism. At best this could be used to permanently switch-off redundant or useless parts of the system such as orphaned actors \cite{linux-book} \ie actors that are no longer in use but for some reason did not terminate. At worst, however, killing actors in a non-adaptive system might cause it to become unresponsive. 

We must therefore look into other adaptation techniques that could enable us to mitigate erroneous actors though high-level actor-manipulation techniques, that can be applied regardless of the design and implementation of the system under scrutiny. In other words our adaptations should be applicable to \emph{any} system actor irrespective of the code that it executes, and of the way that it is linked with other actors. However, implementing such complex adaptations which are agnostic of the actor's code, is quite challenging due to certain restrictions imposed by the actor model and the host language (in our case Erlang).  

\subsection{Identifying ideal Runtime Adaptation Actions} \label{sec:syn-asy-adapt-classes}
An RA framework for actor systems should provide generic adaptations that allow the specifier to manipulate a number of \emph{architectural aspects} pertaining to actor systems. To identify what these aspects really are, we recall the basics of the actor model (presented in \secref{sec:erlang}), which we summarise in the following three points:
\begin{enumerate}[label=$(\roman*)$]\itemsep0em \setstretch{1.2}
	\item  An actor is a concurrent \emph{process}.
	\item  An actor has a \emph{mailbox} for \emph{asynchronous interaction} (through message passing).
	\item  Actors can be \emph{structured} (\ie related through \emph{linking}).
\end{enumerate}

As stated in $(i)$, since actors are essentially processes and given that we want to keep our mitigation techniques as generic as possible, the only two types of generic process manipulations that can be applied upon a black-box process (\ie without knowing what it executes) are $(a)$ \emph{Process Terminations} and $(b)$ \emph{Process Restarts}.
\begin{description}\itemsep0em 
	\item[$(a)$ \emph{Process Terminations}:] These operations should be carefully used to \emph{terminate non-crucial} misbehaving actors, such that the actor system is able to carry on free of the erroneous actors, even though some of its functionality might be unavailable. 
\end{description}

\begin{example} \label{ex:saa:1}
	Recall the system illustrated in \figref{fig:sys}, for which we have already formalised RV \Propref{prop:intro:1}, in terms of \detecterGen's logic, as formula \eqref{intro:1} in \exref{ex:background} (restated below). 
	{  	\setlength{\abovedisplayskip}{5pt}
\setlength{\belowdisplayskip}{5pt}
\setlength{\abovedisplayshortskip}{5pt}
\setlength{\belowdisplayshortskip}{5pt}\begin{align*}     
    \hV \;\deftxt\;\;& \mmax{\,\hVarY}{\mnec{\mrecv{i}{\tup{\eatom{inc},x,y}}}
        \begin{mlbrace}(\,\mnec{\msendD{j}{y}{\tup{\eatom{res},\smash{x+1}}}}\,\hVarY)   \; \mand \; (\,\mnec{\msendD{\_\,}{y}{\eatom{err}}}\,\mfls) \end{mlbrace}} \qquad \eqref{intro:1}
  \end{align*}}
	\noindent Based on $(a)$ we reformulate \eqref{intro:1} as \eqref{saa:1} whereby the script now includes a the \textsf{kill} adaptation which terminates the actor that sends the error message \textsf{err} to the connected client. Note that we now map this erroneous actor to term variable $z$ so that we are able to apply the \textsf{kill} adaptation upon it. 
	  {  	\setlength{\abovedisplayskip}{5pt}
\setlength{\belowdisplayskip}{5pt}
\setlength{\abovedisplayshortskip}{5pt}
\setlength{\belowdisplayshortskip}{5pt}
	  \begin{align} \label{saa:1}    
	    \mmax{\,\hVarY}{\mnec{\mrecv{i}{\tup{\eatom{inc},x,y}}}
	        \begin{mlbrace}(\,\mnec{\msendD{j}{y}{\tup{\eatom{res},\smash{x+1}}}}\,\hVarY)   \; \mand \; (\,\mnec{\msendD{z\,}{y}{\eatom{err}}}\,\mkill{z}\,\hVarY) \end{mlbrace}} 
	  \end{align}}
	 \noindent As in our example system (given in \figref{fig:sys}) we do not assume any sort of built-in self-adaptive functionality (\ie no linking, trapping, supervisors, \etc), in formula \eqref{saa:1} we eliminate future occurrences of this error from being made by the same erroneous actor, by permanently switching off this actor. Although this termination restricts the overall functionality of our system, we still retain the rest of the system's functionality as the other concurrent actors are not effected by the applied adaptation. \bqed   
\end{example}

\begin{description}\itemsep0em 
	\item[$(b)$ \emph{Process Restarts}:] These actions should be used to \emph{restart} more crucial components in an attempt to remove/reduce any side-effects (that may cause further errors) incurred due to the detected misbehaviour, by reinitialising them afresh.
\end{description}
\begin{example} \label{ex:saa:2} For the example system in \figref{fig:sys} we can now extend formula \eqref{intro:1} (restated in \exref{ex:saa:1}) with the \textsf{restart} adaptation in \eqref{saa:3}.  
{  	\setlength{\abovedisplayskip}{5pt}
\setlength{\belowdisplayskip}{5pt}
\setlength{\abovedisplayshortskip}{5pt}
\setlength{\belowdisplayshortskip}{5pt}
  \begin{align}\label{saa:3}  \setstretch{1.1}   
    \mmax{\,\hVarY}{\mnec{\mrecv{i}{\tup{\eatom{inc},x,y}}}
	        \begin{mlbrace}(\,\mnec{\msendD{j}{y}{\tup{\eatom{res},\smash{x+1}}}}\,\hVarY)   \; \mand \; (\,\mnec{\msendD{z\,}{y}{\eatom{err}}}\,\mrestart{i,z}\mtru) \end{mlbrace}} 
  \end{align} }  
\noindent We use the \textsf{restart} adaptation to reinitialise both the common-interface actor $i$, and also the actor mapped to term variable $z$, in order to mitigate the invalid behaviour which lead to producing error \textsf{err}. In this way we still retain the full functionality of the system while we reduced the possible side-effects inflicted on the mitigated actors, by restarting them from their point of initialization. \bqed
\end{example}

Furthermore from architectural aspect $(ii)$, we know that actors are not just concurrent process, but they are also \emph{interactive processes}. In fact actors are able to interact asynchronously by depositing messages in the private mailbox of other actors. Due to this method of interaction we identify another generic adaptation, namely:
\begin{description}\itemsep0em 
	\item[$(c)$ \emph{Message Interception}:] These adaptations should be used to intercept and manipulate messages that might have been erroneously exchanged due to the detected violation, thereby preventing them from causing further problems.
\end{description}
\begin{example} \label{ex:saa:3} Assuming that we can use adaptation \textsf{prg} to purge all the messages in an actor's mailbox, we now develop \eqref{saa:4} as an extension of \eqref{intro:1}. 
  {  	\setlength{\abovedisplayskip}{5pt}
\setlength{\belowdisplayskip}{5pt}
\setlength{\abovedisplayshortskip}{5pt}
\setlength{\belowdisplayshortskip}{5pt}
\begin{align}\label{saa:4}     
    \mmax{\,\hVarY}{\mnec{\mrecv{i}{\tup{\eatom{inc},x,y}}}
        \begin{mlbrace}(\,\mnec{\msendD{j}{y}{\tup{\eatom{res},\smash{x+1}}}}\,\hVarY)   \; \mand \; (\,\mnec{\msendD{z\,}{y}{\eatom{err}}}\,\mclearMailbox{z}\hVarY)\end{mlbrace}}
  \end{align} }
  In \eqref{saa:4} we use the \textsf{prg} adaptation to flush the mailbox contents of the actor mapped to $z$, with the aim of removing other conflicting messages that the erroneous actor (mapped to $z$) might have received prior to sending the error message to the client. \exqed
\end{example}

Furthermore, as by $(iii)$ we know that these communicating actors can also be structured and related, we identify yet another generic adaptation that allows us to manipulate this aspect of actor systems:
\begin{description}\itemsep0em 
	\item[$(c)$ \emph{Process Restructuring}:] There are various ways how actor systems can be structured including: how the actors are \emph{linked} together \eg actor $A$ is linked to actor $B$ but not to C; the type of linking between actors \eg unidirectional or bidirectional, trapped or untrapped; and the way actors are identified \eg some actors are registered with a unique name while others are not. Restructuring should therefore allow us to manipulate these structural aspects.
\end{description}
\begin{example} \label{ex:saa:4} Recall from \secref{sec:erlang} that when a process dies, every process linked to it dies as well (unless trapping is used). Assuming that the common-interface (from \figref{fig:sys}) is linked to the decrementor and incrementor components, in \eqref{saa:5} we use the \textsf{unlink} restructuring adaptation to remove the link between the interface, $i$, and the actor mapped to $z$.
  {  	\setlength{\abovedisplayskip}{5pt}
\setlength{\belowdisplayskip}{5pt}
\setlength{\abovedisplayshortskip}{5pt}
\setlength{\belowdisplayshortskip}{5pt}
\begin{align}\label{saa:5}     
    \mmax{\,\hVarY}{\mnec{\mrecv{i}{\tup{\eatom{inc},x,y}}}
        \begin{mlbrace}(\,\mnec{\msendD{j}{y}{\tup{\eatom{res},\smash{x+1}}}}\,\hVarY)   \; \mand \; (\,\mnec{\msendD{z\,}{y}{\eatom{err}}}\,\munlink{i,z}\hVarY)\end{mlbrace}}
  \end{align} }
In this way, if the actor mapped to $z$ crashes as a result of producing error \textsf{err}, at least the other linked components can still operate. Hence rather than losing all the system's functionality, we allow for the system to remain partially operational. \exqed
\end{example}

The identified adaptations $(a)$, $(b)$, $(c)$ and $(d)$ are essentially \emph{categories} of adaptations rather than actual adaptation actions. We also conjecture that \emph{instances} of these categories can be implemented in various ways, but ultimately provide the same overall effect \eg one instance of the restructuring category is used to \emph{add links} between actors, while another instance is used to \emph{remove links} instead; these reside in the same category as they ultimately both manipulate the structure of an actor system, even if they do it in a different way.

\begin{remark}\label{remark:silent-kill} The overall effects of different categories can also be combined to obtain even complex adaptations. A case in point is the \emph{silent-kill} adaptation which despite being an instance of category $(a)$ (\ie process terminations), it is also able to manipulate the structural aspects of an actor system. In fact whenever the \emph{silent-kill} is applied to an actor $i$, it removes any link related to $i$ prior to terminating it. It is therefore able to suppress the system's built-in self-adaptive functionality (if any) by removing the links that alert the other actors about the terminated actor, and thus prevent the system from attempting to apply its own mitigation. \exqed
\end{remark}

\begin{example} \label{ex:adapt}
With the identified adaptation categories we are now in a position to formally specify RA \propref{prop:intro:2} (see \exref{ex:intro}) as an extension of RV \propref{prop:intro:1} which we have already formalised as formula \eqref{intro:1} in \exref{ex:background}. We thus obtain formula \eqref{intro:2}: 
{  	\setlength{\abovedisplayskip}{5pt}
\setlength{\belowdisplayskip}{5pt}
\setlength{\abovedisplayshortskip}{5pt}
\setlength{\belowdisplayshortskip}{5pt}
\begin{align}
    \!\!\!\label{intro:2}
   & \mmax{\,\hVarY}{\mnec{\mrecv{i}{\tup{\eatom{inc},x,y}}}\begin{mlbrace}  (\,\mnec{\msendD{j}{y}{\tup{\eatom{res},\smash{x+1}}}}\,\hVarY)   
        \;\mand \; (\,\mnec{\msendD{z\,}{y}{\eatom{err}}}\,\mrestart{i}\,\mclearMailbox{z}\,\hVarY) 
       \end{mlbrace}}
  \end{align}}  
The specifier presumes that the error (which may arise after a number of correct interactions) is caused by the interface actor $i$ (as shown in \figref{fig:sys}, where an \eatom{inc} request is erroneously forwarded to the decrementor actor $k$) ---  the specifier may, for instance, have prior knowledge that actor $i$ is a newly-installed, untested component.  The monitor thus restarts actor $i$ by using adaptation \mrestart{i}, and empties the mailbox of the backend server through adaptation \mclearMailbox{z}, as this may contain more erroneously forwarded messages. Note that the actor to be purged is determined at runtime, where term variable $z$ is bound to identifier $k$ from the previous action $\mnec{\msendD{z\,}{y}{\eatom{err}}}$).  Importantly, note that in the above execution (where $k$ is the actor sending the error message), actor $j$'s execution is \emph{not affected} by any adaptation action taken. \exqed
\end{example}

\subsection{The Implementation Constraints of Adaptation Actions} \label{ra:impl-issues}
\begin{figure}[ht!]
	\centering
	\includegraphics[width=0.85\textwidth]{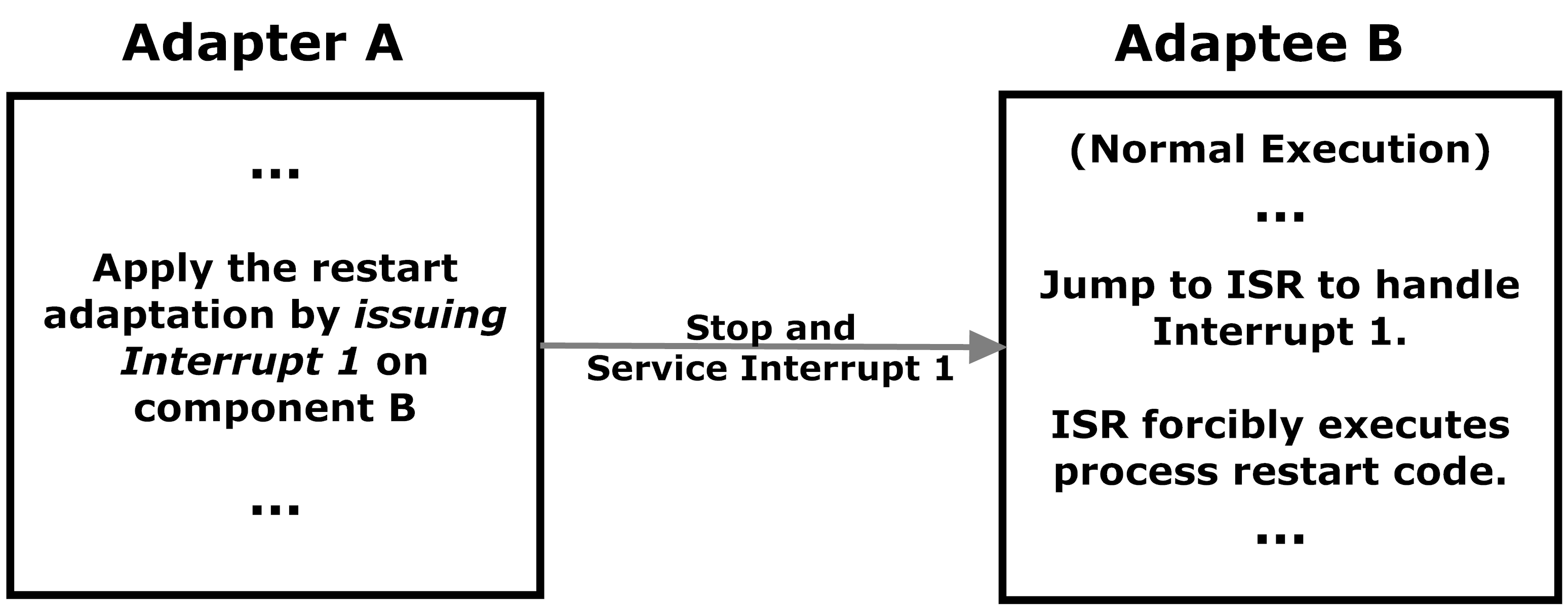}
	\caption{Applying adaptations using interrupts.}
	\label{fig:adapt-ISR}
\end{figure}

Although some adaptations are supported by the host language, other adaptations need to be implemented. Implementing adaptations pertaining to some of the identified categories is not always straight forward, as the actor-model and the host language restrict the way that actors can externally effect other actors. In the case of Erlang, as illustrated in \figref{fig:adapt-ISR}, an actor $A$ is \emph{not allowed} to externally interrupt the operation of another actor $B$, thereby forcing it to execute interrupt service routine (ISR) which invokes the required adaptation function. Although interrupts and ISRs are commonly used in concurrent and multithreaded systems \cite{linux-book}, they are strictly abolished by the actor-model and Erlang.

\begin{figure}[ht!]
	\centering
	\includegraphics[width=0.85\textwidth]{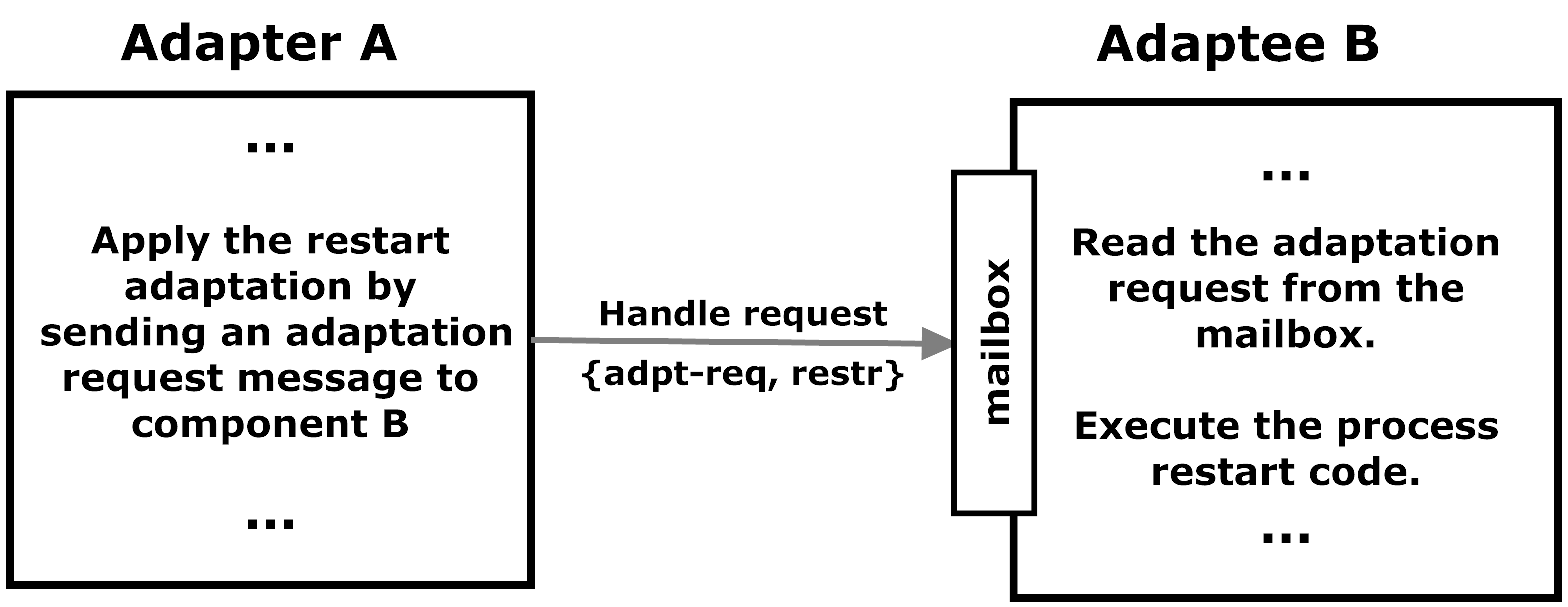}
	\caption{Applying adaptations encoded as messages.}
	\label{fig:adapt-messages}
\end{figure}

Instead, as shown in \figref{fig:adapt-messages}, in the actor-model the adapter actor $A$ must encode and adaptation request as a message and deposit it in the mailbox of adaptee $B$, requesting it to execute the required adaptation function. The recipient actor, $B$, must then read the message from its private mailbox, interpret it and willingly execute the required function (it might choose not to). 

The only type of external actions that are natively supported by the Erlang OTP libraries \cite{ErlangOTP} are actor \emph{killing} and \emph{linking}. This allows us to augment \detecterGen's RV monitors with flavours of adaptation categories $(a)$ and $(d)$ by enabling them to execute the respective OTP functions over the necessary system actors. \bigskip

\begin{remark}	\label{remark:1} Although these natively supported adaptations can also be used with asynchronous monitoring, recall that this suffers from \emph{late detections}. This means that even though we are able to apply these adaptations, it does not necessarily mean that they can be \emph{effectively} applied to prevent or properly mitigate any possible side-effects. For instance, recall that in formula \eqref{saa:1} we use the \textsf{kill} operation to terminate the actor that sends the \textsf{err} message to the client. However, if this actor is not terminated in a timely manner it would still be able to progress and perform further operations, that may lead to more serious consequences, prior to being killed. Similarly, in the case of formula \eqref{saa:5}, if the interface actor $i$ and the erroneous actor mapped to $z$ are not unlinked on-time, actor $z$ might progress and crash (thereby causing the linked actors to crash as well) before the monitor manages to remove the links. \exqed 
\end{remark} \bigskip

The Erlang OTP libraries, however, still do not provide native mechanisms for restarting actors (\ie implementing $(b$)) or for reading and modifying the mailbox contents of the of another actor (\ie to implement $(c)$; in fact this is strictly abolished). Hence as outlined in \figref{fig:adapt-messages}, to implement these adaptations we have to somehow `\textsl{trick}' the adaptee into providing the monitor with enough control to willingly perform these more complex operations.

It turns our that \emph{synchronisation} not only enables us to address the issue of applying adaptations effectively, outlined in Remark~\ref{remark:1} (as explored in \Cref{chp:syn-asyn}), but also provides the monitor with a tighter control over the system actors. This allows for the implementation of more complex adaptations such as those pertaining to categories $(b)$ and $(c)$. The general idea for implementing these adaptations is to use synchronisation as a form of \emph{local anesthetic} which places the necessary system actors in a state of \emph{controlled dormancy} that allows the monitor to operate upon them. 

\begin{figure}[ht!]
	\centering
	\includegraphics[width=0.85\textwidth]{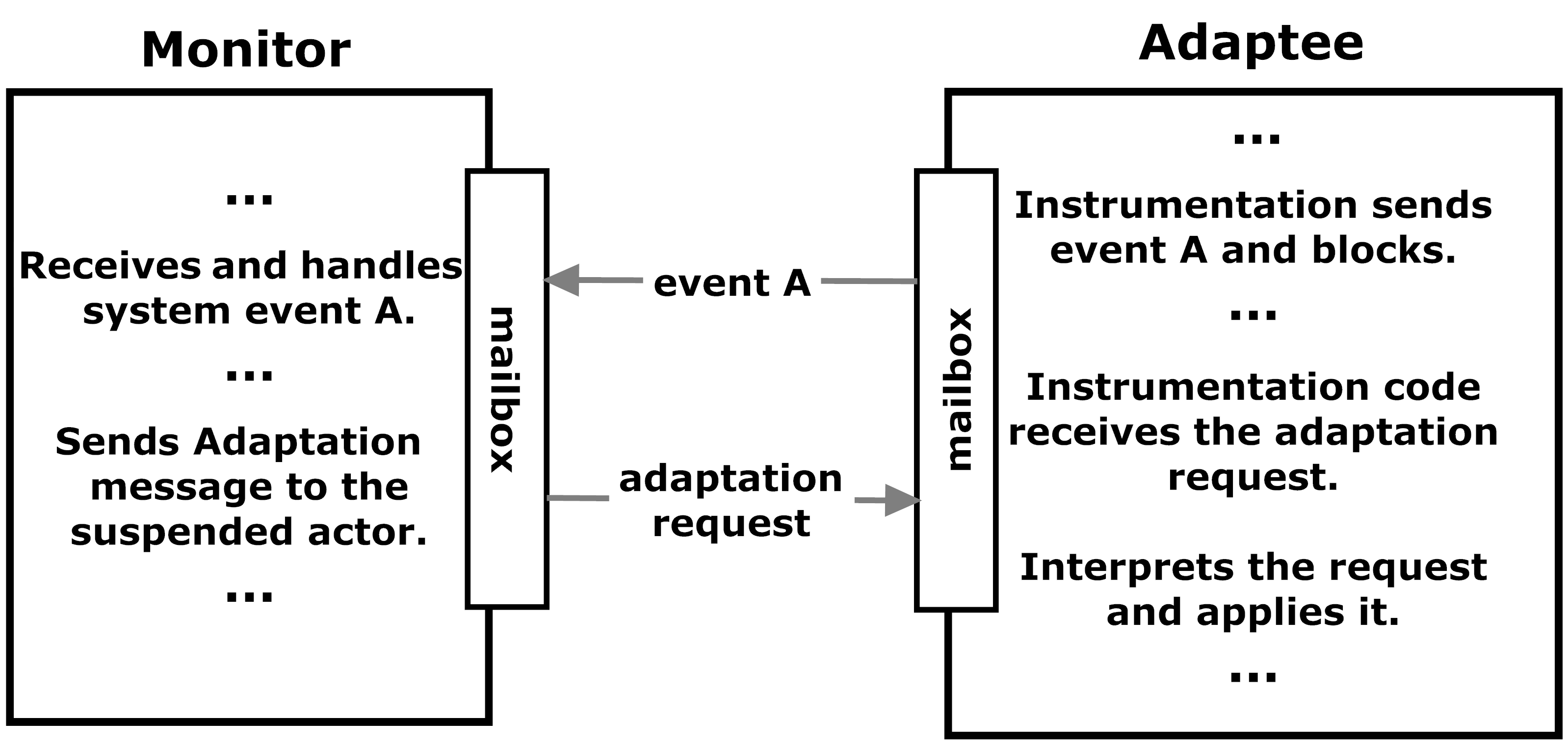}
	\caption{A conceptual protocol for applying synchronous adaptations.}
	\label{fig:adapt-msgs-det}
\end{figure}

More specifically, as illustrated in \figref{fig:adapt-msgs-det} whenever a system actor is \emph{blocked} by the instrumented synchronisation protocol, this does not only wait for the monitor's acknowledgement message, but also waits for \emph{adaptation requests} encoded as messages. Upon receiving an adaptation request, \eg \textsf{restr}, the code instrumented in the system actor can then analyse the request and carry out the necessary procedure for applying the requested pre-instrumented adaptations. As the adaptations are executed within the system actor's own thread of execution, these adaptation functions would also have full access to the actor's private mailbox and state (\ie its process dictionary), thereby allowing for more intrusive operations, such as message interception, to be performed. The required adaptations can therefore be implemented as regular functions which are then instrumented in the system such that they can be invoked by the instrumented synchronisation mechanism.

However, if an adaptation message is sent to an \emph{active} system component (\ie an actor which was not blocked by our instrumentation), the actor might suffer the following consequences:
\begin{itemize}\itemsep0em \setstretch{1.2}
	\item  it might not even read the request message, hence the adaptation is \emph{not conveyed};
	\item  it might read the message while it is performing other operations and \emph{ignore} it, as it would not know what it means;
	\item  the message might be read during the actor's normal execution and cause \emph{conflicts} leading to a \emph{crash}.
\end{itemize}
It is therefore imperative to make sure that the respective actors are \emph{suspended by our instrumentation prior to applying these adaptations}.

Hence, due to these implementation constraints, we classify the implementable adaptations into the following two classes:
\begin{description}\itemsep0em
	\item[$\bullet$\; Asynchronous Adaptations:] these can be administered on the respective actors while they are executing, typically these include natively supported actions.
	\item[$\bullet$\; Synchronous Adaptations:] the actors requiring these adaptations must be suspended by our instrumentation prior to applying them.	
\end{description}
Based on the adaptation categories identified in \secref{sec:syn-asy-adapt-classes} and the adaptation classes presented above we create Table~\ref{tab:syn-asyn-adapts}. This presents a matrix of adaptations denoting the categories that can be applied synchronously, asynchronously or both. In general, asynchronous adaptations \eg actor linking and killing, can also be applied synchronously.\medskip

\begin{table}[ht!]
	\centering
	\small
	\noindent\begin{tabular}{c|c|c|c|c|c|}
		\cline{2-6}
			 & \multicolumn{5}{c|}{Categories} \\ 
		\cline{2-6}
			 & & 	$(a)$ Terminations & $(b)$ Restarts & $(c)$ Interceptions  & $(d)$ Restructuring\\
		\hline
			\multicolumn{1}{|c|}{\multirow{2}{*}{\begin{sideways}Classes\,\,\end{sideways}}} & Synchronous & $\accept$ & $\accept$ & $\accept$ & $\accept$	\\[2mm]
		\cline{2-6}
			\multicolumn{1}{|c|}{} & Asynchronous & $\accept$ &  & & $\accept$\\[2mm]
		\hline
	\end{tabular}
	\caption{A classification for Synchronous and Asynchronous Adaptation Categories.}
	\label{tab:syn-asyn-adapts}
\end{table}
\vspace{-5mm}

\section{Designing the Runtime Adaptation Mechanism}
\label{sec:lang-design}
As synchronous adaptations require suspending the adaptee prior to applying the adaption upon it, we employ an incremental synchronisation mechanism for \emph{gradually} suspending \emph{only} the actor executions of interest as the property is being monitored for, \ie \emph{without} suspending any of the other actors. We opt for incremental synchronisation based on the results we obtained from our impact assessment carried out in \Cref{chp:syn-asyn}. This demonstrated that incremental synchronisation is able to keep synchronisation overheads under control by applying synchrony on a \emph{by-need} basis, thereby also allowing for feasible runtime adaptation. Hence, inspired from synchronous necessities (\ie $\msnec{\patE}$) in our hybrid monitoring scripts (see \secref{sec:hybrid-logic-extensions}) we 
extend necessity formulas with a \emph{synchronisation modality}, $\mattrNec{\patE}{\attr}{
}{\hV}$, where \attr\ ranges over either \agg\ (blocking): stating that the subject of the action (\ie the actor that committed the action) is suspended if its pattern \patE is matched; or \norm\ (asynchronous): stating that the action subject is allowed to continue executing asynchronously, even when the pattern \patE\ is matched.   

We also recall that if the necessity formula \mnec{\patE}{\hV} mismatches with a trace action, its observation terminates due to the \emph{trivial satisfaction} of the formula. 
In our case this also means that the synchronous adaptations contained the continuation \hV\ are never administered. Hence, any actor executions that were previously blocked so as to make these synchronous adaptations applicable are no longer required to remain blocked. We thus provide a mechanism for \emph{releasing} the actors blocked so far by further extending necessity formulas with a list of actor references, \textsf{RL}. This list denotes the (blocked) actors to be released in case the necessity pattern \patE\ mismatches, $\mattrNec{\patE}{\attr}{\mathsf{RL}}{\hV}$.  

\begin{example} Recall formula \eqref{intro:2} from \exref{ex:adapt} (restated below).
	{  	\setlength{\abovedisplayskip}{5pt}
\setlength{\belowdisplayskip}{5pt}
\setlength{\abovedisplayshortskip}{5pt}
\setlength{\belowdisplayshortskip}{5pt}\begin{align*}
    \mmax{\,\hVarY}{\mnec{\mrecv{i}{\tup{\eatom{inc},x,y}}}\begin{mlbrace}  (\,\mnec{\msendD{j}{y}{\tup{\eatom{res},\smash{x+1}}}}\,\hVarY)   
        \;\mand \; (\,\mnec{\msendD{z\,}{y}{\eatom{err}}}\,\mrestart{i}\,\mclearMailbox{z}\,\hVarY) 
       \end{mlbrace}}  \quad \eqref{intro:2}
  	\end{align*}}
  	We augment the necessity operations in \eqref{intro:2} with the required synchronisation modalities and release lists so to obtain \eqref{intro:2:sync}.   
  	{ \setlength{\abovedisplayskip}{5pt}
\setlength{\belowdisplayskip}{5pt}
\setlength{\abovedisplayshortskip}{5pt}
\setlength{\belowdisplayshortskip}{5pt}
\begin{align}\label{intro:2:sync}
\!\!\!\mmax{\,\hVarY}{\mBNec{\mrecv{i}{\tup{\eatom{inc},x,y}}}{\varepsilon}\!\begin{mlbrace}  (\mANec{\msendD{j}{y}{\tup{\eatom{res},\smash{x+1}}}}{\varepsilon}\hVarY)   
        \mand (\mBNec{\msendD{z\,}{y}{\eatom{err}}}{i}\mrestart{i}\,\mclearMailbox{z}\,\hVarY) 
       \end{mlbrace}}  
  	\end{align}}
  	Formula \eqref{intro:2:sync} states that since actor $i$ might require being adapted (by $\mrestart{i}$), then this must be blocked by setting the modality of necessity declared \emph{prior} to the adaptation to \emph{blocking}. In this case we block actor $i$ whenever it receives an increment request \footnote{Note that this might cause the entire system to deadlock; we however explain this later on in Remark~\ref{remark:internal-msgs} and \exref{ex:adapt-and-block}.}, such as $\tup{\eatom{inc},4,h}$, \ie by using blocking necessity $\mBNec{\mrecv{i}{\tup{\eatom{inc},x,y}}}{\varepsilon}$. In case a result, \eg $\tup{\eatom{res},5}$, is sent to the client mapped to $y$ instead of an error message $\eatom{err}$, this means that this event would match the pattern of necessity $\mANec{\msendD{j\,}{y}{\tup{\eatom{res},\smash{x\!+\!1}}}}{\varepsilon}$. 
  	
  	However, this also implies that necessity $\mBNec{\msendD{z\,}{y}{\eatom{err}}}{i}$ would become \emph{trivially satisfied} and hence would not proceed to adapting actor $i$ with $\mrestart{i}$, as this is no longer necessary. Hence, our formula must release actor $i$ upon its trivial satisfaction thereby allowing it to receive further increment requests. We therefore add $i$ to the release list of this necessity, \ie $\mBNec{\msendD{z\,}{y}{\eatom{err}}}{\scriptstyle\pmb{i}}$. Moreover we set the synchronisation modality of this necessity to blocking, \ie $\mattrNec{\msendD{z\,}{y}{\eatom{err}}}{\scriptstyle\pmb{\agg}}{i}{}$ in order to suspend the actor mapped to variable $z$ prior to purging its mailbox. \bqed
\end{example}

Since adaptations 
in a script may be followed by further observations, we also require a similar release mechanism for adaptation actions, \ie $\textsf{adaptation}(\textsf{AL})_{\mathsf{RL}}$, where the actors defined in the \emph{release list}, \textsf{RL}, are unblocked after the adaptation is administered to the actors defined in the \emph{adaptation list}, \textsf{AL}.

\medskip
\begin{remark} \label{remark:internal-msgs}
  Although minimally intrusive, the expressivity of our mechanism for incremental synchronisation relies on what system actions can be observed (\ie the level of abstraction at which the system is monitored).  For instance, recall the server system depicted in \figref{fig:sys}.   If monitored from an external viewpoint, the communications sent from the interface actor, \idV, to the backend incrementor and decrementor actors, \idVV and \idVVVV, are not visible (according to \cite{FraSey14}, they are seen as internal $\tau$-actions).  However, for observations required by properties such as \eqref{intro:2}, we would need to 
 block actor \idV only \emph{after} it sends a message to either of the backend actors---otherwise the entire system blocks (such as in \eqref{intro:2:sync}). This requires observing the system at a \emph{lower level of abstraction}, where certain $\tau$-actions are converted into  visible ones \eg the instrumentation used by  \detecterGen allows us to observe internal actions such as function calls or internal messages sent between actors as discussed for  \figref{fig:sys}. 
\end{remark}

\begin{example}
  \label{ex:adapt-and-block} In \eqref{eq:ra:4} we extend \eqref{intro:2:sync} with the observation of an \emph{internal communication action} (\ie $\msendD{\idV}{\_}{\{\eatom{inc},x,y\}}$) for blocking purposes (as explained in Remark~\ref{remark:internal-msgs}), and also with \emph{release lists} for adaptations \textsf{restr} and \textsf{prg}. 
   \begin{equation} \setstretch{1.1}
    \label{eq:ra:4}
    \begin{array}{l}
    \hV'\;\deftxt\;\; \mmax{\,\hVarY}{\mANec{\mrecv{\idV}{\tup{\eatom{inc},x,y}}}{\epsilon} \\[-2mm]
      \qquad\qquad\qquad\quad\mBNec{\msendD{\idV}{\_}{\{\eatom{inc},x,y\}}}{\varepsilon}
                \begin{mlbrace} 
        (\,\mANec{\msendD{j}{y}{\tup{\eatom{res},\smash{x+1}}}}{\epsilon}\,\hVarY)   \;\mand \; \\[0mm] 
        (\,\mBNec{\msendD{z\,}{y}{\eatom{err}}}{i}\,\mrestart{i}_{\varepsilon}\,\mclearMailbox{z}_{i,z}\,\hVarY) 
      \end{mlbrace}}
    \end{array}
 \end{equation} 
After \emph{asynchronously} observing \mrecv{\idV\!}{\tup{\eatom{inc},\vV,\idVVV}} (for some \vV, \idVVV), the respective monitor \emph{synchronously} listens for an \emph{internal communication} action (\ie $\mBNec{\msendD{\idV\,}{\_}{\{\eatom{inc},x,y\}}}{\varepsilon}$) from \idV to some actor with the same data, \tup{\eatom{inc},\vV,\idVVV}; if this action is observed, the subject of the action (\ie \idV) is blocked. If the subsequently observed action is an error reply, $\msendD{z\;}{\idVVV}{\eatom{err}}$ originating from an actor bound at runtime to $z$, we block this and start applying the necessary synchronous adaptation actions, \ie $\mrestart{i}_{\varepsilon}$ and $\mclearMailbox{z}_{i;z}$.  Note that the last adaptation action releases the two blocked actors \idV and $z$ before recurring; similarly the necessity formula for the error reply releases the blocked actor \idV if the respective action, $\mBNec{\msendD{z\,}{y}{\eatom{err}}}{i}$, is not matched. \exqed
\end{example}

\section{Implementing the Runtime Adaptation Protocol}
\label{sec:lang-implementation}

We implement our runtime adaptation protocol (illustrated in \figref{fig:adaptation} below) as an \emph{extension} of the incremental synchronisation technique introduced in our hybrid monitoring protocol (see \secref{sec:hybr-instr}). The system instrumentation for the new protocol remains more or less the same as in the hybrid protocol. In fact \emph{asynchronous actions} (\ie $\mANec{\patE}{\mathsf{RL}}{}$) inject advice functions that send monitoring messages, containing the event details and a null nonce, without blocking the action's subject. By contrast, \emph{blocking actions} (\ie $\mBNec{\patE}{\mathsf{RL}}{}$) generate and send a fresh nonce along with the event details to the monitor, and subsequently cause the instrumented actor to \emph{block} waiting for a releasing acknowledgement message from the monitor. While waiting, an instrumented blocked actor can now also receive \emph{synchronous adaptation requests}. 

Upon receiving an adaptation request message containing: a \emph{valid nonce}, and an \emph{adaptation tag} denoting a valid pre-instrumented adaptation function (\eg \textsf{restr} to denote the \emph{restart} function), the system side instrumentation sets the requested adaptation as \emph{next-in-line} for execution. The requested adaptations are only applied if and when the blocked system actor receives the releasing acknowledgement. 

\begin{figure}[ht!]
  \centering
  \begin{tikzpicture}[>=latex,auto,thick]
    \begin{scope}[draw=blue!50,fill=blue!20,minimum size=0.8cm]
      \node (sys) at ( -1,1) [shape=rectangle,draw,fill,text width=4.8cm, minimum height = 3cm, text centered] 
          {...\\ Actor $i$ commits event e1;\\ $i$ gets data d1 from e1;\\$i$ blocks on nonce(e1);\\...\\Actor $j$ commits event e2;\\ $j$ gets data d2 from e2;\\$j$ blocks on nonce(e3);\\...\\ Actor $k$ commits event e3;\\ $k$ gets data d3 from e3;\\...\\ $i$ sets restart; \\ $j$ sets purge;\\[4mm] $i$ unblocks with nonce1; \\ $j$ unblocks with nonce2;\\ \ldots\; };
       \node (mon) at ( 8,1) [shape=rectangle,draw,fill,text width=6.4cm, minimum height = 3cm] {\\[1.5mm]$\; $loop(Map)\;$\rightarrow$\\$\quad$\{Evt,Id,Nonce\} = recv\_event(), \\ $\quad$if Nonce $\neq$ null, \\$\qquad$ Map2 = Updt(Id,Nonce,Map); \\$\quad$else\\$\qquad$Map2 = Map;\\ \quad end, \\ $\quad$\{PtrnMtch,AL,RL\} = handle(Evt), \\ $\quad$if(PtrnMtch) $\rightarrow$ \\ $\qquad$adapt(AL,Map2), \\$\qquad$release(RL,Map2), \\$\qquad$loop(Map2);  \\ $\quad$end.\\[15mm]\; };  
        \node (sysN) at (-1,5.8) {\textbf{System}}; 
        \node (monN) at (8,5.8) {\textbf{Monitor}}; 
    \end{scope}
       \node (sys1) at (1.4,4.1){};
       \node (mon1) at (4.81,4.1){};
       \node (sys2) at (1.4,2){};
       \node (mon2) at (4.81,2){};
       \node (sys3) at (1.4,0.2){};
       \node (mon3) at (4.81,0.2){};
       \node (sys4) at (1.4,-0.8){};
       \node (mon4) at (4.8,-0.8){};
       \node (sys5) at (1.4,-1.1){};
       \node (mon5) at (4.8,-1.1){};
       \node (sys6) at (1.4,-2.3){};
       \node (mon6) at (4.81,-2.3){};
       \node (sys7) at (1.4,-2.6){};
       \node (mon7) at (4.81,-2.6){};
    \begin{scope}[draw=red,fill=red, dashed]
      \draw[->] (sys1) to node{\small\textbf{evt}(d1,\{i,nonce1\})} (mon1); 
      \draw[->] (sys2) to node{\small\textbf{evt}(d2,\{j,nonce2\})} (mon2); 
      \draw[->] (sys3) to node{\small\textbf{evt}(d3,null)} (mon3); 
      \draw[<-] (sys4) to node{\small\textbf{adpt}(\textsf{restr},nonce1)} (mon4); 
      \draw[->] (mon5) to node{\small\textbf{adpt}(\textsf{prg},nonce2)} (sys5); 
      \draw[<-] (sys6) to node{\small\textbf{ack}(nonce1)} (mon6); 
      \draw[->] (mon7) to node{\small\textbf{ack}(nonce2)} (sys7);
    \end{scope}
  \end{tikzpicture}
  \caption{A high-level depiction of the Runtime Adaptation protocol. 
  }
       \label{fig:adaptation}
\end{figure}
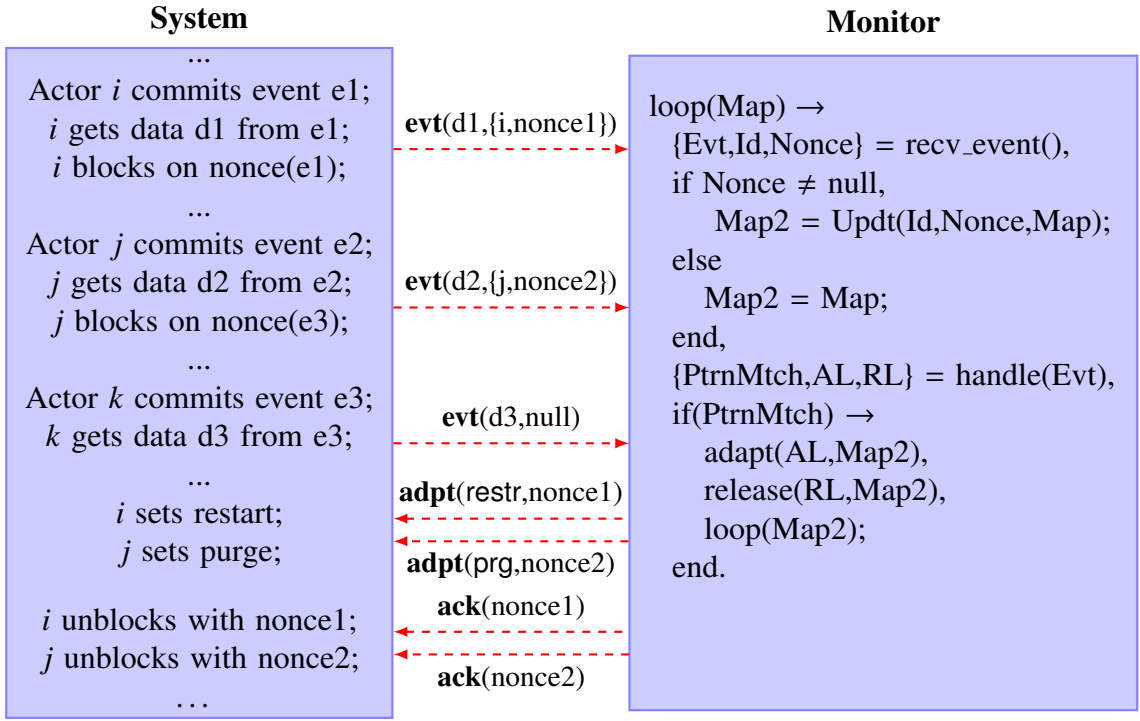

To allow for the monitor to adapt and release system actors, even after receiving further event messages, the monitoring loop now maintains a \emph{map of nonces}. Whenever the monitor receives an event message containing a \emph{non-null} nonce and the subject's process id, it adds the received nonce as a new entry to its map by using the subject's id as key; it then proceeds by handling the event. As event handling now also includes applying adaptations (\ie $\textsf{adaptation}(\textsf{AL})_{\mathsf{RL}}$), 
the monitor may proceed by issuing the necessary adaptation actions denoted in the adaptation list \textsf{AL} (if any), and then releases the actors enlisted in the release-list \textsf{RL}. 
To issue synchronous adaptations and release acknowledgements, the monitor uses the process ids enlisted in \textsf{AL} and \textsf{RL} as keys for retrieving the associated nonces from its map. After applying the necessary adaptions and releases, the monitor removes the map entries associated with the released actors. It finally reiterates and waits for the next system event.  

\subsection{Implementing Pre-instrumented Synchronous Adaptations} \label{sec:syn-action-impl}
Although Erlang already natively implements some of the adaptations that we categorised in Table~\ref{tab:syn-asyn-adapts}, as part of our prototype we implement more complex and effective synchronous adaptation actions. More specifically, we tried to implement a representative adaptation action for each synchronous adaptation category presented in Table~\ref{tab:syn-asyn-adapts}. These implementations can be applied over \emph{instrumented system actors} using the RA protocol implementation outlined in \figref{fig:adaptation}. We therefore implement the following pre-instrumented synchronous adaptation functions:
\begin{itemize}[leftmargin=5mm]\itemsep0em
	\item\textbf{\emph{silent-kill}} (Terminations category $-$ $\msKill{\mathsf{AL}}_{\mathsf{RL}}$): As outlined in Remark~\ref{remark:silent-kill} (in the beginning of \secref{sec:syn-asy-adapt-classes}), this action instructs a blocked actor (using an adaptation request message) to willingly relinquish its supervision links prior to terminating itself. By doing so it refrains from alerting any of its supervisors or linked actors about its termination. This enables the monitor to terminate actors without possibly triggering any of the system's built-in self-adaptation mechanisms. Unlike asynchronous killing, since the \emph{silent-kill} requires the recipient actor to be blocked, it ensures that actors are terminated \emph{on-time}, as blocking forbids an actor from progressing after committing an action.
	\item\textbf{\emph{purge}} (Interception category $-$ $\mclearMailbox{\mathsf{AL}}_{\mathsf{RL}}$): Upon interpreting a purge request, the blocked actor uses a loop of non-blocking receives \cite{Armstrong07} to continuously read and ignore the messages in its mailbox. The loop stops when no more messages are left, \ie the mailbox is emptied. In this case, synchronisation was essential given that the actor model strictly prohibits external actors (like our monitor) from accessing and modifying the mailbox contents of another actor. A variant of this adaptation is \textbf{\emph{intercept}}, in which only the messages matching a specific pattern are purged from the mailbox, while the rest remain intact.	
	\item\textbf{\emph{sync-link}} (Interception category $-$ $\mslink{\mathsf{AL}}_{\mathsf{RL}}$): The monitor sends a link request message to an instrumented blocked system actor containing a list of process ids. When the instrumented code interprets this request, it links the host actor with \emph{all the process ids} enlisted in the request message (if not already linked) in a timely manner. The same mechanism applies for the reverse function \textbf{\emph{sync-unlink}}, which unlinks the instrumented actor from all the enlisted actors if a link exists.	
	\item\textbf{\emph{restart}} (Restarts category $-$ $\mrestart{\mathsf{AL}}_{\mathsf{RL}}$): Restarting a process in Erlang is normally performed by first killing the process and then respawning it. However, since the adaptee may be linked to other actors, killing it might also terminate (or inform) other linked actors as a side-effect which could cause further complications. Moreover, even if the actor is terminated silently with a \emph{silent-kill} action, we still run into the following two problems:
	\begin{itemize}
		\item the links removed by the \emph{silent-kill} must be relinked to the newly respawned process, in the same way as they were prior to killing the original actor;
		\item as respawning an adaptee actor changes its process id, we need to somehow inform the other actors about this change, otherwise they would not be able to contact the restarted process. 		
	\end{itemize}
\end{itemize}
	To avoid these problems with our restart implementation, a monitor restart request causes the blocked actor to: empty its mailbox (same as in \emph{purge}), refresh its internal state (\ie it removes any non-instrumentation data residing in its process dictionary \cite{Armstrong07}) and finally restarts the actor by calling the original initial function used by the spawn operation that created the actor in question. In other words, the adaptee is not terminated and respawned; instead we call its \emph{initialisation function} from \emph{within} the actor's own thread of control. This is better explained in \exref{ex:restart}.
	\begin{example} \label{ex:restart}		
		Recall the system given in \figref{fig:sys}, and assume that the common-interface $i$ was at some point during system initialisation spawned by a parent process which called the function \textsf{spawn(interface:start())}. This spawn operation therefore creates a new actor process which internally calls the function \textsf{interface:start()}, \ie the function which starts the common-interface from the very beginning. Whenever $i$ is sent a restart request by our monitor, it empties its mailbox, refreshes its internal state and finally calls \textsf{interface:start()} to start afresh.	\bqed
	\end{example}
	By using this mechanism, a process is able to \emph{retain its process identifier} even after restarting. This is crucial to avoid having to inform other system actors about the new process id; failing to do so would lead to having other system actors unable to contact the restarted actor. Implementing this adaptation was not as straight forward as the previous ones, as this required a further extension to the AOP instrumentation framework. The extended AOP is now also able to save the details (module, function and arguments) of the initial function used in spawn operation, within the internal state (\ie its process dictionary) of the newly spawned actor. This information is then retrieved if and when the instrumented actor receives a restart adaptation request while it is blocked. 

\section{Strengthening Yaws through Runtime Adaptation}
\label{sec:yaws-ra-casestudy}
To provide a better insight on how runtime verification properties can easily be converted into runtime adaptation properties, we incrementally extend Yaws property \eqref{prop:yaws:2} (given in \secref{sec:monit-safety-prop}). More specifically we strengthen the Yaws webserver with runtime adaptations for mitigating invalid HTTP requests that might contribute to a \emph{dot-dot-slash} attack which exploits the directory traversal vulnerability. Hence we show that by introducing \emph{minimal} constructs in the currently available RV scripts we can use our RA framework to mitigate third-party applications such as the Yaws webserver.

 \vspace{-5mm}
\begin{equation} \setstretch{1.1}
\begin{small}
	\begin{array}{l}
	 \mmax{\hVarX}{\Bigl(
	 	\\\indent \mnec{\mrecv{\eatom{acceptor}}{\{hId,\eatom{next},\_\}}}
	 	 \\\indent\mnec{\mret{hId}{\{\eatom{yaws}, \eatom{do\_recv},3,\{\texttt{ok},\{\eatom{http\_req},\eatom{'GET'},\{\eatom{abs\_path},path\},\_\}\}\}}} 
		 \\\indent \mnec{\mret{hId}{\etuple{\eatom{yaws}, \eatom{do\_recv},\eatom{3},\etuple{\eatom{ok},h1}}}} 
		 \\\indent \indent \vdots	 
		 \\\indent \mnec{\mret{hId}{\etuple{\eatom{yaws}, \eatom{do\_recv},\eatom{3},\etuple{\eatom{ok},h6}}}}
		 \\\indent \mnec{\mret{hId}{\etuple{\eatom{yaws}, \eatom{do\_recv},\eatom{3},\etuple{\eatom{ok},\eatom{http_eoh}}}}}
		 \\\indent \mboolE{path == ``\eatom{/pic.png}'' \texttt{ orelse } path == ``\eatom{/site.html}''}{
	 	\\\indent\indent \mnec{\mcall{hId}{\{\eatom{yaws\_sendfile}, \eatom{send},[\_,path,\_,\_]\}}} \hVarX\\\indent }{ \\\indent\indent\msKill{hId}\;\mclearMailbox{\eatom{acceptor}}\;\hVarX} 	
	\\\Bigr)}
	\end{array}\label{prop:v2}
\end{small}
\end{equation}

Contrasting to \eqref{prop:yaws:2}, instead of issuing a violation ($\mfls$), property \eqref{prop:v2} applies a \emph{silent-kill} action (\ie $\msKill{..}$) to immediately isolate and terminate the handler actor assigned to the requesting client, thus preventing the termination of the handler from possibly affecting other linked actors. Note that we are able to apply a \emph{silent-kill} mitigation over the assigned handler actor (bound to $hId$), as our RA framework also supports applying adaptations over actors which are only known at runtime. This adaptation is then followed by a \emph{purge} adaptation (\ie $\mclearMailbox{\eatom{..}}$) that empties the mailbox of the \eatom{acceptor}.

As both the \emph{silent-kill} and \emph{purge} adaptations are synchronous adaptations (as explained in \secref{sec:syn-action-impl}) we need to specify the necessary synchronisation operations. As these adaptations are applied on system actors $hId$ and \eatom{acceptor}, we must identify a set of necessities that must be converted into blocking necessities to ensure that the adaptations are applied correctly. The rule of thumb for identifying such candidates is that whenever we have an adaptation which applies on a system actor $\idV$, \eg $\mclearMailbox{\idV}{}$, then a candidate necessity should:
\begin{enumerate}[label=(\roman*)] \itemsep0em \setstretch{1.1}
	\item be defined \emph{prior} to the adaptation;
	\item have $i$ defined as its \emph{subject}; and
	\item be the \emph{closest} candidate necessity lying in the path leading to the adaptation $-$ this is important so that an actor $i$ is not blocked earlier than required.
\end{enumerate}

For instance, to correctly apply the first $\msKill{hId}$ action declared in \eqref{prop:v2}, we convert the end-of-headers necessity (\ie $\mnec{\mret{hId}{\etuple{\eatom{yaws}, \eatom{do\_recv},\eatom{3},\etuple{\eatom{ok},\eatom{http_eoh}}}}}$ into a blocking one. This candidate was chosen as it satisfies rules $(i)$, $(ii)$ and $(iii)$. Using the same argument we add the blocking attribute to the first necessity, allowing the \mclearMailbox{\eatom{acceptor}} adaptation to be applied. The rest of the necessities are converted into asynchronous necessities. Hence by using these 3 rules we augment \eqref{prop:v2} with blocking necessities to obtain \eqref{prop:v3}. 

 \vspace{-5mm}
\begin{equation} \setstretch{1.1}
\begin{small}
	\begin{array}{l}
	 \mmax{\hVarX}{\Bigl(
	 	\\\indent \mBNec{\mrecv{\eatom{acceptor}}{\{hId,\texttt{next},\_\}}}{}{}
		 \\\indent \mANec{\mret{hId}{\{\eatom{yaws}, \eatom{do\_recv},3,\{\texttt{ok},\{\eatom{http\_req},\eatom{'GET'},\{\eatom{abs\_path},path\},\_\}\}\}}}{}{} 
		 \\\indent \mANec{\mret{hId}{\etuple{\eatom{yaws}, \eatom{do\_recv},\eatom{3},\etuple{\eatom{ok},h1}}}}{}{} 
		 \\\indent \indent \vdots	 
		 \\\indent \mANec{\mret{hId}{\etuple{\eatom{yaws}, \eatom{do\_recv},\eatom{3},\etuple{\eatom{ok},h6}}}}{}{}
		 \\\indent \mBNec{\mret{hId}{\etuple{\eatom{yaws}, \eatom{do\_recv},\eatom{3},\etuple{\eatom{ok},\eatom{http_eoh}}}}}{}{}
		 \\\indent \mboolE{path = ``\eatom{/pic.png}'' \texttt{ orelse } path = ``\eatom{/site.html}''}{
	 	\\\indent\indent \mANec{\mcall{hId}{\{\eatom{yaws\_sendfile}, \eatom{send},[\_,path,\_,\_]\}}}{}{} \hVarX\\\indent }{ \\\indent\indent\msKill{hId}\; \mclearMailbox{\eatom{acceptor}}\;\hVarX} 
	 \\\Bigr) }
	\end{array}\label{prop:v3}
\end{small}
\end{equation}

Thus far in \eqref{prop:v3}, we have only added synchronisation to block the adaptees without ever releasing them. We must therefore release these blocked actors by specifying them in the appropriate \emph{release-lists}. As shown by \eqref{prop:v4} we add the \eatom{acceptor} to the release list of every necessity operation that lies between the blocking necessity which blocks the acceptor actor ($\mBNec{\mrecv{\eatom{acceptor}}{\{hId,\texttt{next},\_\}}}{}{}$), and the adaptation (\ie \mclearMailbox{\eatom{acceptor}}). This ensures that the \eatom{acceptor} is released in case the property is trivially satisfied before reaching the point where the adaptation needs to be applied. Furthermore, we also add the handler and the acceptor to the release list of the last adaptation; this ensures that the actors are released after applying the adaptations. 

 \vspace{-5mm}
\begin{equation} \setstretch{1.1}
\begin{small}
	\begin{array}{l}
	 \mmax{\hVarX}{\Bigl(
	 	\\\indent \mBNec{\msend{\eatom{acceptor}}{\{hId,\texttt{next},\_\}}}{\varepsilon}{}
		 \\\indent \mANec{\mret{hId}{\{\eatom{yaws}, \eatom{do\_recv},3,\{\texttt{ok},\{\eatom{http\_req},\eatom{'GET'},\{\eatom{abs\_path},path\},\_\}\}\}}}{}{}_{\hspace{-2mm}\eatom{acceptor}} 
		 \\\indent \mANec{\mret{hId}{\etuple{\eatom{yaws}, \eatom{do\_recv},\eatom{3},\etuple{\eatom{ok},h1}}}}{}{}_{\hspace{-2mm}\eatom{acceptor}}  
		 \\\indent \indent \vdots	 
		 \\\indent \mANec{\mret{hId}{\etuple{\eatom{yaws}, \eatom{do\_recv},\eatom{3},\etuple{\eatom{ok},h6}}}}{}{}_{\hspace{-2mm}\eatom{acceptor}} 
		 \\\indent \mBNec{\mret{hId}{\etuple{\eatom{yaws}, \eatom{do\_recv},\eatom{3},\etuple{\eatom{ok},\eatom{http_eoh}}}}}{}{}_{\hspace{-2mm}\eatom{acceptor}} 
		 \\\indent \mboolE{path == \text{``}\eatom{/pic.png}\text{''} \texttt{ orelse } path == \text{``}\eatom{/site.html}\text{''}}{
	 	\\\indent\indent \mANec{\mcall{hId}{\{\eatom{yaws\_sendfile}, \eatom{send},[\_,path,\_,\_]\}}}{}{}_{\hspace{-2mm}\eatom{acceptor}} \; \hVarX\\\indent }{ \\\indent\indent\msKill{hId}_{\varepsilon}\;\mclearMailbox{\eatom{acceptor}}_{hId,\eatom{acceptor}}\;\hVarX} 
	 \\\Bigr) }
	\end{array}\label{prop:v4}
\end{small}
\end{equation}

\noindent Therefore the rule of thumb for releasing a blocked actor is to add its identifier to the release list of:
\begin{enumerate}[label=(\roman*)] \itemsep0em \setstretch{1.1}
	\item every necessity lying between the point where it was blocked and its adaptation action; and to 
	\item the release list of the adaptation which is applied to it (or to those defined after it within the same path).
\end{enumerate}

\section{A Formal Model for Runtime Adaptation}
\label{sec:ra-model}

In this section we develop a formal operational model of our runtime adaptation framework. The primary reasons for developing this formal model are:
\begin{itemize}[leftmargin=5mm,itemindent=0mm]\itemsep0em
	\item A formal model enables us to understand how the monitors specified by an arbitrary script are to behave at runtime without having to actually compile the script and run the synthesised monitors. This also relieves us from having to deal with the complexities of the implementation.
	\item As during the course of this research we developed the model in parallel to the implementation, the model also served to guide the implementation (and vice-versa).
	\item The model allows us to prove that the operational behaviour of our framework corresponds to the semantics defined in earlier work \cite{FraSey14}.
	\item Formalisation also enables us to better understand the kind of errors that our RA scripts may introduce, thereby allowing us to look into static analysis techniques in \Cref{chp:typ-sys} for detecting erroneous scripts prior to deployment.	
\end{itemize}

We develop this model in two subsequent phases. As part of the first phase, in \secref{sec:derivative-semantics} we develop and present a preliminary operational model for \detecterGen's original (unextended) logic (as presented in \figref{fig:logic} in \secref{sec:detecter-primer-2}), given that currently such a model does not exist. We then prove that our operational model \emph{corresponds} to \detecterGen's semantics developed in previous work \cite{FraSey14}.  

For the second phase, in \secref{sec:ra-derivative-semantics}, we formalise the mechanisms for applying adaptation and incremental synchronisation, in a way which reflects the implementation constraints (explained \secref{sec:syn-asyn-adaptations}). We then augment these formalised mechanisms in our preliminary formal model. 

\subsection{An Operational Model for \detecterGen}
\label{sec:derivative-semantics}
\begin{figure}[ht!]
  \begin{small}
	\textbf{Syntax}\vspace{-2mm}
	  \begin{align*}
		    \hV,\hVV \in \FRM & \;\bnfdef\; \mtru \;\bnfsep\; \mfls \;\bnfsep\; \hV\mand\hVV \;\bnfsep\; \mnec{\patE}{\hV}  \;\bnfsep\; \hVarX \;\bnfsep\; \mmax{\hVarX}{\hV} \;\bnfsep\;  \mboolE{\bV}{\hV\,}{\,\hVV} 
	  \end{align*} 
	\textbf{Structural Equivalence Semantics}\vspace{-2mm}
		\begin{mathpar}
	        \hV_1 \mand \hV_2 \steq \hV_2 \mand \hV_1   
	        \and 
	        \hV_1 \mand (\hV_2 \mand \hV_3) \steq  (\hV_1 \mand \hV_2) \mand \hV_3 
	        \and
	        \mtru \mand \hV \steq \hV
	        \and
	        \mfls \mand \hV \steq \mfls        
	  \end{mathpar} 
	  \textbf{Reduction Semantics}
	  \begin{mathpar}	       
	        \inference[\rtit{rIdem1}]{}{\mfls \traS{\actE} \mfls} 
	        \and
	        \inference[\rtit{rIdem2}]{}{\mtru \traS{\actE} \mtru}
	        \\ 
		\inference[\rtit{rTru}]{\mcondEval{\bV}{\textsf{true}}}{\mboolE{\bV}{\hV}{\hVV} \traS{\tau} \hV}
	        \quad
	        \inference[\rtit{rFls}]{\mcondEval{\bV}{\textsf{false}}}{\mboolE{\bV}{\hV}{\hVV} \traS{\tau} \hVV}
	        \quad
	        \inference[\rtit{rStr}]{\hV\steq\hV'\traS{\act}\hVV'\steq\hVV}{\hV\traS{\act}\hVV} 
	        \\
	        \inference[\rtit{rCn1}]{\hV\traS{\actE}\hV' & \hVV\traS{\actE}\hVV'}{\hV\mand\hVV \traS{\actE} \hV'\mand\hVV'}  
	        \and
	        \inference[\rtit{rCn2}]{\hV\traceEvent{\tau} \hV' }{\hV\mand\hVV \traS{\tau} \hV'\mand\hVV} 
	        \and
	        \inference[\rtit{rCn3}]{\hVV\traceEvent{\tau} \hVV' }{\hV\mand\hVV \traS{\tau} \hV\mand\hVV'}\\
	         \inference[\rtit{rMax}]{}{\mmax{X}{\hV} \traS{\tau} \hV\sub{\mmax{X}{\hV}}{X}} 
	         \quad
	        \inference[\rtit{rNc1}]{\match{\patE}{\actE}{\sigma}}{\mnec{\patE}{\hV} \traS{\actE} \hV\sigma} 
	        \quad
	        \inference[\rtit{rNc2}]{\match{\patE}{\actE}{\bot}}{\mnec{\patE}{\hV} \traS{\actE} \mtru}
	\end{mathpar}
\end{small}
  \caption{The Logic and its LTS Semantics}
  \label{fig:logic-transition-rules}
\end{figure}

%

A semantics \cite{SyncVSAsync:Rosu:2005} for the \emph{closed} logic formulas (\ie no free term or formula variables) of the core logic presented in \secref{sec:detecter:semantics}, is given as a Labelled Transition System (LTS). This is defined by the transition rules in \figref{fig:logic-transition-rules}. Note that for convenience we restate the logic's syntax in \figref{fig:logic-transition-rules} as well. 

\subsubsection{The Preliminaries}
The LTS semantics models abstractly the monitoring of the respective property, and assumes: a set of (visible) actions $\actE,\actEE\in\Act$, and a distinguished \emph{silent} action, $\tau$; we let \act range over $\Act \cup \sset{\tau}$. Visible actions represent system operations which may contain values $\vV,\vVV \in \Val$, that range over either actor identifiers, $\idV,\idVV,\idVVV \in \Pid$, or generic data $p\in\Data$ such as integers $n$, lists $p_{1}\!:\!\ldots\!:\!p_{n}$, and tuples $\{p_{1},\ldots,p_{n}$\}. 
We also use metavariable $\varid\in(\Pid \cup  \Vars)$ to represent either a process identifier or a term variable. 

In this semantics we work up-to structural equivalence of formulas $\hV\steq\hVV$ (see Equivalence Semantics in \figref{fig:logic-transition-rules}) for commutativity, associativity \etc The semantics also assumes a partial function  $\mtch{\patE}{\actE}$, defined in Def.~\ref{def:match}, for matching action patterns, \patE, with visible system actions, \actE. When a match is successful, the function returns a substitution from the term variables found in the pattern, to the corresponding values of the matched action, $\sigma:: \Vars\rightharpoonup \Val$. \medskip

\begin{definition}[\textbf{Pattern Matching}]\label{def:match}
Function $\mtch{\patE}{\patEE}$ defines how a pattern $\patE$ can be compared to another pattern $\patEE$. When a match occurs this function returns a mapping from term variables to actual values, while it returns \undef when the patterns do not match. 

\vspace{-2mm}
{\small\setstretch{1.2}  	\setlength{\abovedisplayskip}{0pt}
\setlength{\belowdisplayskip}{0pt}
\setlength{\abovedisplayshortskip}{0pt}
\setlength{\belowdisplayshortskip}{0pt}
\begin{align*}
	\mtch{\patE}{\patEE} &\defEquals \begin{cases} 
				\cmtch{\pmtch{\varid^{s}_{1}}{\varid^{s}_{2}}}{\cmtch{\pmtch{\varid^{\,r}_{1}}{\varid^{\,r}_{2}}\\[-2mm] \;\;}{\;\pmtch{p_{1}}{p_{2}}}} & \text{if }\patE\eq \msendD{\varid^{s}_{1}}{\varid^{\,r}_{1}}{p_{1}}\,\land\,\patEE\eq \msendD{\varid^{s}_{2}}{\varid^{\,r}_{2}}{p_{2}}\\
				\cmtch{\pmtch{\varid_{1}}{\varid_{2}}}{\pmtch{p_{1}}{p_{2}}} & \text{if }\patE\eq \mrecv{\varid_{1}}{p_{1}}\,\land\,\patEE\eq \mrecv{\varid_{2}}{p_{2}}\\
				\cmtch{\pmtch{\varid_{1}}{\varid_{2}}}{\pmtch{p_{1}}{p_{2}}} & \text{if }\patE\eq \mcall{\varid_{1}}{p_{1}}\,\land\,\patEE\eq \mcall{\varid_{2}}{p_{2}}\\
				\cmtch{\pmtch{\varid_{1}}{\varid_{2}}}{\pmtch{p_{1}}{p_{2}}} & \text{if }\patE\eq \mret{\varid_{1}}{p_{1}}\,\land\,\patEE\eq \mret{\varid_{2}}{p_{2}}\\
				\{\} & \text{if }\patE\eq \_ \text{ or } \patEE\eq \_\\
				\undef & \text{otherwise} \\
\end{cases}\\
	\pmtch{p'}{p''} &\defEquals \begin{cases} 
					    \{\} & \text{if }p'\eq n \,\land\, p''\eq n\\
					    \{\} & \text{if }p'\eq x \,\land\, p''\eq y\\
					    \{x\mapsto v\} & \text{if }(p'\eq x \text{ and }p''\eq n) \text{ or } (p'\eq n \text{ and }p''\eq x)\\
					    \bigoplus^{n}_{i\eq 1}\sigma_{\varid_{2}} & \text{if } p'\eq \{p'_{1},\ldots,p'_{n}\}, p''\eq \{p''_{1},\ldots,p''_{n}\} \\& \quad \text{ where }\pmtch{p'_{i}}{p''_{i}}\eq \sigma_{i} \\
					    \cmtch{\pmtch{p'_{1}}{p''_{1}}}{\pmtch{p'_{2}}{p''_{2}}}  & \text{if } p'\eq p'_{1};p'_{2} \,\land\, p''\eq p''_{1}:p''_{2} \\
					    \undef & \text{otherwise} \\
					  \end{cases}\\
	\cmtch{\sigma_{1}}{\sigma_{2}} &\defEquals \begin{cases} 
					    \sigma_{1}\cup\sigma_{2} & \text{if }\forall p\in\dom(\sigma_{1})\cap\dom(\sigma_{2})\cdot\, \sigma_{1}(p) \eq  \sigma_{2}(p) \\
					    \undef & \text{if }\sigma_{1}\eq \undef \text{ or } \sigma_{2}\eq \undef\\
					    \undef & \text{otherwise} \\
					  \end{cases}  
\end{align*} }

\noindent The $\mtch{\patE}{\patEE}$ function starts by checking whether the patterns define the same kind action \eg whether both patterns define an output action. The $\textsf{mtch}$ function also supports the \emph{universally matching} pattern ``\_'', this matches any action pattern. If the action types match, this function uses the $\pmtch{p'}{p''}$ function to inspect the actions' subjects and the data, where the latter includes values $v$, term variables $x$, tuples $\{p'_{1},\ldots,p'_{n}\}$ and lists $p'_{1};p'_{2}$. The $\pmtch{p'}{p''}$ function checks that the data patterns correspond in both patterns \eg a tuple of three matching entries are present in both patterns. If the data patterns match it returns a substitution environment $\sigma$ which maps the term variables declared in one pattern, to the values of the other pattern.

Note that although the $\textsf{mtch}$ function is defined for comparing necessity patterns \ie $\mtch{\patE}{\patEE}$, it can still be used to compare a necessity pattern $\patE$ with a system action $\actE$ \ie $\mtch{\patE}{\actE}$, as well as to compare two system actions $\mtch{\actE}{\actEE}$. This is possible as system actions are equivalent to \emph{closed patterns} \ie patterns that do not define any term variables. This pattern matching mechanism is further explained in the following example.

\begin{example}
We consider a number of scenarios where two patterns are pattern matched using the \textsf{mtch} function. \vspace{-3mm}

{\small\setstretch{1.1} \setlength{\abovedisplayskip}{5pt}
\setlength{\belowdisplayskip}{5pt}
\setlength{\abovedisplayshortskip}{5pt}
\setlength{\belowdisplayshortskip}{5pt}
\begin{align}
  \label{eq:mtch:1}
  &\match{\msendD{\eatom{srv}}{\eatom{clnt}}{\etuple{x,\eatom{ack},y}}\, }{\, \msendD{\eatom{srv}}{\eatom{clnt}}{\etuple{5,\eatom{ack},z}}}{\etuple{x\mapsto 5}} \\
  \label{eq:mtch:2}
  &\match{\msendD{\eatom{srv}}{\eatom{clnt}}{\etuple{5,\eatom{ack},z}}\, }{\, \msendD{\eatom{srv}}{\eatom{clnt}}{\etuple{x,\eatom{ack},y}}}{\etuple{x\mapsto 5}}   
\end{align}
}

\noindent For instance in (\ref{eq:mtch:1}) the output (open) necessity pattern $\msendD{\eatom{srv}\,}{\,\eatom{clnt}}{\etuple{x,\eatom{ack},y}}$ is successfully matched with another (open) output pattern $\msendD{\eatom{srv}\,}{\,\eatom{clnt}}{\etuple{5,\eatom{ack},z}}$. Note that term variable $x$ is pattern matched with the value $5$ and thus returned in the resultant substitution environment. Although variable $y$ is also matched to variable $z$, the match function does not return an entry in the resultant substitution environment (remember that the substitution only maps variables to values). Also note that the match returns the same substitution environment if the arguments are inverted as shown in (\ref{eq:mtch:2}).\vspace{-3mm}

{\small\setstretch{1.1} \setlength{\abovedisplayskip}{5pt}
\setlength{\belowdisplayskip}{5pt}
\setlength{\abovedisplayshortskip}{5pt}
\setlength{\belowdisplayshortskip}{5pt}
\begin{align}
  \label{eq:mtch:3}
  &\match{\msendD{\eatom{srv}}{\eatom{clnt}}{\etuple{x,\eatom{ack},y}}}{\msendD{\eatom{srv}}{\eatom{clnt}}{\etuple{5,\eatom{ack},\eatom{joe}}}}{\etuple{x\mapsto 5, y\mapsto \eatom{joe}}} \\ 
 \label{eq:mtch:4}
  &\match{\msendD{\eatom{srv}}{\eatom{clnt}}{\etuple{5,\eatom{ack},\eatom{joe}}}}{\msendD{\eatom{srv}}{\eatom{clnt}}{\etuple{5,\eatom{ack},\eatom{joe}}}}{\etuple{}}
\end{align}
}%

\noindent In (\ref{eq:mtch:3}) the necessity pattern $\msendD{\eatom{srv}}{\eatom{clnt}}{\etuple{x,\eatom{ack},y}}$ is now matched with a \emph{closed patten} \ie a system action $\msendD{\eatom{srv}}{\eatom{clnt}}{\etuple{5,\eatom{ack},\eatom{joe}}}$, where $x$ and $y$ are pattern matched with the values $5$ and \eatom{joe} respectively. Moreover, in (\ref{eq:mtch:4}) the two closed actions are matched (exactly), thereby returning the empty substitution.\vspace{-3mm}

{\small\setstretch{1.1} \setlength{\abovedisplayskip}{5pt}
\setlength{\belowdisplayskip}{5pt}
\setlength{\abovedisplayshortskip}{5pt}
\setlength{\belowdisplayshortskip}{5pt}
\begin{align} 
 \label{eq:mtch:5}
  &\match{\msendD{\eatom{clnt}}{\eatom{srv}}{\etuple{x,\eatom{ack},y}}}{\msendD{\eatom{srv}}{\eatom{clnt}}{\etuple{5,\eatom{ack},\eatom{joe}}}}{\undef} \\
  \label{eq:mtch:6}
  &\match{\mrecv{\eatom{clnt}}{\etuple{x,\eatom{ack},y}}}{\msendD{\eatom{srv}}{\eatom{clnt}}{\etuple{5,\eatom{ack},\eatom{joe}}}}{\undef}
\end{align}
}%

\noindent The mismatch in (\ref{eq:mtch:5}) is caused by the mismatching subjects of the output actions \ie \eatom{srv} versus \eatom{clnt}, whereas the mismatch in (\ref{eq:mtch:6}) is caused since the different type of actions cannot pattern match, \ie input action $\mrecv{\eatom{clnt}}{\etuple{x,\eatom{ack},y}}$ and output action $\msend{\eatom{srv}}{\eatom{clnt}}{\etuple{5,\eatom{ack},\eatom{joe}}}$ do not match. \bqed 
\end{example}\vspace{-2mm}
\end{definition}

\subsubsection{The Modelled Reductions}
In \figref{fig:logic-transition-rules}, we give two kinds of reductions namely $\actE$-reductions and $\tau$-reductions, which are respectively represented using judgements $\hV\traS{\actE}\hV'$ and $\hV\traS{\tau}\hV'$. The first judgement, \ie $\hV\traS{\actE}\hV'$, represents cases where a monitor $\hV$ evolves (progresses) into a different monitor $\hV'$ after observing an \emph{external} system event $\actE$. This models the implementation mechanism where a monitor reads a system event from its mailbox and progresses accordingly. The second judgement, \ie $\hV\traS{\tau}\hV'$, models an \emph{internal} (silent) action that a monitor $\hV$ can perform so to evolve into $\hV'$. Internal actions include operations such as unfolding recursive definitions and evaluating if-statement conditions. Using these judgements we can represent a monitor execution as a sequence of internal ($\tau$) and external ($\actE$) reductions as shown in \exref{ex:mon-exec}.

\begin{example}\label{ex:mon-exec} Reduction sequence \eqref{redseq:1} models the execution of a monitor $\hVV$ \wrt the following sequence of observable system events: $\actE_{1};\ldots; \actE_{n}$.
  {\setlength{\abovedisplayskip}{5pt}
\setlength{\belowdisplayskip}{5pt}
\setlength{\abovedisplayshortskip}{5pt}
\setlength{\belowdisplayshortskip}{5pt}\begin{align}\label{redseq:1} \hVV\, (\traS{\tau})^\ast\!\traS{\actE_{1}}\!(\traS{\tau})^\ast\, \cdots \, (\traS{\tau})^\ast\!\traS{\actE_{n}}\!(\traS{\tau})^\ast\, \hVV'  \end{align}}
As shown by \eqref{redseq:1}, a monitoring reduction over an observation $\actE_{i}$ may be interposed by internal (silent) $\tau$-reductions, \ie $\hVV''(\traS{\tau})^\ast\!\traS{\actE_{i}}\!(\traS{\tau})^\ast\,\hVV'''$.
\bqed
\end{example}

More specifically, in \figref{fig:logic-transition-rules},  formulas \mtru\ and \mfls\ are idempotent \wrt external transitions and interpreted as final states (verdicts).  Conditional formulas silently (internally) branch to the respective subformula depending on the evaluation of a decidable boolean expression (\rtit{rTru} and \rtit{rFls}), whereas rule \rtit{rMax} internally unfolds a recursive formula.  Necessity formulas, \mnec{\patE}{\hV},  transition only with a visible action, \actE. If the action \actE matches the necessity's pattern (\ie $\match{\patE}{\actE}{\sigma}$), it transitions to the continuation subformula \hV\, where the variables bound by the matched pattern are substituted with the respective matched values obtained from the action, \ie it reduces into $\hV\sigma$ where $\sigma$ contains the data mappings. Otherwise, the necessity formula transitions to \mtru\ in case of a mismatch (\rtit{rNc2}).

The rules for conjunction formulas model the parallel execution of subformulas as described in \cite{FraSey14}, whereby subformulas are allowed to perform \emph{independent} silent $\tau$-transitions (\rtit{rCn2} and \rtit{rCn3}) but transition \emph{together} for external ($\actE$) actions depending on their individual transitions (\rtit{rCn1}). Finally, \rtit{rStr} allows us to abstract over structurally equivalent formulas. We write $\hV \wtraS{\act} \hVV$ in lieu of $\hV (\traS{\tau})^\ast\traS{\act}(\traS{\tau})^\ast \hVV$. We let $\sV
\in \Act^\ast$ range over \emph{lists} of visible actions and write $\hV \wtraS{\sV} \hVV$ to denote $\hV \wtraS{\actE_1}\ldots\wtraS{\actE_n} \hVV$ where $\sV = \actE_1\ldots\actE_n$.

\begin{example} \label{ex:oper-semantics}
  Recall property \hV\ from \eqref{intro:1} of \exref{ex:background} (restated below). 
  {\setlength{\abovedisplayskip}{5pt}
\setlength{\belowdisplayskip}{5pt}
\setlength{\abovedisplayshortskip}{5pt}
\setlength{\belowdisplayshortskip}{5pt}
\begin{align*}     
    \hV \;\deftxt\;\;& \mmax{\,\hVarY}{\mnec{\mrecv{i}{\tup{\eatom{inc},x,y}}}
        \begin{mlbrace}(\,\mnec{\msendD{j}{y}{\tup{\eatom{res},\smash{x+1}}}}\,\hVarY)   \; \mand \; (\,\mnec{\msendD{\_\,}{y}{\eatom{err}}}\,\mfls) \end{mlbrace}} \qquad \eqref{intro:1}
  \end{align*}}
  Using the semantics of \figref{fig:logic-transition-rules}, we can express how formula $\hV$ reduces to a violation when executed \wrt to the following trace: 
  	{\setlength{\abovedisplayskip}{3pt}
\setlength{\belowdisplayskip}{3pt}
\setlength{\abovedisplayshortskip}{4pt}
\setlength{\belowdisplayshortskip}{4pt}
  	\begin{align*}
		\mrecv{i}{\tup{\eatom{inc},5,h}};\;\msendD{j}{h}{\tup{\eatom{res},6}};\;\mrecv{i}{\tup{\eatom{inc},3,h'}};\;\msendD{k}{h'}{\eatom{err}}
	\end{align*}} The runtime derivation is given as follows: \vspace{-2mm}
  
  {\setlength{\abovedisplayskip}{3pt}
\setlength{\belowdisplayskip}{5pt}
\setlength{\abovedisplayshortskip}{3pt}
\setlength{\belowdisplayshortskip}{5pt}
\setstretch{1.1}
  \begin{align*}
    \hV &\wtraS{\scriptstyle\mrecv{i}{\tup{\eatom{inc},5,h}}} \bigl( (\,\mnec{\msendD{j}{y}{\tup{\eatom{res},\smash{x+1}}}}\,\hV)   \,\mand \, (\,\mnec{\msendD{\_\,}{y}{\eatom{err}}}\,\mfls)\bigr) \sset{x\!\mapsto\!5, y\!\mapsto\! h} \\
     & \qquad \equiv \bigl( (\,\mnec{\msendD{j}{h}{\tup{\eatom{res},\smash{5+1}}}}\,\hV)   \,\mand \, (\,\mnec{\msendD{\_\,}{h}{\eatom{err}}}\,\mfls)\bigr) 
\intertext{(\hV\, reduced with rules \rtit{rMax} and \rtit{rNc1}, where {\small$\match{\mrecv{i}{\tup{\eatom{inc},x,y}}}{\mrecv{i}{\tup{\eatom{inc},5,h}}}{\!\sset{x\!\mapsto\!5, y\!\mapsto\! h}}$})}
           & \wtraS{\scriptstyle\msendD{j\;}{\;h}{\tup{\eatom{res},6}}} \hV
             \wtraS{\scriptstyle\mrecv{i}{\tup{\eatom{inc},3,h'}}} \bigl((\,\mnec{\msendD{j}{h'}{\tup{\eatom{res},\smash{3+1}}}}\,\hV)   \mand  (\,\mnec{\msendD{\_\,}{h'}{\eatom{err}}}\,\mfls)\bigr) \wtraS{\scriptstyle\msendD{k\;}{\;h'}{\eatom{err}}}  \mfls \qquad \exqed
  \end{align*}
  }
\end{example}

\subsubsection{The Correspondence Evaluation}
Although we have developed LTS semantics which describe the runtime behaviour of the formulas expressed in \detecterGen's logic, we still do not have any guarantees that this semantics corresponds to other semantics already available for \detecterGen \cite{FraSey14}.

As outlined in \secref{sec:detecter-primer-2}, in \cite{FraSey14} the authors provide \emph{satisfaction semantics} describing how an actor system satisfies a formal property, along with \emph{violation semantics} which describe how a \emph{finite trace} of system events, violates a formal property expressed in \detecterGen's logic.

Similar to the way that the authors of \cite{FraSey14} prove semantic correspondence between the satisfaction and violation semantics, in Thm.~\ref{thm:1} we prove that our LTS semantics corresponds to the \emph{violation semantics}. The primary reason for proving correspondence \wrt the violations semantics (and not the satisfaction semantics), is that violations are defined \wrt a finite trace $t$, which in some sense is analogous to the way a formula \hV\ reduces into \hV' over a trace $t$, \ie $\hV\wtra{\sV}\hV'$.

\begin{theorem}[\textbf{Semantic Correspondence}] \label{thm:1}
	$\\$\indent\indent$\vsat{\actV}{\sV}{\varphi} \quad\text{iff}\quad (\varphi\traceE{\sV}\mfls  \;\text{ and }\; \actV \wtra{\sV})$
\end{theorem}

\noindent In Thm.~\ref{thm:1} we prove that if an arbitrary actor system $A$ generates a trace $t$ which violates $\hV$, \ie $\vsat{\actV}{\sV\!}{\!\hV}$, then for the same trace $t$, the formula \hV\ should reduce to $\mfls$ (\ie to a violation) with our LTS semantics. We also prove the converse, \ie if \hV\ reduces to $\mfls$ \wrt some trace $t$ with our LTS semantics, then the respective violation relation should also hold.

We here give an outline of the main cases of the proof for Thm.~\ref{thm:1}. The reader may safely skip the details of the proof and proceed with \secref{sec:ra-derivative-semantics}.

\begin{proof} During this proof we make reference to \lemmaref{lemmaA} (stated below).\\\bigskip
\noindent {\bfseries\itshape\lemmaref{lemmaA}.} $\\[-5mm]$\indent$\quad\hV\traS{\act}\hV' \;\text{ and }\; \vsat{\actVV}{\sV}{\hV'} \;\text{ and }\; \actV\wtraS{\act}\actVV  \;\imp\; \vsat{\actV}{\act\sV}{\hV}$\\
\noindent This Lemma is proved by rule induction on $\hV\traS{\act}\hV'$. We provide a proof outline of this lemma in Appendix \secref{sec:corr-lemma}. \bigskip
	
	\noindent We prove the \emph{only-if} case for Thm.~\ref{thm:1} by rule induction on $\vsat{\actV}{\sV}{\hV}$. We here outline the main cases.	
	
           \begin{description}\setstretch{0.5}
    \item[\underline{Case $\vsat{\actV}{\sV}{\mnec{\patE}{\hV}}$:}] We know  
             \begin{gather}
                        \sV = \actE\sV' \label{T1:2:0}\\
			\actV\wtraS{\actE}\actVV \label{T1:2:1}\\
			\match{\patE}{\actE}{\sigma} \label{T1:2:2} \\
			\vsat{\actVV}{\sV'}{\hV\sigma} \label{T1:2:3} 
		\end{gather}
		By \eqref{T1:2:3} and IH we know 
		\begin{gather}
			\hV\sigma\traceE{\sV'}\mfls \label{T1:2:4}\\
			\actVV \wtra{\sV'} \label{T1:2:5}
		\end{gather}
		Thus, by \eqref{T1:2:1} and \eqref{T1:2:5} we obtain $\actV\wtra{\actE\sV'}$, \ie $\actV\wtra{\sV}$ by (\ref{T1:2:0}), and by  \eqref{T1:2:2}, \eqref{T1:2:4} and \textsc{rNc1} we derive $\mnec{\patE}{\hV}\wtra{\actE\sV'}\mfls$, and by (\ref{T1:2:0}) this means  $\mnec{\patE}{\hV}\wtra{\sV}\mfls$ as required.\\
		
   \item[\underline{Case $\vsat{\actV}{\sV}{\mmax{\hVarX}{\hV}}$:}] We know 
              \begin{gather}
			\vsat{\actV}{\sV}{\hV\sub{\hVarX}{\mmax{\hVarX}{\hV}}} \label{T1:3:1}		
		\end{gather}
		By \eqref{T1:3:1} and IH we know 
		\begin{gather}
			\hV\sub{\hVarX}{\mmax{\hVarX}{\hV}}\traceE{\sV}\mfls \label{T1:3:2}\\
			\actV \wtra{\sV} \label{T1:3:3}
		\end{gather}
		By  \textsc{rMax} we know  $\mmax{\hVarX}{\hV}\tra{\tau}\hV\sub{\hVarX}{\mmax{\hVarX}{\hV}}$, and hence, by\eqref{T1:3:2}, we obtain  	$\mmax{\hVarX}{\hV}\wtra{\sV}\mfls$ as required.\\	
		
	\item[\underline{Case $\vsat{\actV}{\sV}{\hV\mand\hVV}$:}] We know
		\begin{align}
			& \vsat{\actV}{\sV}{\hV} \label{T1:4:1}	 \\
			\text{or} \quad &\vsat{\actV}{\sV}{\hVV} \label{T1:4:2}	
		\end{align}
		We must consider the following cases:
		\item[Subcase $\sV=\varepsilon$:\;]
			By \eqref{T1:4:1}, \eqref{T1:4:2} and since $\sV=\varepsilon$ we know
			\begin{align}
				& \hV=\mfls \label{T1:4:3}	 \\
				\text{or} \quad &\hVV=\mfls \label{T1:4:4}	
			\end{align}
			By \eqref{T1:4:3} and since $\mfls\mand\hVV\equiv\mfls$ we know $\mfls\mand\hVV\wtra{\varepsilon}\mfls$ as required. Similarly, by \eqref{T1:4:4} and since $\hV\mand\mfls\equiv\mfls$ we know $\hV\mand\mfls\wtra{\varepsilon}\mfls$.\\
	
		\item[Subcase $\sV=\actE\sV$:\;]
			Since $\sV=\actE\sV$, by \eqref{T1:4:1} and IH we know 
			\begin{gather}
				 \hV\wtra{\actE\sV}\mfls \label{T1:4:5}	 \\
				 \actV \wtra{\actE\sV} \label{T1:4:6}	
			\end{gather}
			Similarly, since $\sV=\actE\sV$, by \eqref{T1:4:2} and IH we know 
			\begin{gather}
				 \hV\wtra{\actE\sV}\mfls \label{T1:4:7}	 \\
				 \actV \wtra{\actE\sV} \label{T1:4:8}	
			\end{gather}
			From \eqref{T1:4:5} and \eqref{T1:4:7} we can deduce
			\begin{gather}
				 \hV\traS{\actE}\hV' \label{T1:4:9}	 \\
				 \hV'\wtra{\sV}\mfls \label{T1:4:10}	 \\
				 \hVV\traS{\actE}\hVV' \label{T1:4:11}	 \\
				 \hVV'\wtra{\sV}\mfls \label{T1:4:12}	
			\end{gather}
			By \eqref{T1:4:9}, \eqref{T1:4:11} and \textsc{rCn1} we know
			\begin{gather}
				\hV\mand\hVV\traS{\actE}\hV'\mand\hVV' \label{T1:4:13}
			\end{gather}
			By \eqref{T1:4:10}, \eqref{T1:4:12} and \eqref{T1:4:13} we know
			\begin{gather}
				\hV\mand\hVV\traS{\actE}\hV'\mand\hVV'\wtra{\sV}\mfls\mand\mfls \quad \equiv \quad \hV\mand\hVV\wtra{\actE\sV}\mfls \label{T1:4:14} 
			\end{gather}
			Hence this subcase holds by \eqref{T1:4:8} and \eqref{T1:4:14}.\\
		
		\item[Subcase $\sV=\tau\sV\;$ ($\hV$ makes the $\tau$ transition):\;]
			Since $\sV=\tau\sV$, by \eqref{T1:4:2} and IH we know 
			\begin{gather}
				 \hV\wtra{\tau\sV}\mfls \label{T1:4:15}	 \\
				 \actV\wtra{\tau\sV} \quad \equiv \quad \actV\wtra{\sV} \label{T1:4:16}	
			\end{gather}			
			From \eqref{T1:4:15} we can deduce
			\begin{gather}
				 \hV\traS{\tau}\hV' \label{T1:4:17}	 \\
				 \hV'\wtra{\sV}\mfls \label{T1:4:18}	  
			\end{gather}
			By \eqref{T1:4:17} and \textsc{rCn2} we know
			\begin{gather}
				\hV\mand\hVV\traS{\tau}\hV'\mand\hVV \label{T1:4:19}
			\end{gather}
			We know that for trace $\sV$, $\hVV$ reduces to some $\hVV'$ such that by \eqref{T1:4:19} we can conclude
			\begin{gather}
				\hV\mand\hVV\traS{\tau}\hV'\mand\hVV\wtra{\sV}\mfls\mand\hVV' \quad \equiv \quad \hV\mand\hVV\wtra{\sV}\mfls \label{T1:4:20} 
			\end{gather}
			Hence this subcase holds by \eqref{T1:4:16} and \eqref{T1:4:20}.\\
				
		\item[Subcase $\sV=\tau\sV\;$ ($\hVV$ makes the $\tau$ transition):\;]
			$\\[3mm]$The proof for this subcase is synonymous to that of the previous subcase.
	\end{description} \bigskip

\noindent We prove the \emph{if} case by numerical induction on the number of transitions used in $\hV\wtraS{\sV}\mfls$.\smallskip
\begin{description}\setstretch{0.5} \itemsep0em
   \item[$n=0$:] We know that $\hV=\mfls$ and $\sV=\varepsilon$ and we trivially have $\vsat{\actV}{\varepsilon}{\mfls}$ for arbitrary \actV.
   \item[$n=k+1$:] For some $\hV'$ we know $t=\act t'$ such that	
		\begin{gather}
			\hV\traS{\act}\hV' \label{T1:6:3}\\
			\hV'\wtra{\sV'}\mfls \quad \text{ where } |\,\wtra{\;\sV'\;}\!| = k.\label{T1:6:4} 			
		\end{gather}
		Similarly we also know 	
		\begin{gather}			
			\actV\wtraS{\act}\actVV \label{T1:6:5} \\
			\actVV\wtra{\sV'} \label{T1:6:6} 					
		\end{gather}
		By \eqref{T1:6:4}, \eqref{T1:6:6} and IH we know 
		\begin{gather}			
			\vsat{\actVV}{\sV'}{\hV'} \label{T1:6:7} 					
		\end{gather}	
		By \eqref{T1:6:3}, \eqref{T1:6:5}, \eqref{T1:6:7} and \lemmaref{lemmaA}\! we know $\vsat{\actV}{\act\sV'}{\hV'}$ as required.			
\end{description}	\vspace{-10mm}
\end{proof}\vspace{-3mm}

\subsection{Extending the Operational Model with Runtime Adaptation}
\label{sec:ra-derivative-semantics}
\figref{fig:rt-semantics} (below) describes 
a semantics for the extended logic with adaptations as discussed in \secref{sec:syn-asyn-adaptations} to \ref{sec:lang-implementation}.  Recall that in Table~\ref{tab:syn-asyn-adapts} (given in \secref{ra:impl-issues}) we distinguished between synchronous and asynchronous classes of adaptations. Given that adaptations pertaining to the same class are applied (to the adaptee) in the same manner, in this extended logic we work at this level of abstraction.

We therefore represent asynchronous and synchronous classes of adaptations as $\macor{\vLstB}{\vLstA}$ and $\mscor{\vLstB}{\vLstA}$ respectively: they both take two lists of actor references as argument --- $\vLstB,\vLstA \in \bigl(\Pid \cup \Vars\bigr)^\ast$ --- which are synonymous to the adaptation and release lists. Furthermore, the extended logic also includes the extended necessity, $\mattrNec{\patE}{\attr}{\vLstB}$, where in this case $\vLstB$ is analogous to the release list, \textsf{RL}, of the necessity. 

\begin{figure}[ht!]\vspace{-2mm}
\begin{small} \setstretch{0.1}
	\textbf{Extended Logic Syntax with Adaptations and Synchronisations} \bigskip
	\begin{mathpar}
		\hV, \hVV \in \FRM \bnfdef \; \ldots \quad \bnfsep \quad \mattrNec{\patE}{\attr}{\vLstB}{\hV}   \quad \bnfsep \quad \macor{\vLstA}{\vLstB}\hV \quad \bnfsep \quad \mscor{\vLstA}{\vLstB}\hV \quad \bnfsep \quad \mblock{\vLstA}\hV  \quad \bnfsep \quad \mrelease{\vLstA}\hV 
	\end{mathpar} 
	\bigskip
	
	\textbf{Monitor Transition Rules}\bigskip
	\begin{mathpar}
	   \inference[\rNecA]{\match{\patE}{\actE}{\sigma} & \id{\actE}=\idV}{\mBNec{\patE}{\vLstA}{\varphi} \traS{\actE} \mblock{\idV}(\varphi\sigma)}  \qquad
	  \inference[\rNecB]{\match{\patE}{\actE}{\sigma}}{\mANec{\patE}{\vLstA}{\hV} \traS{\actE} \hV\sigma} \\
	  \inference[\rNecC]{\match{\patE}{\actE}{\bot}}{\mattrNec{\patE}{\attr}{\vLstA}{\varphi} \traS{\actE} \mrelease{\vLstA}\mtru} \quad 
	  \inference[\rAAct]{}{\macor{\vLstB}{\vLstA}\hV \traS{\cora{\vLstB}} \mrelease{\vLstA}\hV} \quad
	  \inference[\rSAct]{}{\mscor{\vLstB}{\vLstA}\hV \traS{\cors{\vLstB}} \mrelease{\vLstA}\hV} \\
	  \inference[\rRel]{}{\mrelease{\vLstA}\hV \traS{\rel{\vLstA}} \hV} \qquad 
	  \inference[\rBlk]{}{\mblock{\vLstA}\hV \traS{\blk{\vLstA}} \hV} \qquad
	  \inference[\rAndD]{\hV \traS{\actC} \hV'}{\hV\mand\hVV \traS{\actC} \hV'\mand\hVV} 
	\end{mathpar} 
	\bigskip
	
	\textbf{System Transition Rules}\bigskip
		\begin{mathpar}	
			\inference[\sNew]{}{\stV \traS{\tau}\stV, \mapstate{\idV}{\unblocked}} \qquad 
	                \inference[\sAct]{\id{\actE} = \idV & \ids{\actE} \subseteq \dom(\stV)}{\stV,\mapstate{\idV}{\unblocked} \traSS{\actE} \stV,\mapstate{\idV}{\unblocked}}  \qquad 
	                \inference[\rtit{sAdA}]{\vLstB \subseteq \dom(\stV)}{\stV \traS{\cora{\vLstB}} \stV}\\
			\inference[\sBlk]{}{\stV,\mapstate{\vLstB}{\unblocked} \traS{\blk{\vLstB}} \stV,\mapstate{\vLstB}{\blocked}} 
	                \qquad
	                \inference[\sRel]{}{\stV,\mapstate{\vLstB}{\blocked} \traS{\rel{\vLstB}} \stV,\mapstate{\vLstB}{\unblocked}} 
	                \qquad
	                \inference[\rtit{sAdS}]{}{\stV,\mapstate{\vLstB}{\blocked} \traS{\cors{\vLstB}} \stV,\mapstate{\vLstB}{\blocked}} 	
		\end{mathpar}
	 \bigskip
	 
	\textbf{Instrumentation Transition Rules} \bigskip
	\begin{mathpar}
	   \inference[\rtit{iAda}]{\hV \traS{\actC} \hV' & \stV \traS{\actC} \stV'}{\instr{\stV}{\hV} \traS{\tau} \instr{\stV'}{\hV'}} \qquad 
	   \inference[\rtit{iAct}]{  \hV  \notTraceEvent{\;\actC\;}  &   \stV \traS{\actE} \stV'  & \hV \traS{\actE} \hV'}{\instr{\stV}{\hV} \traS{\actE} \instr{\stV'}{\hV'}} \\
	   \inference[\rtit{iTrm}]{\hV  \notTraceEvent{\;\actC\;}  &   \stV \traS{\actE} \stV'  &  \hV  \notTraceEvent{\;\actE\;}}{\instr{\stV}{\hV} \traS{\actE} \instr{\stV'}{\mtru}} \qquad
	   \inference[\rtit{iSys}]{\stV \traS{\tau} \stV}{\instr{\stV}{\hV} \traS{\tau} \instr{\stV'}{\hV}} \qquad
	   \inference[\rtit{iMon}]{\hV \traS{\tau} \hV'}{\instr{\stV}{\hV} \traS{\tau} \instr{\stV}{\hV'}}		
	\end{mathpar}
\end{small}
\caption{A Runtime Semantics for Instrumented Properties with Adaptations}
\label{fig:rt-semantics}
\end{figure}

The extended logic also uses two additional constructs, $\mblock{\vLstA}\hV$  and $\mrelease{\vLstA}\hV$; these are not meant to be part of the specification scripts, but are used as part of the \emph{runtime syntax}.  Since the extended logic can also affect the system being monitored, through \emph{adaptations} and \emph{synchronisations}, the operational semantics is given in terms of \emph{configurations}, $\instr{\stV}{\hV}$, where $\stV$ is an \emph{abstract representation} of the system and $\hV$ is a \emph{closed formula}. Note that in the previous operational model (given in \figref{fig:logic-transition-rules}), giving an abstract representation of the system was not necessary as the formulas were \emph{passive}, \ie unable to effect the system. By contrast in the extended logic we now include adaptations and synchronisation mechanisms that can effect the system; an abstract representation $s$ therefore allows us to model some of these effects at a high level of abstraction.

In addition to closed formulas $\hV$, configurations thus include the monitored system represented abstractly as a partial map, $\stV :: \Pid \rightharpoonup \sset{\blocked,\unblocked}$, describing whether an actor (through its unique identifier) is currently blocked (suspended), $\blocked$, or executing, $\unblocked$.    We occasionally 
write $\mapstate{\vLstB}{\blocked}$ to denote the list of mappings $\mapstate{\idV_1}{\blocked},\ldots,\mapstate{\idV_n}{\blocked}$ where $\vLstB = \idV_1,\ldots,\idV_n$  (similarly for $\mapstate{\vLstB}{\unblocked}$).

In the operational model given in \figref{fig:rt-semantics}, we also distinguish between \emph{two} kinds of reductions that can be carried out by a \emph{configuration}, namely $\actE$-reductions, formalised as $\instr{\stV}{\hV}\traS{\actE}\instr{\stV'}{\hV'}$, and $\tau$-reductions \ie $\instr{\stV}{\hV}\traS{\tau}\instr{\stV'}{\hV'}$. The former defines cases where a system $s$ commits an \emph{external} monitorable event $\actE$ which is perceived by monitor $\hV$; due to this event, both the system and the monitor may evolve together thus forming a new configuration $\instr{\stV'}{\hV'}$. The latter (\ie $\tau$-reductions) formalise cases where the whole configuration $\instr{\stV}{\hV}$ performs an \emph{internal} transition into another configuration $\instr{\stV'}{\hV'}$ where $\stV'$ and $\hV'$ are the evolved versions of $\stV$ and $\hV$ respectively. These internal transitions can be caused in the following cases:
\begin{itemize}\setstretch{1.1}
	\item System $\stV$ performs an internal operation which is unobservable by $\hV$, and thus $\stV$ evolves into some $\stV'$ independently of formula $\hV$.
	\item Formula $\hV$ performs the internal transition thereby reducing into some formula $\hV'$, independently from the system $\stV$.
	\item System $\stV$ and formula $\hV$ \emph{interact} over a \emph{synchronisation} operation or an \emph{adaptation} action $-$ this causes both $\stV$ and $\hV$ to evolve together into some new $\stV'$ and $\hV'$, according to the exchanged interaction.  
\end{itemize}

More specifically, to describe adaptation interactions between the monitor and the system, the LTS semantics of \figref{fig:rt-semantics} employs four additional labels, ranged over by the variable $\actC$.  These include the asynchronous and synchronous adaptation labels, \cora{\vLstB} and \cors{\vLstB}, to denote respectively that an asynchronous and synchronous action affecting actors \vLstB, has been executed.  They also include a blocking action, \blk{\vLstB},  and an unblocking (release) action, \rel{\vLstB},  affecting the execution of actors with identifiers in the list  \vLstB.

The semantics is defined in terms of three LTSs: one for logical formulas (\ie monitors), one for systems, and one for configurations which is based on the other two LTSs.  

\subsubsection{The Monitor Transition Rules}
These extend the rules in \figref{fig:logic-transition-rules} with the exception of those for the necessity formulas, which are replaced by rules \rNecA, \rNecB\ and \rNecC.  Whereas the new \rNecB\ follows the same format as that of \rtit{rNc1} from \figref{fig:logic-transition-rules}, the rule for \emph{blocking} necessity formulas, \rNecA, transitions into a blocking construct,  $\mblock{\idV}\hV$, for blocking the subject of the action (\ie actor \idV) in case a pattern match is successful. Note that the action's subject is retrieved using the $\id{\patE}$ function as defined in \defref{def:subj} (below). \noindent Furthermore, in case an action mismatches a necessity pattern, \rNecC\ transitions the necessity formula to a release construct,  $\mrelease{\vLstA}\hV$, with the specified release list of actors, $\vLstA$. \medskip

\begin{definition}[The subject Function]\label{def:subj}
{\setstretch{1}\setlength{\abovedisplayskip}{0pt}
\setlength{\belowdisplayskip}{0pt}
\setlength{\abovedisplayshortskip}{0pt}
\setlength{\belowdisplayshortskip}{0pt}\begin{align*}
\id{\gamma} &\defEquals \begin{cases} 
					   \varid_{1} & \text{if }\patE=\msendD{\varid_{1}}{\varid_{2}}{p}\\
					   \varid & \text{if }\patE=\mrecv{\varid}{p}\\
					   \varid & \text{if }\patE=\mcall{\varid}{p}\\
					   \varid & \text{if }\patE=\mret{\varid}{p}\\
					  \end{cases}
\end{align*}}
This function inspects a necessity pattern $\patE$ and extracts the process identifier of the actor that committed the action. In particular note that in output actions \eg $\msendD{i}{j}{3}$, the subject is $i$, as this actor is sending the message to $j$, and hence it is the one performing the output action. \bqed
\end{definition}

Asynchronous and synchronous adaptation formulas, $\macor{\vLstB}{\vLstA}\hV$ and $\mscor{\vLstB}{\vLstA}\hV$, transition with the respective labels (\ie $\cora{\vLstB}$ and $\cors{\vLstB}$), using rules \rtit{rAdA} and \rtit{rAdS}, to a release construct, $\mrelease{\vLstA}$, thereby communicating the respective adaptations to $\vLstB$. Similarly the block and release constructs reduce over labels (\ie \blk{\vLstA} and \rel{\vLstA} with rules \rtit{rRel} and \rtit{rBlk} respectively.  Finally, rule \rAndD\ allows monitor actions affecting the system, \ie $\actC$, to be carried out under a conjunction formula, independent of the other branch; we elide the obvious symmetric rule \rAndE. 

\subsubsection{The System Transition Rules}
These rules allow further actor spawning (rule \rtit{sNew} in \figref{fig:rt-semantics}) but restrict external actions $(\actE)$ to those whose subject is currently active, \ie unblocked  $\mapstate{\idV}{\unblocked}$ (rule \rtit{sAct}); external actions model the system events that can be reported to the monitor.  Whereas asynchronous adaptations can be applied to any actor list, irrespective of their status \ie can be either $\mapstate{\vLstB\!}{\!\unblocked}$ or $\mapstate{\vLstB\!}{\!\blocked}$ (rule \rtit{sAdA}), synchronous ones require the adaptees to be blocked \ie must be $\mapstate{\vLstB\!}{\!\blocked}$ (rule \rtit{sAdS}). This transpires from our implementation constraints where synchronous adaptations require the respective system actors to be blocked by our instrumented code in some prior stage, so that the monitor can later on communicate synchronous adaptations as messages, which are then received and interpreted by the instrumented actor. By contrast, this is not required by asynchronous adaptations.  

Finally rules \rtit{sBlk} and \rtit{sRel} model the blocking and releasing mechanism used by the RA protocol code instrumented in the system. In fact, blocking is modelled in terms of a transition which changes the status of the respective adaptees from active $\mapstate{i}{\unblocked}$, to blocked $\mapstate{i}{\blocked}$; the converse applies for releasing.

\subsubsection{The Instrumentation Rules}
The \emph{instrumentation rules} for configurations describe how system (visible) actions, \actE, affect the monitors, and dually, how the monitor adaptation and synchronisation actions, \actC, affect the system. For instance, if the monitor instigates  action \actC\ and the system allows it, they both transition together as a silent (internal) transition (rule \rtit{iAda}).  Dually, if the system generates external action \actE and the monitor can observe it, they also transition together (rule \rtit{iAct}); this essentially models the event reporting mechanism whereby monitorable events are forwarded to the monitor by the system. 

If the monitor cannot observe an external $\actE$-action (rule \rtit{iTrm}), it terminates as formula \mtru. Importantly note that both rules \rtit{iAct} and \rtit{iTrm} require the monitor not to be in a position to perform an adaptation/synchronisation action, \ie premise $\hV  \notTraceEvent{\;\actC\;}$; this gives \emph{precedence} to monitor actions over system ones in our instrumentation. This precedence models the fact that in our RA protocol implementation (refer to \secref{sec:lang-implementation}), the monitoring loop only perceives a system event by reading it from its mailbox \emph{after} it is done applying the necessary handling actions, which may include $\actC$-actions \eg issuing an adaptation. Rules \rtit{iSys} and \rtit{iMon} allow systems and monitors to perform internal transitions (\wrt $\tau$-actions) independently of each other. \medskip

\begin{remark} We point out the subtle difference between rules \rtit{iAct} and \rtit{iAda}. Rule \rtit{iAct} states that whenever the system $s$ performs an external $\actE$-action which is perceived by formula $\hV$, then the \emph{entire configuration} $\instr{\stV}{\hV}$ transitions over the external action, \ie $\instr{\stV}{\hV}\traS{\actE}\instr{\stV'}{\hV'}$. This opposes rule \rtit{iAda} in which whenever the system $\stV$ and formula $\hV$ interact over a synchronisation or adaptation action $\actC$, then the configuration transitions \emph{silently}, \ie $\instr{\stV}{\hV}\traS{\tau}\instr{\stV'}{\hV'}$ instead of $\instr{\stV}{\hV}\traS{\actC}\instr{\stV'}{\hV'}$. This transpires since external $\actE$-actions may be observable by other external systems which interact with the monitored system, and hence these actions must be kept observable. Contrasting this, $\actC$-actions model interactions that are only relevant (\ie internal) to the system and monitor, and hence they are not be made perceivable to external systems. \bqed
\end{remark}

\subsubsection{Formalising Synchronisation Errors}
In the following example we use our extended operational model to demonstrate how adaption formulas reduce at runtime when executed \wrt a system. More importantly we show the subtle errors that erroneously specified synchronisation operations may introduce at runtime.
\begin{example}
  \label{ex:model}  Recall the adaptation formula $\hV'$ defined in (\ref{eq:ra:4}) of \exref{ex:adapt-and-block} (restated below).  
   \begin{align*} \setstretch{1.1}
  \begin{array}{l} 
    \hV'\;\deftxt\;\; \mmax{\,\hVarY}{\mANec{\mrecv{\idV}{\tup{\eatom{inc},x,y}}}{\epsilon}\\[-2mm]
      \qquad\qquad\qquad\quad\mBNec{\msendD{\idV}{\_}{\{\eatom{inc},x,y\}}}{\varepsilon}
                \begin{mlbrace} 
        (\,\mANec{\msendD{j}{y}{\tup{\eatom{res},\smash{x+1}}}}{\epsilon}\,\hVarY)   \;\mand \; \\[0mm] 
        (\,\mBNec{\msendD{z\,}{y}{\eatom{err}}}{i}\,\mrestart{i}_{\varepsilon}\,\mclearMailbox{z}_{i,z}\,\hVarY) 
      \end{mlbrace}} \qquad \eqref{eq:ra:4}
     \end{array}
  	\end{align*}  	
  Now, consider the following system $\smash{\stV=(\mapstate{(\idV,\idVV,\idVVVV,\idVVV)}{\unblocked})}$ where all of the system actors (\ie $i,j,k$ and $h$) are unblocked. We can therefore model the runtime execution of formula $\hV'$ \wrt system $\stV$ with adaptations as follows:
  \begin{align}
  \label{eq:ra:5}
   \instr{\stV}{\hV'} &  \wtraS{\scriptstyle\mrecv{\idV}{\tup{\eatom{inc},1,\idVVV}}}\cdot\wtraS{\scriptstyle\msendD{\idV\;\,}{\;\idVVVV}{\{\eatom{inc},1,\idVVV\}}} \instr{\stV}{\mblock{\idV} 
    \left(\begin{array}{l}
\mANec{\msendD{j}{\idVVV}{\tup{\eatom{res},2}}}{\varepsilon}\,\hV'   \;\mand\; \\
        \mBNec{\msendD{z\,}{\idVVV}{\eatom{err}}}{i}\,\mrestart{\idV}_{\varepsilon}\,\mclearMailbox{z}_{\idV,z}\,\hV' 
    \end{array}\right)
}\\
\label{eq:ra:3}
  & \traS{\scriptstyle\tau} \instr{\bigl(\mapstate{(\idVV,\idVVV,\idVVVV)}{\unblocked},\mapstate{\idV}{\blocked}\bigr)}{
    \left(
\mANec{\msendD{j}{\idVVV\,}{\tup{\eatom{res},2}}}{\varepsilon}\hV'   \mand  
        \mBNec{\msendD{z\,}{\idVVV}{\eatom{err}}}{i}\mrestart{\idV}_{\varepsilon}\mclearMailbox{z}_{\idV,z}\hV' 
    \right)
}\\
  \label{eq:ra:6}
  & \traS{\scriptstyle\msendD{\idVVVV\;}{\;\idVVV}{\eatom{err}}}  \instr{\bigl(\mapstate{(\idVV,\idVVV,\idVVVV)}{\unblocked},\mapstate{\idV}{\blocked}\bigr)}{\mblock{\idVVVV}\,\mrestart{\idV}_{\varepsilon}\,\mclearMailbox{\idVVVV}_{i,\idVVVV}\,\hV'}\\
  \label{eq:ra:7}
  & \traS{\scriptstyle\tau} \instr{\bigl(\mapstate{(\idVV,\idVVV)}{\unblocked},\mapstate{\idV}{\blocked},\mapstate{\idVVVV}{\blocked}\bigr)}{\mrestart{\idV}_{\varepsilon}\,\mclearMailbox{\idVVVV}_{i,\idVVVV}\,\hV'} \\ 
   \label{eq:ra:8}
   & \wtraS{\scriptstyle\tau} \instr{\bigl(\mapstate{(\idVV,\idVVV)}{\unblocked},\mapstate{\idV}{\blocked},\mapstate{\idVVVV}{\blocked}\bigr)}{\mrelease{\idV,\idVVVV}\hV'} \traSS{\tau} \instr{\stV}{\hV'}
  \end{align}
 In particular, the blocking necessities matched in (\ref{eq:ra:5}) and (\ref{eq:ra:6}) yield the actor blocking constructs which block actors $i$ and $k$ in (\ref{eq:ra:3}) and (\ref{eq:ra:7}) respectively.  This allows for the synchronous adaptations in \eqref{eq:ra:8} to be performed, thereby proceeding to release of the respective actors. 

 Erroneous blocking directives however may lead to stuck synchronous adaptations \ie rule \rtit{sAdS} is inapplicable on the system.  For instance, if we change the first blocking necessity in   $\hV'$ of (\ref{eq:ra:4}) to an asynchronous one, $\mattrNec{\msendD{\idV}{\_}{\{\eatom{inc},x,y\}}}{\scriptstyle\pmb{\norm}}{\varepsilon}$, it yields the execution below, where $\hV''$ is the erroneous formula:
  \begin{align*} \setstretch{1.1}
   \begin{array}{l}
    \hV''\;\deftxt\;\; \mmax{\,\hVarY}{\mANec{\mrecv{\idV}{\tup{\eatom{inc},x,y}}}{\epsilon}
   		 \mattrNec{\msendD{\idV}{\_}{\{\eatom{inc},x,y\}}}{\scriptstyle\pmb{\norm}}{\varepsilon}
                \begin{mlbrace} 
        (\,\mANec{\msendD{j}{y}{\tup{\eatom{res},\smash{x+1}}}}{\epsilon}\,\hVarY)   \;\mand \; \\[0mm] 
        (\,\mBNec{\msendD{z\,}{y}{\eatom{err}}}{i}\,\mrestart{i}_{\varepsilon}\,\mclearMailbox{z}_{i,z}\,\hVarY) 
      \end{mlbrace}}
      \end{array} 
  	\end{align*}
  	\vspace{-15mm}
   \begin{align}\label{eq:ra:9}
     \instr{\stV}{\hV''} &  \wtraS{\scriptstyle\mrecv{\idV\,}{\tup{\eatom{inc},1,\idVVV}}}\cdot\wtraS{\scriptstyle\msendD{\idV\;}{\;\idVVVV}{\{\eatom{inc},1,\idVVV\}}}\cdot \wtraS{\scriptstyle\msendD{\idVVVV\;}{\;\idVVV}{\eatom{err}}}  \instr{\bigl(\mapstate{(\idV,\idVV,\idVVV)}{\unblocked},\mapstate{\idVVVV}{\blocked}\bigr)}{\mrestart{\idV}_{\varepsilon}\,\mclearMailbox{\idVVVV}_{i,\idVVVV}\,\hV''} \vspace{-3mm}
   \end{align}
 The final configuration in \eqref{eq:ra:9} (\ie $\instr{\bigl(\mapstate{(\idV,\idVV,\idVVV)}{\unblocked},\mapstate{\idVVVV}{\blocked}\bigr)}{\mrestart{\idV}_{\varepsilon}\,\mclearMailbox{\idVVVV}_{i,\idVVVV}\,\hV''}$) is stuck because the synchronous adaptation on \idV (\ie $\mrestart{\idV}$) cannot be carried out since \idV is \emph{not} blocked (\ie since $\mapstate{\idV}{\unblocked}$). A similar situation is reached if a blocked actor is \emph{released prematurely} by a concurrent branch.  For instance, if we erroneously change the release list of the necessity subformula $\mANec{\msendD{j}{y}{\tup{\eatom{res},\smash{x+1}}}}{\varepsilon}$ from $\varepsilon$ to $\idV$, this releases $\idV$ upon mismatch, interfering with adaptation actions along the other branch of the conjunction. Hence this erroneous formula ($\hV'''$) exhibits a \emph{race condition} whereby it can non-deterministically yield a number of executions, some of which may lead to stuck configurations. Here we present two possible executions, where one of them applies the $\mrestart{\idV}$ adaptation erroneously:
 \begin{align*} \setstretch{1.1} 
 	 \begin{array}{l}
    \hV'''\;\deftxt\;\; \mmax{\,\hVarY}{\mANec{\mrecv{\idV}{\tup{\eatom{inc},x,y}}}{\epsilon}
     	\mBNec{\msendD{\idV}{\_}{\{\eatom{inc},x,y\}}}{\varepsilon}
                \begin{mlbrace} 
        (\,\mANec{\msendD{j}{y}{\tup{\eatom{res},\smash{x+1}}}}{\scriptstyle\pmb{i}}\,\hVarY)   \;\mand \; \\[0mm] 
        (\,\mBNec{\msendD{z\,}{y}{\eatom{err}}}{i}\,\mrestart{i}_{\varepsilon}\,\mclearMailbox{z}_{i,z}\,\hVarY) 
      \end{mlbrace}}
     \end{array}
 \end{align*}
 \vspace{-13mm}
 \begin{align}
	    \instr{\stV}{\hV'''} &  \wtraS{\scriptstyle\mrecv{\idV}{\tup{\eatom{inc},1,\idVVV}}}\cdot\wtraS{\scriptstyle\msendD{\idV\;}{\;\idVVVV}{\{\eatom{inc},1,\idVVV\}}} \instr{\bigl(\mapstate{(\idVV,\idVVV,\idVVVV)}{\unblocked},\mapstate{\idV}{\blocked}\bigr)}{
	    \left(\begin{array}{l}
	\mANec{\msendD{\idVV}{\idVVV}{\tup{\eatom{res},2}}}{\idV}\,\hV'   \;\mand\; \\
	        \mBNec{\msendD{z\,}{\idVVV}{\eatom{err}}}{\idV}\,\mrestart{\idV}_{\varepsilon}\,\mclearMailbox{z}_{\idV,z}\,\hV''' 
	    \end{array}\right)
	} \nonumber\\	
	  & \traS{\scriptstyle\msendD{\idVVVV\;}{\;\idVVV}{\eatom{err}}}  \instr{\bigl(\mapstate{(\idVV,\idVVV,\idVVVV)}{\unblocked},\mapstate{\idV}{\blocked}\bigr)}{\big(\mrelease{\idV}\mtru\big) \,\mand\, \big(\mblock{\idVVVV}\,\mrestart{\idV}_{\varepsilon}\,\mclearMailbox{\idVVVV}_{\idV,\idVVVV}\,\hV'''\big) \label{eq:ra:10}
	}
\end{align}	
\noindent\underline{\textbf{Execution 1: }The adaptation is \emph{erroneously} applied.}\\
The concurrent branch $\mrelease{i}\mtru$ in \eqref{eq:ra:10} is evaluated first, thereby releasing actor $i$ \emph{prior} to applying the restart action $\mrestart{\idV}$ defined in the opposing branch (\ie $\mblock{\idVVVV}\,\mrestart{\idV}_{\varepsilon}\ldots$). This leads to the stuck configuration shown in \eqref{eq:ra:11} where adaptation $\mrestart{\idV}_{\varepsilon}$ cannot be applied (with rule \rtit{iAda}) as actor $i$ is active, \ie $\mapstate{\idV}{\unblocked}$. \vspace{-5mm}
\begin{align}
	  \; & \wtraS{\scriptstyle\tau} \instr{\bigl(\mapstate{(\idV,\idVV,\idVVV)}{\unblocked},\mapstate{\idVVVV}{\blocked}\bigr)}{\mrestart{\idV}_{\varepsilon}\,\mclearMailbox{\idVVVV}_{\idV,\idVVVV}\,\hV'''} \label{eq:ra:11}
\end{align}	

\noindent\underline{\textbf{Execution 2: }The adaptation is \emph{validly} applied.}\\
The adaptation can however be validly applied in case the concurrent branch in \eqref{eq:ra:10} containing the restart adaptation (\ie branch $\mblock{\idVVVV}\,\mrestart{\idV}_{\varepsilon}\ldots$) is evaluated before the releasing branch as shown in \eqref{eq:ra:12} below.  
{\setlength{\abovedisplayskip}{5pt}
\setlength{\belowdisplayskip}{5pt}
\setlength{\abovedisplayshortskip}{5pt}
\setlength{\belowdisplayshortskip}{5pt}
\begin{align}
	  \; & \traS{\scriptstyle\tau} \instr{\bigl(\mapstate{(\idVV,\idVVV)}{\unblocked},\mapstate{(\idV,\idVVVV)}{\blocked}\bigr)}{\big(\mrelease{\idV}\mtru\big) \,\mand\, \big(\mrestart{\idV}_{\varepsilon}\,\mclearMailbox{\idVVVV}_{\idV,\idVVVV}\,\hV'''\big)} \nonumber \\ 
	  \; & \traS{\scriptstyle\tau} \instr{\bigl(\mapstate{(\idVV,\idVVV)}{\unblocked},\mapstate{(\idV,\idVVVV)}{\blocked}\bigr)}{\big(\mrelease{\idV}\mtru\big) \,\mand\, \big(\mclearMailbox{\idVVVV}_{\idV,\idVVVV}\,\hV'''\big)} \label{eq:ra:12}
\end{align}	}
\noindent Synchronisation errors caused due to concurrency can be more difficult for the specifier to notice without any assistance. Similarly, synchronisation errors can also be introduced in cases where the monitor attempts to release an unblocked actor. For instance, if we continue with the second execution (\ie from \eqref{eq:ra:12}), the configuration progresses to a point where \emph{both} concurrent branches attempt to release actor $i$. In the derivation below we consider two possible continuations, starting from \eqref{eq:ra:12}, both of which lead to a stuck configuration.\\[2mm]
\noindent\underline{\textbf{Continuation 1: }}\vspace{-9mm}
{\setlength{\abovedisplayskip}{5pt}
\setlength{\belowdisplayskip}{5pt}
\setlength{\abovedisplayshortskip}{5pt}
\setlength{\belowdisplayshortskip}{5pt}
\begin{align}
	  \; & \traS{\scriptstyle\tau} \instr{\bigl(\mapstate{(\idV,\idVV,\idVVV)}{\unblocked},\mapstate{\idVVVV}{\blocked}\bigr)}{\mtru \,\mand\, \big(\mclearMailbox{\idVVVV}_{\idV,\idVVVV}\,\hV'''\big)} \nonumber \\ 
	  \; & \traS{\scriptstyle\tau} \instr{\bigl(\mapstate{(\idV,\idVV,\idVVV)}{\unblocked},\mapstate{\idVVVV}{\blocked}\bigr)}{\big(\mrelease{\idV,\idVVVV}\,\hV'''\big)} \nonumber \\ 
	  \; & \traS{\scriptstyle\tau} \instr{\bigl(\mapstate{(\idV,\idVV,\idVVV,\idVVVV)}{\unblocked}\bigr)}{\big(\mrelease{\idV}\,\hV'''\big)} \label{eq:ra:13} 
\end{align}	}
\noindent\underline{\!\textbf{Continuation 2: }}\vspace{-9mm}
{\setlength{\abovedisplayskip}{5pt}
\setlength{\belowdisplayskip}{5pt}
\setlength{\abovedisplayshortskip}{5pt}
\setlength{\belowdisplayshortskip}{5pt}
\begin{align}
	  \; & \traS{\scriptstyle\tau} \instr{\bigl(\mapstate{(\idVV,\idVVV)}{\unblocked},\mapstate{(\idV,\idVVVV)}{\blocked}\bigr)}{\big(\mrelease{\idV}\mtru\big) \,\mand\, \big(\mrelease{\idV,\idVVVV}\,\hV'''\big)} \nonumber\\
	  \; & \wtraS{\scriptstyle\tau} \instr{\bigl(\mapstate{(\idV,\idVV,\idVVV,\idVVVV)}{\unblocked}\bigr)}{\big(\mrelease{\idV}\mtru\big) \,\mand\, \hV'''} \label{eq:ra:14}
\end{align}	}
\noindent Once again in both \eqref{eq:ra:13} and \eqref{eq:ra:14} we have reached a stuck configuration because a second release action on $\idV$ could not be carried out as this was already released.\exqed 
\end{example}

The semantics of \figref{fig:rt-semantics} allows us to formalise configurations in an erroneous state (as identified in \exref{ex:model}) \ie when a monitor wants to release or apply synchronous adaptations that the system prohibits, as follows:

\begin{definition}
  \label{def:err-conf}
  $\mErrorC{\instr{\stV}{\hV}} \;\deftxt\; \Bigl(\hV \traS{\scriptstyle\cors{\vLstB}} \;\text{ and }\; \stV\notTraceEvent{\;\scriptstyle\cors{\vLstB}\;}\Bigr)\; \text{ or }\; \Bigl(\hV \traS{\scriptstyle\rel{\vLstB}} \;\text{ and }\; \stV\notTraceEvent{\scriptstyle\;\rel{\vLstB}\;}\Bigr) \\$ \indent $\qquad \text{for some } \vLstB \in\dom(\stV)$
   
   \medskip   
  \noindent This definition states that a synchronisation error can be caused in two ways. The first way occurs when a formula $\hV$ communicates a synchronous adaptation action $\hV \traS{\scriptstyle\cors{\vLstB}}$ to the system actors defined in $\vLstB$, but one (or more) of these actors are active (\ie there exists some $i\!\in\!\vLstB$ such that $\mapstate{i}{\unblocked}$), and is thus not in a position to correctly react to the reported adaptation. Similarly, the second way occurs when a formula $\hV$ communicates a release action instead, \ie $\hV \traS{\scriptstyle\rel{\vLstB}}$. 
\end{definition} \vspace{-2mm}


\section{Conclusion}

In this chapter we addressed our second objective (\ie objective $(ii)$ in \secref{sec:intro:obj}) by designing and implementing an effective Runtime Adaptation framework for actor systems as an extension of an existing RV setup. In fact we have identified two classes of adaptation actions, namely \emph{asynchronous} and \emph{synchronous} adaptations, where the latter require a degree of synchronisation in order to be applied over the respective adaptees, as opposed to the former. The required synchronisation was therefore introduced by building upon the efficient incremental synchronous (Hybrid) monitoring approach explored in \Cref{chp:syn-asyn}. In fact the prototype implementation of our Runtime Adaptation framework was developed as an extension of \detecter\ \ie the extended version of \detecterGen\ that we created in \Cref{chp:syn-asyn}. This RA framework for actor systems is now called \adapter\footnote{The prototype tool implementation is accessible from \texttt{https://bitbucket.org/casian/adapter}.}.

Furthermore, to be able to better understand the subtle errors that our RA synchronisation protocol may introduce, we developed a \emph{formal operational model}. This allowed us to identify situations where our runtime adaptation framework may introduce errors in the monitored system, if synchronisation is not properly defined. This model thus enabled us to identify a notion of synchronisation errors which we formalised as \defref{def:err-conf}. This therefore enables us to address our third objective by showing that these errors can be detected using static analysis techniques, particularly by developing a type system. In \Cref{chp:typ-sys} we therefore investigate ways how to develop a type system for assisting the specifier into writing valid scripts, by rejecting scripts containing invalid synchronisation commands that lead to the identified error.

\chapter{Filtering Out Invalid Adaptation Scripts}
\label{chp:typ-sys}
\newcommand{\lemmaA}{\lemmaref[(Type Safety)]{lemma1}\xspace}
\newcommand{\lemmaB}{\lemmaref[(Subject Reduction)]{lemma2}\xspace}
\newcommand{\lemmaX}{\lemmaref[(System Reduction)]{lemmaX}\xspace}
\newcommand{\lemmaY}{\lemmaref[(Monitor Reduction)]{lemmaY}\xspace}
\newcommand{\lemmaC}{\lemmaref[(Formula Substitution)]{lemma3}\xspace}
\newcommand{\lemmaD}{\lemmaref[(Term Substitution)]{lemma4}\xspace}
\newcommand{\lemmaI}{\lemmaref[(Weakening)]{lemmaI}{}\xspace}

As we seen in the previous chapter, by contrast to RV scripts, our runtime adaptation scripts are no longer passive. In fact our RA framework permits the specifier to define blocking necessities and synchronous adaptations that could possibly introduce synchronisation errors. We conjecture that there are various types of errors that erroneously defined scripts may introduce. Such errors include:
\begin{enumerate}[label=$(\arabic*)$]
	\item Applying a synchronous adaptation or a release operation on an actor which was not priorly suspended by a blocking necessity.
	\item Blocking an actor earlier then required, therefore disabling subsequent events from occurring; a case in point is shown in \exref{ex:adapt-and-block} where we had to explicitly specify the internal communication action $\msendD{\idV\,}{\_}{\{\eatom{inc},x,y\}}$ to be able to block actor $i$ after it forwards the increment request to some other internal component.
	\item Adaptations may be applied to external systems interacting with the system which might not be have been instrumented by our RA framework. For instance, we could specify an RA script for our running example system in \figref{fig:sys} in which we apply an adaptation to the connected external client mapped to term variable $y$.	
\end{enumerate}
\noindent A static analysis phase can be employed to detect errors in our RA scripts and reject them at compile time, thereby serving as a tool for assisting the specifier into creating valid adaptation scripts. As a proof of concept, in this chapter we focus on developing a static type system for ruling out scripts that are erroneous \wrt to $(1)$ which we have already formalised as \defref{def:err-conf} in \secref{sec:ra-derivative-semantics}. However, we conjecture that similar static analysis techniques may be employed to detect other errors such as $(2)$ and $(3)$.

This chapter is structured in three sections. In \secref{sec:static} we develop a static type system for ruling out scripts that are erroneous \wrt error $(1)$, prior to deployment. Since static typechecking is performed \wrt certain assumptions, that might not always be honoured by a system execution, in \secref{sec:dynamic} we augment our runtime monitors with dynamic checks that halt monitoring in case the system deviates from our assumptions. This further prevents synchronisation errors from being introduced by our RA monitors. Finally, in \secref{sec:soundness} we evaluate our typing strategies by proving \emph{type soundness}, \ie we prove that at no point during execution can an accepted typed script introduce synchronisation errors in the monitored system.

\section{Static Type Checking}
\label{sec:static}
Recall that in \exref{ex:model} we presented cases where invalid synchronisation leads to erroneous (stuck) configurations in which synchronous adaptations or release operations are incapable of being validly applied on an active system actor. Particularly, we showed how RA scripts containing concurrent conjunction branches can introduce errors which can be difficult for the specifier to notice prior to deployment, without any form of assistance. 

In the previous chapter we formalised these synchronisation errors as \defref{def:err-conf} (restated below). In this section we therefore focus on developing a static type system that assists the specifier into writing valid scripts by rejecting the scripts which they may violate this formalised notion of synchronisation errors, and are thus \emph{unsafe}.  

\medskip
\noindent\textbf{\defref{def:err-conf}.}  $\,\mErrorC{\instr{\stV}{\hV}} \;\deftxt\; \Bigl(\hV \traS{\scriptstyle\cors{\vLstB}} \;\text{ and }\; \stV\notTraceEvent{\;\scriptstyle\cors{\vLstB}\;}\Bigr)\; \text{ or }\; \Bigl(\hV \traS{\scriptstyle\rel{\vLstB}} \;\text{ and }\; \stV\notTraceEvent{\scriptstyle\;\rel{\vLstB}\;}\Bigr) \\$ \indent $\qquad \text{for some } \vLstB \in\dom(\stV)$
\medskip

\noindent From this definition note that these synchronisation errors are caused due to (one or more) invalid entries specified in the list of actor references, used for applying a synchronous adaptation or release operations. We note that an invalid entry $i$, defined in a list of actor references $\vLstB$, can have the following three forms:
\begin{itemize}[leftmargin=4mm] \itemsep0em
	\item it can either refer to an actor which is currently active, \ie an actor which is not blocked by the instrumented code, $\mapstate{i}{\unblocked}$;
	\item it might not be referring to a process identifier pertaining to the system, \ie it might refer to some generic data, $i\notin\stV$; or else
	\item it might be referring to some generic data, $i\!=\!3\notin\stV$.
\end{itemize}
\noindent Hence to begin with, our type system must be able to make a distinction between \emph{generic data} and \emph{actual process identifiers}. Moreover, the type system must also be able to distinguish between two \emph{generic classes} of process identifiers specified within the RA script it is analysing.

The first class includes the process identifiers that are used in the synchronous adaptations and synchronisation operations (\ie blocking and releasing) of the RA script being analysed. As these operations may potentially contribute to synchronisation errors, the type system must ensure that the identifiers pertaining to this class are carefully used within the script. Hence our type system must employ a mechanism for keeping track of how these identifiers are being blocked and released at runtime. For instance the type system must be able to statically track that these identifiers are being blocked by a blocking necessity that specified \emph{prior} to the application of a synchronous adaptation or release operation.

More importantly, the type system must ensure that these identifiers are \emph{carefully} specified within concurrent branches, thereby being able to reject cases where two (or more) branches exhibit a \emph{race condition}. Race conditions such as those exhibited by $\hV'''$ in \exref{ex:model}, include for instance cases where a branch may only apply a synchronous adaptation correctly over an actor $i$ if it (non-deterministically) executes \emph{before} another concurrent branch which releases $i$. Ensuring careful use of these identifiers requires introducing an element of \emph{linearity}, whereby the use of a process id (pertaining to this class) must be \emph{restricted} to a \emph{single monitoring branch}. We thus refer to process identifiers pertaining to this class as \emph{linear}.

The second class includes other process identifiers which are not involved with any kind of operation that requires synchronisation, \ie these typically used as the subjects of \emph{asynchronous necessities} and thus never used as subjects of blocking necessities or as part of release lists or adaptation lists. As these process identifiers are not used in such sensitive operations, they can safely be specified amongst concurrent branches (unlike linear process ids). Hence we refer to the process ids pertaining to this class as \emph{unrestricted}. The type system must however make sure that processes that are assumed to be unrestricted are in fact not being used in synchronisation operations and synchronous adaptations.

We present our static type system in the four subsections: in \secref{sec:type-annotations} we explain the type annotations introduced by our type system; followed by \secref{sec:type-environments} in which we explain the type environments containing the required assumptions for type checking a script. Finally in \secref{sec:type-rules} and \ref{sec:type-rules-excl} we present details explaining the typing rules forming our type system.

\subsection{The Type Annotations} \label{sec:type-annotations}
\begin{figure}[!ht]
\begin{small}	\setstretch{0.8}
	\textbf{Type Structure}\medskip
	\begin{align*}
	  \tV,\tVV \in \Typ \;\bnfdef
	  \; \dat \quad\textsl{(generic)}\;\bnfsep
	  \quad \upid \quad\textsl{(unrestricted)}\;\bnfsep
	  \quad  \lpid \quad\textsl{(linear)}\;\bnfsep
	  \quad \lpidb \quad\textsl{(blocked)}\\
	\end{align*}
	
	\textbf{Type Environment Splitting}\smallskip
	\begin{mathpar}
	  \inference[\rtit{sE}]{}{\emptyset + \emptyset = \emptyset} \and
	  \inference[\rtit{sU}]{\envV_1 + \envV_2 = \envV_3 & \tV \in \sset{\dat, \upid}}{(\envV_1,\typ{\varid}{\tV})  + (\envV_2,\typ{\varid}{\tV}) = (\envV_3,\typ{\varid}{\tV})} \and
	  \inference[\rtit{sL}]{\envV_1 + \envV_2 = \envV_3 & \tV \in \sset{\lpid, \lpidb}}{(\envV_1,\typ{\varid}{\tV})  + \envV_2 = (\envV_3,\typ{\varid}{\tV})} \and
	  \inference[\rtit{sR}]{\envV_1 + \envV_2 = \envV_3 & \tV \in \envV_1  +  \sset{\lpid, \lpidb}}{(\envV_2,\typ{\varid}{\tV}) = (\envV_3,\typ{\varid}{\tV})} \\
	\end{mathpar}
	
	\textbf{Adaptation-Script Typing Rules}\medskip
	\begin{mathpar}
	  	\inference[\rtit{tNcA}]{
	          \typeRule{\env{\envVV}{(\envV,\tbnd{\patE})}}{\hV}\qquad \typeRule{\env{\envVV}{\envV}}{\mrelease{\vLstA}\mtru} 
			}{\typeRule{\env{\envVV}{\envV}}{\mANec{\patE}{\vLstA}{\hV}}}  
	                \and
	           \inference[tFls]{}{\typeRule{\env{\envVV}{\envV}}{\mfls}} 
	           \and
		\inference[tTru]{}{\typeRule{\env{\envVV}{\envV}}{\mtru}}\\
	        \inference[\rtit{tNcB}]{ \id{\patE}\!=\!\varid \qquad
	          \typeRule{\env{\envVV}{(\envV,\tbnd{\patE})}}{\mblock{\varid}\hV}
	          \qquad \typeRule{\env{\envVV}{\envV}}{\mrelease{\vLstA}\mtru} 
			}{\typeRule{\env{\envVV}{\envV}}{\mBNec{\patE}{\vLstA}{\hV}}}
	              \quad
	        \inference[\rtit{tIf}]{\typeRule{\env{\envVV}{\envV}}{\hV} \qquad \typeRule{\env{\envVV}{\envV}}{\hVV}}{\typeRule{\env{\envVV}{\envV}}{\mboolE{\bV}{\hV}{\hVV}}} 	
	                \\
	          \inference[\rtit{tBlk}]{\envV = \envV', \typ{\vLstB}{\lpid} \qquad\typeRule{\env{\envVV}{(\envV', \typ{\vLstB}{\lpidb})}}{\hV} }{\typeRule{\env{\envVV}{\envV}}{\mblock{\vLstB}\hV}} 
	          \and
		\inference[\rtit{tRel}]{\envV = \envV', \typ{\vLstB}{\lpidb} \qquad\typeRule{\env{\envVV}{(\envV', \typ{\vLstB}{\lpid})}}{\hV}}{\typeRule{\env{\envVV}{\envV}}{\mrelease{\vLstB}\hV}}
	\\ 
	      	\inference[\rtit{tAdA}]{\envV = \envV',\typ{\vLstB}{\lpid} \qquad\typeRule{\env{\envVV}{\envV}}{\mrelease{\vLstA}\hV} }{\typeRule{\env{\envVV}{\envV}}{\macor{\vLstB}{\vLstA}\,\hV}}  \and
		\inference[\rtit{tAdS}]{\envV = \envV',\typ{\vLstB}{\lpidb} \qquad \typeRule{\env{\envVV}{\envV}}{\mrelease{\vLstA}\hV} }{\typeRule{\env{\envVV}{\envV}}{\mscor{\vLstB}{\vLstA}\,\hV}} 
	        \\
	       \inference[\rtit{tCn1}]{\excl{\hV,\hVV}= \bot \qquad\typeRule{\env{\envVV}{\envV_1}}{\hV} \qquad \typeRule{\env{\envVV}{\envV_2}}{\hVV}}{\typeRule{\env{\envVV}{(\envV_1 + \envV_2)}}{\hV\mand\hVV}}
	       \and 
	       \inference[\rtit{tMax}]{\typeRule{\env{(\envVV, \hVarX\mapsto\envV)}{\envV}}{\hV}}{\typeRule{\env{\envVV}{\envV}}{\mmax{\hVarX}{\hV}}} 
	       \and
	       \inference[\rtit{tCn2}]{\excl{\hV,\hVV}=\langle\vLstA[\hV],\vLstA[\hVV]\rangle\qquad  \typeRule{\env{\envVV}{\eff{\envV}{\vLstA[\hVV]}}}{\hV} \quad \typeRule{\env{\envVV}{\eff{\envV}{\vLstA[\hV]}}}{\hVV}  }{\typeRule{\env{\envVV}{\envV}}{\hV\mand\hVV}}
	       \and 
	       \inference[\rtit{tVar}]{\envVV(\hVarX)\subseteq\envV}{\typeRule{\env{\envVV}{\envV}}{\hVarX}}
	         \end{mathpar}
\end{small}
\caption{A Type System for Adaptation \SHML scripts }
\label{fig:static-rules}
\end{figure}

The type system relies on the type structure defined in \figref{fig:static-rules} (below) where values are partitioned into either generic data, \dat, or actor identifiers, where the latter are further subdivided into unrestricted, \upid, and linear, \lpid.    The type system is substructural \cite{ATTAPL}, in the sense that it uses linear types to \emph{statically track} how the linear actor identifiers used for synchronous adaptations, are blocked and released by the parallel branches (\ie conjunctions) of the respective script. 

To be able to statically keep track of how the linear identifiers specified in a given script, are blocked and released, type checking (internally) uses a sub-category for linear types, \lpidb, to denote a \emph{blocked} linear identifier. 
The general idea behind this tracking mechanism is that the type system uses a \emph{look-up table} to statically approximate how a system actor is specified to blocked and released at runtime; this is achieved by switching between types \lpid and \lpidb. In fact, whenever the type system analyses an identifier $i$ that is used in a blocking operation (\ie a blocking necessity), it consults its look-up table and checks that $i$ is linear typed, \ie $\typ{i}{\lpid}$; if this is true, then it changes the type associated with the identifier in its look-up table, from \lpid to \lpidb. In this way the type system can detect whether a synchronous adaptation or a release operation is being correctly applied to $i$, by consulting its look-up table and checking that the associated type is \lpidb. This signifies that at runtime $i$ would have been blocked in some prior operation as required.    

Type checking works on \emph{typed scripts}, where the syntax of \figref{fig:rt-semantics} is extended so that the binding term variables used in action patterns are annotated by the types \dat, \upid or \lpid. Note that the type \lpidb is not part of the external syntax, \ie this cannot be specified as a type annotation in a script, because this type is \emph{reserved} for internal use by the type system, for tracking purposes.

\begin{example}\label{ex:typed-script}  The adaptation-script (\ref{eq:ra:4}) of \exref{ex:adapt-and-block} would be annotated as follows: 
  \begin{equation}\setstretch{1.1}
    \label{eq:static:9}
    \begin{array}{l}
    \hV'\;\deftxt\;\; \mmax{\,\hVarY}{\mANec{\mrecv{\idV}{\tup{\eatom{inc},\typ{x}{\dat},\typ{y}{\upid}}}}{\epsilon}}\\[-1mm]
      \qquad\qquad\qquad\qquad\mBNec{\msendD{\idV}{\_}{\{\eatom{inc},x,y\}}}{\varepsilon}
                \begin{mlbrace} 
        (\,\mANec{\msendD{j}{y}{\tup{\eatom{res},\smash{x+1}}}}{\epsilon}\,\hVarY)   \;\mand \; \\[0mm] 
        (\,\mBNec{\msendD{\typ{z}{\lpid}\,}{y}{\eatom{err}}}{i}\,\mrestart{i}_{\varepsilon}\,\mclearMailbox{z}_{i,z}\,\hVarY) 
      \end{mlbrace}
    \end{array}
 \end{equation} 
In (\ref{eq:static:9}) above, pattern variables $x,y$ and $z$ are associated to types \dat, \upid and \lpid respectively.  \exqed 
\end{example}

Note that explicit type annotations are only required to be specified for \emph{binding term variables}, \ie they are only used within necessities that \emph{first bind} a variable. For example, in \eqref{eq:static:9}, we do not add the type annotations for variables $x$ and $y$ in the second necessity, \ie $\mBNec{\msendD{\idV}{\_}{\{\eatom{inc},x,y\}}}{\varepsilon}$, as these variables are bound (and were thus annotated) in the prior necessity, \ie $\mANec{\mrecv{\idV}{\tup{\eatom{inc},\typ{x}{\dat},\typ{y}{\upid}}}}{\epsilon}$.

In addition, note that we do not include explicit type annotations for \emph{statically known} process ids such as $i$ and $j$; these typically represent process identifiers that are \emph{registered} and can thus be know at compile time. The necessary type annotations for these registered processes are passed over to the type system as type assumptions represented in the form of a type environment.

\subsection{The Type Environments} \label{sec:type-environments}
Our type system for (typed) adaptation-scripts is defined as the least relation satisfying the rules in \figref{fig:static-rules}. The type system judgements take the form $\typeRule{\env{\envVV}{\envV}}{\hV}$ where
\begin{itemize}[leftmargin=4mm]
\item Value  environments,
  $\envV \in \Env :: (\Pid \cup \Vars) \rightharpoonup \Typ$, map 
  identifiers and variables to types --- these are used as a look-up table containing the necessary type assumptions, which the system consults and updates, while it compositionally analyses a given typed script. 
\item  Formula  environments,
  $\envVV \in \LVars \rightharpoonup \Env$, map 
  formula variables
  to value  environments, \eg \envVVN{X}{\envV} --- these are used to analyse recursive formulas in a compositional way. (The use of this environment will be clarified when we discuss rules \rtit{tMax} and \rtit{tVar} in \secref{sec:type-rules}).
\end{itemize}
\noindent Note that we sometimes write $\typeRule{\envV}{\hV}$ in lieu of  $\typeRule{\env{\emptyset}{\envV}}{\hV}$. We also write $\typ{\vLstB}{\tV}$ to denote the list of mappings $\typ{\varid_1}{\tV},\ldots,\typ{\varid_n}{\tV}$ where $\vLstB=\varid_1,\ldots,\varid_n$ (recall that metavariable $\varid\!\in\!(\Pid\!\cup\!\Vars)$ ranges over identifiers and variables).

\subsubsection{Extending the Type Environments}
The rules  in \figref{fig:static-rules} assume standard \emph{environment extensions}, $(\envV,\envV')$. This allows the type system to add new type assumptions by extending its current value environment with new information. Since value environments are \emph{maps}, they are therefore implicitly restricted such that a domain value of the map 
(\eg $i$ or $x$) can only map to \emph{one} type, \eg value environment $\envV\!=\!\{\typ{i}{\upid},\typ{i}{\lpid}\}$ is invalid as $i$ is mapped to two different types. This restriction also reflects in the way a value environment is extended. 

For instance, if a valid value environment $\envV\!=\!\{\typ{i}{\upid},\typ{j}{\upid}\}$ is extended by $\envV'\!=\!\{\typ{i}{\upid},\typ{z}{\lpid}\}$, we get the resultant environment $(\envV,\envV')\!=\!\{\typ{i}{\upid},\typ{j}{\upid},\typ{z}{\lpid}\}$. Although identifier $i$ is present in both $\envV$ and $\envV'$, the resultant extended environment $(\envV,\envV')$ is still valid as both environments map $i$ to the \emph{same type}, \ie to $\upid$. Hence the implicit mapping restriction is not breached by the extension. By contrast if $\envV$ is extended by $\envV''=\{\typ{i}{\lpid},\typ{z}{\lpid}\}$, the resultant environment $(\envV,\envV'')$ would not constitute a valid map. 

A similar environment extension is also used for formula environments which map formula variables, \eg $\hVarX$, onto a value environment \eg $\envV$. 

\subsubsection{Splitting Value Environments}
The rules in \figref{fig:static-rules} also use \emph{environment splitting}, $\envV_1 + \envV_2$, as defined by the environment splitting rules \rtit{sE}, \rtit{sU}, \rtit{sL} and \rtit{sR}. These rules are used to distribute linearly the mappings amongst two (possibly) smaller environments. Rule \rtit{sE} states that an empty value environment $\envV\!=\!\emptyset$ can be divided into two empty environments \ie $\emptyset + \emptyset$. Rule \rtit{sU} dictates that generic data and unrestricted typed references, \eg $\typ{x}{\upid}$, can be divided amongst the two resultant value environments. For instance, if $\typ{x\!}{\!\upid}\!\in\!\envV$ and $\envV$ is split into $\envVA$ and $\envVB$, \ie $\envV=\envVA+\envVB$, then $\typ{x}{\upid}$ is distributed to both $\envVA$ and $\envVB$, such that $\typ{x}{\upid}\in\envVA$ and $\typ{x}{\upid}\in\envVB$. 

This contrasts rules \rtit{sL} and \rtit{sR} which state that linear typed references, \eg $\typ{i}{\lpid}$ or $\typ{i}{\lpidb}$, can only be given to \emph{one} of the two resultant environments. For example, if $\typ{z}{\lpid}\in\envV$ and this is split into $\envVA$ and $\envVB$ then $\typ{x}{\lpid}$ can only be distributed to either $\envVA$ (with rule \rtit{sL}), \ie $\typ{z}{\lpid}\in\envVA$ but $\typ{z}{\lpid}\notin\envVB$, or else to $\envVB$ (with rule \rtit{sR}). 

This splitting mechanism is essential for the type system to make sure that linear typed references are carefully used amongst the concurrent conjunction branches, by restricting their use to only one branch. 
   
\subsection{The Typing Rules} \label{sec:type-rules}
The typing rules presented in \figref{fig:static-rules} serve to statically analyse a typed script in a \emph{compositional} manner. Hence these rules regularly consult and update the type assumptions, stored in the type environments, such that subsequent type judgements are made \wrt the information gained by previous judgements. 

For instance in formula \eqref{eq:static:9} (stated in \exref{ex:typed-script}), necessity $\mBNec{\msendD{\idV}{\_}{\{\eatom{inc},x,y\}}}{\varepsilon}$ and other subsequent constructs are typechecked \wrt the information gained from analysing prior constructs. In fact, from the prior necessity $\mANec{\mrecv{\idV}{\tup{\eatom{inc},\typ{x}{\dat},\typ{y}{\upid}}}}{\epsilon}$ the type system can deduce that $x$ and $y$ were bound and were given the types \dat and \upid respectively. 

Rules \rtit{tTru}, \rtit{tFls} and \rtit{tIf} are quite straightforward, \ie truth and falsities always typecheck, while the analysis of if-statements is approximated by requiring both branches to typecheck. Furthermore, typing rules that update the type environments of subsequent judgements include rules \rtit{tNcA}, for analysing \emph{asynchronous necessities}, and rules \rtit{tNcB} for \emph{blocking necessities}. 

These rules are very similar: \rtit{tNcA} updates the value environment $\envV$ with the \emph{process bindings} (\ie variables annotated with either $\upid$ or $\lpid$) introduced by the necessity pattern $\patE$. This permits for the continuation formula $\hV$ (\eg $\hV$ is the continuation formula of $\mattrNec{\patE}{\attr}{\vLstA}\hV$) to be typechecked \wrt these new process bindings (\ie $\typeRule{\env{\envVV}{(\envV,\envV')}}{\hV}$, where $\envV'$ contains the new process bindings). 

It also checks that in case of action mismatch, the resultant \emph{actor releases}, defined the release list $\vLstA$ of the respective necessity, also typecheck (\ie $\typeRule{\env{\envVV}{\envV}}{\mrelease{\vLstA}\mtru}$).  Rule \rtit{tNcB} performs similar checks, but the continuation formula typechecking is prefixed by the blocking of the subject of the pattern, (\ie $\typeRule{\env{\envVV}{(\envV,\envV')}}{\mblock{\varid}\hV}$).  \bigskip

\begin{remark} Both \rtit{tNcA} and \rtit{tNcB} utilise the auxiliary function $\tbnd{\patE}$ that extracts a map of process bindings from a necessity pattern as shown below:

{\setstretch{1.2} \setlength{\abovedisplayskip}{-5pt}\setlength{\belowdisplayskip}{-10pt} \setlength{\abovedisplayshortskip}{-5pt} \setlength{\belowdisplayshortskip}{-10pt} 
	\begin{gather} 
		\tbnd{\mrecv{i}{\etuple{\eatom{inc},\typ{x}{\dat},\typ{y}{\upid}}}}=\{\typ{y}{\upid}\} \label{eq:bnd:1} \\
		\tbnd{\msendD{\typ{x}{\lpid}}{\typ{y}{\upid}}{\etuple{\eatom{res},z}}}=\{\typ{x}{\lpid},\typ{y}{\upid}\} \label{eq:bnd:2} 
	\end{gather}
}
	
\noindent From \eqref{eq:bnd:1} note that this function only returns term variable bindings that are only annotated with a process type \ie \lpid or \upid. The reason for not considering generic data types is that the type system does not track the use of these variables. In fact the main focus of our type system is to detect cases where a synchronous adaptation or a release operation is applied to an invalid actor id, and not about how the data variables are being used and inspected by if-statements. 

Furthermore, from \eqref{eq:bnd:2} note that variable $z$ was not included in the resultant map as it does \emph{not} have a type annotation. As our analyses assumes closed formulas, by the point the type system inspects necessity pattern $\msendD{\typ{x}{\lpid}}{\typ{y}{\upid}}{\etuple{\eatom{res},z}}$, variable $z$ would have already been bound and added to the respective value environment. 

Rule \rtit{tNcA} then uses the map extracted by this function from a pattern $\patE$, to typecheck the continuation formula, \eg $\hV$, \wrt this extended type environment, \ie $\typeRule{\env{\envVV}{(\envV,\tbnd{\patE})}}{\hV}$. The same applies for rule \rtit{tNcB}, with the difference that the continuation formula is prefixed by a block of the pattern's subject, \ie $\typeRule{\env{\envVV}{(\envV,\tbnd{\patE})}}{\mblock{\varid}\hV}$). \bqed
\end{remark}\bigskip
 
The typing rules \rtit{tBlk} and \rtit{tRel} for actor blocking (\ie $\mblock{\vLstB}$) and releasing (\ie $\mrelease{\vLstB}$), enable the system to track which actor references are marked as blocked or not. For instance, whenever rule \rtit{tBlk} is used to analyse script $\mblock{\vLstB}\hV$ \wrt a value environment $\envV$, this rule consults $\envV$ and checks that all actor references in $\vLstB$ are present in $\envV$ and are typed as linear, \ie $\envV\!=\!\envV''\!,\typ{\vLstB\!}{\!\lpid}$. If this condition is satisfied, it updates the $\envV$ environment by changing the type of the identifiers enlisted in $\vLstB$ to \lpidb, thereby creating a new environment $\envV'\!=\!\envV''\!,\typ{\vLstB\!}{\!\lpidb}$. Rule \rtit{tBlk} then typechecks the continuation formula $\hV$ \wrt $\envV'$. In this way the typesystem is able to check whether the synchronous adaptations or release operations declared in $\hV$, are being applied to actor references that were priorly blocked.

In fact, rule \rtit{tRel} uses a mechanism similar to that of \rtit{tBlk}, whereby it checks that all actor references, defined in a release list $\vLstB$, have the type \lpidb in \envV. Hence the type system is able to reject cases where release operations are applied to active processes by consulting the value environment and checking whether their type was changed to \lpidb by a previously defined blocking operation. If all actor references enlisted in $\vLstB$ are typed as linear blocked (\lpidb) in the $\envV$ value environment, then rule \rtit{tRel} changes their type back to \lpid, thus creating a new value environment which is used to typecheck the continuation formula.

Typechecking asynchronous adaptations, $\macor{\vLstB}{\vLstA}$, with \rtit{tAdA}, requires the adaptees specified in $\vLstB$, to be linearly typed, \ie $\typ{\vLstB}{\lpid}$, whereas typechecking synchronous adaptations, $\mscor{\vLstB}{\vLstA}$, with rule \rtit{tAdS}, requires the adaptees to be linearly blocked, \ie $\typ{\vLstB}{\lpidb}$. In this way the type system ensures that synchronous adaptations are only applied on actor references which were blocked by a \emph{prior} blocking operation (\ie their type was changed from \lpid to \lpidb). Both rules consider the respective released actors when typechecking the continuations, (\ie $\typeRule{\env{\envVV}{\envV}}{\mrelease{\vLstA}\hV}$), thus also checking that the actors defined in the release list $\vLstA$ are also applied in a valid way, prior to checking the continuation formula $\hV$. 

We have two rules for typechecking conjunction formulas, $\hV\mand\hVV$, namely rules \rtit{tCn1} and \rtit{tCn2}; for now we only consider the former, as the latter in covered in more detail in \secref{sec:type-rules-excl}. Recall that since conjunction subformulas may be executing in parallel (as defined by the reduction rules \rAndA, \rAndB, \rAndC\ and \rAndD\ from \figref{fig:logic-transition-rules} and \figref{fig:rt-semantics}), synchronous adaptations and release actions may lead to race conditions as explained in \exref{ex:model} (see \eqref{eq:ra:11} and \eqref{eq:ra:12}). Rule \rtit{tCn1} eliminates such cases by typechecking each subformula $\hV$ and $\hVV$ \wrt a \emph{split environment}, $\envV_1 + \envV_2$, as is standard in linear type systems \cite{ATTAPL} (refer to \secref{sec:type-environments}). This ensures that linear typed actor references (\eg $\typ{\varid}{\lpid}$ or $\typ{\varid}{\lpidb}$) are only distributed to a \emph{single concurrent branch}, thereby eliminating cases where multiple branches attempt to release or apply a synchronous adaptation on the same actor references. Note that conjunction formulas are inspected with rule \rtit{tCn1}, if the side-condition $\excl{\hV,\hVV}=\undef$ holds (otherwise \rtit{tCn2} applies); this side-condition is explained in more depth in \secref{sec:type-rules-excl}.

While analysing recursive definitions, $\mmax{\hVarX}{\hV}$, formula environment $\envVV$ also plays a crucial role in preserving linearity. In fact this environment permits rule \rtit{tMax} to save a copy of the value environment $\envV$ at the point where a formula variable $X$ is bound by $\mmax{X}{\hV}$ (\ie $\envV$ becomes $\envVVN{X}{\envV}$). This then enables rule \rtit{tVar} to compare the saved copy to the contents of the (possibly different) value environment $\envVP$ at the point where $X$ is used (\ie when the formula recurs), by checking that $\envVV(X)\subseteq\envVP$. Note that we check whether the saved copy $\envVV(X)$ is a subset of (or equal to) the current value environment $\envVP$, as during the type derivation the value environment might be extended with new bindings while inspecting necessity operations with rules \rtit{tNcA} and \rtit{tNcB}. Furthermore, this condition together with environment splitting, allow the type system to eliminate cases where a formula that applies synchronous adaptations (or releases), \emph{recurs} along two (or more) concurrent branches which may eventually lead to race conditions. 

Hence, checking formula variables is imperative to eliminate cases in which synchronisation errors may appear at a later stage during runtime due to \emph{concurrency and recursion}. For instance consider formula \eqref{eq:Sigma} which may appear to be free of potential synchronisation errors, yet \exref{ex:SigmaCE} proves otherwise. 
\begin{dmath}   \label{eq:Sigma}
	   \hV_{err} = \mmax{\hVarX}{\mBNec{\mrecv{i}{3}}{\varepsilon}\Bigl((\mscor{i}{i}\hVarX)\mand(\hVarX)\Bigr)}
\end{dmath}

\begin{example} \label{ex:SigmaCE} Assume that when formula $\hV_{err}$ is used to verify trace $t=\mrecv{i}{3};\,\mrecv{i}{3};\,\ldots$ \wrt $\stV\!=\!(\mapstate{i}{\unblocked})$, their configuration yields the following execution (there may be others).
 \begin{align}\setlength{\abovedisplayskip}{-5pt}
\setlength{\belowdisplayskip}{-5pt}
\setlength{\abovedisplayshortskip}{-5pt}
\setlength{\belowdisplayshortskip}{-5pt}
\instr{\stV}{\hV_{err}}
	 &  \traS{\scriptstyle\;\;\tau\;\;} \instr{\bigl(\mapstate{\idV}{\unblocked}\bigr)}{ \mBNec{\mrecv{i}{3}}{\varepsilon}\Bigl((\mscor{i}{i}\,\hV_{err})\mand \hV_{err}\Bigr)} \label{eq:err:0} \\	
	  &  \traS{\scriptstyle\mrecv{i}{3}} \instr{\bigl(\mapstate{\idV}{\unblocked}\bigr)}{ \mblock{i}\Bigl((\mscor{i}{i}\,\hV_{err})\mand \hV_{err}\Bigr)} \label{eq:err:1} \\	
	  &  \traS{\scriptstyle\;\;\tau\;\;} \instr{\bigl(\mapstate{\idV}{\blocked}\bigr)}{ (\mscor{i}{i}\,\hV_{err})\mand \hV_{err}  } \label{eq:err:2}\\	
	  &  \traS{\scriptstyle\;\;\tau\;\;} \instr{\bigl(\mapstate{\idV}{\blocked}\bigr)}{ (\mrelease{i}\,\hV_{err})\mand \hV_{err}  } \label{eq:err:3}\\		  
	  & \wtraS{\scriptstyle\;\;\tau\;\;}  \instr{\bigl(\mapstate{\idV}{\unblocked}\bigr)}{ \Bigl(\mBNec{\mrecv{i}{3}}{\varepsilon}\bigl((\mscor{i}{i}\,\hV_{err})\mand\hV_{err}\bigr)\Bigr) \mand \Bigl(\mBNec{\mrecv{i}{3}}{\varepsilon}\bigl((\mscor{i}{i}\,\hV_{err})\mand\hV_{err}\bigr)\Bigr)  }\label{eq:err:4}\\
	  &  \wtraS{\scriptstyle\mrecv{i}{3}}  \instr{\bigl(\mapstate{\idV}{\blocked}\bigr)}{ \Bigl((\mscor{i}{i}\,\hV_{err})\mand\hV_{err}\Bigr) \mand \Bigl((\mscor{i}{i}\,\hV_{err})\mand\hV_{err}\Bigr)  }\label{eq:err:5}\\
	   &  \wtraS{\scriptstyle\;\;\tau\;\;}  \instr{\bigl(\mapstate{\idV}{\unblocked}\bigr)}{ \Bigl(\hV_{err}\mand\hV_{err}\Bigr) \mand \Bigl((\mscor{i}{i}\,\hV_{err})\mand\hV_{err}\Bigr)  } \label{eq:err:6} 
\end{align}
\end{example}\vspace{-1mm}

The above derivation shows that although formula $\hV_{err}$ does not immediately yield a stuck configuration, it still manages to get stuck after unfolding formula variable $\hVarX$ twice as a result of receiving system event $\mrecv{i}{3}$ twice in a row. More specifically, in the above derivation notice that after unfolding the recursive definition in \eqref{eq:err:0}, the resultant formula perceives system action $\mrecv{i}{3}$ which matches with necessity $\mBNec{\mrecv{i}{3}}{\varepsilon}$ thus causing the subject actor $i$ to block in \eqref{eq:err:1}. In \eqref{eq:err:2} we see that a synchronous adaptation could validly be applied on actor $i$ as it was already blocked (\ie $\mapstate{\idV}{\blocked}$) in the previous step. This actor is then released in \eqref{eq:err:3} as it was specified in the release list of the synchronous adaptation applied in \eqref{eq:err:2}. In \eqref{eq:err:4} we see that after $i$ was released, the recursive definitions of the resultant conjunction formula (\ie $\hV_{err}\mand\hV_{err}$) were unfolded. Once again in \eqref{eq:err:5}, action $\mrecv{i}{3}$ matches blocking necessities $\mBNec{\mrecv{i}{3}}{\varepsilon}$ thus causing actor $i$ to block. Finally from \eqref{eq:err:5} to \eqref{eq:err:6} we see that the synchronous adaptation of the left branch was applied on $i$, and then $i$ was released, \ie $\mapstate{\idV}{\unblocked}$. Hence the last configuration in \eqref{eq:err:6} is \emph{stuck} since the synchronous adaptation in the right branch is unable to apply on $i$.

This shows how concurrency and recursion can induce synchronisation errors that are very hard to notice without assistance. Therefore, as shown in Derivation~\ref{drv:Phi4}, our typesystem uses the linearity imposed by type environment splitting in \rtit{tCn1}, together with the formula environment, $\envVV$, to rule out such cases by using rules \rtit{tMax} and \rtit{tVar}. 

\begin{derivation} \label{drv:Phi4} \vspace{-3mm}
\begin{small}
	\begin{mathpar}
		\inferrule*[Left=tMax]{
			\inferrule*[Left=tNcB]{
				\inferrule*[Left=tBlk]{
					\inferrule*[Left=tCn1]{
						\inferrule*[Left=tAdS]{						
								\inferrule*[Left=tRel]{
										\inferrule*[Left=tVar]{
											(\envVVP(X)\eq\{\typ{i}{\lpid}\})\subseteq(\typ{i}{\lpid})
										}{\typeRule{\env{\envVVP}{(\typ{i}{\lpid})}}{\hVarX}} \;	
									}{\typeRule{\env{\envVVP}{\envVP}}{\mrelease{i}\hVarX}}	\;	
							}{\typeRule{\env{\envVVP}{\envVP}}{(\mscor{i}{i}\hVarX)}} \;
						\inferrule*[Right=tVar]{
								\textbf{Rejected!!} \\\\ (\envVVP(X)=\{\typ{i\,}{\,\lpid}\})\not{\subseteq}\emptyset
							}{\typeRule{\env{\envVVP}{\emptyset}}{\hVarX}}		
					}{\typeRule{\env{\envVVP}{\envVP\eq\sset{\typ{i}{\lpidb}}}}{((\mscor{i}{i}\hVarX)\mand(\hVarX))}}
				}{\typeRule{\env{\envVVP}{\envVP\eq\sset{\typ{i}{\lpidb}}}}{\mblock{i}((\mscor{i}{i}\hVarX)\mand(\hVarX))}}
					\inferrule*[Right=tRel]{
							\inferrule*[Right=tTru]{\;}{\typeRule{\env{\envVV}{\envV}}{\mtru}}
						}{\typeRule{\env{\envVV}{\envV}}{\mrelease{\varepsilon}\mtru}}
			}{\typeRule{\env{(\envVVP\eq\sset{\hVarX\smapsto\envV})}{\envV}}{\mBNec{\mrecv{i}{3}}{\varepsilon}((\mscor{i}{i}\hVarX)\mand(\hVarX))}}
		}{\typeRule{\env{(\envVV\eq\emptyset)}{(\envV\eq\sset{\typ{i}{\lpid}})}}{\mmax{\hVarX}{\mBNec{\mrecv{i}{3}}{\varepsilon}((\mscor{i}{i}\hVarX)\mand(\hVarX))}}}
	\end{mathpar}
\end{small}
\end{derivation}
\noindent The above type derivation rejects formula $\hV_{err}$ by combining the use of environment splitting with formula environment $\envVV$. In particular, when the typesystem analyses the recursive binder $\mmax{X}{}$ with rule \rtit{tMax}, it saves a copy of $\envV\eq\sset{\typ{i}{\lpid}}$, identified by $X$, in $\envVV$. Then while checking conjunction formula $((\mscor{i}{i}\hVarX)\mand(\hVarX))$ with rule \rtit{tCn1}\footnote{This rule was chosen over \rtit{tCn2} as the side-condition $\excl{(\mscor{i}{i}\hVarX), (\hVarX)}=\undef$ holds; this is discussed in \secref{sec:type-rules-excl}.}, the value environment $\envVP\eq\sset{\typ{i}{\lpidb}}$ is split as $\envV'+\emptyset$. The left branch\footnote{Note that the formula still gets rejected even if the right branch is typechecked \wrt $\envVP$ while the left branch is typechecked \wrt $\emptyset$. } is thus typechecked \wrt $\envVP$, while the right branch is typechecked \wrt $\emptyset$. The right branch then gets rejected when rule \rtit{tVar} is applied to verify that the saved copy \ie $\envVP\eq\sset{\typ{i}{\lpidb}}$, is a subset of (or equal to) the assigned value environment \ie $\emptyset$; however this is not the case. Hence the formula gets rejected.  \bqed

\subsection{Introducing Mutual Exclusion}\label{sec:type-rules-excl}
Unfortunately, ensuring that linearity is preserved by using environment splitting (in rule \rtit{tCn1}) is however too coarse of an analysis and rejects useful adaptation-scripts such as \eqref{eq:static:9} from \exref{ex:typed-script} as shown below. 

\begin{example}  \label{ex:mutually-exclusive}The conjunction formula used in \eqref{eq:static:9} from \exref{ex:typed-script} has the form:
  \begin{dmath*}\setstretch{0.5}
    (\,\mANec{\msendD{j}{y}{\tup{\eatom{res},\smash{x+1}}}}{\epsilon}\,\hVarY)   \;\mand \; 
        (\,\mBNec{\msendD{\typ{z}{\lpid}\,}{y}{\eatom{err}}}{i}\,\mrestart{i}_{\varepsilon}\,\mclearMailbox{z}_{i,z}\,\hVarY) 
  \end{dmath*}
  where the subformulas are necessity formulas with \emph{mutually exclusive} patterns \ie there is no action satisfying both patterns $\msendD{j}{y}{\tup{\eatom{res},\smash{x+1}}}$ and $\msendD{\typ{z}{\lpid}\,}{y}{\eatom{err}}$.  In such cases, a conjunction formula operates more like an \emph{external choice} construct rather than a parallel composition \cite{Milner:Concurrency}, where only one branch continues monitoring. Hence these branches can never exhibit a race condition not even due to recursion (as opposed to \eqref{eq:Sigma} in \exref{ex:SigmaCE}). However, if the type system uses rule \rtit{tCn1} to analyse this conjunction formula, this gets unnecessarily rejected when rule \rtit{tVar} is applied to check the formula variable $Y$ which is used in both branches (\ie in a similar manner as in \exref{ex:SigmaCE}). \exqed   
\end{example}\medskip

To refine our analysis, in \defref{def:excl} (below), we define an \emph{approximating} function $\excl{\hV,\hVV}$ that syntactically analyses subformulas to determine whether they are \emph{mutually exclusive} or not.  When this can be \emph{determined statically} (as in the case of \eqref{eq:static:9}), it means that at runtime \emph{at most one branch} will continue, whereas the others will terminate due to \emph{trivial satisfaction}, thereby releasing the actors specified by the respective necessity formulas (recall \rtit{rNc3} from \ref{fig:rt-semantics}). We refer to these releases as \emph{side-effects} (formally defined in \defref{def:eff} below).

Accordingly,  $\excl{\hV,\hVV}$ denotes that mutual exclusion can be determined by returning a tuple consisting of two \emph{release sets}, $\langle\vLstA[\hV],\vLstA[\hVV]\rangle$, containing the actors released by the respective subformulas when an action is mismatched.  Rule \rtit{tCn2} then typechecks each subformula \wrt the \emph{entire} value environment $\envV$, adjusted to take into consideration the side-effects imposed on one branch by the other branch, \ie it considers actors released by the other (defunct) branch, \eg  $\typeRule{\env{\envVV}{\eff{\envV}{\vLstA[\hVV]}}}{\hV}$, where $\vLstA[\hVV]$ signifies the actors that are released in case the other branch $\hVV$ gets trivially satisfied. 

\subsubsection{Defining Side-effects}
At runtime, whenever a mutually exclusive branch $\hVV$ terminates trivially due to mismatching necessities, as a side-effect, it releases a number of actors specified in the release-lists of the terminated necessities. These side-effects may therefore effect the matching branch $\hV$ by for instance releasing a specific linear actor on which this branch applies a synchronous adaptation, \ie in a similar way to what we have previously shown in \exref{ex:model} (see \eqref{eq:ra:11} and \eqref{eq:ra:12}). 

The $\eff{\envV}{\vLstA}$ function defined in \defref{def:eff}, therefore permits the type system to take into consideration these side-effect releases while typecheck a branch $\hV$. This means that a branch $\hV$ is typechecked as if the other exclusive branch $\hVV$ has terminated and thus released the respective actors. 

\begin{definition}[Side-effects]\label{def:eff} 
{
	$\\$\indent\indent$\qquad\eff{\envV}{\vLstA}\defEquals\set{\typ{\varid}{\lpid}}{\varid\!\in\!\vLstA\,\land\,\typ{\varid}{\lpidb}\!\in\!\envV}\cup\set{\typ{\varid}{t}}{\varid\!\notin\!\vLstA\,\land\,\typ{\varid}{t}\!\in\!\envV}$
}
\end{definition}

This function therefore takes as input the value environment $\envV$ associated with one branch $\hV$, together with the release set derived from the opposing branch, \ie $\vLstA[\hVV]$. For each process reference $\varid$ residing in release set $\vLstA[\hVV]$, this function searches for $\varid$ in $\envV$ and changes its type from \lpidb to \lpid. Each branch (\eg $\hV$) can thus be analysed separately with the assumption that the processes defined in the release set of the opposing branch (\eg $\vLstA[\hVV]$) will actually be released at runtime in case a system event causes this opposing branch (\ie $\hVV$) to terminate trivially.
 
\subsubsection{Defining Mutual Exclusion} 
The $\excl{\hV,\hVV}$ function, defined in \defref{def:excl}, returns tuple $\langle\vLstA[\hV],\vLstA[\hVV]\rangle$ in case both functions $\vexcl{\hV,\hVV}$ and $\vexcl{\hVV,\hV}$ respectively return release sets $\vLstA[\hV]$ and $\vLstA[\hVV]$. These release sets denote the actors released by branches $\hV$ and $\hVV$ respectively, in case these get trivially satisfied at runtime. If either one of these two \textsf{vexcl} functions return $\undef$, then the \textsf{excl} function also returns $\undef$, thus denoting that branches $\hV$ and $\hVV$ are \emph{not} mutually exclusive. 
 
\begin{definition}[Mutual Exclusion] \label{def:excl}
{\small  \setstretch{1.1}
	\begin{align*}
		\excl{\hV,\hVV} &\defEquals \begin{cases} 
								\langle\vLstA[\hV],\vLstA[\hVV]\rangle  & \text{if } (\vexcl{\hV,\hVV}\eq\vLstA[\hV]\!\neq\!\bot)\, \land\, (\vexcl{\hVV,\hV}\eq\vLstA[\hVV]\!\neq\!\bot) \\
								\bot & \text{otherwise} \\
							\end{cases}\\
		\vexcl{\hV,\hVV} &\defEquals \begin{cases} 
						    	\vexcl{\hV_{1},\hVV}\cup\vexcl{\hV_{2},\hVV} & \text{if } \hV\eq\hV_{1}\mand\hV_{2} \\
						    	\vexcl{\mrelease{\vLstA}\mtru,\hVV} & \text{if }\hV=\mattrNec{\patE}{\attr}{\vLstA}{\hV'}\, \land\, \hVV\!\neq\!\trivSat\\ & \quad \land\, \forall\patE_{1}\in\fps{\hVV}\cdot\match{\patE}{\patE_{1}}{\bot} \\
						    	\vLstA & \text{if } \hV\eq\mboolE{c}{\hV_{1}}{\hV_{2}}\,\land\,\hVV\!\neq\!\trivSat \\& \quad \land \, \vexcl{\hV_{1},\hVV}\eq\vexcl{\hV_{2},\hVV}\eq\vLstA \\
						    	\vexcl{\hV',\hVV} & \text{if } \hV=\mmax{X}{\hV'}\,\land\,\hVV\!\neq\!\trivSat \\
						    	\vLstA & \text{if } \hV\eq\mrelease{\vLstA}\mtru \\
						    	\varepsilon & \text{if } \hV\neq\mrelease{\vLstA}\mtru \,\land\, \hVV\eq\trivSat\\
						    	\bot & \text{otherwise} \\
						  \end{cases}\\
		\fps{\hV} &\defEquals \begin{cases} 
						    \patE & \text{if }\hV=\mattrNec{\patE}{\attr}{\vLstA}{\hV'}  \\
						    \fps{\hV_{1}},\fps{\hV_{2}} & \text{if }\hV\!\in\!\sset{\mboolE{c}{\hV}{\hVV},\, \hV\mand\hVV}\\
						    \fps{\hV'} & \text{if }\hV=\mmax{X}{\hV'}\\
						    \_ & \text{otherwise} \\
						  \end{cases}
	\end{align*}
	}
\end{definition}

The $\vexcl{\hV,\hVV}$ function incrementally inspects the contents of a branch $\hV$ \wrt the opposing branch $\hVV$ which is always kept constant during analysis. This function returns the release set of $\hV$ only if this is \emph{mutually exclusive} to $\hVV$. We now highlight the three main cases of this function.

The first case we consider is when $\hV$ is also a conjunction $\hV_{1}\mand\hV_{2}$. In this case the \textsf{vexcl} is recursively applied for both $\hV_{1}$ and $\hV_{2}$ \wrt the same opposing branch $\hVV$. The idea is that if $\hV_{1}$ and $\hV_{2}$ are both mutually exclusive to $\hVV$ then their conjunction, \ie $\hV_{1}\mand\hV_{2}$ is also mutually exclusive to $\hVV$. In this case the function returns a \emph{union} of the release sets of both $\hV_{1}$ and $\hV_{2}$, returned by the respective recursive applications. Note however that the inner branches $\hV_{1}$ and $\hV_{2}$ need not be mutually exclusive to each other, as long as they are mutually exclusive to $\hVV$ (we explain this in more detail in \exref{ex:excl}). 

The second case we consider is when  $\hV$ starts with a necessity, \ie $\mattrNec{\patE}{\attr}{\vLstA}\hV'$. In this case the function internally calls the $\fps{\hVV}$ function to retrieve \emph{all} the \emph{top patterns} defined in the opposing branch $\hVV$. To better understand what these top patterns are, as an example assume that we use the \textsf{fps} function to analyse $\hVV\!=\!(\mattrNec{\patE_{1}}{\attr}{\vLstA[1]}\hVV_{1}) \mand (\mattrNec{\patE_{2}}{\attr}{\vLstA[2]}\hVV_{2}) \mand (\mscor{\vLstB[3]}{\vLstA[3]}\hVV_{3})$. In this case, $\fps{\hVV}$ returns the pattern set $\{\patE_{1},\patE_{2},\_\}$, where $\patE_{1}$ and $\patE_{2}$ refer to the patterns of the guarding necessities of branches $(\mattrNec{\patE_{1}}{\attr}{\vLstA[1]}\hVV_{1})$ and $(\mattrNec{\patE_{2}}{\attr}{\vLstA[2]}\hVV_{2})$, while ``\_ ''is a universally matching pattern which is returned due to the branch starting with $\mscor{\vLstB[3]}{\vLstA[3]}$ --- we sometimes refer to these branches as \emph{unguarded}. 

The \textsf{vexcl} function then uses standard pattern matching (as defined by the \textsf{mtch} function in \defref{def:match}) to check whether the pattern $\patE$ of the necessity being analysed matches any of the top patterns defined in the opposing branch $\hVV$. If it does \emph{not} match with any of them (\ie \textsf{mtch} returns $\undef$ for each top pattern), then this means that branch $\hV$  is mutually exclusive to branch $\hVV$. This function is therefore recursively applied for $\mrelease{\vLstA}\mtru$ (\ie the function internally calls $\vexcl{\mrelease{\vLstA}\mtru,\hVV}$), thereby denoting that whenever a necessity $\mattrNec{\patE}{\attr}{\vLstA}$ gets trivially satisfied at runtime, it releases the actors enlisted in $\vLstA$ prior to reducing into $\mtru$; this recursive application thus returns $\vLstA$.

Until now we have ignored the side condition $\hVV\!\neq\!\trivSat$, which states that for this case to apply, the opposing branch $\hVV$ must not form a collection of trivially satisfied branches \eg $\hV\equiv(\mrelease{\vLstA[1]}\mtru)\mand\cdots\mand(\mrelease{\vLstA[n]}\mtru)$. This condition is used throughout most of the cases in the \textsf{vexcl} function to navigate its inspection to the penultimate case. This case states that if the formula being inspected, $\hV$, is not a release operation and the opposing branch $\hVV$ forms a collection of trivially satisfied branches, \ie $\hVV\eq\trivSat$, then the function returns an empty release set, $\varepsilon$. This condition is thus used to signify that \emph{any branch} $\hV$ is mutually exclusive to a collection of trivially satisfied branches. For instance, if $\hV\eq\mscor{\vLstB}{\vLstA}\hV'$ and $\hVV\eq\trivSat$, then $\vexcl{\hV,\hVV}$ returns $\varepsilon$ and not $\undef$.\medskip

\begin{remark} These side-conditions therefore allow us to show that two mutually exclusive branches (\eg $\Bigl(\mBNec{\mrecv{i}{3}}{\varepsilon}\mscor{i}{i}\mtru\Bigr)\mand\Bigl(\mBNec{\mrecv{i}{4}}{\varepsilon}\mscor{i}{i}\mfls\Bigr)$), remain mutually exclusive even when one of the branches gets trivially satisfied at runtime. For instance, \lemmaref{lemmaXXXB} (stated below) proves that if two mutually exclusive branches $\hV$ and $\hVV$ reduce over a $\tau$-action, then the resultant branches $\hV'$ and $\hVV$ remain mutually exclusive to each other. \medskip

\textbf{\lemmaref{lemmaXXXB}} {\small\setstretch{0.5} \begin{math}\excl{\hV,\hVV}=(\vLstA[\hV],\vLstA[\hVV]) \text{ and } \hV\traceEventTau\hV'\, \imp \,\excl{\hV',\hVV}=(\vLstA[\hV],\vLstA[\hVV])\end{math} }\medskip

\noindent This Lemma is proved by rule induction on $\hV\traceEventTau\hV'$ in Appendix \secref{sec:other-aux-lemmas}, along with two other similar lemmas, \ie \lemmaref{lemmaXXXA} and \lemmaref{lemmaH} which respectively prove that mutual exclusion is also preserved in the case of $\actE$ and $\actC$ reductions. \bqed
\end{remark}

\begin{example}\label{ex:excl}
	To better understand how the \textsf{excl} function works, consider the following partial script having 4 branches.	\medskip
	
	\begin{dmath*}\Bigl(\underbrace{(\mANec{\mrecv{i}{3}}{j}{\hV_{1}}\, \mand\, \mANec{\mrecv{\typ{x}{\upid}}{3}}{k}{\hV_{2}})}_{\displaystyle\hV}\, \mand\, \underbrace{(\mANec{\mrecv{i}{4}}{\varepsilon}{\hVV_{1}}\, \mand\, \mANec{\mrecv{\typ{y}{\upid}}{5}}{l}{\hVV_{2}})}_{\displaystyle\hVV}\Bigr)\end{dmath*}
	
	\noindent As shown by the \textsf{excl} function derivation below, if we check for mutual exclusion between the topmost two branches, $\hV$ and $\hVV$ we can notice that the patterns in the right topmost branch (\ie $\mrecv{i}{3}$ and $\mrecv{\typ{x}{\upid}}{3}$) clearly do not match any pattern in the left topmost branch (\ie $\mrecv{i}{4}$ and $\mrecv{\typ{x}{\upid}}{5}$). Hence this function returns tuple $\langle (j;\!k) , l\rangle$ denoting that if $\hV$ gets trivially satisfied at runtime, then the actors contained in release set $(j;\!k)$ would be released; similarly actor $l$ is released in case $\hVV$ gets trivially satisfied.
	
	{\small\setlength{\abovedisplayskip}{-5pt}\setlength{\belowdisplayskip}{-10pt} \setlength{\abovedisplayshortskip}{-5pt} \setlength{\belowdisplayshortskip}{-10pt} 
		 \begin{align*}
		\hspace{-3mm} \excl{&\hspace{-3mm}&\overbrace{\Bigl(\mANec{\mrecv{i}{3}}{j}{\hV_{1}}\, \mand\, \mANec{\mrecv{\typ{x}{\upid}}{3}}{k}{\hV_{2}}\Bigr)}^{\displaystyle\hV}\;,&\;\overbrace{\Bigl(\mANec{\mrecv{i}{4}}{\varepsilon}{\hVV_{1}}\, \mand\, \mANec{\mrecv{\typ{y}{\upid}}{5}}{l}{\hVV_{2}}\Bigr)}^{\displaystyle\hVV}& \hspace{-3mm}} \\[0mm]
		\hspace{-8mm} = \Big\langle&& \hspace{-3mm}\vexcl{\mANec{\mrecv{i}{3}}{j}{\hV_{1}}\, \mand\, \mANec{\mrecv{\typ{x}{\upid}}{3}}{k}{\hV_{2}},\,\hVV}\;,&\; \vexcl{\mANec{\mrecv{i}{4}}{\varepsilon}{\hVV_{1}}\, \mand\, \mANec{\mrecv{\typ{y}{\upid}}{5}}{l}{\hVV_{2}},\,\hV}& \hspace{-3mm}\Big\rangle  \\[-2mm]		
		\hspace{-8mm} = \Big\langle&& \hspace{-3mm}\vexcl{\mANec{\mrecv{i}{3}}{j}{\hV_{1}},\,\hVV}\cup\vexcl{\mANec{\mrecv{\typ{x}{\upid}}{3}}{k}{\hV_{2}},\,\hVV} \;,&\; \vexcl{\mANec{\mrecv{i}{4}}{\varepsilon}{\hVV_{1}},\,\hV}\cup\vexcl{\mANec{\mrecv{\typ{y}{\upid}}{5}}{l}{\hVV_{2}},\,\hV}& \hspace{-3mm}\Big\rangle \\[3mm]
		\shortintertext{\normalsize\centering (since $\mrecv{i}{3}$ and $\mrecv{\typ{x\,}{\,\upid}}{3}$ do \emph{not} match \wrt $\mrecv{i}{4}$ and $\mrecv{\typ{y\,}{\,\upid}}{5}$)}  \\[-6.5mm]
		 \hspace{-8mm} = \Big\langle&& \hspace{-3mm}\vexcl{\mrelease{j}\mtru,\,\hVV}\cup\vexcl{\mrelease{k}\mtru,\,\hVV} \;,&\; \vexcl{\mrelease{\varepsilon}\mtru,\,\hV}\cup\vexcl{\mrelease{k}\mtru,\,\hV}& \hspace{-3mm}\Big\rangle  \\[-2mm]
		\hspace{-8mm}  = \Big\langle&& \hspace{-3mm} \{j\}\cup\{k\} \;,&\; \varepsilon\cup\{l\}& \hspace{-3mm}\Big\rangle  \\[-2mm]		
	    \hspace{-8mm}  = \Big\langle&& \hspace{-3mm} (j;\!k) \;,&\; l& \hspace{-3mm}\Big\rangle
		\end{align*}	}	
	
	Therefore with this mechanism we can statically infer that if an event matches either $\mrecv{i}{3}$ or $\mrecv{\typ{x}{\upid}}{3}$, it would certainly \emph{not match} patterns $\mrecv{i}{4}$ and $\mrecv{\typ{x}{\upid}}{5}$ (and vice-versa). However, this does not mean that the internal branches are exclusive amongst themselves as well. In fact $\excl{\mANec{\mrecv{i}{3}}{j}{\hV_{1}},\mANec{\mrecv{\typ{x}{\upid}}{3}}{k}{\hV_{2}}}=\undef$ as both patterns may match system action $\mrecv{i}{3}$. 

	Therefore, as shown in the type derivation (below), while typechecking this partial formula the type system can be less restrictive about the topmost branches $\hV$ and $\hVV$ by using type rule \rtit{tCn2} (as they are mutually exclusive). However, in the case of inner branches $\mANec{\mrecv{i}{3}}{j}{\hV_{1}}$ and $\mANec{\mrecv{\typ{x}{\upid}}{3}}{k}{\hV_{2}}$, the type system takes a more stringent approach and preserves linearity by splitting the value environment with rule \rtit{tCn1}.
	
	{\footnotesize 
		\noindent\begin{derivation} 
			{\setstretch{1.4} \setlength{\abovedisplayskip}{0pt}\setlength{\belowdisplayskip}{0pt} \setlength{\abovedisplayshortskip}{0pt} \setlength{\belowdisplayshortskip}{0pt}  
			\begin{mathpar}
				\inferrule*[Left=tCn2]{
					\inferrule*[Left=tCn1]{ 					
						\inferrule*[Left=tNcA]{ \cdots }{\typeRule{\env{\envVV}{\envVA}}{\mANec{\mrecv{i}{3}}{j}{\hV_{1}}}}\and							
						\inferrule*[Right=tNcA]{ \cdots }{\typeRule{\env{\envVV}{\envVB}}{\mANec{\mrecv{\typ{x}{\upid}}{3}}{k}{\hV_{2}})}}
					}{\typeRule{\env{\envVV}{\eff{\envV}{l}}}{(\mANec{\mrecv{i}{3}}{j}{\hV_{1}}\, \mand\, \mANec{\mrecv{\typ{x}{\upid}}{3}}{k}{\hV_{2}})}} \and					
					\inferrule*[Right=tCn2]{ \cdots
					}{\typeRule{\env{\envVV}{\eff{\envV}{(j;k)}}}{(\mANec{\mrecv{i}{4}}{\varepsilon}{\hVV_{1}}\, \mand\, \mANec{\mrecv{\typ{y}{\upid}}{5}}{l}{\hVV_{2}})}}					
				}{\typeRule{\env{\envVV}{\envV}}{(\mANec{\mrecv{i}{3}}{j}{\hV_{1}}\, \mand\, \mANec{\mrecv{\typ{x}{\upid}}{3}}{k}{\hV_{2}}) \mand (\mANec{\mrecv{i}{4}}{\varepsilon}{\hVV_{1}}\, \mand\, \mANec{\mrecv{\typ{y}{\upid}}{5}}{l}{\hVV_{2}})}} \\ 	
			\qquad (\text{where } \eff{\envV}{l}=\envVA+\envVB) $\bqed$
			\end{mathpar}}
		\end{derivation}		
	}	
\end{example}

\subsubsection{Example Typechecking Derivations}
The better understand how our type system accepts valid scripts and rejects invalid ones, we present a number of example type derivations for some of the properties that we have seen so far.  For instance in Derivation~\ref{drv:ex:1} (below) we show how \eqref{eq:static:9} typechecks (\ie is accepted by the typesystem) \wrt value environment $\envV_{0}=\{\typ{\idV}{\lpid},\typ{\idVV}{\upid}\}$.

	{\footnotesize 
		\begin{derivation} \label{drv:ex:1} $\\[2mm]$
			\noindent\fbox{%
		    \parbox{\textwidth}{%
		    	\textbf{Formula Environments:} $\envVV\eq(\hVarY\mapsto\envV_{0})$\\
		    	\textbf{Value Environments:} $\mathbf{\envV_{0}}\eq\sset{\typ{i}{\lpid},\typ{j}{\upid}}$; \hfill $\mathbf{\envV_{1}}\eq\envV_{0},\sset{\typ{y}{\upid}}$; \hfill $\mathbf{\envV_{2}}\eq\sset{\typ{i}{\lpidb},\typ{j}{\upid},\typ{y}{\upid}}$; \hfill $\mathbf{\envV_{3}}\eq\eff{\envV_{2}}{i}\eq\envV_{1}$; \hfill $\mathbf{\envV_{4}}\eq\eff{\envV_{2}}{\varepsilon}\eq\envV_{2}$; \hfill $\mathbf{\envV_{5}}\eq\envV_{4},\sset{\typ{z}{\lpid}}$; \hfill $\mathbf{\envV_{6}}\eq\envV_{4},\sset{\typ{z}{\lpidb}}$; \hfill $\mathbf{\envV_{7}}\eq\sset{\typ{i}{\lpid},\typ{j}{\upid},\typ{y}{\upid},\typ{z}{\lpid}}$.}} \vspace{3mm}
			{\setstretch{1.4} \begin{mathpar}
				\inferrule*[left=tMax, narrower=5]{
					\inferrule*[Left=tNcA]{
						\inferrule*[Left=tNcB]{
							\inferrule*[Left=tBlk]{
								\inferrule*[Left=tCn2]{
									\inferrule*[Left=tNcA]{	
										\inferrule*[Left=tVar]{
													\textbf{Accepted}\\\\(\envVV(\hVarY)\eq\envV_{0})\subset\envV_{3}
											}{\typeRule{\env{\envVV}{\envV_{3}}}{\hVarY}} \;													
										}{\typeRule{\env{\envVV}{\envV_{3}}}{\mANec{\msendD{j}{y}{\tup{\eatom{res},\smash{x+1}}}}{\varepsilon}\,\hVarY}} \and						
								\inferrule*[Right=tNcB]{	
										\inferrule*[Left=tBlk]{
											\inferrule*[Left=tAdS]{
													\inferrule*[Left=tRel]{
															\inferrule*[Left=tVar]{
																\textbf{Accepted}\\\\{(\envVV(\hVarY)\!=\!\envV_{0})\subseteq\envV_{7}}
															}{\typeRule{\env{\envVV}{\envV_{7}}}{\hVarY}}
														}{\typeRule{\env{\envVV}{\envV_{6}}}{\mrelease{i;z}\,\hVarY}}
												}{\typeRule{\env{\envVV}{\envV_{6}}}{\mrestart{i}_{\varepsilon}\,\mclearMailbox{z}_{i;z}\,\hVarY}}
											}{\typeRule{\env{\envVV}{\envV_{5}}}{\mblock{z}\mrestart{i}_{\varepsilon}\,\mclearMailbox{z}_{i;z}\,\hVarY}} 	\;												
										\inferrule*[Right=tRel]{
												\inferrule*[Right=tTru]{\textbf{Accepted}}{\typeRule{\env{\envVV}{\envV_{4}}}{\mtru}}
											}{\typeRule{\env{\envVV}{\envV_{4}}}{\mrelease{i;z}\mtru}}
										}{\typeRule{\env{\envVV}{\envV_{4}}}{\mBNec{\msendD{\typ{z}{\lpid}\,}{y}{\eatom{err}}}{i}\,\mrestart{i}_{\varepsilon}\,\mclearMailbox{z}_{i;z}\,\hVarY}}
								}{\typeRule{\env{\envVV}{\envV_{2}}}{((\mANec{\msendD{j}{y}{\tup{\eatom{res},\smash{x+1}}}}{\varepsilon}\,\hVarY) \,\mand \, (\mBNec{\msendD{\typ{z}{\lpid}\,}{y}{\eatom{err}}}{i}\,\mrestart{i}_{\varepsilon}\,\mclearMailbox{z}_{i;z}\,\hVarY))}}							
							}{\typeRule{\env{\envVV}{\envV_{1}}}{\mblock{i}((\mANec{\msendD{j}{y}{\tup{\eatom{res},\smash{x+1}}}}{\varepsilon}\,\hVarY) \,\mand \, (\mBNec{\msendD{\typ{z}{\lpid}\,}{y}{\eatom{err}}}{i}\,\mrestart{i}_{\varepsilon}\,\mclearMailbox{z}_{i;z}\,\hVarY))}}							
						}{\typeRule{\env{\envVV}{\envV_{1}}}{\mBNec{\msendD{i}{\_}{\{\eatom{inc},x,y\}}}{\varepsilon}((\mANec{\msendD{j}{y}{\tup{\eatom{res},\smash{x+1}}}}{\varepsilon}\,\hVarY) \,\mand \, (\mBNec{\msendD{\typ{z}{\lpid}\,}{y}{\eatom{err}}}{i}\,\mrestart{i}_{\varepsilon}\,\mclearMailbox{z}_{i;z}\,\hVarY))}}
					}{\typeRule{\env{\envVV}{\envV_{0}}}{\mANec{\mrecv{i}{\{\eatom{inc},\typ{x}{\dat},\typ{y}{\upid}\}}}{\varepsilon}\mBNec{\msendD{i}{\_}{\{\eatom{inc},x,y\}}}{\varepsilon}{\setstretch{1.2}\begin{mlbrace}\setstretch{1}(\mANec{\msendD{j}{y}{\tup{\eatom{res},\smash{x+1}}}}{\varepsilon}\,\hVarY) \cr \; \mand (\mBNec{\msendD{\typ{z}{\lpid}\,}{y}{\eatom{err}}}{i}\,\mrestart{i}_{\varepsilon}\,\mclearMailbox{z}_{i;z}\,\hVarY)\end{mlbrace}}}}
				}{\typeRule{\env{\emptyset}{\envV_{0}}}{\mmax{\,\hVarY}{\mANec{\mrecv{i}{\{\eatom{inc},\typ{x}{\dat},\typ{y}{\upid}\}}}{\varepsilon}\mBNec{\msendD{i}{\_}{\{\eatom{inc},x,y\}}}{\varepsilon}{\setstretch{1.2}\begin{mlbrace}(\mANec{\msendD{j}{y}{\tup{\eatom{res},\smash{x+1}}}}{\varepsilon}\,\hVarY) \\ \;\mand (\mBNec{\msendD{\typ{z}{\lpid}\,}{y}{\eatom{err}}}{i}\,\mrestart{i}_{\varepsilon}\,\mclearMailbox{z}_{i;z}\hVarY)\end{mlbrace}}}}} 	\vspace{-3mm} \;\; \exqed
			\end{mathpar}}
		\end{derivation}		
	}
	
	\noindent Formula \eqref{eq:static:9} is accepted by the type system since all possible three branches in Derivation~\ref{drv:ex:1} managed to reach (and satisfy) a base case typing rule (\ie \rtit{tVar} and \rtit{tTru}). 
	
	By contrast, for any value environment $\envV$, the typesystem is unable to typecheck erroneous script $\hV''$ (augmented with the necessary type annotations) from \exref{ex:model}, and thus rejects it. 
	The derivation below shows how this erroneous script is rejected when typechecked \wrt $\envV_{0}$: 

{\footnotesize \setstretch{1.4}
		\begin{derivation} $\\[2mm]$
			\noindent\fbox{%
		    \parbox{\textwidth}{%
		    	\textbf{Formula Environments:} \quad $\envVV\eq(\hVarY\mapsto\envV_{0})$\\
		    	\textbf{Value Environments:} \quad $\mathbf{\envV_{0}}\eq\sset{\typ{i}{\lpid},\typ{j}{\upid}}$; \hfill $\mathbf{\envV_{1}}\eq\envV_{0},\sset{\typ{y}{\upid}}$; \hfill $\mathbf{\envV_{2}}\eq\eff{\envV_{1}}{i}\eq\envV_{1}$; \\ \indent \hfill $\mathbf{\envV_{3}}\eq\eff{\envV_{1}}{\varepsilon}\eq\envV_{1}$; \hfill $\mathbf{\envV_{4}}\eq\envV_{3},\sset{\typ{z}{\lpid}}$; \hfill $\mathbf{\envV_{5}}\eq\sset{\typ{i}{\lpid},\typ{j}{\upid},\typ{y}{\upid},\typ{z}{\lpidb}}$.\indent}}\vspace{-1mm}
			\begin{mathpar}
				\inferrule*[left=tMax, narrower=5]{
					\inferrule*[Left=tNcA]{
						\inferrule*[Left=tNcA]{
								\inferrule*[Left=tCn2]{
									\inferrule*[Left=tNcA]{	
										\inferrule*[Left=tVar]{
													(\envVV(\hVarY)\eq\envV_{0})\subset\envV_{2}
											}{\typeRule{\env{\envVV}{\envV_{2}}}{\hVarY}} \;													
										}{\typeRule{\env{\envVV}{\envV_{2}}}{\mANec{\msendD{j}{y}{\tup{\eatom{res},\smash{x+1}}}}{\varepsilon}\,\hVarY}} \and						
								\inferrule*[Right=tNcB]{	
										\inferrule*[Left=tBlk]{
											\inferrule*[Left=tAdS]{		
													\textbf{Rejected!!}\\\\\typ{i}{\lpidb}\notin\envV_{5}
												}{\typeRule{\env{\envVV}{\envV_{5}}}{\mrestart{i}_{\varepsilon}\,\mclearMailbox{z}_{i;z}\,\hVarY}}
											}{\typeRule{\env{\envVV}{\envV_{4}}}{\mblock{z}\mrestart{i}_{\varepsilon}\,\mclearMailbox{z}_{i;z}\,\hVarY}} 	\;												
										\inferrule*[Right=tRel]{
												\inferrule*[Right=tTru]{\;}{\typeRule{\env{\envVV}{\envV_{3}}}{\mtru}}
											}{\typeRule{\env{\envVV}{\envV_{3}}}{\mrelease{i;z}\mtru}}
										}{\typeRule{\env{\envVV}{\envV_{3}}}{\mBNec{\msendD{\typ{z}{\lpid}\,}{y}{\eatom{err}}}{i}\,\mrestart{i}_{\varepsilon}\,\mclearMailbox{z}_{i;z}\,\hVarY}}
								}{\typeRule{\env{\envVV}{\envV_{1}}}{((\mANec{\msendD{j}{y}{\tup{\eatom{res},\smash{x+1}}}}{\varepsilon}\,\hVarY) \,\mand \, (\mBNec{\msendD{\typ{z}{\lpid}\,}{y}{\eatom{err}}}{i}\,\mrestart{i}_{\varepsilon}\,\mclearMailbox{z}_{i;z}\,\hVarY))}}	
						}{\typeRule{\env{\envVV}{\envV_{1}}}{\mattrNec{\msendD{i}{\_}{\{\eatom{inc},x,y\}}}{\scriptstyle\pmb{\norm}}{\varepsilon}((\mANec{\msendD{j}{y}{\tup{\eatom{res},\smash{x+1}}}}{\varepsilon}\,\hVarY) \,\mand \, (\mBNec{\msendD{\typ{z}{\lpid}\,}{y}{\eatom{err}}}{i}\,\mrestart{i}_{\varepsilon}\,\mclearMailbox{z}_{i;z}\,\hVarY))}}
					}{\typeRule{\env{\envVV}{\envV_{0}}}{\mANec{\mrecv{i}{\{\eatom{inc},\typ{x}{\dat},\typ{y}{\upid}\}}}{\varepsilon}\mattrNec{\msendD{i}{\_}{\{\eatom{inc},x,y\}}}{\scriptstyle\pmb{\norm}}{\varepsilon}{\setstretch{1.2}\begin{mlbrace}\setstretch{1}(\mANec{\msendD{j}{y}{\tup{\eatom{res},\smash{x+1}}}}{\varepsilon}\,\hVarY) \cr \; \mand (\mBNec{\msendD{\typ{z}{\lpid}\,}{y}{\eatom{err}}}{i}\,\mrestart{i}_{\varepsilon}\,\mclearMailbox{z}_{i;z}\,\hVarY)\end{mlbrace}}}}
				}{\typeRule{\env{\emptyset}{\envV_{0}}}{\mmax{\,\hVarY}{\mANec{\mrecv{i}{\{\eatom{inc},\typ{x}{\dat},\typ{y}{\upid}\}}}{\varepsilon}\mattrNec{\msendD{i}{\_}{\{\eatom{inc},x,y\}}}{\scriptstyle\pmb{\norm}}{\varepsilon}{\setstretch{1.2}\begin{mlbrace}(\mANec{\msendD{j}{y}{\tup{\eatom{res},\smash{x+1}}}}{\varepsilon}\,\hVarY) \\ \;\mand (\mANec{\msendD{\typ{z}{\lpid}\,}{y}{\eatom{err}}}{i}\,\mrestart{i}_{\varepsilon}\,\mclearMailbox{z}_{i;z}\hVarY)\end{mlbrace}}}}} 	\vspace{-3mm} \; \exqed
			\end{mathpar}
		\end{derivation}		
	}

	\noindent Our type system thus rejects formula $\hV''$ as it detects that actor $i$ is not being blocked by some blocking necessity prior to applying $\mrestart{i}_{\varepsilon}$. In fact, this error is detected, and thus rejected, by rule \rtit{tAds} which detects that the adaptee $i$ had the type $\lpid$ in the value environment $\envV_{5}$; if $i$ had been blocked by a prior construct, then its type would have been changed to \lpidb and thus the script would have been accepted (as we have previously shown in Derivation~\ref{drv:ex:1} for formula \eqref{eq:static:9}). 

	Similarly, the typesystem also rejects scripts that introduce synchronisation errors due to concurrency such as $\hV'''$ from \exref{ex:model}. Recall that $\hV'''$ suffers from a race condition whereby necessity $\mANec{\msendD{j}{y}{\tup{\eatom{res},\smash{x+1}}}}{i}$ erroneously releases actor $i$ if it mismatches a system action. As discussed in \exref{ex:model}, this results in a premature release of actor \idV, which may interfere with the synchronous adaptation $\mrestart{i}_{\varepsilon}$ along the other branch.  However, as shown in Derivation~\ref{drv:Phi3}, rule \rtit{tCn2} allows for this interference to be statically detected by considering the \emph{side-effects} of the respective mutually exclusive branches.  

	{\footnotesize 
		\begin{derivation}\label{drv:Phi3} $\\[2mm]$
			\noindent\fbox{%
		    \parbox{\textwidth}{%
		    	\textbf{Formula Environments:} $\envVV\eq(\hVarY\mapsto\envV_{0})$\\
		    	\textbf{Value Environments:} $\mathbf{\envV_{0}}\eq\sset{\typ{i}{\lpid},\typ{j}{\upid}}$; \hfill $\mathbf{\envV_{1}}\eq\envV_{0},\sset{\typ{y}{\upid}}$; \hfill $\mathbf{\envV_{2}}\eq\sset{\typ{i}{\lpidb},\typ{j}{\upid},\typ{y}{\upid}}$; \\ \indent $\mathbf{\envV_{3}}\eq\eff{\envV_{2}}{i}\eq\envV_{1}$; \hfill $\mathbf{\envV_{4}}\eq\eff{\envV_{2}}{i}\eq\envV_{1}$; \hfill $\mathbf{\envV_{5}}\eq\envV_{4},\sset{\typ{z}{\lpid}}$; \hfill $\mathbf{\envV_{6}}\eq\sset{\typ{i}{\lpid},\typ{j}{\upid},\typ{y}{\upid},\typ{z}{\lpidb}}$.}} 
			{\setstretch{1.4} \begin{mathpar}
				\inferrule*[left=tMax, narrower=5]{
					\inferrule*[Left=tNcA]{
						\inferrule*[Left=tNcB]{
							\inferrule*[Left=tBlk]{
								\inferrule*[Left=tCn2]{
									\inferrule*[Left=tNcA]{	
										\inferrule*[Left=tVar]{
													(\envVV(\hVarY)\eq\envV_{0})\subset\envV_{3}
											}{\typeRule{\env{\envVV}{\envV_{3}}}{\hVarY}} \;													
										}{\typeRule{\env{\envVV}{\envV_{3}}}{\mANec{\msendD{j}{y}{\tup{\eatom{res},\smash{x+1}}}}{\scriptstyle\pmb{i}}\,\hVarY}} \and						
								\inferrule*[Right=tNcB]{	
										\inferrule*[Left=tBlk]{
											\inferrule*[Left=tAdS]{											
													\textbf{Rejected!!}\\\\\typ{i}{\lpidb}\notin\envV_{6}
												}{\typeRule{\env{\envVV}{\envV_{6}}}{\mrestart{i}_{\varepsilon}\,\mclearMailbox{z}_{i;z}\,\hVarY}}
											}{\typeRule{\env{\envVV}{\envV_{5}}}{\mblock{z}\mrestart{i}_{\varepsilon}\,\mclearMailbox{z}_{i;z}\,\hVarY}} 	\;												
										\inferrule*[Right=tRel]{
												\inferrule*[Right=tTru]{\;}{\typeRule{\env{\envVV}{\envV_{4}}}{\mtru}}
											}{\typeRule{\env{\envVV}{\envV_{4}}}{\mrelease{i;z}\mtru}}
										}{\typeRule{\env{\envVV}{\envV_{4}}}{\mBNec{\msendD{\typ{z}{\lpid}\,}{y}{\eatom{err}}}{i}\,\mrestart{i}_{\varepsilon}\,\mclearMailbox{z}_{i;z}\,\hVarY}}
								}{\typeRule{\env{\envVV}{\envV_{2}}}{((\mANec{\msendD{j}{y}{\tup{\eatom{res},\smash{x+1}}}}{\scriptstyle\pmb{i}}\,\hVarY) \,\mand \, (\mBNec{\msendD{\typ{z}{\lpid}\,}{y}{\eatom{err}}}{i}\,\mrestart{i}_{\varepsilon}\,\mclearMailbox{z}_{i;z}\,\hVarY))}}							
							}{\typeRule{\env{\envVV}{\envV_{1}}}{\mblock{i}((\mANec{\msendD{j}{y}{\tup{\eatom{res},\smash{x+1}}}}{\scriptstyle\pmb{i}}\,\hVarY) \,\mand \, (\mBNec{\msendD{\typ{z}{\lpid}\,}{y}{\eatom{err}}}{i}\,\mrestart{i}_{\varepsilon}\,\mclearMailbox{z}_{i;z}\,\hVarY))}}							
						}{\typeRule{\env{\envVV}{\envV_{1}}}{\mBNec{\msendD{i}{\_}{\{\eatom{inc},x,y\}}}{\varepsilon}((\mANec{\msendD{j}{y}{\tup{\eatom{res},\smash{x+1}}}}{\scriptstyle\pmb{i}}\,\hVarY) \,\mand \, (\mBNec{\msendD{\typ{z}{\lpid}\,}{y}{\eatom{err}}}{i}\,\mrestart{i}_{\varepsilon}\,\mclearMailbox{z}_{i;z}\,\hVarY))}}
					}{\typeRule{\env{\envVV}{\envV_{0}}}{\mANec{\mrecv{i}{\{\eatom{inc},\typ{x}{\dat},\typ{y}{\upid}\}}}{\varepsilon}\mBNec{\msendD{i}{\_}{\{\eatom{inc},x,y\}}}{\varepsilon}{\setstretch{1.2}\begin{mlbrace}\setstretch{1}(\mANec{\msendD{j}{y}{\tup{\eatom{res},\smash{x+1}}}}{\scriptstyle\pmb{i}}\,\hVarY) \cr \; \mand (\mBNec{\msendD{\typ{z}{\lpid}\,}{y}{\eatom{err}}}{i}\,\mrestart{i}_{\varepsilon}\,\mclearMailbox{z}_{i;z}\,\hVarY)\end{mlbrace}}}}
				}{\typeRule{\env{\emptyset}{\envV_{0}}}{\mmax{\,\hVarY}{\mANec{\mrecv{i}{\{\eatom{inc},\typ{x}{\dat},\typ{y}{\upid}\}}}{\varepsilon}\mBNec{\msendD{i}{\_}{\{\eatom{inc},x,y\}}}{\varepsilon}{\setstretch{1.2}\begin{mlbrace}(\mANec{\msendD{j}{y}{\tup{\eatom{res},\smash{x+1}}}}{\scriptstyle\pmb{i}}\,\hVarY) \\ \;\mand (\mBNec{\msendD{\typ{z}{\lpid}\,}{y}{\eatom{err}}}{i}\,\mrestart{i}_{\varepsilon}\,\mclearMailbox{z}_{i;z}\hVarY)\end{mlbrace}}}}} \vspace{-3mm} \; \exqed
			\end{mathpar}} 
		\end{derivation}		
	} 
	
	\noindent In fact, when rule \rtit{tCn2} is applied in Derivation~\ref{drv:Phi3}, it uses the \textsf{excl} function to analyse the respective branches. Since these branches are mutually exclusive, the function returns the tuple $\langle i, i\rangle$, \ie the release lists of the top necessities $\mANec{\msendD{j}{y}{\tup{\eatom{res},\smash{x+1}}}}{\scriptstyle\pmb{i}}$ and $\mBNec{\msendD{\typ{z}{\lpid}\,}{y}{\eatom{err}}}{i}$. The release sets in this tuple are then used to construct environments $\envV_{3}$ and $\envV_{4}$, from $\envV_{2}$, \wrt the side-effects imposed by one branch on the other branch in case of trivial satisfaction. Since both branches release $i$ upon being trivially satisfied, then both $\envV_{3}$ and $\envV_{4}$ are constructed in the same way, \ie $\envV_{3}\!=\!\eff{\envV_{2}}{i}\!=\!\{\typ{i}{\lpid},\typ{j}{\upid},\typ{y}{\upid}\}\!=\!\envV_{4}$.
	
	Note that $\typ{i}{\lpidb}$ in $\envV_{2}$ was changed to $\typ{i}{\lpid}$ in $\envV_{3}$ and $\envV_{4}$. This allows for the type system to detect that if at runtime the synchronous adaptation $\mrestart{i}_{\varepsilon}$ is applied, then the trivially satisfied opposing branch (\ie $\mANec{\msendD{j}{y}{\tup{\eatom{res},\smash{x+1}}}}{\scriptstyle\pmb{i}}$) could have already released actor $i$, thereby leading to a synchronisation error.  

\section{Dynamic Analysis of Typed Scripts}
\label{sec:dynamic}
The typed adaptation-scripts of \secref{sec:static} need to execute \wrt the systems described in \secref{sec:ra-model}.  Crucially, however, we cannot expect that a monitored system observes the type discipline assumed by the script. This, in turn, may create \emph{type incompatibilities} that need to be detected and handled \emph{at runtime} by the monitor.

\begin{example}
  \label{ex:dynamic-typechecking} Recall that in Derivation~\ref{drv:ex:1}, the typed script (\ref{eq:static:9}) (restated below) was accepted by the type system when typechecked \wrt value environment $\envV_{0}=\{\typ{\idV}{\lpid},\typ{\idVV}{\upid}\}$.  
		  \begin{equation*}\setstretch{1.1}
		    \begin{array}{l}
		    \hV'\;\deftxt\;\; \mmax{\,\hVarY}{\mANec{\mrecv{\idV}{\tup{\eatom{inc},\typ{x}{\dat},\typ{y}{\upid}}}}{\varepsilon}}\\[-1mm]
		      \qquad\qquad\qquad\qquad\mBNec{\msendD{\idV}{\_}{\{\eatom{inc},x,y\}}}{\varepsilon}
		                \begin{mlbrace} 
		        (\,\mANec{\msendD{j}{y}{\tup{\eatom{res},\smash{x+1}}}}{\varepsilon}\,\hVarY)   \;\mand \; \\[0mm] 
		        (\,\mBNec{\msendD{\typ{z}{\lpid}\,}{y}{\eatom{err}}}{i}\,\mrestart{i}_{\varepsilon}\,\mclearMailbox{z}_{i,z}\,\hVarY) 
		      \end{mlbrace} \quad \eqref{eq:static:9}
		    \end{array}
		 \end{equation*} 
  There are two classes of type incompatibilities that may arise during runtime monitoring:
  \begin{itemize}[leftmargin=5mm]
  \item When listening for a pattern, \eg $\mrecv{\idV}{\tup{\eatom{inc},\typ{x}{\dat},\typ{y}{\upid}}}$, the system may generate the action $\mrecv{\idV}{\tup{\eatom{inc},5,\idV}}$; pattern matching the two would incorrectly map the identifier variable $y$ (of type \upid) to the linear identifier value $i$ (which is already typed as \lpid); we call this a \emph{type mismatch} incompatibility, as an identifier $i$ can only be assigned one type.
  \item When listening for pattern $\msendD{\typ{z\!}{\!\lpid}\,}{y}{\eatom{err}}$, the system may generate a matching action $\msendD{\idV\,}{\idVVV}{\eatom{err}}$ mapping 
    variable $z$ to \idV.  Aliasing $z$ with $\idV$ violates the linearity assumption associated with $z$, (\ie denoted by its type \lpid), which assumes it to be distinct from any other identifier mentioned in the script \cite{ATTAPL}; we call this an \emph{aliasing} incompatibility. 
  \end{itemize}
Such incompatibilities would not occur if a script is used to monitor a \emph{typed system} that typechecks \wrt the same value environment used for typechecking the script. For instance, in the case of formula \eqref{eq:static:9} which typechecks \wrt $\envV_{0}$, the system in \figref{fig:sys} would also need to typecheck \wrt to the same environment. By employing static typechecking for both the monitor and the system, we would therefore be able to statically detect cases where a system event, such as $\mrecv{\typ{i}{\lpid}}{3}$, may erroneously match a necessity pattern such as $\mrecv{\typ{x}{\upid}}{3}$, thereby detecting a type mismatch. A similar technique can also be used to statically detect that a linear process identifier $i$, may at runtime be mapped to other linear typed variables in the adaptation formula; thereby detecting cases of aliasing. \exqed
\end{example}

In the absence of system typing, our monitors need to perform \emph{dynamic type checks} (at runtime) and \emph{abort monitoring} as soon as a type incompatibility is detected. This is important since any violations to the type discipline assumed by the script, may potentially render the adaptations specified in the script \emph{unsafe}, and thus these should \emph{not} be administered on the system. 

In order to perform dynamic type checks, the monitor should be able to start with an \emph{initial set} of type assumptions and then \emph{extend} these assumptions from the information that it learns about the system while handling system events. Hence, the operational semantics of typed scripts is defined \wrt the value type environment, $\envV$, with which they are statically typechecked. 
This environment is then extended at runtime with new type assumptions, that the monitor infers whenever a value received from a system event, is bound to a typed variable defined in a matching necessity pattern. The new type assumption would thus consist in the matched value mapped to the type of the variable it was bound to. 

For instance, if a monitor which executes \wrt environment $\envV$, receives a system event $\mrecv{i}{3}$ which matches necessity pattern $\mrecv{\typ{x}{\upid}}{3}$, it would therefore infer a new type assumption in which it associates identifier $i$ to type \upid (which is the type of variable $x$ to which $i$ was bound). The monitor therefore extends $\envV$ with this new assumption, \ie $\typ{i}{\upid}$, thereby creating environment $(\envVN{\typ{i}{\upid}})$.

\begin{example} \label{ex:dynamic-env-use} As shown in the derivation below, the execution of the typed script $\hV'$ in (\ref{eq:static:9}) \wrt system $\stV=\mapstate{(i,j,k,h)}{\unblocked}$, would use the type environment $\envV\!=\!\{\typ{\idV}{\lpid},\typ{\idVV}{\upid}\}$ from \exref{ex:typed-script} as a look-up table which its consults in order to detect type incompatibilities. For instance the typed script is able determine that an action such as $\mrecv{\idV}{\tup{\eatom{inc},5,\idV}}$ cannot be pattern matched with $\mrecv{\idV}{\tup{\eatom{inc},\typ{x}{\dat},\typ{y}{\upid}}}$, as this would lead to a \emph{type mismatch} when $\typ{y}{\upid}$ is mapped as $\typ{\idV}{\upid}$ (note the mismatching types, \ie $\typ{y}{\upid}\mapsto\idV$ conflicts with the assumption $\typ{i}{\lpid}$ which is already present in $\envV$).    

  {\small\setstretch{1.1}\setlength{\abovedisplayskip}{-8pt}
\setlength{\belowdisplayskip}{2pt}
\setlength{\abovedisplayshortskip}{-8pt}
\setlength{\belowdisplayshortskip}{2pt}
  \begin{align*}  	 
   &\instr{\stV}{\langle\envV,\hV'\rangle}  \wtraS{\scriptstyle\mrecv{\idV}{\tup{\eatom{inc},5,\idV}}} \instr{\stV}{\langle \{\envV,\typ{y}{\upid}\},\mBNec{\msendD{\idV}{\_}{\{\eatom{inc},x,y\}}}{\varepsilon}\cdots
		\rangle}\, \{x\mapsto 5,y\mapsto i\} \\
				        &\qquad \text{ (By applying substitution }  \{x\!\mapsto\!5,y\!\mapsto\!i\} \text{, environment }\{\envV,\typ{y}{\upid}\} \text{ becomes } \{\typ{\idV}{\lpid},\typ{\idVV}{\upid},\typ{i}{\upid}\} \text{)}\\
				        & \qquad\Rightarrow \text{ Type mismatch!! }	\vspace{-3mm}
  \end{align*}} \vspace{-5mm}
  
\noindent The above reduction leads to a stuck configuration caused by the type mismatch of $i$, as this yields an invalid extension to the value environment \ie the resultant environment is no longer a map (as previously explained in \secref{sec:type-environments}).

Conversely as shown in \eqref{eq:dt1:1} of the below derivation, matching pattern $\mrecv{\idV}{\tup{\eatom{inc},\typ{x}{\dat},\typ{y}{\upid}}}$ with action $\mrecv{\idV}{\tup{\eatom{inc},5,\idVVV}}$ would \emph{not only} constitute a \emph{valid match}, but also allow monitoring to \emph{extend} the assumed knowledge of the system from $\envV$ to $\smash{\envV'\!=\!\{\envV,\typ{\idVVV}{\upid}\}}$, where $\idVVV$ is associated to the type of the matched pattern variable $y$.  The extended environment $\envV'$ would however allow the monitor to detect a type mismatch between pattern  $\msendD{\typ{z}{\lpid}\,}{\idVVV}{\eatom{err}}$ and action $\msendD{\idVVV\,}{\idVVV}{\eatom{err}}$ as shown in \eqref{eq:dt1:3}.

  {\small\setstretch{1.1}\setlength{\abovedisplayskip}{1pt}
\setlength{\belowdisplayskip}{1pt}
\setlength{\abovedisplayshortskip}{1pt}
\setlength{\belowdisplayshortskip}{1pt}
  \begin{align}  	 
   	\instr{\stV}{\langle\envV,\,\hV'\rangle}  
   		&\wtraS{\scriptstyle\mrecv{\idV}{\tup{\eatom{inc},5,h}}} \instr{\stV}{\langle \{\envV,\typ{h}{\upid}\},\mBNec{\msendD{\idV}{\_}{\{\eatom{inc},5,h\}}}{\varepsilon}
		           \begin{mlbrace} 
				        (\mANec{\msendD{j}{h}{\tup{\eatom{res},\smash{5+1}}}}{\varepsilon}\,\hV')   \;\mand \; \\[0mm] 
				        (\mBNec{\msendD{\typ{z}{\lpid}\,}{h}{\eatom{err}}}{i}\,\mrestart{i}_{\varepsilon}\,\mclearMailbox{z}_{i,z}\,\hV')
				   \end{mlbrace} \;
		\rangle} \label{eq:dt1:1}\\
		&\wtraS{\scriptstyle\msendD{\idV}{h}{\tup{\eatom{inc},5,h}}} \instr{(\mapstate{(j,k,h)}{\unblocked},\mapstate{i}{\blocked})}{\langle \{\envV,\typ{h}{\upid}\},
		           \begin{mlbrace} 
				        (\mANec{\msendD{j}{h}{\tup{\eatom{res},\smash{5+1}}}}{\varepsilon}\,\hV')   \;\mand \; \\[0mm] 
				        (\mBNec{\msendD{\typ{z}{\lpid}\,}{h}{\eatom{err}}}{i}\,\mrestart{i}_{\varepsilon}\,\mclearMailbox{z}_{i,z}\,\hV')
				   \end{mlbrace} \; \rangle} \\
		&\wtraS{\scriptstyle\msendD{h}{h}{\eatom{err}}} \instr{(\mapstate{(j,k,h)}{\unblocked},\mapstate{i}{\blocked})}{\langle \{\envV,\typ{h}{\upid},\typ{z}{\lpid}\},
		          \mrestart{i}_{\varepsilon}\,\mclearMailbox{z}_{i,z}\,\hV'\rangle}\,  \{z\mapsto h\} \label{eq:dt1:3}\\			   
				  &\quad \text{ (If we apply substitution }\{z\mapsto h\}\text{ we get environment }\{\envV,\typ{h}{\upid},\typ{h}{\lpid}\} \text{)} \nonumber \\
				  & \qquad\Rightarrow \text{ Type mismatch!! } \nonumber	\vspace{-3mm}
  \end{align}} \vspace{-5mm}
  
Importantly, the derivation below shows that the monitor is also enabled to detect an aliasing violation between the same pattern, $\msendD{\typ{z}{\lpid}\,}{\idVVV}{\eatom{err}}$, and action $\msendD{\idV\,}{\idVVV}{\eatom{err}}$, since variable $z$ cannot be mapped to $\idV$, as this linear identifier is already present in the respective value environment. 

{\small\setstretch{1.1}\setlength{\abovedisplayskip}{1pt}
\setlength{\belowdisplayskip}{1pt}
\setlength{\abovedisplayshortskip}{1pt}
\setlength{\belowdisplayshortskip}{1pt}
  \begin{align*}  	 
   	\instr{\stV}{\langle\envV,\,\hV'\rangle}  
   		&\wtraS{\scriptstyle\mrecv{\idV}{\tup{\eatom{inc},5,h}}} \cdot \wtraS{\scriptstyle\msendD{\idV}{h}{\tup{\eatom{inc},5,h}}} \cdot \wtraS{\scriptstyle\msendD{i}{h}{\eatom{err}}} \\
   		&\instr{(\mapstate{(j,k,h)}{\unblocked},\mapstate{i}{\blocked})}{\langle \{\envV,\typ{h}{\upid},\typ{z}{\lpid}\},
		          \mrestart{i}_{\varepsilon}\,\mclearMailbox{z}_{i,z}\,\hV'\;\rangle}\,  \{z\mapsto i\}\\			   
				  &\text{ (If we apply substitution }\{z\mapsto i\}\text{ we get environment }\envV'=\{\typ{i}{\lpid},\typ{j}{\upid},\typ{h}{\upid},\typ{i}{\lpid}\} \text{)} \nonumber \\
				  &\quad\text{ Linear identifier }\typ{i}{\lpid}\in\envV\text{ was remapped to }\typ{z}{\lpid} \Rightarrow\text{ Aliasing!! } \nonumber	\vspace{-3mm} 
  \end{align*}} \vspace{-5mm}

\noindent It would however allow $\msendD{\typ{z}{\lpid}\,}{\idVVV}{\eatom{err}}$ to be matched to action $\msendD{\idVVVV\,}{\idVVV}{\eatom{err}}$, which would (again) extend the current type environment to $(\envV', \typ{\idVVVV}{\lpid})$ using the script type association $\typ{z}{\lpid}$. \exqed			
 \end{example}
 
\subsubsection{The Preliminaries for Dynamic Typechecking}

Although safe, the mechanism discussed in \exref{ex:dynamic-env-use} turns out to be rather restrictive for recursive properties (using maximal fixpoints). By this we mean that we end up unnecessarily aborting monitoring for the fear of unsafety. Note that, by alpha-conversion, the variable bindings made by a necessity formula under a recursive formula is \emph{different} for every unfolding of that formula:  \eg unfolding script (\ref{eq:static:9})  twice yields 
\begin{align*}
  & \mANec{\mrecv{\idV}{\tup{\eatom{inc},\typ{x}{\dat},\typ{y}{\upid}}}}{\varepsilon}\left(\ldots \mBNec{\msendD{\typ{z}{\lpid}\,}{y}{\eatom{err}}}{i}\,\ldots\, 
    \left(\begin{array}{l}
\mANec{\mrecv{\idV}{\tup{\eatom{inc},\typ{x'}{\dat},\typ{y'}{\upid}}}}{\varepsilon}\\
\qquad\quad      (\ldots \mBNec{\msendD{\typ{z'}{\lpid}\,}{y}{\eatom{err}}}{i}\,\ldots\,\hV' )
    \end{array}\right)
\right)
\end{align*}
where the outer bindings $x,y,z$ are distinct from the inner bindings $x',y',z'$. More importantly, however, the scope of these bindings extends until the next fixpoint unfolding, and are not used again beyond that point. For instance, $x,y,z$ above are not used beyond the respective adaptations of the first unfolding. Thus, one possible method for allowing a finer dynamic analysis for adaptation-scripts, especially relating to linearity and aliasing violations, is to employ a mechanism that keeps track of which linear bindings \emph{are still in use}.  In the example above, this would allow us to  bind \idVVVV twice --- once with $z$ during the first iteration, and another time with $z'$. This is safe as we know that by the time the second binding occurs ($z'$), the first binding ($z$)  is not in use anymore, \ie there is no aliasing. \medskip

\noindent To implement this mechanism, our formalisation uses three additional components.\vspace{-1mm}
\begin{enumerate}[label=(\arabic*),leftmargin=7mm,itemindent=0mm]
	\item  The operational semantics for  adaptation-scripts uses an extra scoping environment, $\Used \in \Pid \rightharpoonup \pset{\LVars}$, keeping track of the \emph{recursion  variables under which an identifier binding is introduced}, by associating that identifier to a set of formula variables, $\stk \in \pset{\LVars}$. Environment \Used\ keeps track of the \emph{linear identifiers} that are currently in use only. For instance, if $\envV=\{i\!:\!\upid\}$, $\Used=\emptyset$, $\hV=\mmax{\hVarY}{(\mANec{\mrecv{i}{\typ{x}{\lpid}}}{\varepsilon}\mBNec{\mrecv{x}{3}}{\varepsilon}\mscor{x}{x}\hVarY)}$, the script would reduce as follows for event $\mrecv{i}{j}$:
	
	{\small\setstretch{1.1}\setlength{\abovedisplayskip}{0pt}
\setlength{\belowdisplayskip}{-5pt}
\setlength{\abovedisplayshortskip}{0pt}
\setlength{\belowdisplayshortskip}{-5pt}
	  \begin{align}  	 
	   	\instr{\stV}{\cmon{\envV}{\Used}{\hV}} \wtraS{\scriptstyle\mrecv{i}{j}}\instr{\stV}{ \cmon{\{\envV,\typ{j}{\lpid}\}}{\UsedP=\{\typ{j}{\{\hVarY\}}\}}{\mBNec{\mrecv{j}{3}}{\varepsilon}\mscor{j}{j}\hV}} \label{drv:used:1:1} \vspace{-3mm} 
	  \end{align}} \vspace{-5mm}	
	  
	\noindent Hence as identifier $j$ was bound to $\typ{x}{\lpid}$, the monitor extended its type assumptions in $\envV$ with the assumption that $i$ is a linear process id \ie $\typ{i}{\lpid}$. It also added an entry in the $\UsedP$ environment showing that linear identifier $j$ is \emph{scoped} under formula variable $\hVarY$ and is thus currently in use. Therefore, if $i$ is rebound to a linear variable while it is still in use, we can detect aliasing by consulting the scoping environment. 	
	\item To facilitate updates to  environment \Used, the patterns in necessity formulas are \emph{decorated by sets of formula variables}, denoting their respective recursion scope. For instance, if we apply the decoration process on $\mmax{\hVarX}{(\mattrNec{\patE}{\attr}{\vLstA}{\ldots \mmax{\hVarY}{(\mattrNec{\patE'}{{\attr'}}{\vLstB}{\mfls})} })}$ we get $\mmax{\hVarX}{(\mattrNec{\patE_{\{\hVarX\}}}{\attr}{\vLstA}{\ldots \mmax{\hVarY}{(\mattrNec{\patE'_{\{\hVarX,\hVarY\}}}{{\attr'}}{\vLstB}{\mfls})} })}$. Decoration can easily be performed using a linear scan of the script before executing it.
	\item The runtime syntax uses an additional construct, $\clr{\hVarY}\hV$, when unfolding a recursive formula. This is required: to mark the \emph{end of an unfolding}; and upon execution, to remove all identifier entries in \Used\ that have the recursion variable \hVarY\ declared in their respective set of formula variables, $\stk$, so as to record that they are not in use anymore. For instance, when using this additional construct in \eqref{drv:used:1:1} we obtain the following derivation:
	
	{\small\setstretch{1.1}\setlength{\abovedisplayskip}{-8pt}
\setlength{\belowdisplayskip}{-2pt}
\setlength{\abovedisplayshortskip}{-8pt}
\setlength{\belowdisplayshortskip}{-2pt}
	  \begin{align}  	 
	   	\instr{\stV}{\cmon{\envV}{\Used}{\hV}} &\wtraS{\scriptstyle\mrecv{i}{j}}\instr{\stV}{ \cmon{\{\envV,\typ{j}{\lpid}\}}{\{\typ{j}{\{\hVarY\}}\}}{\mBNec{\mrecv{j}{3}_{\{\hVarY\}}}{\varepsilon}\mscor{j}{j}\clr{\hVarY}\hV}} \nonumber \\
			&\wtraS{\scriptstyle\mrecv{j}{3}}\instr{\stV}{ \cmon{\{\envV,\typ{j}{\lpid}\}}{\{\typ{j}{\{\hVarY\}}\}}{\clr{\hVarY}\hV}} \nonumber \\
			&\traS{\scriptstyle\tau}\instr{\stV}{ \cmon{\{\envV,\typ{j}{\lpid}\}}{\emptyset}{\hV}} \label{drv:used:1:2} 			\vspace{-3mm} 
	  \end{align}} \vspace{-5mm}	
	  
	 Note that in \eqref{drv:used:1:2} the $\clr{\hVarY}$ was used to clear the contents in $\UsedP$ which were associated with formula variable $\hVarY$.
\end{enumerate}
Moreover, the $\Used$ environment is first initialised by associating \emph{every} statically know linear process identifier $\typ{i}{\lpid}$ residing in $\envV$ to $\stk\!=\!\emptyset$, in the following manner: {\setlength{\abovedisplayskip}{2pt}
\setlength{\belowdisplayskip}{2pt}
\setlength{\abovedisplayshortskip}{2pt}
\setlength{\belowdisplayshortskip}{2pt}\begin{align*}
	\Used=\set{i:\emptyset}{\envV(i)=\lpid}
\end{align*}} For example, if a script typechecks statically \wrt $\envV=\{\typ{i}{\lpid},\typ{j}{\upid}\}$, the $\Used$ environment is initialised as $\Used=\{i:\emptyset\}$. These linear identifiers are therefore \emph{universally scoped} and thus cannot be cleared. \medskip

\begin{remark} In this scoping mechanism, the identifiers bound to linear variables from different concurrent branches are added to a conceptually global scoping environment $\Used$, in which they are associated to the formula variable $\hVarX$ under which they are currently scoped \ie in use. The bindings associated to formula variable $\hVarX$ are then cleared from $\Used$ whenever $\clr{\hVarX}$ is called. This approach may appear to be problematic in the following situations:
\begin{enumerate}[label=(\roman*), leftmargin=7mm]
	\item  A formula multiply defines the same recursive variable, \eg $(\mmax{\hVarX}{\hV_{1}})\mand(\mmax{\hVarX}{\hV_{2}})$. Our clearing mechanism may erroneously clear the bindings used in branch $\mmax{\hVarX}{\hV_{2}}$ when branch $\mmax{\hVarX}{\hV_{1}}$ recurs (and vice-versa).
	\item  A formula recurs on the \emph{same} recursive variable in two (or more) coexisting concurrent branches, \eg $\mmax{\hVarX}{\mBNec{\mrecv{i}{3}}{\varepsilon}\Big((\mscor{i}{i}\hVarX)\mand\hVarX\Big)}$. In such cases the clearing mechanism could clear the bindings or one recurring branch \eg $(\mscor{i}{i}\hVarX)$ by using $\clr{\hVarX}$, which could erroneously clear the bindings scoped under the other branch.
\end{enumerate}
We solve issue (i) as we work up to $\alpha$-conversion \ie we rename one of the variables by conducting a linear scan of the script, \eg the renamed script becomes: {\setlength{\abovedisplayskip}{2pt}
\setlength{\belowdisplayskip}{2pt}
\setlength{\abovedisplayshortskip}{2pt}
\setlength{\belowdisplayshortskip}{2pt}\begin{align*}(\mmax{\hVarX}{\hV_{1}})\mand(\mmax{\hVarY}{\hV_{2}\sub{X}{Y}})\end{align*}} In the case of issue (ii), recall that in $\Used$ we only track \emph{linear identifiers}, \eg $\typ{i}{\lpid}$, which means that this issue is problematic whenever a formula that \emph{uses a linear identifier} $\typ{i}{\lpid}$, recurs over the \emph{same} formula variable in \emph{multiple non-mutually exclusive branches}. However from \exref{ex:SigmaCE} in \secref{sec:type-rules}, recall that these cases are \emph{statically rejected} by our typing rules. Hence, a \emph{well-typed} script can never introduce such an issue. \bqed
\end{remark} \bigskip

\begin{figure}[!ht]
\begin{small}
\begin{mathpar}
	 \inference[\rtit{rIdem1}]{}{\cmon{\envV}{\Used}{\mtru} \traS{\actE} \cmon{\envV}{\Used}{\mtru}} \and
     \inference[\rtit{rIdem2}]{}{\cmon{\envV}{\Used}{\mfls} \traS{\actE} \cmon{\envV}{\Used}{\mfls}} \and
	\inference[\rtit{rMax}]{}{\cmon{\envV}{\Used}{\mmax{\hVarX}{\hV}} \traS{\tau} \cmon{\envV}{\Used}{\hV\sub{\bigl(\clr{\hVarX}\mmax{\hVarX}{\hV}\bigr)}{\hVarX}}}
         \and
	\inference[\rtit{rClr}]{}{\cmon{\envV}{\Used}{\clr{\hVarX}\hV} \traS{\tau} \cmon{\envV}{\set{\typ{\idV}{\stk}}{\typ{\idV}{\stk}\in\Used\land\hVarX\notin\stk}}{\hV}} 
        \and		
	 \inference[\rtit{rNc1}]{
           \match{\patE}{\actE}{\sigma} & \id{\actE}=\idV  & \envV' = \tbnd{\patE}\sigma 
            \\
              \dom(\Used) \cap \dom(\envV') = \emptyset & \forall \typ{(x,y)}{\lpid}\!\in\!\tbnd{\patE}\cdot x\neq y \implies x\sigma\neq y\sigma 
           }{\cmon{\envV}{\Used}{\mBDNec{\patE}{\stk}{\vLstB}{\hV}} \traSS{\actE} \cmon{\bigl(\envV,\envV'\bigr)\,}{\,\bigl(\Used \cup \set{\typ{\idV}{\stk}}{\,\smash{\envV'(\idV)\! =\! \lpid}}\bigr)\,}              {\,\mblock{i}\hV\sigma}
           } 
        \and
	 \inference[\rtit{rNc2}]{
              \match{\patE}{\actE}{\sigma}   & \envV' = \tbnd{\patE}\sigma  
              \\
              \dom(\Used) \cap \dom(\envV') = \emptyset & \forall \typ{(x,y)}{\lpid}\!\in\!\tbnd{\patE}\cdot x\neq y \implies x\sigma\neq y\sigma 
            }{
              \cmon{\envV}{\Used}{\mADNec{\patE}{\stk}{\vLstB}{\hV}} \traSS{\actE} \cmon{\bigl(\envV,\envV'\bigr)\,}{\,\bigl(\Used \cup \set{\typ{\idV}{\stk}}{\,\smash{\envV'(\idV)\! =\! \lpid}}\bigr)\,}{\,\hV\sigma}
            }
       \and	
    \inference[\rtit{rNc3}]{\match{\patE}{\actEE}{\bot}}{\cmon{\envV}{\Used}{\mattrDNec{\patE}{\attr}{\stk}{\vLstB}{\hV}}\traS{\actE} \cmon{\envV}{\Used}{\mrelease{\vLstB}\mtru}}
    \and
    \inference[\rtit{rStr}]{\cmon{\envV}{\Used}{\hV\steq\hV'}\traS{\act}\cmon{\envV}{\Used}{\hVV'\steq\hVV}}{\cmon{\envV}{\Used}{\hV}\traS{\act}\cmon{\envV}{\Used}{\hVV}} 
       \and
    \inference[\rtit{rTru}]{\mcondEval{c}{true}}{\cmon{\envV}{\Used}{\mboolE{c}{\varphi}{\psi}} \traS{\tau} \cmon{\envV}{\Used}{\varphi}} 
        \;\;
	\inference[\rtit{rFls}]{\mcondEval{c}{false}}{\cmon{\envV}{\Used}{\mboolE{c}{\varphi}{\psi}} \traS{\tau} \cmon{\envV}{\Used}{\psi}}         
	\and   			
    \inference[\rtit{rCn1}]{
			  \cmon{\envV}{\Used}{\hV} \!\traS{\actE}\! \cmon{\envV'}{\UsedP}{\hV'} 
			& \cmon{\envV}{\Used}{\hVV} \!\traS{\actE}\! \cmon{\envV''}{\UsedPP}{\hVV'} 
                        & \dom(\Used) \!=\! \bigl(\dom(\UsedP) \!\cap\! \dom(\UsedPP)\bigr) 
		}{\cmon{\envV}{\Used}{\hV\mand\hVV} \traS{\actE} \cmon{(\envV',\envV'')}{\UsedP\!\cup\!\UsedPP}{\hV'\mand\hVV'}} 
	\and
	\inference[\rtit{rCn2}]{\cmon{\envV}{\Used}{\varphi} \traS{\tau}  \cmon{\envV}{\UsedP}{\varphi'}}{\cmon{\envV}{\Used}{\varphi\mand\psi}  \traS{\tau}  \cmon{\envV}{\UsedP}{\varphi'\mand\psi}}
        \and
	\inference[\rtit{rCn3}]{\cmon{\envV}{\Used}{\varphi} \traS{\actC} \cmon{\envV}{\Used}{\varphi'}}{\cmon{\envV}{\Used}{\varphi\mand\psi} \traS{\actC} \cmon{\envV}{\Used}{\varphi'\mand\psi}}
        \and	   
	\inference[\rtit{rAdA}]{
	}{\cmon{\envV}{\Used}{\macor{\vLstA}{\vLstB}\varphi} \traS{\cora{\vLstA}} \cmon{\envV}{\Used}{\mrelease{\vLstB}\varphi}}	
        \and
	\inference[\rtit{rAdS}]{
	}{\cmon{\envV}{\Used}{\mscor{\vLstA}{\vLstB}\varphi} \traS{\cors{\vLstA}} \cmon{\envV}{\Used}{\mrelease{\vLstB}\varphi}}	
        \and
	\inference[\rtit{rRel}]{}{\cmon{(\envV,\typ{\vLstB}{\lpidb})}{\Used}{\mrelease{\vLstB}\varphi} \traS{\rel{\vLstB}} \cmon{(\envV,\typ{\vLstB}{\lpid})}{\Used}{\varphi}}
        \and
	\inference[\rtit{rBlk}]{}{\cmon{(\envV,\typ{\vLstB}{\lpid})}{\Used}{\mblock{\vLstB}\varphi} \traS{\blk{\vLstB}} \cmon{(\envV,\typ{\vLstB}{\lpidb})}{\Used}{\varphi}}     
\end{mathpar}
\end{small}
\caption{Dynamically Typed Adaptation-Script Rules}
\label{fig:dt-mon-rules-1}
\end{figure}

\begin{figure}[!ht]
\begin{small} 
	\begin{mathpar}
	   \inference[\rtit{iAct}]{  \cmon{\envV}{\Used}{\hV}  \notTraceEvent{\;\actC\;}  &   \stV \traS{\actE} \stV'  & \cmon{\envV}{\Used}{\hV} \traS{\actE} \cmon{\envV'}{\Used'}{\hV'}}{\instr{\stV}{\cmon{\envV}{\Used}{\hV}} \traS{\actE} \instr{\stV'}{\cmon{\envV'}{\Used'}{\hV'}}} \and
	   \inference[\rtit{iTrm}]{\cmon{\envV}{\Used}{\hV}  \notTraceEvent{\;\actC\;}  &   \stV \traS{\actE} \stV'  &  \cmon{\envV}{\Used}{\hV} \notTraceEvent{\;\actE\;}}{\instr{\stV}{\cmon{\envV}{\Used}{\hV}} \traS{\actE} \instr{\stV'}{\mtru}} \and
	    \inference[\rtit{iAda}]{\cmon{\envV}{\Used}{\hV} \traS{\actC} \cmon{\envV'}{\Used'}{\hV'} & \stV \traS{\actC} \stV'}{\instr{\stV}{\cmon{\envV}{\Used}{\hV} } \traS{\tau} \instr{\stV'}{\cmon{\envV'}{\Used'}{\hV'}}} \and 	  
	   \inference[\rtit{iSys}]{\stV \traS{\tau} \stV}{\instr{\stV}{\cmon{\envV}{\Used}{\hV}} \traS{\tau} \instr{\stV'}{\cmon{\envV}{\Used}{\hV}}} \and
	   \inference[\rtit{iMon}]{\cmon{\envV}{\Used}{\hV} \traS{\tau} \cmon{\envV'}{\Used'}{\cmon{\envV'}{\Used'}{\hV'}}}{\instr{\stV}{\cmon{\envV}{\Used}{\hV}} \traS{\tau} \instr{\stV}{\cmon{\envV'}{\Used'}{\hV'}}}		
	\end{mathpar}
\end{small}
\caption[The Instrumentation Transition Rules]{The Instrumentation Transition Rules (adapted from \figref{fig:rt-semantics})}
\label{fig:dt-mon-instr}
\end{figure}

\subsubsection{The Dynamically Typed Model}

\figref{fig:dt-mon-rules-1} describes the transition rules for typed adaptation-scripts, defined over triples $\cmon{\envV}{\Used}{\hV}$. 
Along with the system rules of \figref{fig:rt-semantics} and the instrumentation rules (adapted to triples $\cmon{\envV}{\Used}{\hV}$ from \figref{fig:rt-semantics}) of \figref{fig:dt-mon-instr}, they form the complete operational semantics.   
By contrast to 
\figref{fig:logic-transition-rules}, \rtit{rMax} in \figref{fig:dt-mon-rules-1} unfolds a recursive formula to one prefixed by a clear construct, (\ie $\clr{\hVarX}\mmax{\hVarX}{\hV}$), where rule \rtit{rClr} is then used to remove all entries in \Used\ associated with \hVarX. 

The new version of \rtit{rNc1} in \figref{fig:dt-mon-rules-1} implicitly checks for \emph{type mismatch} incompatibilities by requiring that the environment extension, $\bigl(\envV,\envV'\bigr)$, is still a map, \ie conflicting entries \eg $\typ{\idV}{\upid},\typ{\idV}{\lpid}$ would violate this condition.  It also uses the predicate $\smash{\forall \typ{(x,y)}{\lpid}\!\in\!\tbnd{\patE}\cdot x\!\neq\!y\!\implies\! x\sigma\!\neq\!y\sigma}$, to check that the substitution environment, $\sigma$, derived from matching a necessity pattern $\patE$ to a system event, does not map the same process identifier, $i$, to two or more \emph{distinct} linear typed variables such as $\typ{x}{\lpid}$ and $\typ{y}{\lpid}$. For instance, if a system event $\{\eatom{id},i,i\}$ is matched with necessity pattern $\{\eatom{id},\typ{x}{\lpid},\typ{y}{\lpid}\}$ this predicate detects that $x\neq y$ but $x\sigma\!=\!y\sigma\!=\!i$, and thus detects aliasing. Furthermore, this new version of \rtit{rNc1} checks that the new bindings, \ie $\dom(\envV')$, are distinct from the linear identifiers that are currently in use, \ie $\dom(\Used)$, as these would constitute aliasing incompatibilities. If no type incompatibilities are detected, the rule transitions by updating \Used\ accordingly. Rule \rtit{rNc2} is analogous to \rtit{rNc1}.

Rule \rtit{rCn1} performs similar checks by ensuring that linear aliasing introduced along separate branches do not overlap. In fact this rule uses predicate $\dom(\Used)\!=\!\bigl(\dom(\UsedP) \cap \dom(\UsedPP)\bigr)$ to detect cases where the same identifier is \emph{simultaneously} mapped to two distinct linear typed variables pertaining to the two concurrent branches $\hV$ and $\hVV$ (\eg identifier $i$ is simultaneously mapped to $\typ{x}{\lpid}$ defined in $\hV$, and also to $\typ{y}{\lpid}$ in $\hVV$). 

If any of the conditions for \rtit{rNc1}, \rtit{rNc2} and \rtit{rCn1} are not satisfied, the adaptation-script blocks and is terminated in an instrumented setup using rule \rtit{iTrm} from \figref{fig:dt-mon-instr}, \ie it aborts as soon as type incompatibilities are detected. 

\subsubsection{An example driven explanation for Detecting Aliasing}
As shown by our newly introduced dynamic checks in rules \rtit{rNc1}, \rtit{rNc2} and \rtit{rCn1}, aliasing can be introduced in two ways: $(i)$ either by remapping a linear process id while it is already in use, or else $(ii)$ by having two or more branches that simultaneously map the same linear process id to different linear typed variables.
\begin{example} For the first case consider the following derivation for formula \eqref{eq:static:9} (restated below), \wrt system $\stV=(\mapstate{(i,j,k,h)}{\unblocked})$, $\envV=\{\typ{i}{\lpid},\typ{j}{\upid}\}$, $\Used=\{i:\emptyset\}$. Note that decorations were added accordingly while deriving \eqref{eq:static:9}. \medskip
	\begin{equation*}\setstretch{1.1}
		    \begin{array}{l}
		    \hV'\;\deftxt\;\; \mmax{\,\hVarY}{\mANec{\mrecv{\idV}{\tup{\eatom{inc},\typ{x}{\dat},\typ{y}{\upid}}}}{\varepsilon}}\\[-3mm]
		      \qquad\qquad\qquad\qquad\quad\mBNec{\msendD{\idV}{\_}{\{\eatom{inc},x,y\}}}{\varepsilon}
		                \begin{mlbrace} 
		        (\,\mANec{\msendD{j}{y}{\tup{\eatom{res},\smash{x+1}}}}{\varepsilon}\,\hVarY)   \;\mand \; \\[0mm] 
		        (\,\mBNec{\msendD{\typ{z}{\lpid}\,}{y}{\eatom{err}}}{i}\,\mrestart{i}_{\varepsilon}\,\mclearMailbox{z}_{i,z}\,\hVarY) 
		      \end{mlbrace} \quad \eqref{eq:static:9}
		    \end{array}
	\end{equation*} 
	{\small\setstretch{1.2}\setlength{\abovedisplayskip}{-1pt}
\setlength{\belowdisplayskip}{-8pt}
\setlength{\abovedisplayshortskip}{-1pt}
\setlength{\belowdisplayshortskip}{-8pt}
	  \begin{align*}  	
	   	\instr{\stV}{\cmon{\envV}{\Used}{\hV'}} 
	   		&\traS{\scriptstyle\tau}\instr{\stV}{\cmon{\envV}{\Used}{\mANec{(\mrecv{\idV}{\tup{\eatom{inc},\typ{x}{\dat},\typ{y}{\upid}}})_{\{\hVarY\}}}{\varepsilon}\cdots \begin{mlbrace} 
		        (\,\mANec{(\msendD{j}{y}{\tup{\eatom{res},\smash{x+1}}})_{\{\hVarY\}}}{\varepsilon}\,\clr{\hVarY}\hV')  \mand  \\[0mm] 
		        (\,\cdots\,\mrestart{i}_{\varepsilon}\,\mclearMailbox{z}_{i,z}\clr{\hVarY}\hV') 
		      \end{mlbrace}\;}} \\
		   &\traS{\scriptstyle\mrecv{i}{\etuple{\eatom{inc},3,h}}}\instr{\stV}{\cmon{\envV}{\Used}{\mBNec{(\msendD{\idV}{\_}{\{\eatom{inc},3,h\}})_{\{\hVarY\}}}{\varepsilon} \begin{mlbrace} 
		        (\,\mANec{(\msendD{j}{h}{\tup{\eatom{res},\smash{3+1}}})_{\{\hVarY\}}}{\varepsilon}\,\clr{\hVarY}\hV') \mand \\[0mm] 
		        (\,\cdots\,\mrestart{i}_{\varepsilon}\,\mclearMailbox{z}_{i,z}\clr{\hVarY}\hV') 
		      \end{mlbrace}\;}} \\  
		  &\traS{\scriptstyle\msendD{i\,}{\,i}{\etuple{\eatom{inc},3,h}}}\instr{(\mapstate{i}{\blocked},\mapstate{(j,k,h)}{\unblocked})}{\cmon{\envV}{\Used}{ \begin{mlbrace} 
		        (\,\mANec{(\msendD{j}{h}{\tup{\eatom{res},\smash{3+1}}})_{\{\hVarY\}}}{\varepsilon}\,\clr{\hVarY}\hV') \mand  \\[0mm] 
		        (\,\mBNec{(\msendD{\typ{z}{\lpid}\,}{h}{\eatom{err}})_{\{\hVarY\}}}{i}\,\mrestart{i}_{\varepsilon}\,
		        \ldots)
		      \end{mlbrace}\;}}  \\  
		 &\traS{\scriptstyle\msendD{i\,}{\,h}{\eatom{err}}}\instr{(\mapstate{i}{\blocked},\mapstate{(j,k,h)}{\unblocked})}{\cmon{(\envV,\envV')}{\UsedP}{ \begin{mlbrace} 
		      \mrestart{i}_{\varepsilon}\,\mclearMailbox{i}_{i,i}\clr{\hVarY}\hV'
		      \end{mlbrace}\;}} \\ & \qquad\qquad\qquad \text{ where } \envV'=\{\typ{i}{\lpid}\} \text{  and  }  \UsedP=\{i:\emptyset,i:\hVarY\}\\
		&\text{But }\dom(\Used)\cap\dom(\envV')=\{i\}\neq\emptyset\, \Rightarrow \text{ Aliasing Detected! }		      
	  \end{align*}} 	  
	\noindent Aliasing is therefore detected since in the penultimate reduction of the above derivation, linear process id $i$ was remapped by necessity $\mBNec{\msendD{\typ{z}{\lpid}\,}{y}{\eatom{err}}}{i}$ to linear typed variable $\typ{z}{\lpid}$, while this was still in use as stated by the entry $\{i:\emptyset\}$ in $\Used$. More specifically, aliasing was detected by the dynamic check $\dom(\Used) \cap \dom(\envV') = \emptyset$ performed by \rtit{rNc2}. 
	
	\noindent For the second case, consider the following script: 
	{\setlength{\abovedisplayskip}{5pt}
\setlength{\belowdisplayskip}{5pt}
\setlength{\abovedisplayshortskip}{5pt}
\setlength{\belowdisplayshortskip}{5pt}\begin{equation*}
	  \hV_{alias}\;\deftxt\;\; \begin{mlbrace}(\mBNec{\mrecv{\typ{x}{\lpid}}{3}}{\varepsilon}\mscor{x}{x}\mtru) \;\; \mand \\ (\mBNec{\mrecv{\typ{y}{\lpid}}{\typ{z}{\dat}}}{\varepsilon}\;\mbool{z>0\;}{\;\mscor{x}{x}\;\mtru})\end{mlbrace}		   
	\end{equation*} }
	\noindent Although $\hV_{alias}$ observes the linearity imposed by the static typing rules given in \figref{fig:static-rules}, aliasing across concurrent branches may however lead to unsafety, particularly in the case a system actor \eg $i$, commits an input operation of $3$, \ie upon event $\mrecv{i}{3}$. The monitor reduction below shows how the global dynamic check made by rule \rtit{rCn1} is used to detect aliasing amongst the two concurrent branches (note that we have added the necessary decorations to $\hV_{alias}$ in the reductions). \smallskip
	{\small \begin{align*}\vspace{-2mm}\inference[\rtit{rCn1}]{
			 {\bigl(\dom(\UsedP) \!\cap\! \dom(\UsedPP)\!=\!\{i\}\neq\dom(\Used)\bigr)} \; \imp \; \text{Aliasing!!} \\
			  \cmon{\emptyset}{\Used\!=\!\emptyset}{\mBNec{(\mrecv{\typ{x}{\lpid}}{3})_{\emptyset}}{\varepsilon}\mscor{x}{x}\mtru} \!\traS{\mrecv{i}{3}}\! \cmon{\{\typ{i}{\lpid}\}}{\UsedP\!=\!\{\typ{i}{\emptyset}\}}{\mscor{i}{i}\mtru} \\
			 \cmon{\envV}{\Used\!=\!\emptyset}{\mBNec{\mrecv{\typ{y}{\lpid}}{\typ{z}{\dat}}_{\emptyset}}{\varepsilon}\,\mbool{z\!>\!0\,}{\,\mscor{x}{x}\,\mtru}} \!\traS{\mrecv{i}{3}}\! \cmon{\{\typ{i}{\lpid}\}}{\UsedPP\!\!=\!\{\typ{i}{\emptyset}\}}{\mbool{3\!>\!0\,}{\,\mscor{i}{i}\,\mtru}} 
		}{\cmon{\emptyset}{\Used\!=\!\emptyset}{\hV_{alias}} \traS{\mrecv{i}{3}} \text{Aliasing Detected!!}} \end{align*} }\smallskip
	\noindent Aliasing is thus detected by rule \rtit{rCn1} since $\bigl(\dom(\UsedP) \cap \dom(\UsedPP)\bigr)=\{i\}\neq\dom(\Used)$, meaning that both branches have simultaneously mapped the same linear process id $i$ to two different linear typed variables, \ie $\typ{x}{\lpid}$ and $\typ{y}{\lpid}$. \bqed
\end{example}

\subsubsection{Defining Type Soundness}
Using a straightforward extension we redefine the definition of synchronisation errors (given in \defref{def:err-conf}), \wrt our dynamically typed operational model given in \figref{fig:dt-mon-rules-1}, as:
\begin{definition}[Synchronisation Error] \label{def:error-dt}
	{\small$\\$\indent$\mErrorC{\instr{\stV}{\cmon{\envV}{\Used}{\hV}}} \;\deftxt\; \Bigl(\cmon{\envV}{\Used}{\hV} \traS{\scriptstyle\cors{\vLstB}} \;\text{ and }\; \stV\notTraceEvent{\;\scriptstyle\cors{\vLstB}\;}\Bigr)\; \text{ or } \; \Bigl(\cmon{\envV}{\Used}{\hV} \traS{\scriptstyle\rel{\vLstB}} \;\text{ and }\; \stV\notTraceEvent{\scriptstyle\;\rel{\vLstB}\;}\Bigr) \\$\indent\indent\indent$ {\normalsize\text{ for some } \vLstB \in\dom(\stV)}$}
\end{definition}
\noindent Based on \defref{def:error-dt}, we prove \emph{Type Soundness} \wrt the semantics of typed adaptation scripts. 

\begin{theorem}[\textbf{Type Soundness}] \label{thm:soundness} 
\setlength{\abovedisplayskip}{0pt}
\setlength{\belowdisplayskip}{0pt}
\setlength{\abovedisplayshortskip}{0pt}
\setlength{\belowdisplayshortskip}{0pt}
Whenever $\typeRule{\envV}{\hV}$ then:
  \begin{align*}
     &\instr{\stV}{\cmon{\envV}{\sset{{\typ{\idV}{\emptyset}}\;|{\;\,\smash{\envV(\idV)\! =\! \lpid}}}}{\hV}} \wtraS{\,\sV\,} \instr{\stV'}{\cmon{\envV'\!}{\Used'\!}{\hV'}} \quad \text{implies}\quad \neg\mErrorC{\instr{\stV'}{\cmon{\envV'\!}{\Used'\!}{\hV'}}}
  \end{align*}
\end{theorem}
\noindent This theorem states that if a dynamically checked monitor $\cmon{\envV}{\Used}{\hV}$ (where $\Used$ is initialised as $\Used=\sset{{\typ{\idV}{\emptyset}}\;|{\;\,\smash{\envV(\idV)\! =\! \lpid}}}$), is executed \wrt a system $\stV$ forming configuration $\instr{\stV}{\cmon{\envV}{\Used}{\hV}}$, then no matter how this configuration evolves at runtime over some trace $t$, it can \emph{never} reach an \emph{unsafe configuration}, \ie a configuration that satisfies the error definition given in \defref{def:error-dt}.

\section{Proving Type Soundness}
\label{sec:soundness}
\noindent So far we have relied on examples to show that our typing mechanisms are able to accept valid scripts while rejecting erroneous ones (either statically or dynamically). However, since there are an infinite number of possible scripts and executions that one can express, we want to provide \emph{soundness} guarantees. We want to show that a synchronisation error can never be introduced by any arbitrary script $\hV$ which is: statically accepted by our type system, and which executes \wrt a system $\stV$ and the respective $\envV$ and $\Used$ environments, \ie as configuration $\instr{\stV}{\cmon{\envV}{\Used}{\hV}}$. We therefore prove \emph{type soundness} for dynamically typed scripts to ensure that at no point during execution can an accepted script yield a synchronisation error (as defined by \defref{def:error-dt}). A few comments are in order prior to discussing the proofs.

\subsection{Preliminaries}
When proving type soundness (and the supporting lemmas) we assume \emph{closed formulas}. This is required as our type system is compositional, \ie it decomposes a formula into subparts and analyses them separately. For this reason, certain parts of the formula may be taken out context and become open formulas as shown in the example below:

	$\\\typeCheckRule[tMax]{\typeRule{\env{\envVVN{\hVarX}{\envV}}{\envV}}{\overbrace{\mattrNec{\patE_{1}}{\norm}{\vLstA}{\hVarX}\; \mand\; \mattrNec{\patE_{2}}{\norm}{\vLstB}{\mfls}}^{\text{Open Formula}^{\star}}}}{\typeRule{\env{\envVV}{\envV}}{\underbrace{\mmax{\hVarX}{\mattrNec{\patE_{1}}{\norm}{\vLstA}{\hVarX}\; \mand\; \mattrNec{\patE_{2}}{\norm}{\vLstB}{\mfls}}}_{\text{Closed Formula}}}} 
	\begin{aligned}
		\; ^{\star}&\footnotesize \text{But at runtime, }X\text{ is substituted by}\\[-2mm]
		\; \; &\footnotesize \mmax{\hVarX}{(\mattrNec{\patE_{1}}{\norm}{\vLstA}{\hVarX} \mand \mattrNec{\patE_{2}}{\norm}{\vLstB}{\mfls})}\text{, thus keeping the}\\[-2mm]
		\; \; & \footnotesize\text{formula closed.}\\[0mm]
		\; &\text{\footnotesize(The same argument applies for Data Variables.)}
	\end{aligned}$	\medskip

\noindent Furthermore, as dynamic typechecking introduces a new runtime construct, \ie $\clr{\hVarX}$, we make the following additions:
\begin{itemize}
	\item  We extend our static typesystem (see \figref{fig:static-rules}) with type rule \rtit{tClr} that we define as follows:
		\[\inference[\rtit{tClr}]{\typeRule{\env{\envVV}{\envV}}{\hV}}{\typeRule{\env{\envVV}{\envV}}{\clr{\hVarX}{\hV}}}\]
	\item  We also extend function \textsf{vexcl} (defined in \defref{def:excl}) with an extra case which handles the clear construct in the same way as per the maximal fixed points, \ie we modify function \textsf{vexcl} as follows:
		
		{\small \setstretch{0.8} \[\vexcl{\varphi,\psi} \defEquals \begin{cases} 
						    	\quad\dots \\
						    	\;\vexcl{\varphi',\psi} & \text{if } \varphi\in\{\mmax{X}{\varphi'},\clr{X}\varphi'\}\,\land\,\psi\!\neq\!\trivSat \\
						    	\quad\dots
						  \end{cases}\]}
\end{itemize}

Furthermore, our proof of type soundness makes use of a well-formedness judgement for systems \wrt type environments \envV.  It that states that when a linear actor identifier is annotated as blocked in \envV, then it must indeed be blocked in \stV\ for it to be well-formed.
\begin{definition}[System Well-formedness]
  \label{def:wf-sys}
  \begin{math}
     \textbf{wf}(\stV,\envV) \;\deftxt\;\; \envV(\idV)=\lpidb \;\textsl{ implies }\; \stV(\idV) = \blocked
  \end{math}
\end{definition}

\noindent Note that \defref{def:wf-sys} is a technical device used to decompose soundness into the respective individual transitions.   However, it does not affect our premise that the system itself may not adhere to the typing discipline of the typing environment.  Recall also that  linear actors annotated as \lpidb are not allowed in initial environments. Thus, by   \defref{def:wf-sys}, any system is well-formed \wrt an environment \envV where $\lpidb \not\in\cod(\envV)$.

\subsection{The Top Level Proof}
As is standard, type soundness relies on the following two properties, namely Type Safety (\lemref{lemma1}) and Subject Reduction (\lemref{lemma2}). 

\begin{description}\itemsep0em
	\item[\lemmaA.]  $\\$\indent\indent$\text{For all } \Used\cdot\wf{\envV}{\stV}, \typeRule{\env{\envVV}{\envV}}{\hV}\; \text{ implies }\; \neg\mErrorC{\instr{\stV}{\cmon{\envV\!}{\Used\!}{\hV}}}$
\end{description}
\noindent Type Safety (\lemref{lemma1}) ensures that any typed script cannot \emph{immediately} be in error \wrt well-formed systems, as defined in \defref{def:error-dt}. This however does not imply that it will not reach an error after some number of transitions.  

For this reason, we also need Subject Reduction (\lemref{lemma2}), which states that a typed script and a well-formed system remain typed and well-formed after a single transition. 
\begin{description}\itemsep0em
	\item[\lemmaB.] $\\$\indent\indent$\wf{\envV}{\stV} \,\text{ and }\, \typeRule{\envV}{\hV}, \,\text{ and }\, \instr{\stV\!}{\cmon{\envV}{\Used}{\hV}}\!\traS{\act}\!\instr{\stV'\!}{\cmon{\envV'\!}{\Used'\!}{\hV'}}\;\text{ implies }\\\qquad\wf{\envV'}{\stV'}\,\text{ and }\,\typeRule{\envV'}{\hV'}$
\end{description}

\noindent Thus by repeatedly applying  \lemref{lemma2} over a sequence of transitions we know that starting from a typed script and a well-formed system will always lead to an eventual typed script and well-formed system.  From this latter fact, we can then apply \lemref{lemma1} to deduce that the resulting configuration cannot be in error. \bigskip

\noindent\textbf{To prove Theorem \ref{thm:soundness} (Type Soundness). }
	Whenever $\typeRule{\envV}{\hV}$ then:\\
	\begin{math}
   		 \instr{\stV}{\cmon{\envV}{\set{\typ{\idV}{\emptyset}}{\,\smash{\envV(\idV)\! =\! \lpid}}}{\hV}} \wtraS{\,\sV\,} \instr{\stV'}{\cmon{\envV'\!}{\Used'\!}{\hV'}} \quad \text{implies}\quad \neg\mErrorC{\instr{\stV'}{\cmon{\envV'\!}{\Used'\!}{\hV'}}}
  \end{math}

\begin{proof}  We prove the following statement from which the theorem follows:
\begin{equation*}\setstretch{1.1} 
  \begin{rcases}
    \wf{\envV}{\stV} & \text{ and }\\
    \typeRule{\env{\emptyset}{\envV}}{\hV} & \text{ and }\\
    \instr{\stV}{\cmon{\envV}{\Used}{\hV}}\!\!\! &  \wtraS{\,\sV\,} \instr{\stV'}{\cmon{\envV'\!}{\Used'\!}{\hV'}} \;
  \end{rcases}
\quad \text{implies}\quad \neg\mErrorC{\instr{\stV'}{\cmon{\envV'\!}{\Used'\!}{\hV'}}}
\end{equation*}

\noindent This is proved by numerical induction on the number of transitions in $\wtraS{\sV}$, \ie $|\,\wtraS{\sV}\!|\,=n$ using Type Safety and Subject Reduction as follows: \medskip

	\begin{Case}[$n\!=\!0$] Since $n=0$ implies that no transitions were performed, we know
		\begin{gather}
			\wf{\envV}{\stV} \label{snd:1:1} \\
			\typeRule{\env{\envVV}{\envV}}{\hV} \label{snd:1:2} \\
			\stV'\!=\!\stV \,\text{ and }\, \envV'\!=\!\envV \,\text{ and }\, \Used'\!=\!\Used \,\text{ and }\, \hV'\!=\!\hV, \label{snd:1:3} 
		\end{gather}	
	By \eqref{snd:1:1}, \eqref{snd:1:2} and \lemmaA we know that for all $\Used$
		\begin{gather}
			\neg\mErrorC{\instr{\stV}{\cmon{\envV}{\Used}{\hV}}} \label{snd:1:4}
		\end{gather}
	By \eqref{snd:1:3} and \eqref{snd:1:4} we can deduce
		\begin{gather}
			\neg\mErrorC{\instr{\stV'}{\cmon{\envV'}{\Used'}{\hV'}}} \label{snd:1:5}
		\end{gather}
	\noindent$\therefore$ Case holds by \eqref{snd:1:5}.
	\end{Case}
	
	\begin{lastcase}[$n\!=\!k\!+\!1$] Since $n=k+1$, and the rule premises we know 
		\begin{gather}
			\wf{\envV}{\stV} \label{snd:2:1} \\
			\typeRule{\env{\envVV}{\envV}}{\hV} \label{snd:2:2} \\
			\instr{\stV}{\cmon{\envV}{\Used}{\hV}}\traS{\act}\cdot\wtraS{\sV'}\instr{\stV'}{\cmon{\envV'}{\Used'}{\hV'}} \;\; \text{  where } \vert\,\wtraS{\sV'}\vert\,=k \label{snd:2:3} 
		\end{gather}	
	By \eqref{snd:2:3} and defn of $t$ we know 
		\begin{gather}
			\instr{\stV}{\cmon{\envV}{\Used}{\hV}}\traS{\act}\instr{\stV''}{\cmon{\envV''}{\Used''}{\hV''}}  \label{snd:2:4}\\
			\instr{\stV''}{\cmon{\envV''}{\Used''}{\hV''}}\wtraS{\sV'}\instr{\stV'}{\cmon{\envV'}{\Used'}{\hV'}}  \label{snd:2:5}
		\end{gather}
	By \eqref{snd:2:1}, \eqref{snd:2:2}, \eqref{snd:2:4} and \lemmaB we know
		\begin{gather}
			\wf{\envV''}{\stV''} \label{snd:2:6} \\
			\typeRule{\env{\envVV''}{\envV''}}{\hV''} \label{snd:2:7} 
		\end{gather}	
	By \eqref{snd:2:5}, \eqref{snd:2:6}, \eqref{snd:2:7} and IH we know 
		\begin{gather}
			\neg\mErrorC{\instr{\stV'}{\cmon{\envV'}{\Used'}{\hV'}}} \label{snd:2:8}
		\end{gather}
	\noindent$\therefore$ Case holds by \eqref{snd:2:8}.
	\end{lastcase}
\end{proof}

\subsection{Proving Type Safety}
We now consider proving type safety; here we detail the proof for \lemmaref{lemma1}.
\begin{lemma}[Type Safety] \label{lemma1}$\\$\indent\indent$\wf{\envV}{\stV} \;\text{  and  }\; \typeRule{\env{\envVV}{\envV}}{\hV}$\imp$\lnot(\mErrorC{\instr{\stV}{\cmon{\envV}{\Used}{\hV}}})$
\begin{proof}\emph{By Rule induction on } $\typeRule{\env{\envVV}{\envV}}{\hV}$.\medskip

\begin{Case}[\rtit{tFls}] From the rule premises we know that for all $\Used$  
\begin{gather}
 \typeRule{\env{\envVV}{\envV}}{\mfls} \label{A:1:1} \\ 
 \wf{\envV}{\stV} \label{A:1:2} 
\end{gather}
From the monitor reduction rules in \figref{fig:dt-mon-rules-1} we know that $\cmon{\envV}{\Used}{\mfls}\notTraceEvent{\scriptstyle\cors{\vLstB}}$ and also $\cmon{\envV}{\Used}{\mfls}\notTraceEvent{\scriptstyle\rel{\vLstB}}$. Hence, this case holds as by \defref{def:error-dt} (Synchronisation Error) we know $\lnot(\mErrorC{\instr{\stV}{\cmon{\envV}{\Used}{\mfls}}})$ as required.
\end{Case}

\noindent\textbf{Note:} The proofs for cases \rtit{tTru}, \rtit{tVar}, \rtit{tIf}, \rtit{tAdA}, \rtit{tMax}, \rtit{tClr}, \rtit{tNcA} and \rtit{tNcB} are analogous to that of case \rtit{tFls}.\medskip

\begin{Case}[\rtit{tCN1}] From the rule premises we know that for all $\Used$
\begin{align}
\typeRule{\env{\envVV}{\envV}}{\hV\mand\hVV} \label{A:2:1} 
\end{align}
because
\begin{gather}
	\envV=(\envVA+\envVB)  \label{A:2:2} \\ 
	\typeRule{\env{\envVV}{\envVA}}{\hV}  \label{A:2:3} \\ 
 	\typeRule{\env{\envVV}{\envVB}}{\hVV} \label{A:2:4} 
\end{gather} 	
and
\begin{align}
 \wf{\envV}{\stV} \label{A:2:5} 
\end{align} 
By \eqref{A:2:5} and \emph{system well-formedness} (\defref{def:wf-sys}) we know 
\begin{align}
 \scond{\envV}{\stV} \label{A:2:6} 
\end{align} 
As by \eqref{A:2:2} we know that $\envVA\subseteq\envV$ and $\envVB\subseteq\envV$, then from \eqref{A:2:6} we can deduce 
\begin{gather}
 \scond{\envVA}{\stV} \label{A:2:7} \\
 \scond{\envVB}{\stV} \label{A:2:8} 
 \end{gather} 
By applying the defn of well-formedness (\defref{def:wf-sys}) on both \eqref{A:2:7} and \eqref{A:2:8} we know 
\begin{align}
 \wf{\envVA}{\stV} \label{A:2:9} \\
 \wf{\envVB}{\stV} \label{A:2:10} 
\end{align} 
By \eqref{A:2:3}, \eqref{A:2:9} and IH we know 
\begin{align}
 \lnot(\mErrorC{\instr{\stV}{\cmon{\envVA}{\Used}{\hV}}}) \label{A:2:11} 
\end{align}  
By \eqref{A:2:4}, \eqref{A:2:10} and IH we know 
\begin{align}
 \lnot(\mErrorC{\instr{\stV}{\cmon{\envVB}{\Used}{\hVV}}}) \label{A:2:12} 
\end{align}
By \eqref{A:2:11} and \eqref{A:2:12} we know that both branches \emph{do not} yield errors. Hence we can conclude 
\begin{align}
 \lnot(\mErrorC{\instr{\stV}{\cmon{\envV}{\Used}{\hV\mand\hVV}}}) \label{A:2:13} 
\end{align} 
\noindent $\therefore$ Case holds by \eqref{A:2:13}. 
\end{Case}

\begin{Case}[\rtit{tCN2}] From the rule premises we know that for all $\Used$
\begin{align}
\typeRule{\env{\envVV}{\envV}}{\hV\mand\hVV} \label{A:2.5:1} 
\end{align}
because
\begin{gather}
	\excl{\hV,\hVV}=(\vLstA[\hV],\vLstA[\hVV])  \label{A:2.5:2} \\ 
	\typeRule{\env{\envVV}{\eff{\envV}{\vLstA[\hVV]}}}{\hV}  \label{A:2.5:3} \\ 
	\typeRule{\env{\envVV}{\eff{\envV}{\vLstA[\hV]}}}{\hVV}  \label{A:2.5:4}
\end{gather} 	
and
\begin{align}
 \wf{\envV}{\stV} \label{A:2.5:5} 
\end{align} 
By \eqref{A:2.5:5} and \emph{system well-formedness} (\defref{def:wf-sys}) we know 
\begin{align}
 \scond{\envV}{\stV} \label{A:2.5:6} 
\end{align} 
{\setstretch{1.2}By defn of \textsf{eff} we know that $\eff{\envV}{\vLstA}$ contains \emph{less} processes of type $\lpidb$ then $\envV$, since \textsf{eff} sets every process reference $\varid\in\vLstA$ to $\lpid$ and then produces a new type environment from $\envV$. This allows us to conclude}
\begin{gather}
	 \forall\typ{j}{\lpidb}\in\eff{\envV}{\vLstA[\hVV]}\,\Rightarrow\,\typ{j}{\lpidb}\in\envV \label{A:2.5:7} \\
	 \forall\typ{j}{\lpidb}\in\eff{\envV}{\vLstA[\hV]}\,\Rightarrow\,\typ{j}{\lpidb}\in\envV \label{A:2.5:8} 
\end{gather} 
By \eqref{A:2.5:6}, \eqref{A:2.5:7}, \eqref{A:2.5:8} and transitivity we can deduce
\begin{gather}
 \scond{(\eff{\envV}{\vLstA[\hVV]})}{\stV} \label{A:2.5:9} \\ 
 \scond{(\eff{\envV}{\vLstA[\hV]})}{\stV} \label{A:2.5:10}  
\end{gather}
By applying the defn of well-formedness (\defref{def:wf-sys}) on both \eqref{A:2.5:9} and \eqref{A:2.5:10} we know 
\begin{gather}
 \wf{\eff{\envV}{\vLstA[\hVV]}}{\stV} \label{A:2.5:11} \\
 \wf{\eff{\envV}{\vLstA[\hV]}}{\stV} \label{A:2.5:12} 
\end{gather} 
By \eqref{A:2.5:3}, \eqref{A:2.5:11} and IH we know 
\begin{align}
 	\lnot(\mErrorC{\instr{\stV}{\cmon{(\eff{\envV}{\vLstA[\hVV]})}{\Used}{\hV}}}) \label{A:2.5:13} 
\end{align}  
By \eqref{A:2.5:4}, \eqref{A:2.5:12} and IH we know 
\begin{align}
 	\lnot(\mErrorC{\instr{\stV}{\cmon{(\eff{\envV}{\vLstA[\hV]})}{\Used}{\hVV}}}) \label{A:2.5:14} 
\end{align}
By \eqref{A:2.5:13} and \eqref{A:2.5:14} we know that both branches \emph{do not} yield errors. Hence we can conclude 
\begin{align}
 \lnot(\mErrorC{\instr{\stV}{\cmon{\envV}{\Used}{\hV\mand\hVV}}}) \label{A:2.5:15} 
\end{align} \noindent $\therefore$ Case holds by \eqref{A:2.5:15}. 
\end{Case}

\begin{Case}[\rtit{tAdS}] From the rule premises we know that for all $\Used$
\begin{align}
	\typeRule{\env{\envVV}{\envV}}{\mscor{\vLstB}{\vLstA}\hV}  \label{A:3:1} 
\end{align}
because
\begin{gather}
 	\envV=\envVPN{\typ{\vLstB}{\lpidb}} \label{A:3:2} \\
 	\typeRule{\env{\envVV}{\envV}}{\mrelease{\vLstA}\hV} \label{A:3:3} 
\end{gather} 	
and
\begin{align}
 	\wf{\envV}{\stV} \label{A:3:4} 
\end{align} 
By \eqref{A:3:4} and \emph{system well-formedness} (\defref{def:wf-sys}) we know 
\begin{align}
 	\scond{\envV}{\stV} \label{A:3:5} 
\end{align} 
From \eqref{A:3:2} and \eqref{A:3:5} we can deduce 
\begin{align}
 	\stV(\vLstB)=\blocked \label{A:3:6} 
\end{align} 
By \eqref{A:3:2} and rule \rtit{rAdS} we know
\begin{align}
 	 \cmon{\envV}{\Used}{\mscor{\vLstB}{\vLstA}\hV}\traceEvent{\scriptstyle\cors{\vLstB}}\cmon{\envV}{\Used}{\mrelease{\vLstA}\hV} \label{A:3:7} 
\end{align} 
By \eqref{A:3:6} and rule \rtit{sAdS} we know 
\begin{align}
 	 \stV\traceEvent{\scriptstyle\cors{\vLstB}}\stV \label{A:3:8} 
\end{align} 
Hence, by \eqref{A:3:7}, \eqref{A:3:8} and \defref{def:error-dt} (Synchronisation Error) we know 
\begin{align}
 	\lnot(\mErrorC{\instr{\stV}{\cmon{\envV}{\Used}{\mscor{\vLstB}{\vLstA}\hV}}}) \label{A:3:9} 
\end{align} 
\noindent $\therefore$ Case holds by $\eqref{A:3:9}$.
\end{Case} 

\begin{lastcase}[\rtit{tRel}] From the rule premises we know that for all $\Used$
\begin{align}
	\typeRule{\env{\envVV}{\envV}}{\mrelease{\vLstB}\hV}  \label{A:4:1} 
\end{align}
because
\begin{gather}
 	\envV=\envVPN{\typ{\vLstB}{\lpidb}} \label{A:4:2} \\
 	\typeRule{\env{\envVV}{\envVPN{\typ{\vLstB}{\lpid}}}}{\hV} \label{A:4:3} 
\end{gather} 	
and
\begin{align}
 	\wf{\envV}{\stV} \label{A:4:4} 
\end{align} 
By \eqref{A:4:4} and \emph{system well-formedness} (\defref{def:wf-sys}) we know 
\begin{align}
 	\scond{\envV}{\stV} \label{A:4:5} 
\end{align} 
From \eqref{A:4:2} and \eqref{A:4:5} we can deduce 
\begin{align}
 	\stV(\vLstA)=\blocked \label{A:4:6} 
\end{align} 
By \eqref{A:4:2} and rule \rtit{rRel} we know
\begin{align}
 	 \cmon{\envVPN{\typ{\vLstB}{\lpidb}}}{\Used}{\mrelease{\vLstB}\hV}\traceEvent{\rel{\vLstA}}\cmon{\envVPN{\typ{\vLstB}{\lpid}}}{\Used}{\hV} \label{A:4:7} 
\end{align} 
By \eqref{A:4:6} and rule \rtit{sRel} we know 
\begin{align}
 	 \state{\mapstate{\vLstA}{\blocked}}{\stV}\traceEvent{\rel{\vLstA}}\state{\mapstate{\vLstA}{\unblocked}}{\stV} \label{A:4:8} 
\end{align} 
Hence, by \eqref{A:4:7}, \eqref{A:4:8} and \defref{def:error-dt} (Synchronisation Error) we know 
\begin{align}
 	\lnot(\mErrorC{\instr{\stV}{\cmon{\envV}{\Used}{\mrelease{\vLstB}\hV}}}) \label{A:4:9} 
\end{align} 
\noindent $\therefore$ Case holds by $\eqref{A:4:9}$.
\end{lastcase}
\end{proof}
\end{lemma} 

\subsection{Proving Subject Reduction} \label{sec:main-proofs}
Before proving subject reduction, in \defref{def:after} we define the \textsf{after} function $-$ this denotes how a value type environment $\envV$ changes based on the execution of a system action $\act$.  
\begin{definition}[The \textsf{after} function]\label{def:after} \smallskip
{\setstretch{1} 
\begin{align*}
	\after{\envV,\act} &\defEquals 
	\begin{cases} 
				\envVPN{\typ{\vLstB}{\lpidb}} & \text{if }\act=\blk{\vLstB}\,\land\,\envV=\envVPN{\typ{\vLstB}{\lpid}}\\
				\envVPN{\typ{\vLstB}{\lpid}} & \text{if }\act=\rel{\vLstB}\,\land\,\envV=\envVPN{\typ{\vLstB}{\lpidb}}\\
				\envVP & \text{if }\act=\actE\,\land\,\exists\envVP\cdot\envV\subseteq\envVP \\
				\envV & \text{otherwise}\\
	\end{cases}
\end{align*}
}
\end{definition}
\noindent This function enables us to make a more precise deduction of how the runtime reductions, over some arbitrary action $\act$, effect and evolve a value environment $\envV$ into a more updated environment $\envV'$ as defined by the function $\after{\envV,\act}$. This opposes proving that a runtime reduction reduces $\envV$ into some arbitrary value environment.  \medskip 

\noindent The proof for Subject Reduction relies on the following auxiliary lemmas: \vspace{-2mm} 
\begin{description}\itemsep0em
	\item[\lemmaX.] $\\\envVP=\after{\envV,\act} \;\text{ and }\; \wf{\envV}{\stV} \;\text{ and }\; s\!\traceEvent{\act}\!s'$\imp$\wf{\envVP}{\stV'}$
\end{description}
		--- This Lemma proves that well-formedness is preserved upon any action $\act$. This is proved by rule induction on $s\!\traceEvent{\act}\!s'$ in \secref{sec:main-aux-lemmas}.
\begin{description}	
	\item[\lemmaY.]  $\\\typeRule{\env{\envVV}{\envV}}{\hV} \;\text{ and }\; \cmon{\envV}{\Used}{\hV}\traceEvent{\act}\cmon{\envVP}{\UsedP}{\hV'}$ \imp $\typeRule{\env{\envVV}{\envVP}}{\hV'}\;\text{ and }\; \envVP=\after{\envV,\act}$ 
\end{description}	
	 --- This Lemma proves that if a well-typed script performs a dynamically checked runtime reduction, it always reduces into a new well-typed script.  This is proved by rule induction on $\cmon{\envV}{\Used}{\hV}\traceEvent{\act}\cmon{\envVP}{\UsedP}{\hV'}$; we show two main cases in \secref{sec:main-mon-red} while the rest are proved in \secref{sec:main-aux-lemmas}. \bigskip

\begin{lemma}[Subject Reduction] \label{lemma2} $\\\wf{\envV}{\stV} \,\text{ and }\, \typeRule{\env{\emptyset}{\envV}}{\hV} \,\text{ and }\, \instr{\stV}{\cmon{\envV}{\Used}{\hV}}\!\traceEvent{\act}\!\instr{\stV'}\cmon{\envVP}{\UsedP}{\hV'}$\imp$\wf{\envVP}{\stV'} \,\text{ and }\, \typeRule{\env{\emptyset}{\envVP}}{\hV'}$. 

\begin{proof}\emph{By Rule induction on } $\instr{\stV}{\cmon{\envV}{\Used}{\hV}}\!\traceEvent{\act}\!\instr{\stV'}\cmon{\envVP}{\UsedP}{\hV'}$.\medskip

\begin{Case}[\rtit{iAct}] From the rule premises we know 
\begin{align}
 \instr{\stV}{\cmon{\envV}{\Used}{\hV}}\!\traceEvent{\actE}\!\instr{\stV}{\cmon{\envVP}{\UsedP}{\hV'}} \label{B:1:1}
\end{align}
because
\begin{gather}
 s\traceEvent{\actE} s \label{B:1:2} \\ 
 \cmon{\envV}{\Used}{\hV}\!\traceEvent{\actE}\!\cmon{\envVP}{\UsedP}{\hV'} \label{B:1:3}
\end{gather}
and
\begin{gather}
	\wf{\envV}{\stV} \label{B:1:4} \\ 
	\typeRule{\env{\emptyset}{\envV}}{\hV} \label{B:1:5}
\end{gather}
By \eqref{B:1:3}, \eqref{B:1:5} and \lemmaY we know 
\begin{gather}
	\typeRule{\env{\emptyset}{\envVP}}{\hV'} \label{B:1:6} \\
	\envVP = \after{\envV,\actE}				\label{B:1:7}
\end{gather}
By \eqref{B:1:2}, \eqref{B:1:4}, \eqref{B:1:7} and \lemmaX we know 
\begin{align}
	\wf{\envVP}{\stV} \label{B:1:8}
\end{align}
\noindent $\therefore$ Case holds by \eqref{B:1:6} and \eqref{B:1:8}.
\end{Case}

\begin{Case}[\rtit{iSys}] From the rule premises we know 
\begin{align}
 \instr{\stV}{\cmon{\envV}{\Used}{\hV}}\!\traceEventTau\!\instr{\stV'}{\cmon{\envV}{\Used}{\hV}} \label{B:2:1}
\end{align}
because
\begin{gather}
 \stV\traceEventTau\stV' \label{B:2:2} 
\end{gather}
and
\begin{gather}
	\wf{\envV}{\stV} \label{B:2:3} \\ 
	\typeRule{\env{\emptyset}{\envV}}{\hV} \label{B:2:4}
\end{gather}
By defn of \textsf{after} (\defref{def:after}) and since $\act=\tau$ we know that 
\begin{align}
	\after{\envV,\tau}=\envV \label{B:2:5}
\end{align}
By \eqref{B:2:2}, \eqref{B:2:3}, \eqref{B:2:5} and \lemmaX we know 
\begin{align}
	\wf{\envV}{\stV'} \label{B:2:6}
\end{align}
\noindent $\therefore$ Case holds by \eqref{B:2:4} and \eqref{B:2:6}. 
\end{Case}

\begin{Case}[\rtit{iMon}] From the rule premises we know 
\begin{align}
 \instr{\stV}{\cmon{\envV}{\Used}{\hV}}\!\traceEventTau\!\instr{\stV}{\cmon{\envVP}{\UsedP}{\hV'}} \label{B:3:1}
\end{align}
because
\begin{gather}
 \cmon{\envV}{\Used}{\hV}\!\traceEventTau\!\cmon{\envVP}{\UsedP}{\hV'} \label{B:3:2} 
\end{gather}
and
\begin{gather}
	\wf{\envV}{\stV} \label{B:3:3} \\ 
	\typeRule{\env{\emptyset}{\envV}}{\hV} \label{B:3:4}
\end{gather}
By \eqref{B:3:2}, \eqref{B:3:4} and \lemmaY we know 
\begin{gather}
	\typeRule{\env{\emptyset}{\envVP}}{\hV'} \label{B:3:5}\\
	\envVP=\after{\envV,\tau}  \label{B:3:6}
\end{gather}
By \eqref{B:3:6} and the defn of \textsf{after} (\defref{def:after}) we know
\begin{align}
	\envVP=\envV \label{B:3:7}
\end{align}
By \eqref{B:3:3} and \emph{system well-formedness} (\defref{def:wf-sys}) we know
\begin{align}
	\scond{\envV}{\stV}  \label{B:3:8}
\end{align}
By \eqref{B:3:7} and \eqref{B:3:8} we know
\begin{align}
	\scond{\envVP}{\stV}  \label{B:3:9}
\end{align}
By \eqref{B:3:9} and \emph{system well-formedness} (\defref{def:wf-sys}) we know
\begin{align}
	\wf{\envVP}{\stV}  \label{B:3:10}
\end{align}
\noindent $\therefore$ Case holds by \eqref{B:3:5} and \eqref{B:3:10}. 
\end{Case}

\begin{Case}[\rtit{iAda}] From the rule premises we know 
\begin{align}
 \instr{\stV}{\cmon{\envV}{\Used}{\hV}}\!\traceEvent{\tau}\!\instr{\stV'}{\cmon{\envVP}{\UsedP}{\hV'}} \label{B:4:1}
\end{align}
because
\begin{gather}
 \stV\!\traceEvent{\actC}\!\stV' \label{B:4:2} \\
 \cmon{\envV}{\Used}{\hV}\!\traceEvent{\actC}\!\cmon{\envVP}{\UsedP}{\hV'} \label{B:4:3} 
\end{gather}
and
\begin{gather}
	\wf{\envV}{\stV} \label{B:4:4} \\ 
	\typeRule{\env{\emptyset}{\envV}}{\hV} \label{B:4:5}
\end{gather}
By \eqref{B:4:3}, \eqref{B:4:5} and \lemmaY we know 
\begin{gather}
	\typeRule{\env{\emptyset}{\envVP}}{\hV'} \label{B:4:6} \\
	\envVP=\after{\envV,\actC}	\label{B:4:7}
\end{gather}
By \eqref{B:4:2}, \eqref{B:4:4}, \eqref{B:4:7} and \lemmaX we know 
\begin{align}
	\wf{\envVP}{\stV'} \label{B:4:8}
\end{align}
\noindent $\therefore$ Case holds by \eqref{B:4:6} and \eqref{B:4:8}.
\end{Case}

\begin{lastcase}[\rtit{iTrm}] From the rule premises we know 
\begin{align}
 \instr{\stV}{\cmon{\envV}{\Used}{\hV}}\!\traceEvent{\actE}\!\cmon{\envV}{\Used}{\mtru} \label{B:5:1}
\end{align}
because
\begin{gather}
 s\traceEvent{\actE} s \label{B:5:2} \\ 
 \cmon{\envV}{\Used}{\hV}\!\notTraceEvent{\actE} \label{B:5:3}
\end{gather}
Given that by \eqref{B:5:3} we know that the monitoring stops applying, then we know that the monitor enters an \emph{idempotent state} (\ie it remains infinitely true) such that it is unable reduce into a script which is \emph{unsound}.\\
 Case holds trivially.
\end{lastcase}
\end{proof}
\end{lemma}

\subsection{Proving the Monitor Reduction Lemma} \label{sec:main-mon-red}
In this section we highlight two main cases pertaining to the proof for \lemmaY; the other cases are proved in \secref{sec:main-aux-lemmas}. The proof for this lemma is conducted by rule induction on the monitor reductions. Particularly, here we show the proofs for cases \rtit{rNc1} and \rtit{rMax} as these cases rely on the following three auxiliary lemmas:
\begin{description}\itemsep0em
	\item[\lemmaC.] $\\\typeRule{\env{\envVV}{\envVP}}{\mmax{\hVarX}{\!\hV}} \text{ and } \typeRule{\env{\envVV}{\envV}}{\hV}\,\imp\,\typeRule{\env{\envVV}{\envV}}{\hV\sub{(\clr{\hVarX}\mmax{\hVarX}{\!\hV})}{\hVarX}}$
\end{description}
		--- The Formula Substitution Lemma proves that whenever a formula variable $\hVarX$ is substituted in a well-typed formula, then the resultant formula is also well-typed formula.
\begin{description}\itemsep0em
	\item[\lemmaI.] $\\\typeRule{\env{\envVVN{\hVarX}{\envVP}}{\envV}}{\hV}, \; \hVarX\notin\fv(\hV)\, \imp \typeRule{\env{\envVV}{\envV}}{\hV}$
\end{description}
		--- The Weakening lemma proves that if a formula $\hV$ which typechecks \wrt formula environment $\envVVN{\hVarX}{\envVP}$ has no free occurrences of formula variable $\hVarX$ (\ie $\hVarX\notin\fv(\hV)$), then we can weaken the formula environment by removing the unnecessary entry $\hVarX\!\mapsto\!\envVP$. 
\begin{description}	
	\item[\lemmaD.]  $\\\typeRule{\env{\envVV}{(\envVN{\tbnd{\patE}}})}{\hV} \text{ and } \match{\patE}{\actE}{\sigma} \text{ and } \dtcheck\\$\indent$\qquad  \text{ and } \dtcheckB \imp\,\typeRule{\env{\envVV}{(\envVN{\tbnd{\patE}}\sigma)}}{\hV\sigma}$
\end{description}	
	 --- The Term Substitution Lemma is used to prove that if a system event $\actE$ \emph{matches} a necessity pattern $\patE$ without introducing any type inconsistencies, then if the substitution environment $\sigma$, created by the match operation, is applied onto a well-typed formula, then the resultant formula is also well-typed.
	 \smallskip
	 
\noindent Note that these three lemmas are proved in \secref{sec:main-aux-lemmas}. \medskip

\noindent{\bfseries\itshape To prove \lemmaY} $\\$\indent$\typeRule{\env{\envVV}{\envV}}{\hV}  \;\; \text{ and } \;\; \cmon{\envV}{\Used}{\hV}\traceEvent{\act}\cmon{\envVP}{\UsedP}{\hV'}\, \imp \,\typeRule{\env{\envVV}{\envVP}}{\hV'}\, \land\, \envVP=\after{\envV,\act}$.
\begin{proof}\emph{By Rule induction on} $\cmon{\envV}{\Used}{\hV}\traceEvent{\act}\cmon{\envVP}{\UsedP}{\hV'}$\medskip

\begin{Case}[\rtit{rMax}] From our rule premises we know 
\begin{gather}
	\cmon{\envV}{\Used}{\mmax{\hVarX}{\!\hV}}\traceEvent{\tau}\cmon{\envV}{\Used}{\hV\sub{(\clr{\hVarX}\mmax{\hVarX}{\!\hV})}{\hVarX}}  \label{Y:1.5:1} \\
 	\typeRule{\env{\envVV}{\envV}}{\mmax{\hVarX}{\!\hV}} \label{Y:1.5:3}
\end{gather} 	
By \eqref{Y:1.5:3} and \rtit{tMax} we know 
\begin{align}
	\typeRule{\env{\envVVN{\hVarX}{\envV}}{\envV}}{\hV} \label{Y:1.5:4}
\end{align}
By \eqref{Y:1.5:3}, \eqref{Y:1.5:4} and \lemmaC we know 
\begin{align}
	\typeRule{\env{\envVVN{\hVarX}{\envV}}{\envV}}{\hV\sub{(\clr{\hVarX}\mmax{\hVarX}{\!\hV})}{\hVarX}} \label{Y:1.5:5}
\end{align}
By \eqref{Y:1.5:5}, due to substitution we know that every free formula variable $\hVarX$ in $\hV$ is substituted by $(\clr{\hVarX}\mmax{\hVarX}{\!\hV})$, and hence we know
\begin{align}
	\hVarX\notin\fv(\hV\sub{(\clr{\hVarX}\mmax{\hVarX}{\!\hV})}{\hVarX}) \label{Y:1.5:6}
\end{align}
By \eqref{Y:1.5:5}, \eqref{Y:1.5:6} and \lemmaI we know 
\begin{align}
	\typeRule{\env{\envVV}{\envV}}{\hV\sub{(\clr{\hVarX}\mmax{\hVarX}{\!\hV})}{\hVarX}} \label{Y:1.5:7}
\end{align}
By the defn of \textsf{after} (\defref{def:after}){} (\defref{def:after}) and since $\act=\tau$ we know
\begin{align}
	\envV=\after{\envV,\tau} \label{Y:1.5:8}
\end{align}
$\therefore$ Case holds by \eqref{Y:1.5:7} and \eqref{Y:1.5:8}. 
\end{Case}

\begin{Case}[\rtit{rNc1}]  From our rule premises we know 
\begin{align}
 \cmon{\envV}{\Used}{\mBDNec{\patE}{\stk}{\vLstA}{\hV}}\traceEventA\cmon{\envVN{\tbnd{\patE}\sigma}}{\UsedP}{\mblock{i}\hV\sigma} \label{Y:7:1} 
\end{align}
because 
\begin{gather}
	\match{\patE}{\actE}{\sigma} \label{Y:7:2} \\
	\dtcheck \label{Y:7:3} \\
	\dtcheckB \label{Y:7:3.5} \\
	\id{\actE}=i 					\label{Y:7:4} 
\end{gather}
and
\begin{gather}
	\typeRule{\env{\envVV}{\envV}}{\mBNec{\patE}{i}{\hV}}  \label{Y:7:5} 
\end{gather}
By \eqref{Y:7:5} and \rtit{tNcB} we know 
\begin{gather}
	\id{\patE}=\varid				\label{Y:7:8} \\
	\typeRule{\env{\envVV}{(\envVN{\tbnd{\patE}})}}{\mblock{\varid}\hV}  \label{Y:7:10} \\
	\typeRule{\env{\envVV}{\envV}}{\mrelease{\vLstA}\mtru} \label{Y:7:11}
\end{gather}
By \eqref{Y:7:2}, \eqref{Y:7:3}, \eqref{Y:7:3.5}, \eqref{Y:7:10} and \lemmaD we know 
\begin{align}
	\typeRule{\env{\envVV}{(\envVN{\tbnd{\patE}\sigma})}}{\mblock{\varid\sigma}\hV\sigma}  \label{Y:7:12} 
\end{align}
By \eqref{Y:7:2}, \eqref{Y:7:4} and \eqref{Y:7:8} we know 
\begin{align}
	\varid\sigma=i  \label{Y:7:13} 
\end{align}
By \eqref{Y:7:12} and \eqref{Y:7:13} we know
\begin{align}
	\typeRule{\env{\envVV}{(\envVN{\tbnd{\patE}\sigma})}}{\mblock{i}\hV\sigma}  \label{Y:7:14} 
\end{align}
Since $\envV\subseteq(\envVN{\tbnd{\patE}\sigma})$, by defn of \textsf{after} (\defref{def:after}){} (\defref{def:after}) we know
\begin{align}
	\after{\envV,\actE}=(\envVN{\tbnd{\patE}\sigma})  \label{Y:7:15} 
\end{align}
$\therefore$ Case holds by \eqref{Y:7:14} and \eqref{Y:7:15}. 
\end{Case}

\begin{Case}[\rtit{rNc2}] The proof for this case is analogous to that of case \rtit{rNc1}. \end{Case}\medskip

\noindent The rest of the cases are proved in \secref{sec:main-aux-lemmas}.
\end{proof}

\section{Conclusion}
In this chapter we have addressed our third objective (\ie objective (iii) in \secref{sec:intro:obj}), whereby we developed a static type system for assisting the specifier into writing valid runtime adaptation scripts that are free from synchronisation errors (as stated in \defref{def:err-conf}). This type system distinguishes between identifiers which are used in synchronisation mechanisms and synchronous adaptations \ie \emph{linear identifiers}, and those which are not \ie \emph{unrestricted identifiers}. By making this distinction, the type system is able to restrict the use of linear identifiers such that errors are not introduce due to race conditions. However it can be less stringent with how the unrestricted identifiers can be used, as these are incapable of introducing synchronisation errors given that they cannot be used in synchronisation operations (if they are the script gets rejected). 

The static type system uses a tracking mechanism, in its analysis, to make sure that linear identifiers are always blocked prior to being used in synchronous adaptations. Furthermore, the typesystem detects erroneous scripts that suffer from race conditions by employing a degree of linearity. In this way, unless two (or more) branches are \emph{mutually exclusive}, the type system allows \emph{only one branch} to operate on the same linear identifier. This therefore eliminates cases where one concurrent monitoring branch may for instance release a blocked system actor before another branch manages to apply a synchronous adaptation. Implementing this tracking mechanism required augmenting RA scripts with the necessary type annotations, thereby also allowing the type system to distinguish between term variables used for binding generic data, and variables that bind linear and unrestricted process identifiers. 

As we monitor untyped systems, process identifiers that are bound to term variables may introduce type inconsistencies such as \emph{type mismatches} and \emph{aliasing}. We thus augmented the monitor's operational LTS semantics with dynamic typechecking so as to eliminate introducing synchronisation errors, due to these type inconsistencies, by aborting monitoring.

Finally we provided a formal proof guaranteeing that our type system is \emph{sound} \wrt to typed adaptation scripts. We thus guarantee that any typed script that typechecks statically, generates dynamically typed monitors that are \emph{safe}, \ie at no point during their execution can they introduce synchronisation errors.

\chapter{Related Work}
\label{chp:related-work}
In this chapter we consider other work available in the current literature which is related to different aspects of this dissertation; we thus structure this chapter in four sections. In \secref{sec:synasyn-rw} we discuss work related to the synchronous and asynchronous monitoring aspects discussed and explored in \Cref{chp:syn-asyn}. Moreover, in relation to the work presented in \Cref{chp:runtime-adaptation}, in \secref{sec:ra-rw} we present other work which employs a form of mitigation mechanisms that relate to our runtime adaptation methodology. In conjunction, in \secref{sec:logicra-rw} we also discuss work which dealt with extending a formal Logic with a notion of adaptations. Finally, in \secref{sec:static-analysis-rw} we present work related to the static and dynamic analysis techniques that we presented in \Cref{chp:typ-sys}.

\section{Synchronous and Asynchronous Monitoring}\label{sec:synasyn-rw}
By and large, most widely used online RV tools employ synchronous instrumentation \cite{java-mac,larva-CPS09,chen-rosu-2007-oopsla, jUnitRv:13,BarringerFHRR12}. There is also a substantial body of work commonly referred to as asynchronous RV \cite{lola:runtime,d'Amorim:2005:ERV:1082983.1083249,Andrews03generaltest}. However, the latter tools and algorithms assume \emph{completed traces}, generated by complete program executions and stored in a database or a log file.  As explained in \secref{sec:off_on}, we term these bodies of work as \emph{offline}, and their aims are considerably different from the work presented in \Cref{chp:syn-asyn}. 

There exist a few tools offering both synchronous and asynchronous monitoring, such as MOP \cite{chen-rosu-2005-tacas,chen-rosu-2007-oopsla}, JPAX \cite{Havelund:2001:MJP:891177,SyncVSAsync:Rosu:2005} and DB-Rover \cite{rover, taxonomy:Delgado:2004}, whereby the specifier can choose whether a property should be monitored for either synchronously or asynchronously. Crucially, however, unlike our newly developed hybrid monitoring approach, these tools do not provide the fine grained facility of supporting both synchronous and asynchronous monitoring at the same time, \ie by switching between the two modes at runtime. Apart from \detecterGen\ \cite{FraSey14}, e\rtit{Larva} \cite{elarva:2012} is another Erlang monitoring tool that uses the EVM tracing mechanism to perform asynchronous monitoring. 

Rosu \etal \cite{SyncVSAsync:Rosu:2005} make a similar distinction to ours in their definitions synchronous and asynchronous online monitoring, based on which they develop formula rewriting algorithms to implement efficient centralised monitors. However, their definitions of synchrony and asynchrony deals with the timeliness of detections made by their monitors \ie that the monitors are efficient enough not to lag behind; however, they provide minimal details on how the system is instrumented. By contrast, we focus on showing how instrumentation is carried out over actor-based systems, and show how hybrid instrumentation can be used to obtain timely detections for certain properties, by blocking strategic parts of the monitored system until the monitor performs the necessary checks. This amounts to the definition of synchronous detection monitoring given in \cite{SyncVSAsync:Rosu:2005}.

A closer work to ours (\wrt this area) is \cite{CP12FF}, whereby the authors allow a decoupling between the system and monitor executions but provide \emph{explicit} mechanisms for pausing the system while the lagging (asynchronous) monitor execution catches up. In our case this mechanism is handled in a more implicit manner as we weave it in a more natural way within our logic. In fact we implement similar mechanisms at a higher level of abstraction by integrating them seamlessly in our logic when switching from asynchronous to synchronous monitoring in our hybrid monitoring approach. Furthermore, in \cite{CP12FF} the authors dealt with introducing asynchrony on top of a setting which supported synchronous monitoring, while in our case we had to do the converse. In fact in \Cref{chp:syn-asyn} we studied ways on how to introduce synchrony in a tool and a setting, which only provide support for asynchrony. 

 Talcott \etal in \cite{Talcott2008} compare and contrast three coordination models for actors which cover a wide spectrum of communication mechanisms and coordination strategies. 
 The comparison focusses on  the level of expressivity of each model, the level of maturity, 
 the level of abstraction of the model,
 and the way user definable coordination behavior is provided. One of the analysed coordination models, \ie the Reo model, is a channel based language in which channels may be either synchronous or asynchronous. This model resembles the way our hybrid monitoring protocol interacts with the monitored system. In fact our monitoring language constructs \texttt{[$\alpha$]} and \texttt{[|$\alpha$|]}, seem analogous to the asynchronous and synchronous channels in the Reo model. The major difference is that these coordination models are used to develop a concurrent system by specifying protocols of communication between its, while our constructs are used for monitoring an existing actor-based system by defining the way that the system and the monitor interact.  

\section{Employing Runtime Mitigation} \label{sec:ra-rw}
In \cite{Rinard04}, Rinard \etal introduce \emph{failure-oblivious-computing}, a tailor made runtime adaptation technique that enables reinforced systems to survive memory errors without memory corruption. The solution consists in the creation of a \emph{safe} C compiler which inserts bounds checks that \emph{dynamically} detect invalid memory accesses. Upon detection, instrumented systems execute recovery code which mitigates the invalid memory access, thus allowing the system to continue with its normal execution instead of crashing. Similarly, in \cite{Rinard12} Rinard \etal present another domain-specific dynamic adaptation technique for \emph{automatic input rectification}. This technique was realized as a prototype SOAP implementation that automatically rectifies potentially dangerous inputs by converting them into typical inputs that the program is more likely to process correctly. Furthermore in \cite{Rinard14}, Rinard \etal implement yet another dynamic adaptation technique called RCV that enables reinforced systems to survive \emph{divide-by-zero} and \emph{null dereference} errors. This technique employs a monitor which attaches itself to the target system whenever an error occurs, mitigates its execution, and keeps on tracking the effects of the applied mitigation until they are flushed from the system's state, in which case the monitor detaches itself. 

These techniques can, to some extent, be considered analogous to our adaptation scripts, in the sense that once an adaptation script is compiled, it generates a monitor which dynamically analyses a system and intervenes whenever it detects the specified erroneous behaviour. This is analogous to what the techniques presented by Rinard \etal are doing in order to reinforce their respective systems. The major differences however are that the adaptation properties presented in Rinard \etal's work target a different domain (\ie not for actor systems), and they were implemented in a tailor-made fashion, \ie they were not specified in some high-level logic and then automatically converted into the required adaptation monitor. By contrast, in our work we provide a logic and a tool for specifying runtime adaptation properties for adapting actor-based systems through actor-level manipulations.

Colombo \etal in \cite{cc-saferAsync,CCPHD2013} integrate the notion of compensations within the \rtit{Larva} runtime monitoring framework to allow for the undoing of undesired system actions. Compensations combined with asynchronous monitoring allow for performance overheads to be minimised. This is done by providing the lagging asynchronous monitor with the ability to \emph{rollback} the system by reverting its state to a point prior to the occurrence of the detected anomaly, even in the case of late detection. This contrasts our adaptations which rather then trying to rollback the system, they acknowledge that an error has occurred and apply mitigation techniques to reduce and possibly eliminate the repercussions of the detected misbehaviour.

Furthermore, compensations are generally \emph{user defined} and related to a specific action; in fact it is generally assumed that a system action has a compensating action which can be used during the rollback procedure once a property violation is detected. This opposes our work as we define and implement a number of adaptations (\ie they are not user defined) which are not related to any kind of system action. In fact our adaptations do \emph{not} assume any information about the code executed by the monitored system. Moreover, Colombo \etal's work differs in the type of systems that they target. Whereas in our work we target long-running (reactive) actor-based systems, in their work the authors target long-lived transaction systems such as the EntroPay transaction system \cite{cc-saferAsync}.

\section{Extending a Logic with a Notion of Adaptation}  \label{sec:logicra-rw}
In \cite{Zhang06} Zhang \etal developed A-LTL, an adaptation-based extension to Linear Temporal Logic (LTL) \cite{Clarke:2000:MC:332656} for formally specifying adaptation requirements of a system in temporal terms. Particularly, the extended logic allows a user to specify the way that an adaptive system can transition from (\ie adapt) one program (\ie the source program) into another program (\ie the target program). This logic was also implemented as a runtime verification tool called \textsc{AMOebA-RT} \cite{Zhang08}, which provides assurance that dynamically adaptive software satisfies its requirements. \textsc{AMOebA-RT} instruments adaptive programs to insert code which forwards runtime state information to a concurrent monitoring process which verifies whether the adaptive system satisfies the adaptation properties specified in A-LTL and LTL. \textsc{AMOebA-RT} therefore allows for continuous monitoring and verification of post-release behavior of a software system, ensuring that whenever an error occurs the adaptive software either flags an error report, or applies an adaption to mitigate the error automatically. 

This opposes our work as our adaptation scripts provide the synthesised monitors with the ability of \emph{actually performing} the required adaptations, thus converting normal systems into adaptive systems where the monitor takes care of applying the respective adaptations when required. This contrasts the monitors synthesised from A-LTL scripts presented in Zhang \etal's work, as these are used to verify that an adaptive system applies the necessary adaptations as specified, and thus their monitors (unlike ours) are incapable of performing the adaptations over the monitored system.

\section{Static Analysis of Monitoring scripts}\label{sec:static-analysis-rw}
In \cite{Bodden09} Bodden \etal present \emph{dependent advices} $-$ an AspectJ language extension containing dependency annotations for preserving crucial domain knowledge. This allows dependent advices to execute only when their dependencies are satisfied, thus allowing for optimizations that exploit this knowledge by removing advice-dispatch code from program locations where these dependencies cannot be satisfied. This concept lead to the creation of \textsc{Clara} \cite{Bodden10Tut,Bodden10} $-$ a static-analysis framework that employs static typestate analyses to automatically convert any AspectJ monitoring aspect into a residual runtime monitor that only monitors events triggered by program locations that the analyses failed to prove safe. 
	
To a certain extent, AspectJ scripts augmented with dependent advices are similar to our typed scripts, as they provide the static compiler with additional information about the environment interacting with the system. In Bodden's case dependencies dictate whether an aspect will execute at runtime or not, thus allowing for static optimisations, while in our case the types provide information which allows us to statically determine whether a script is well formed or not \ie whether it introduces synchronisation errors or not. Therefore, our aims differ from those of  Bodden \etal's work, as in our work we aimed to employ static analysis for our RA scripts so as to ensure a degree of correctness. Conversely, in Bodden \etal's work the main aim was to apply static analysis in order to reduce the overheads imposed by the AspectJ scripts. Furthermore, our work also differs in the way we evaluate our static analysis technique, since we provide type soundness proof as opposed to Bodden \etal which provide empirical performance tests. 
	
Castellani \etal \cite{Castellani14} extend previous work by Dezani \etal \cite{Dezani2014}, whereby they present a model (calculus) of self-adaptive, multiparty communications, in which security violations occur when processes attempt to communicate messages using inappropriate security levels. The model consists in three components: global types, monitors, and processes. Global types represent how processes interact and define reading permissions for each process. Monitors are then obtained by projecting a global type onto a process thus obtaining a \emph{monitored process}; hence global types are in some sense analogous to our typed scripts. Monitors are able to mitigate security violations using local and global adaptations, for handling minor and serious violations, respectively. 

Similar to our work, the authors also establish type soundness by which they ensure that global protocols still execute correctly, even when the system adapts itself to preserve security. 
However, they assume that the processes requiring monitoring are also well-typed, whereas we do not rely on such an assumption. Specifically we consider the system as a black-box and assume minimal information. This is reflected in the way we model the system in our RA operational model, and in the type assumptions used by our type system, whereby the only information that we assume about the monitored system is whether its actors are active or suspended by our instrumentation. Hence our adaptations can apply over any (instrumented) actor regardless of the code it executes.

\chapter{Conclusion}	
\label{chp:conc}

In this thesis we sought to develop a Runtime Adaptation framework for actor systems by \emph{extending} an \emph{existing} Runtime Verification tool, with adaptation functionality that can be  \emph{localised} to individual concurrent system components, thereby allowing for specific actors to be mitigated while leaving the other actors unaffected. This demanded addressing the following challenges:
\begin{enumerate}[label=(\roman*), leftmargin=7mm]
	\item Effective runtime adaptation on actor systems requires introducing a degree of synchrony in our monitors. As synchrony is not natively supported by the actor model, this requires studying ways for introducing synchronous monitoring in an efficient manner.
	\item An effective RA framework must also provide a number of adaptations that can be used to mitigate erroneous system behaviour by modifying several architectural aspects of actor systems. We must therefore identify these adaptations and develop a mechanism by which these can be made applicable to any actor system, \ie regardless of the code that it executes. As RA properties are no longer passive, these may introduce errors at runtime. Formalising the runtime behaviour of our RA scripts thus relieves us from having to deal with the complexities of the implementation. This therefore enables us to identify and better understand the subtle errors that our RA monitors may introduce at runtime.
	\item Formalisation also enables us to study ways how erroneous RA scripts can be detected and rejected prior to being deployed, by employing static analysis techniques. In this way we should therefore provide the specifier with a tool that serves as guidance for writing error-free runtime adaptation scripts. This tool must however provide an element of guarantee about its own correctness.
\end{enumerate}

To address our first objective (\ie (i)), in \Cref{chp:syn-asyn} we have therefore studied different techniques for introducing synchronous monitoring on top of an inherently asynchronous platform (\ie the actor model). After integrating these techniques in \detecterGen, we carried out a \emph{systematic assessment} of the relative overheads incurred by these different monitoring techniques. This assessment allowed us to conclude that although an \emph{incremental synchronisation} approach still introduces a degree of synchronisation overheads, it however permits for these overheads to be minimised by employing synchronisation on a \emph{by-need basis}. 

In light of the obtained performance results, in \Cref{chp:runtime-adaptation} we addressed our second objective, (\ie (ii)). We thus identified a number of adaptation actions that allow for violations to be mitigated by \emph{modifying different aspects} of actor systems, such as in the case of process restarts and message interceptions. We then categorised these adaptations into two classes, namely \emph{asynchronous} and \emph{synchronous} adaptations, where the latter require synchronisation when applied over the respective adaptees. Synchronisation was therefore introduced by building upon the efficient incremental synchronisation methodology identified in the previous chapter. Hence, by adding \emph{minimally intrusive} annotations, we augmented \detecterGen's specification language with synchronisation mechanisms, for blocking and releasing actors, along with implementations of the identified adaptation actions. This allows the specifier to develop adaptation scripts that yield monitors which are capable of mitigating erroneous behaviour by executing adaptations. A \emph{prototype implementation} of our Runtime Adaptation framework called \adapter\footnote{A \emph{prototype} implementation of this tool can be accessed and downloaded from \texttt{https://bitbucket.org/casian/adapter}.}, was thus developed as a result of this extension to \detecterGen.

Furthermore, we also gave a formal specification of the augmented logic (with adaptations) through a formal operational model. This allows us to predict how an adaptation script operates at runtime without actually executing it, and thus enabled us to better understand the subtle errors that our RA monitors may introduce. Hence this model permitted us to identify and \emph{formalise} situations where our runtime adaptation framework may introduce errors if synchronisation is not properly defined.

Based on this formal model, in \Cref{chp:typ-sys} we addressed our third objective (\ie (iii)) by developing a static type system for assisting the specifier into writing valid runtime adaptation scripts. We identified a number of errors that erroneous RA scripts may introduce at runtime in the monitored system. We then focused on showing how one of these errors \ie the synchronisation error formalised in \Cref{chp:runtime-adaptation}, can be statically detected. We finally evaluated our type system by proving \emph{type soundness} \wrt typed adaptation scripts, thereby guaranteeing that the monitors that are synthesised from a \emph{valid} adaptation script (\ie a script which was accepted by our type system), can at no point during execution introduce synchronisation errors.

Once we addressed these three objectives we established that there exists an important interplay between synchronisation and effective adaptation actions. Although this introduces additional overheads, as well as, the possibility of introducing runtime errors, we developed and evaluated \emph{concepts} which enable adaptations to be performed efficiently, and which provide assistance for writing error-free scripts. Despite implementing these concepts \wrt actor systems developed in Erlang, we conjecture that similar challenges would arise when developing a similar RA framework for other actor-based technologies such as Scala \cite{actorsinscala} and Akka \cite{akka}. Furthermore, these concepts may also be applicable to other inherently asynchronous architectures such as service-oriented systems \cite{RA-SOA:2008} and distributed systems \cite{Coulouris:2011}. Therefore despite requiring certain parts of the implementation to be reformulated for the respective technologies, the core concepts studied in this thesis (\eg incremental synchronisation, the identified adaptations, \etc) would still be relevant and applicable.

\section{Future Work}	
In order to further improve our work, we propose the exploration of the following suggestions.

\subsection{Implementing More Adaptations}
In \Cref{chp:runtime-adaptation} we have identified a number of adaptation categories which modify different aspects of an actor system. We have however implemented only a representative adaptation action for each category, as a proof-of-concept. We therefore propose that in future work we implement other adaptation actions thus providing the user with more adaptations that can be used for mitigating the detected violations.    

\subsection{An Algorithmic Type System}
When the type system developed in \Cref{chp:typ-sys}, analyses two non-mutually exclusive branches, it preserves linearity by using an \emph{environment splitting} operation to assign linear identifiers to only one concurrent branch, thereby eliminating cases which could lead to race conditions. Although our type system is implementable, the inherent non-determinism of environment splitting, does not provide a directly implementable algorithm for splitting the type environment. We conjecture that the development of an Algorithmic Type System \cite{ATTAPL} in which linear environment splitting can be performed deterministically, should require minimal alterations to our existing type system. Furthermore, to prove that this algorithmic version is also sound, it suffices proving that it corresponds to our existing type system, for which we have already proven type soundness. 

\subsection{Type inference}
Although we provide a tool which assists the specifier into writing valid scripts, this requires the user to manually include additional annotations and to provide an environment containing the initial type assumptions which are used by the type system. We conjecture that the user should ideally be relieved from this additional effort. We therefore propose the exploration of type inference techniques that could be used to automatically infer the required type assumptions and include the respective annotations. Hence a type inference system could be used to analyse an untyped script in order to generate a type environment and a typed script which typechecks \wrt our type system.

\subsection{Static Detection of Other Errors}
In this thesis we aimed to prove the concept that although erroneous RA scripts introduce runtime errors, we can still provide mechanisms for detecting these errors prior to deployment. Although in \Cref{chp:typ-sys}, we identify a number of errors that our RA monitors may introduce, in order to prove this concept we developed a type system which detects \emph{one} of these errors \ie synchronisation errors which we formalised in \Cref{chp:runtime-adaptation}. 
However we conjecture that in most cases, addressing the other errors would only require minor alterations to our type analysis techniques.

%


\appendix

\chapter{Monitor Optimisations in \detecterGen}
\label{chp:app-mon-opt}
In this chapter we present additional information with regards the different monitor synthesis techniques that are present in the current implementation of the \detecterGen\ RV tool. In \secref{sec:monitor-synthesis} we look into the original monitor synthesis algorithm, presented in \cite{FraSey14} and explain its major causes of inefficiencies. Following this in \secref{sec:proposed-solutions} we give a brief overview on how static and dynamic optimisations where proposed in \cite{CasFraSaid15} and augmented in \detecterGen\ as means for lowering these overheads to a more feasible standard.

\section{Synthesising Unoptimised Monitors}
\label{sec:monitor-synthesis}

The original synthesis algorithm \cite{FraSey14} aims to be \emph{modular} by generating independently executing monitor combinators for each logical construct, interacting with one another through message passing.\footnote{Every combinator is implemented as a (lightweight) Erlang process (actor) \cite{Cesarini:2009}, uniquely identifiable by its process ID.  Messages sent to a process are received in its dedicated mailbox, and may be read selectively using (standard) pattern-matching.}  For instance,  the synthesis  \emph{parallelises} the analysis of the subformulas $\hV_1$ and $\hV_2$ in a conjunction ${\hV_1}\mand{\hV_2}$ by:
\begin{enumerate}[label=(\roman*)] \itemsep0em \setstretch{1.2}
	\item synthesising concurrent monitor systems for $\hV_1$ and $\hV_2$ respectively; and 
	\item creating a conjunction monitor combinator that receives trace events and forks (forwards) them  to the independently-executing monitors of  $\hV_1$ and $\hV_2$. 
\end{enumerate}
Since the submonitor systems for $\hV_1$ and $\hV_2$ may be arbitrarily complex (needing to analyse a stream of events before reaching a verdict), the conjunction monitor is \emph{permanent} in the monitor organisation generated, so as to fork and forward event streams of arbitrary length.  It is also worth noting that the synthesis algorithm assumes formulas to be \emph{guarded}, where recursive variables appear under necessity formulas; this is required to generate monitors that implement \emph{lazy} unrolling of recursive formulas, thereby minimising overheads.

\begin{figure}[ht!]
  \centering
  \[\begin{array}{ccc}
  \includegraphics[scale=0.1]{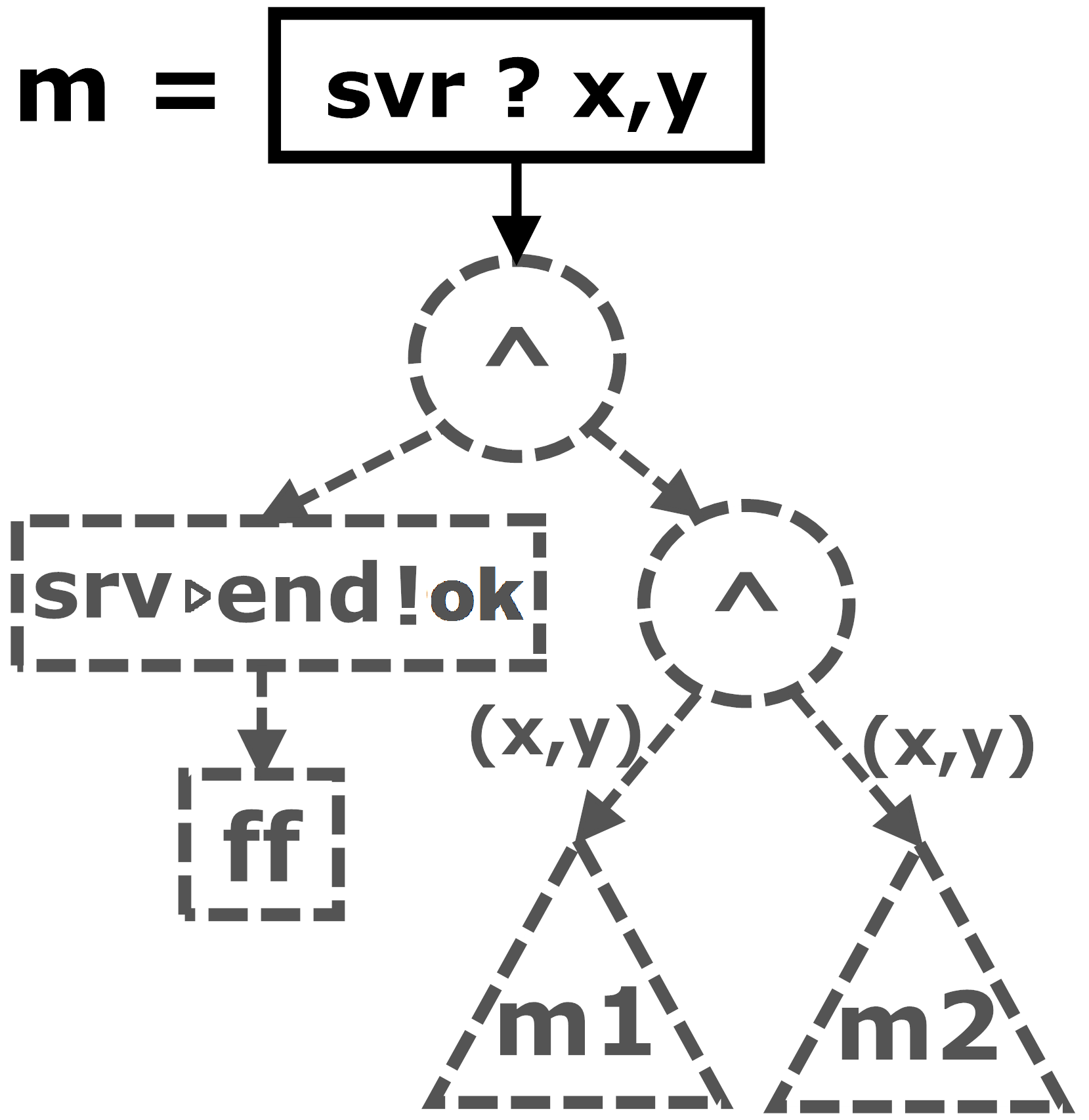} 
&\quad
  \includegraphics[scale=0.1]{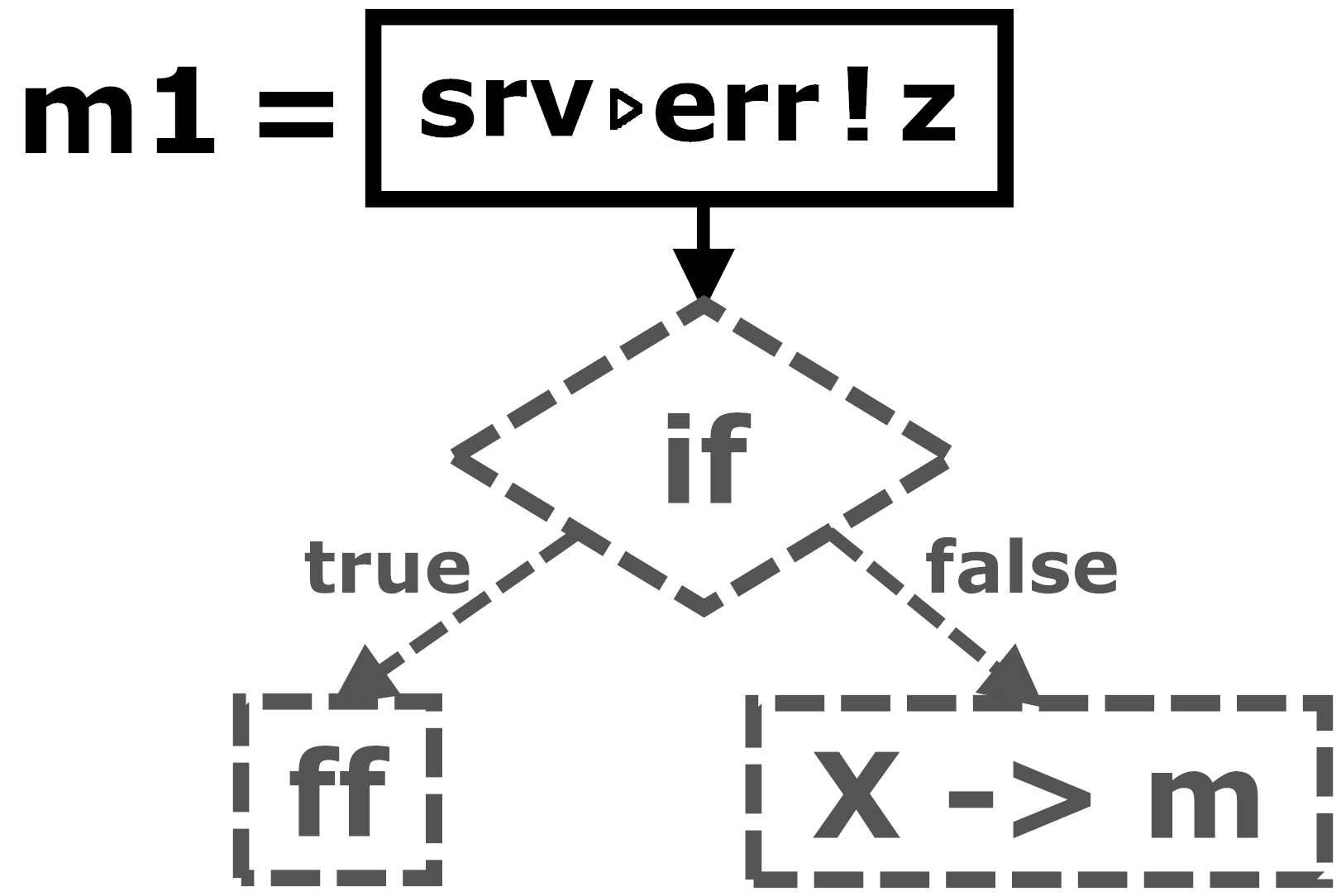}
&\quad
  \includegraphics[scale=0.1]{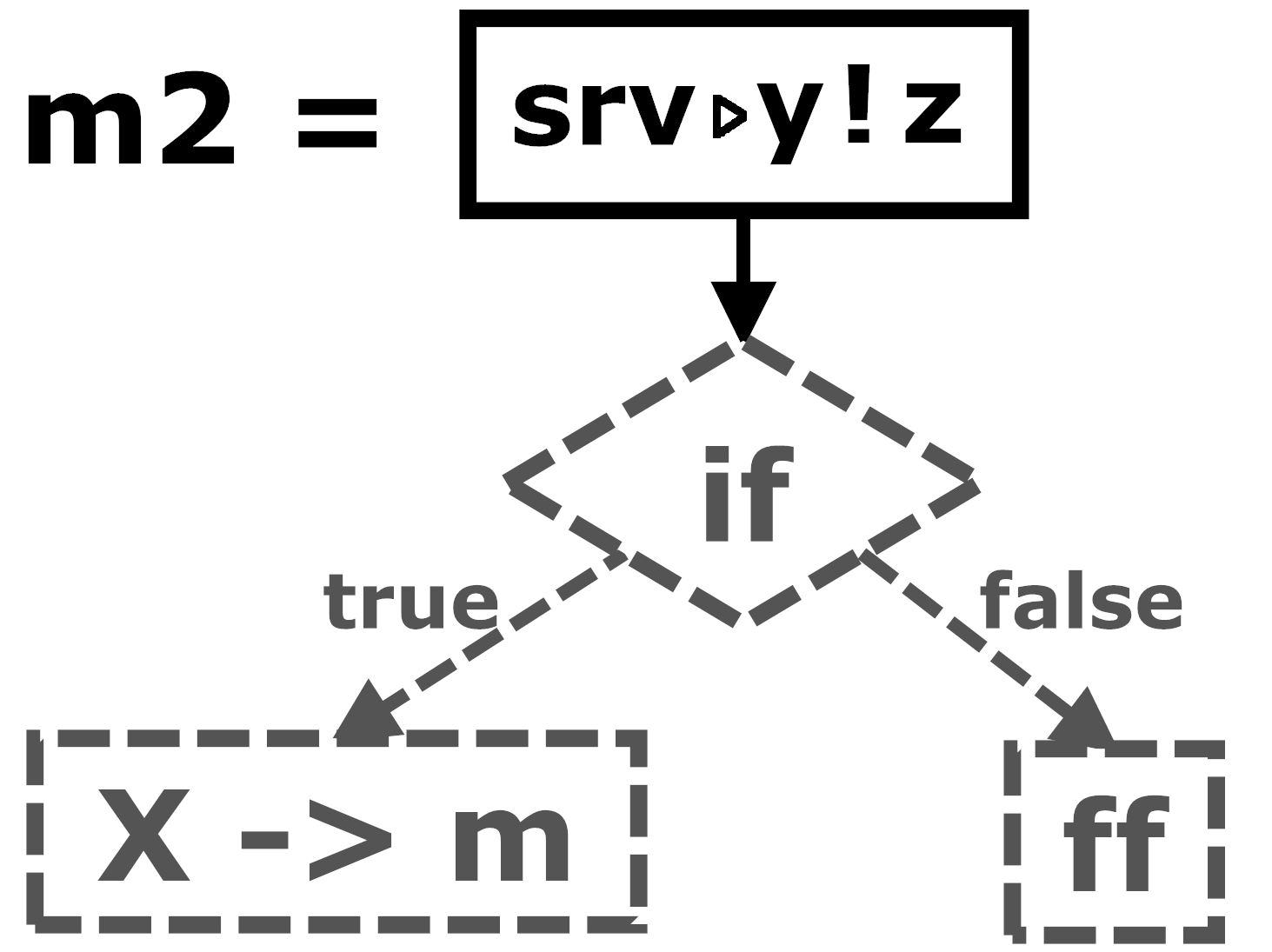}
\end{array}\]
\hrulefill
\[\begin{array}{ccc}
    \begin{array}{l}
\tra{\displaystyle\mrecv{\eatom{srv}}{\etuple{v,c}}}\\
  \quad\includegraphics[scale=0.09]{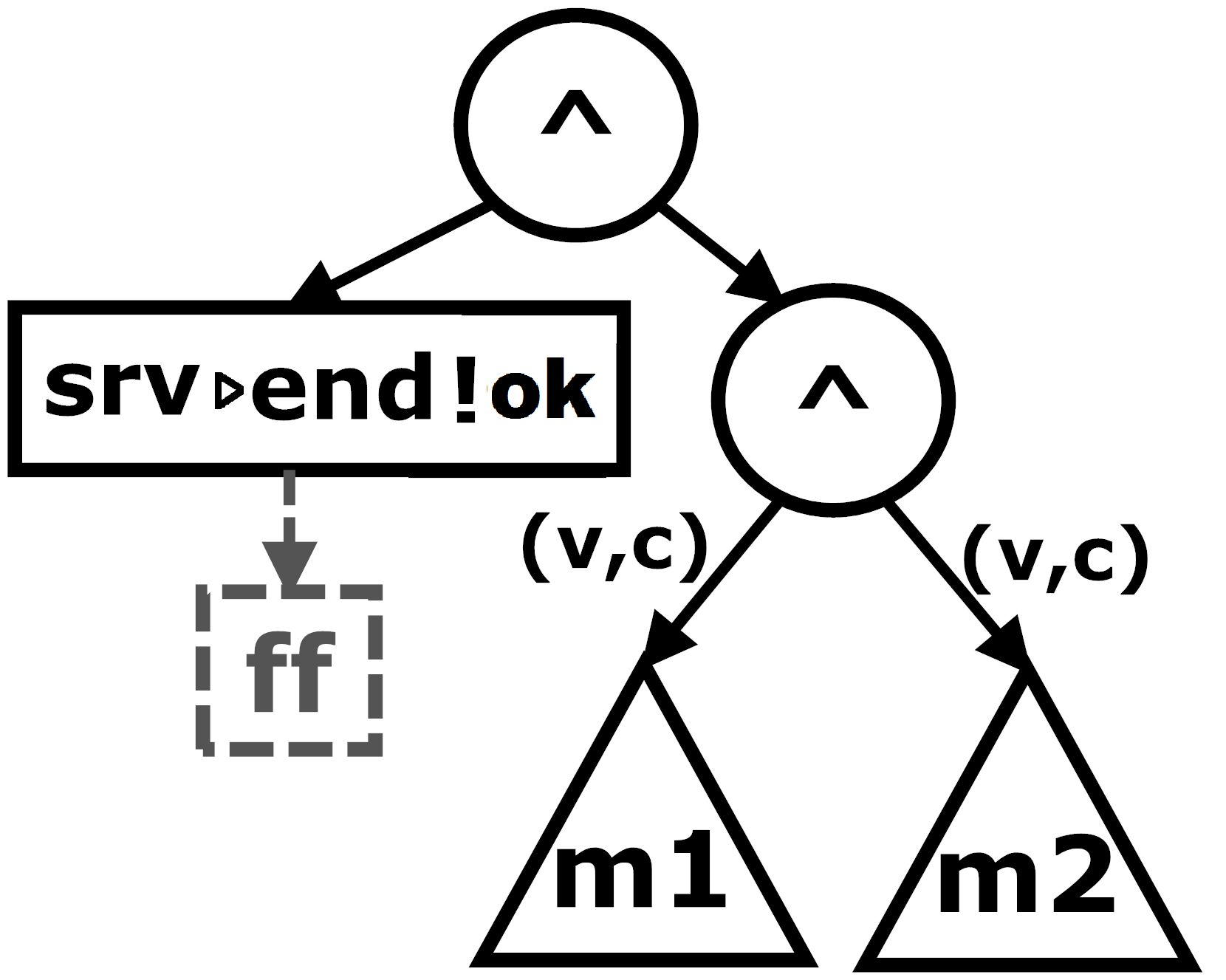}
\end{array}
&\;
\begin{array}{l}
  \tra{\displaystyle\msendD{\eatom{srv}}{c}{(v\!-\!1)}}\\
  \qquad\quad\includegraphics[scale=0.09]{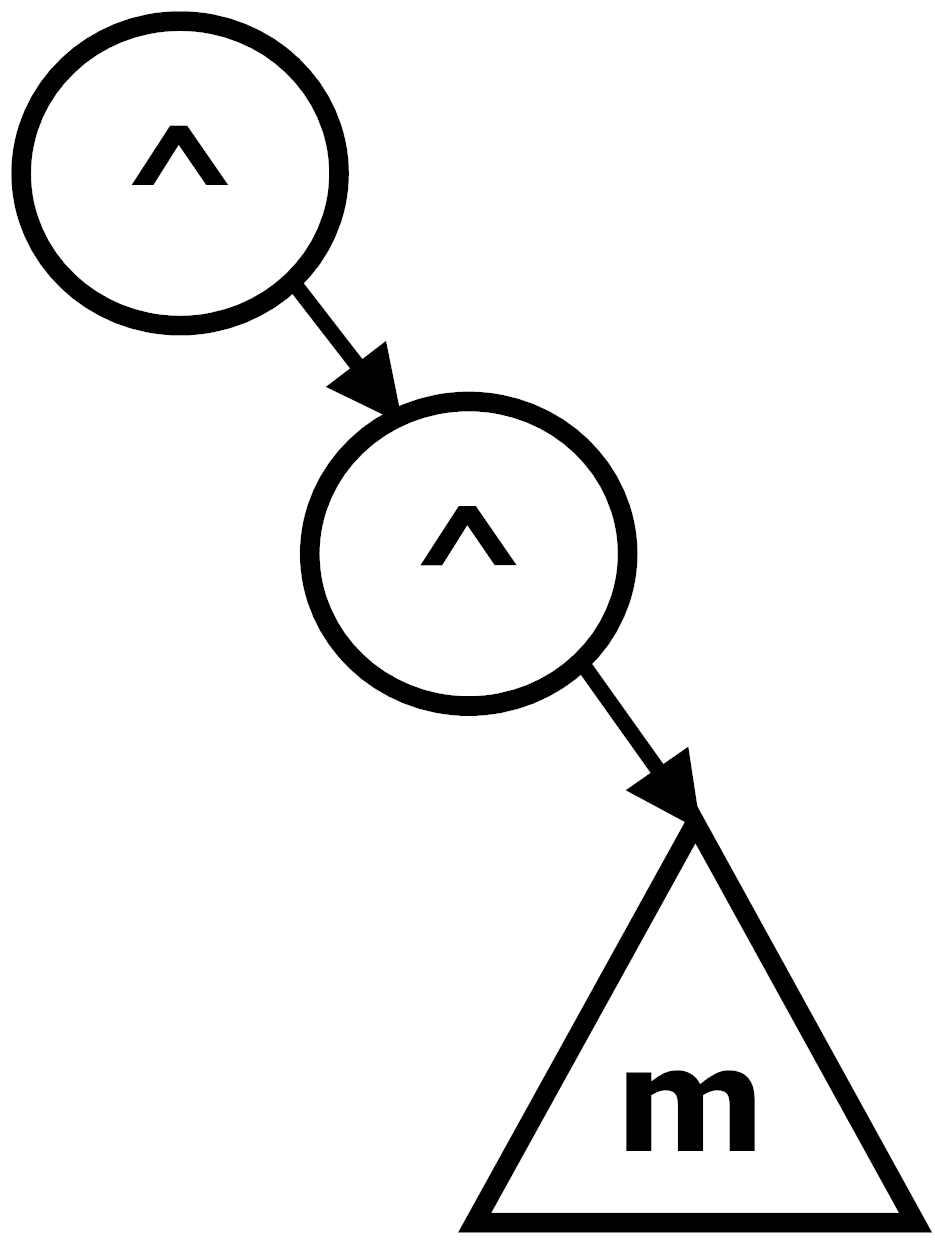}
\end{array}
&\;
\begin{array}{l}
    \tra{\displaystyle\mrecv{\eatom{srv}}{\etuple{v',c'}}}\\
  \quad\qquad\includegraphics[scale=0.09]{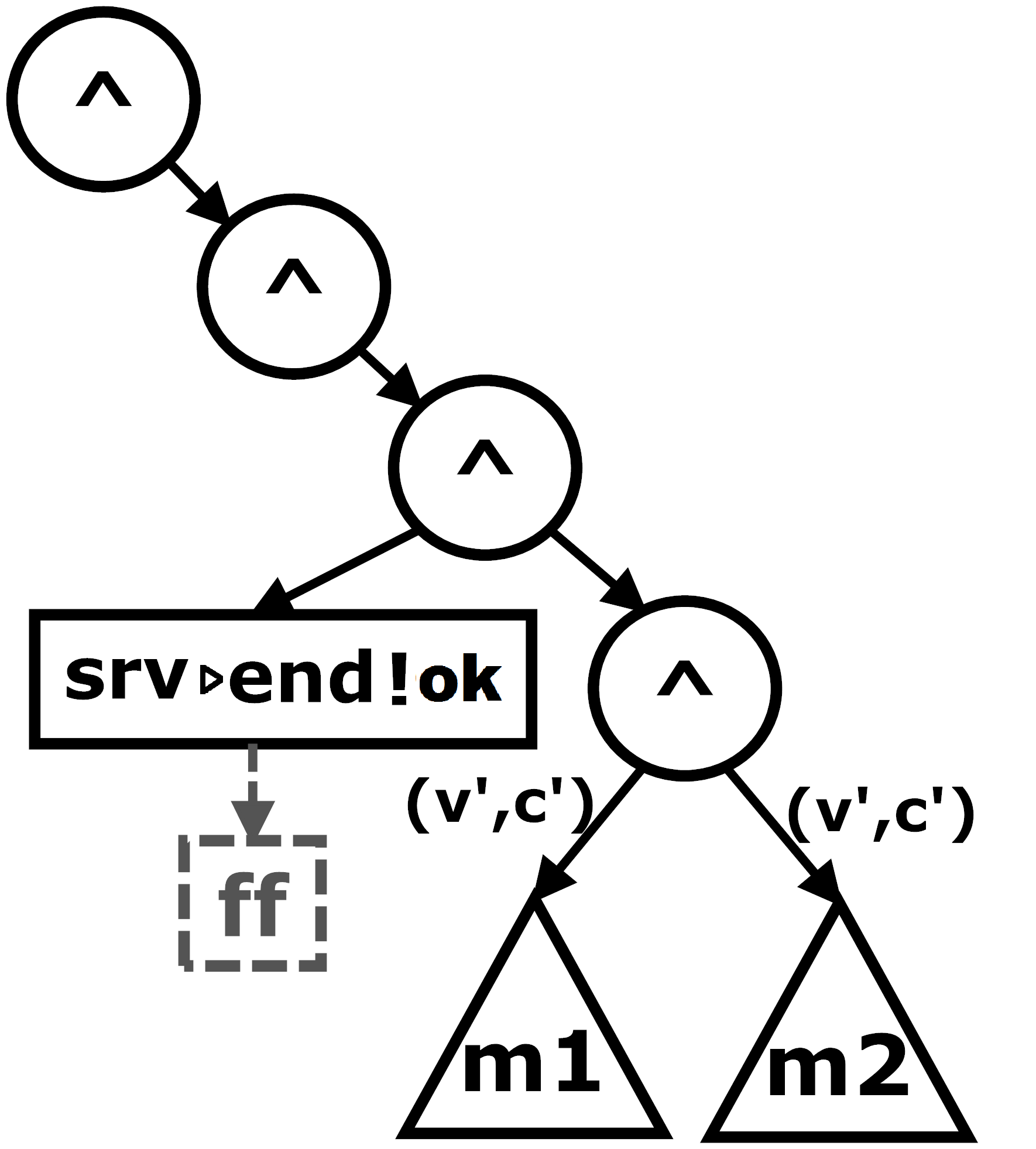}
\end{array}
\end{array}\]
  \caption[Monitor Combinator Generation for Formula~(\ref{eq:7:language})]{Monitor Combinator Generation for Formula~(\ref{eq:7:language}) and its execution \wrt trace $\mrecv{\eatom{srv}}{\etuple{v,c}};\msendD{\eatom{srv}}{c}{(v\!-\!1)};\mrecv{\eatom{srv}}{\etuple{v',c'}}$}
  \label{fig:monit-gener}
\end{figure}

\begin{example}
  \label{ex:server}  
Consider a simple Erlang system implementing a \emph{predecessor server} receiving messages of the form \emph{($n$, clientID)} and returning $n-1$ back to \emph{clientID}  whenever $n>0$, but reporting the offending client to an error handler actor, \eatom{err}, whenever $n=0$.  It may also announce termination of service by sending a message to \eatom{end}. A safety correctness property in our logic would be:
{\small\begin{equation} \label{eq:7:language}
    \mmax{\hVarX}{\mnec{\mrecv{\eatom{srv}}{\etuple{x,y}}}{          
        \left(
          \begin{array}{l}
             {(\mnec{\msendD{\eatom{srv}}{\eatom{end}}{\_\,}}{\mfls})\;}\mand{\;
              (\mnec{\msendD{\eatom{srv}}{\eatom{err}}{z}}{(\mboolE{\;(x\neq 0 \vee y \neq z )\;}{\;\mfls\;}{\;\hVarX}) }) 
             }\\
            \quad\mand{\;\;
              (\mnec{\msendD{\eatom{srv}}{y}{z}}{(\mboolE{\;z=(x-1)\;}{\;\hVarX\;}{\;\mfls} )} )
            }
           \end{array}
        \right)
      }}
\end{equation}}
It is a recursive formula, \mmax{\hVarX}{(\ldots)}, stating that, \emph{whenever} the server implementation receives a request (input action), $\mnec{\mrecv{\eatom{srv}}{\etuple{x,y}}}{\ldots}$, with value $x$ and return (client) address $y$, then it should \emph{not}:
\begin{enumerate} \itemsep0em \setstretch{1.2}
	\item terminate the service (before handing the client request), $\mnec{\msendD{\eatom{srv}}{\eatom{end}}{\_\,}}{\mfls}$.
	\item report an error, $\mnec{\msendD{\eatom{srv}}{\eatom{err}}{z}}{\ldots}$, when $x$ is not $0$, or with a client other than the offending one, $y \neq z$.
	\item service the client request, $\mnec{\msendD{\eatom{srv}}{y}{z}}{\ldots}$, with a value other than $x-1$.
\end{enumerate}
These conditions are \emph{invariant}; maximal fixpoints capture this invariance for server implementations that \emph{may never terminate} as they are considered correct as long as the conditions above are not violated. 

From formula~(\ref{eq:7:language}), the monitor organisation $m$ (depicted in \figref{fig:monit-gener}) is generated, consisting of one process acting as the combinator for the necessity formula $\mnec{\mrecv{\eatom{srv}}{\etuple{x,y}}}{\hV}$.  If an event of the form $\mrecv{\eatom{srv}}{\etuple{v,c}}$ is received, the process pattern matches it with $\mrecv{\eatom{srv}}{\etuple{x,y}}$ (mapping variables $x$ and $y$ to the values $v$ and $c$ \resp) and  spawns the (dashed) monitor system shown underneath it in \figref{fig:monit-gener}, instantiating the variables $x,y$ with $v,c$ \resp  The subsystem consisting of three monitor subsystems, one for each subformula guarded by $\mnec{\mrecv{\eatom{srv}}{\etuple{x,y}}}{}$ in (\ref{eq:7:language}), connected by two conjunction forking monitors.  When the next trace event is received, \eg $\msendD{\eatom{srv}}{c}{v\!-\!1}$ (a server reply to client $c$ with value $v\!-\!1$),  the conjunction monitors replicate and forward this event to the three monitor subtrees. Two of these subtrees do not pattern match this event and terminate, while the third subtree (\ie submonitor $m2$) pattern-matches it and instantiates $z$ for $(v\!-\!1)$, thereby also evaluating the conditional and unfolding the recursive variable $\hVarX$ to monitor $m$.  If another server request event is received, $\mrecv{\eatom{srv}}{\etuple{v',c'}}$ (with potentially different client and value arguments $c'$ and $v'$), the conjunction monitors forward it to $m$, pattern matching it and generating a subsystem with two further conjunction combinators as before. \bqed     
\end{example}

Formula~(\ref{eq:7:language}) is a pathological example, highlighting two inefficiencies of the unoptimised synthesised monitors. First, conjunction  monitors mirror closely their syntactic counterpart and can only handle forwarding to \emph{two} sub-monitors.  As a result, formulas with \emph{nested} conjunctions (as in formula~(\ref{eq:7:language})) translate into organisations of \emph{cascading conjunction monitors} that are inefficient at forwarding trace events.  

Second, the current monitor implementation does not perform any monitor reorganisations at runtime.  When a conjunction formula ${\hV_1}\mand{\hV_2}$ is translated, the  conjunction combinator monitor organisation is kept permanent throughout its execution because it is assumed that the respective sub-monitors for $\hV_1$ and $\hV_2$ are long-lived.  This heuristic however does not apply in the case of  formula~(\ref{eq:7:language}), where two out of the three sub-monitors terminate after receiving a single event.  This feature, in conjunction with recursive unfolding, creates \emph{chains of indirections} through conjunction monitors with only one child, as shown in \figref{fig:monit-gener} (bottom row).

\section{Monitor Optimisations}
\label{sec:proposed-solutions}

\begin{figure}[ht!]
  \centering
\[\begin{array}{ccc}
    \begin{array}{l}
\tra{\displaystyle\mrecv{\eatom{srv}}{\etuple{v,c}}}\\
  \;\includegraphics[scale=0.09]{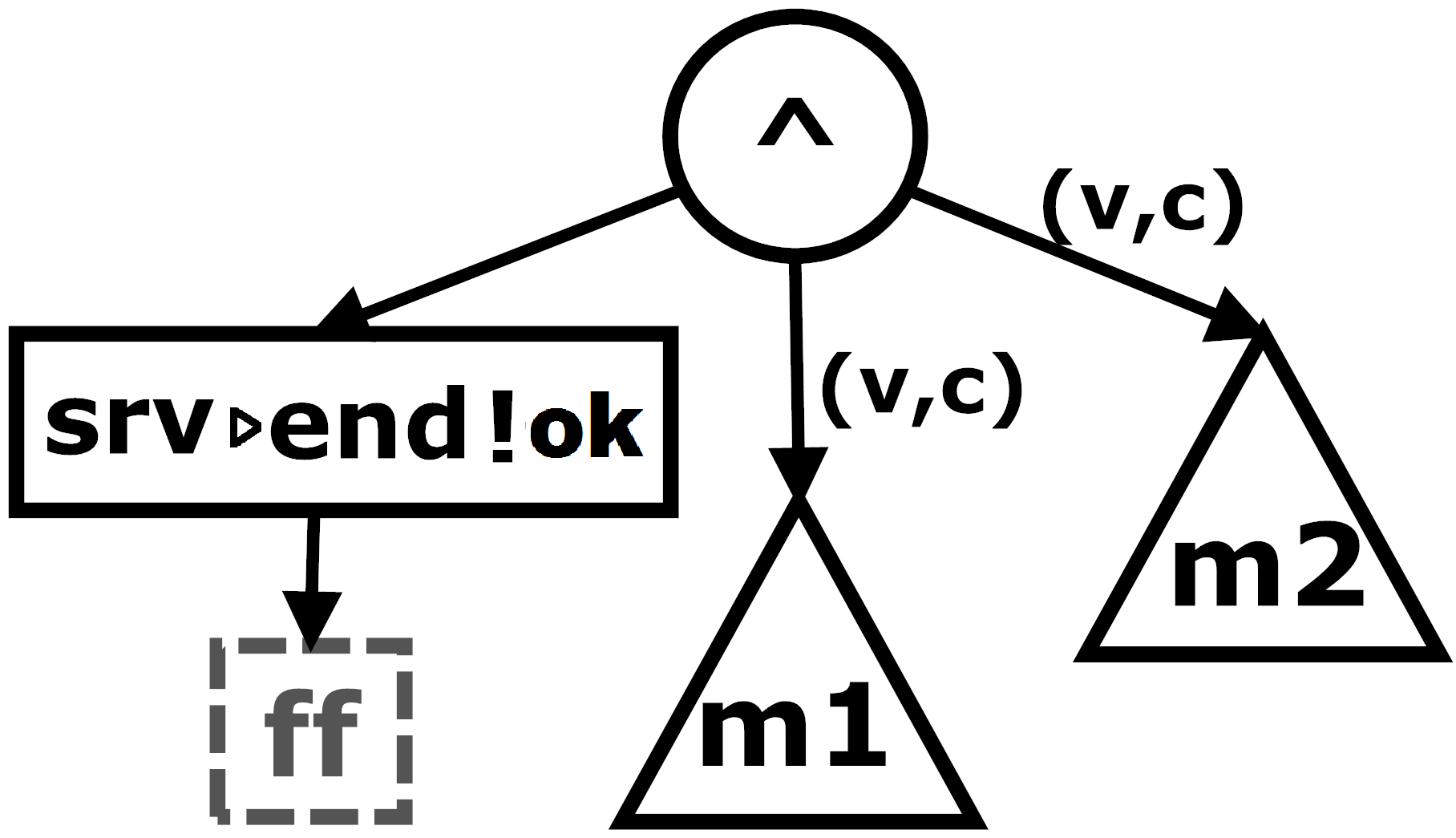}
\end{array}
&\;
\begin{array}{l}
  \tra{\displaystyle\msendD{\eatom{srv}}{c}{(v\!-\!1)}}\\
  \qquad\qquad\includegraphics[scale=0.09]{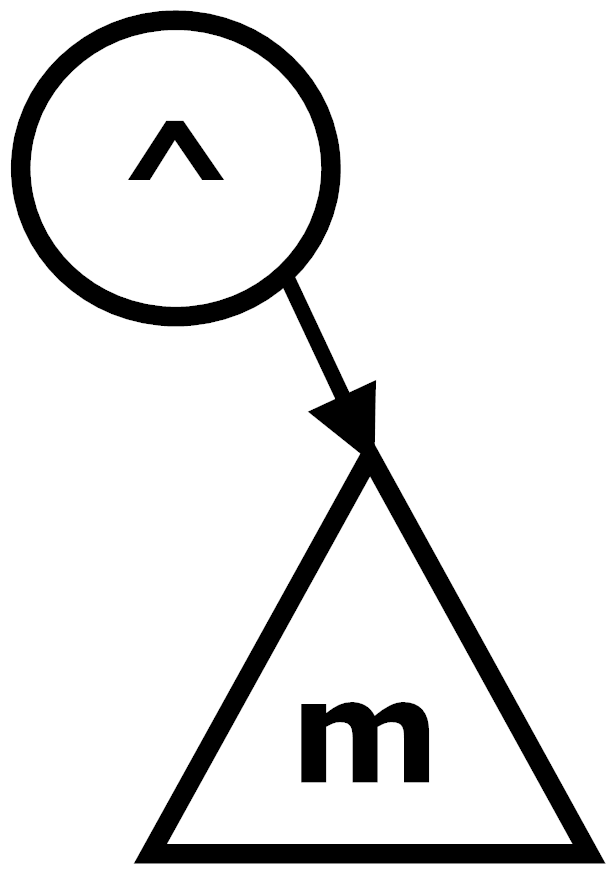}
\end{array}
&\quad
\begin{array}{l}
    \tra{\displaystyle\mrecv{\eatom{srv}}{\etuple{v',c'}}}\\
  \;\includegraphics[scale=0.09]{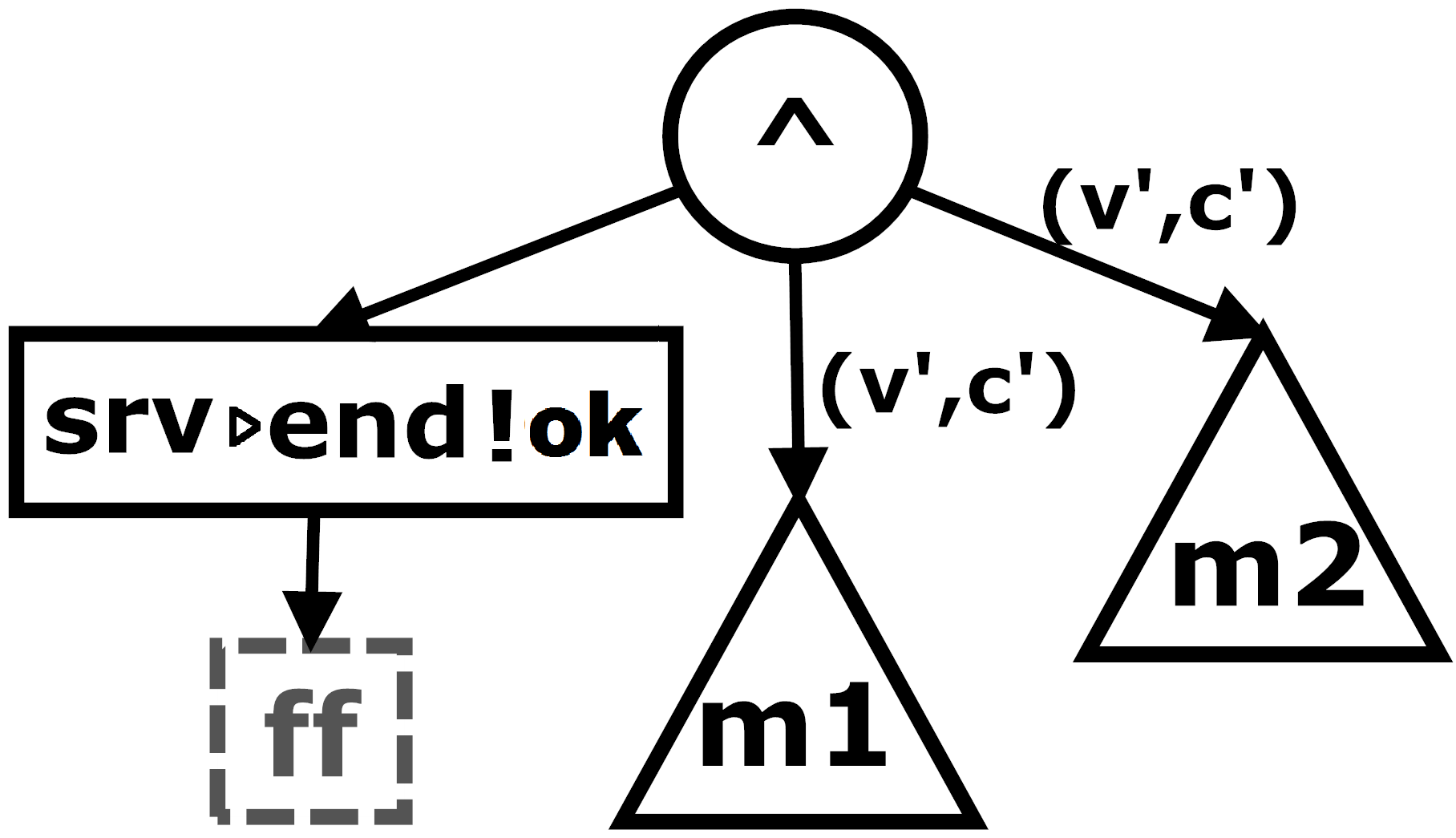}
\end{array}
\end{array}\]
  
  \caption[Optimised Synthesis for Formula~(\ref{eq:7:language}).]{Optimised Synthesis for Formula~(\ref{eq:7:language}) and its execution \wrt trace $\mrecv{\eatom{srv}}{\etuple{v,c}};\msendD{\eatom{srv}}{c}{(v\!-\!1)};\mrecv{\eatom{srv}}{\etuple{v',c'}}$}
  \label{fig:monit-gener-opt}
\end{figure}

The first optimisation introduced in \cite{CasFraSaid15} is that of conjunction monitor combinators that fork-out to an \emph{arbitrary number} of monitor subsystems. For instance, the corresponding monitor formula~(\ref{eq:7:language}) would translate into a monitor organisation consisting of \emph{one} conjunction combinator with \emph{three} children (instead of two combinators with two children each)  as shown in \figref{fig:monit-gener-opt} (left).  This is more efficient from from the point of view of processes created, but also in terms of the number of replicated messages required to perform the necessary event forwarding to monitor subsystems $-$\eg the conjunction combinator of \figref{fig:monit-gener-opt} generates \emph{three} message replications to forward an event to the three sub-monitors, as opposed to the unoptimised approach which requires sending \emph{four} messages (see \figref{fig:monit-gener}). This \emph{static optimisation} is conducted through syntactic manipulation using a pre-processing phase.

The second optimisation \cite{CasFraSaid15} builds on the first one by allowing conjunction monitor combinators to \emph{dynamically reorganise} the monitor configuration  so as to keep the event flow structure as efficient as possible.  In order to keep overheads low, this reconfiguration operation is kept \emph{local}, where unaffected monitor subsystems should continue with their runtime analysis while the restructuring is in process.  Stated otherwise, the monitor reorganisation happens while  trace events are \emph{still being received}, and the operation needs to guarantee that $(i)$ no trace events are lost $(ii)$ trace events are not reordered. Reorganisations are carried out by conjunction combinators, which are provided with the capability to \emph{add} and \emph{delete} monitor subsystems from their internal list of children.  For instance, when an event causes a child sub-monitor  to terminate, the parent (conjunction) monitor is sent a termination message which allows it to remove the terminated sub-monitor  from its child-list. 
   
The dynamic restructuring protocol employed by this optimisation yields monitor organisations with only \emph{one} (eventual) conjunction node at the root, and a list of monitor subsystems processing the forwarded events (a spider-like configuration).  For instance, for the event trace  $\mrecv{\eatom{srv}}{\etuple{v,c}};\msendD{\eatom{srv}}{c}{(v\!-\!1)};\mrecv{\eatom{srv}}{\etuple{v',c'}}$, the synthesised monitor for formula~\eqref{eq:7:language}  yields the evolution shown in \figref{fig:monit-gener-opt}. 

\chapter{Additional Properties and Tabulated Results}
\label{chp:app-syn-asyn}
In this Chapter we present additional information related to the impact assessment experiments conducted in \Cref{chp:syn-asyn}. In \secref{sec:app-aux-properties} we provide an overview of two additional properties written for the Yaws webserver, that we used along with properties \eqref{prop:yaws:1} and \eqref{prop:yaws:2} for obtaining the performance results presented in the graphs in Figures~\ref{fig:unopt_results}, \ref{fig:sopt_results} and \ref{fig:s+dopt_results} given in \secref{sec:eval-results}. To provide an even clearer insight of the results presented in these graphs, in \secref{sec:app-tab-res} we represent the same results as tables.

\section{Additional Runtime Verification Properties} \label{sec:app-aux-properties}
Here we present two other properties developed for the Yaws webserver and used in our experiments while conducting our impact assessment analysis. Here we only present the versions used for asynchronous and synchronous monitoring as the versions for Hybrid monitoring can easily be inferred by converting all the falsities (\mfls) in the scripts to \msfls.

\subsection{Five or more Headers Required Property}
When sending a request to a webserver, HTTP clients also send a number of HTTP headers containing information about themselves and about their request. Typically, an HTTP request should provide about 5 or 6 information headers (\texttt{http\_header}), so that the webserver may correctly handle the request and send the appropriate response. A property may be used to check and ensure that any request sent to the server contains at least 5 headers. This behaviour may be verified using either property \eqref{eq:altProp2} or \eqref{eq:altProp3}. 

{\small\setstretch{1.1}
\begin{equation}
\begin{array}{l}
	 \mmax{\hVarX}{\Big( \\\quad\mnec{\msend{\eatom{AcceptorPid}}{\etuple{hPid,\eatom{next},\_}}}\\
      \quad\mnec{\mret{hPid}{\etuple{\eatom{yaws}, \eatom{do\_recv},\eatom{3},\etuple{\eatom{ok},\etuple{\eatom{http\_req},\eatom{GET},\_,\_}}}}}\\
     \quad(\\\qquad\mnec{\mret{hPid}{\etuple{\eatom{yaws}, \eatom{do\_recv},\eatom{3},\etuple{\eatom{ok},\eatom{http\_eoh}}}}}\mfls
     \\\quad)
     \mand(\\
     \qquad\mnec{\mret{hPid}{\etuple{\eatom{yaws}, \eatom{do\_recv},\eatom{3},\etuple{\eatom{ok},\etuple{\eatom{http\_header},\_}}}}}\\
     \qquad\mnec{\mret{hPid}{\etuple{\eatom{yaws}, \eatom{do\_recv},\eatom{3},\etuple{\eatom{ok},\eatom{http\_eoh}}}}}\mfls
     \\\quad)
     \mand(\\
     \qquad\mnec{\mret{hPid}{\etuple{\eatom{yaws}, \eatom{do\_recv},\eatom{3},\etuple{\eatom{ok},\etuple{\eatom{http\_header},\_}}}}}\\
     \qquad\mnec{\mret{hPid}{\etuple{\eatom{yaws}, \eatom{do\_recv},\eatom{3},\etuple{\eatom{ok},\etuple{\eatom{http\_header},\_}}}}}\\
     \qquad\mnec{\mret{hPid}{\etuple{\eatom{yaws}, \eatom{do\_recv},\eatom{3},\etuple{\eatom{ok},\eatom{http\_eoh}}}}}\mfls
     \\\quad)
     \mand(\\
     \qquad\mnec{\mret{hPid}{\etuple{\eatom{yaws}, \eatom{do\_recv},\eatom{3},\etuple{\eatom{ok},\etuple{\eatom{http\_header},\_}}}}}\\
     \qquad\mnec{\mret{hPid}{\etuple{\eatom{yaws}, \eatom{do\_recv},\eatom{3},\etuple{\eatom{ok},\etuple{\eatom{http\_header},\_}}}}}\\
     \qquad\mnec{\mret{hPid}{\etuple{\eatom{yaws}, \eatom{do\_recv},\eatom{3},\etuple{\eatom{ok},\etuple{\eatom{http\_header},\_}}}}}\\
     \qquad\mnec{\mret{hPid}{\etuple{\eatom{yaws}, \eatom{do\_recv},\eatom{3},\etuple{\eatom{ok},\eatom{http\_eoh}}}}}\mfls
     \\\quad)
     \mand(\\
     \qquad\mnec{\mret{hPid}{\etuple{\eatom{yaws}, \eatom{do\_recv},\eatom{3},\etuple{\eatom{ok},\etuple{\eatom{http\_header},\_}}}}}\\
     \qquad\mnec{\mret{hPid}{\etuple{\eatom{yaws}, \eatom{do\_recv},\eatom{3},\etuple{\eatom{ok},\etuple{\eatom{http\_header},\_}}}}}\\
     \qquad\mnec{\mret{hPid}{\etuple{\eatom{yaws}, \eatom{do\_recv},\eatom{3},\etuple{\eatom{ok},\etuple{\eatom{http\_header},\_}}}}}\\
     \qquad\mnec{\mret{hPid}{\etuple{\eatom{yaws}, \eatom{do\_recv},\eatom{3},\etuple{\eatom{ok},\etuple{\eatom{http\_header},\_}}}}}\\
     \qquad\mnec{\mret{hPid}{\etuple{\eatom{yaws}, \eatom{do\_recv},\eatom{3},\etuple{\eatom{ok},\eatom{http\_eoh}}}}}\mfls
     \\\quad)
     \mand(\\
     \qquad\mnec{\mret{hPid}{\etuple{\eatom{yaws}, \eatom{do\_recv},\eatom{3},\etuple{\eatom{ok},\etuple{\eatom{http\_header},\_}}}}}\\
     \qquad\mnec{\mret{hPid}{\etuple{\eatom{yaws}, \eatom{do\_recv},\eatom{3},\etuple{\eatom{ok},\etuple{\eatom{http\_header},\_}}}}}\\
     \qquad\mnec{\mret{hPid}{\etuple{\eatom{yaws}, \eatom{do\_recv},\eatom{3},\etuple{\eatom{ok},\etuple{\eatom{http\_header},\_}}}}}\\
     \qquad\mnec{\mret{hPid}{\etuple{\eatom{yaws}, \eatom{do\_recv},\eatom{3},\etuple{\eatom{ok},\etuple{\eatom{http\_header},\_}}}}}\\
     \qquad\mnec{\mret{hPid}{\etuple{\eatom{yaws}, \eatom{do\_recv},\eatom{3},\etuple{\eatom{ok},\etuple{\eatom{http\_header},\_}}}}}\\
     \qquad\mmax{Z}{\begin{mlbrace}(\mnec{\mret{hPid}{\etuple{\eatom{yaws}, \eatom{do\_recv},\eatom{3},\etuple{\eatom{ok},\eatom{http\_eoh}}}}} \hVarX)
     \; \\ \mand (\mnec{\mret{hPid}{\etuple{\eatom{yaws}, \eatom{do\_recv},\eatom{3},\etuple{\eatom{ok},\etuple{\eatom{http\_header},\_}}}}}\; Z)\end{mlbrace}}\\\quad) \\ \Big)}
\end{array}\label{eq:altProp2}
\end{equation}
}
\vspace{-3mm}

Although these two properties may be considered equivalent, they do not translate into the same concurrent monitor. In fact the monitor generated from property \eqref{eq:altProp2} is specified to immediately spawn 6 concurrent submonitors for each connected client, upon receiving an HTTP request notification (\texttt{http\_req}) event. Five of these submonitors flag a violation when 4 or less headers are received, otherwise they terminate. The sixth monitor reapplies the property for the connected client, when 5 or more headers are received.

Conversely, in \eqref{eq:altProp3}, submonitors are generated in a lazy manner, meaning that for each event message, only two submonitors are spawned. One of these submonitors detects a violation if the(\texttt{http\_eoh}) message is received prematurely, otherwise it terminates. The other submonitor spawns two further submonitors if an HTTP information header is received. The property is reapplied if the end-of-header message is received after 5 or more information headers. Although \eqref{eq:altProp2} is more readable then \eqref{eq:altProp3}, the latter is less resource intensive, as less submonitors are spawned for each connected client.

{\small\setstretch{1.1}
\begin{equation}
\begin{array}{l}
	 \mmax{\hVarX}{\Big( \\\quad\mnec{\msend{\eatom{AcceptorPid}}{\etuple{hPid,\eatom{next},\_}}}\\
      \quad\mnec{\mret{hPid}{\etuple{\eatom{yaws}, \eatom{do\_recv},\eatom{3},\etuple{\eatom{ok},\etuple{\eatom{http\_req},\eatom{GET},\_,\_}}}}}\\
      \quad(\mnec{\mret{hPid}{\etuple{\eatom{yaws}, \eatom{do\_recv},\eatom{3},\etuple{\eatom{ok},\eatom{http\_eoh}}}}}\mfls) \,\mand\, \\
      \quad\begin{mlbrace}
	    \mnec{\mret{hPid}{\etuple{\eatom{yaws}, \eatom{do\_recv},\eatom{3},\etuple{\eatom{ok},\etuple{\eatom{http\_header},\_}}}}}\\
	    (\mnec{\mret{hPid}{\etuple{\eatom{yaws}, \eatom{do\_recv},\eatom{3},\etuple{\eatom{ok},\eatom{http\_eoh}}}}}\mfls) \,\mand\, \\
	    \begin{mlbrace}
		   \mnec{\mret{hPid}{\etuple{\eatom{yaws}, \eatom{do\_recv},\eatom{3},\etuple{\eatom{ok},\etuple{\eatom{http\_header},\_}}}}} \\
		   (\mnec{\mret{hPid}{\etuple{\eatom{yaws}, \eatom{do\_recv},\eatom{3},\etuple{\eatom{ok},\eatom{http\_eoh}}}}}\mfls) \,\mand\, \\
	       \begin{mlbrace}
			   \mnec{\mret{hPid}{\etuple{\eatom{yaws}, \eatom{do\_recv},\eatom{3},\etuple{\eatom{ok},\etuple{\eatom{http\_header},\_}}}}} \\
			   (\mnec{\mret{hPid}{\etuple{\eatom{yaws}, \eatom{do\_recv},\eatom{3},\etuple{\eatom{ok},\eatom{http\_eoh}}}}}\mfls) \,\mand\, \\
			   \begin{mlbrace}
				  \mnec{\mret{hPid}{\etuple{\eatom{yaws}, \eatom{do\_recv},\eatom{3},\etuple{\eatom{ok},\etuple{\eatom{http\_header},\_}}}}} \\
				  (\mnec{\mret{hPid}{\etuple{\eatom{yaws}, \eatom{do\_recv},\eatom{3},\etuple{\eatom{ok},\eatom{http\_eoh}}}}}\mfls) \,\mand\, \\
				  \begin{mlbrace}
				     \mnec{\mret{hPid}{\etuple{\eatom{yaws}, \eatom{do\_recv},\eatom{3},\etuple{\eatom{ok},\etuple{\eatom{http\_header},\_}}}}} \\    
				     \mmax{Z\!}{\begin{mlbrace}(\mnec{\mret{hPid}{\etuple{\eatom{yaws}, \eatom{do\_recv},\eatom{3},\etuple{\eatom{ok},\eatom{http\_eoh}}}}}\, \hVarX)\quad \\ \mand (\mnec{\mret{hPid}{\etuple{\eatom{yaws}, \eatom{do\_recv},\eatom{3},\etuple{\eatom{ok},\etuple{\eatom{http\_header},\_}}}}}\,Z)\!\!\end{mlbrace}} \\
				 \!\!\! \end{mlbrace}
			   \!\! \end{mlbrace}
		  	\!\! \end{mlbrace} 
	     \!\! \end{mlbrace}
	    \! \! \end{mlbrace}	  
	\\ \Big)}
\end{array}\label{eq:altProp3}
\end{equation}
}
\vspace{-5mm}

\subsection{Valid HTTP Request Format Property}
In the HTTP protocol, every HTTP request must begin with a request notification header (ie,\texttt{http\_req}). This request header must then be followed by a number of HTTP data headers (ie, \texttt{http\_header}) which denote information about the requesting client and other type of information. Finally each request must end with an end-of-header notification (ie, \texttt{http\_eoh}), denoting that the client has finished sending its request headers. The following property was therefore developed to make sure that every HTTP request received by the server follows this protocol.

{\small\setstretch{1.1}
\begin{equation}
\begin{array}{l}
	 \mmax{\hVarX \;}{\Big( \\\quad\mnec{\msend{\eatom{AcceptorPid}}{\etuple{hPid,\eatom{next},\_}}}\\
      \quad(\quad\mnec{\mret{hPid}{\etuple{\eatom{yaws}, \eatom{do\_recv},\eatom{3},\etuple{\eatom{ok},\etuple{\eatom{http\_req},\eatom{GET},\_,\_}}}}}
      \\\qquad\mnec{\mret{hPid}{\etuple{\eatom{yaws}, \eatom{do\_recv},\eatom{3},\etuple{\eatom{ok},\etuple{\eatom{http\_header},\_}}}}}\; \hVarX
     \\\quad
     )\mand(\\     
     \qquad\mnec{\mret{hPid}{\etuple{\eatom{yaws}, \eatom{do\_recv},\eatom{3},\etuple{\eatom{ok},\etuple{\eatom{http\_header},\_}}}}}
      \\\qquad\mnec{\mret{hPid}{\etuple{\eatom{yaws}, \eatom{do\_recv},\eatom{3},\etuple{\eatom{ok},\etuple{\eatom{http\_header},\_}}}}}\; \hVarX
      \\\quad
     )\mand(
     \\\qquad\mnec{\mret{hPid}{\etuple{\eatom{yaws}, \eatom{do\_recv},\eatom{3},\etuple{\eatom{ok},\etuple{\eatom{http\_header},\_}}}}}
      \\\qquad\mnec{\mret{hPid}{\etuple{\eatom{yaws}, \eatom{do\_recv},\eatom{3},\etuple{\eatom{ok},\eatom{http\_eoh}}}}}
      \\\qquad\mnec{\mret{hPid}{\etuple{\eatom{yaws}, \eatom{do\_recv},\eatom{3},\etuple{\eatom{ok},\etuple{\eatom{http\_req},\eatom{GET},\_,\_}}}}}\; \hVarX
      \\\quad
     )\mand(     
     \\\qquad\mnec{\mret{hPid}{\etuple{\eatom{yaws}, \eatom{do\_recv},\eatom{3},\etuple{\eatom{ok},\etuple{\eatom{http\_header},\_}}}}}
      \\\quad\mnec{\mret{hPid}{\etuple{\eatom{yaws}, \eatom{do\_recv},\eatom{3},\etuple{\eatom{ok},\etuple{\eatom{http\_req},\eatom{GET},\_,\_}}}}}\; \mfls
      \\\quad
     )\mand(     
    \\\qquad\mnec{\mret{hPid}{\etuple{\eatom{yaws}, \eatom{do\_recv},\eatom{3},\etuple{\eatom{ok},\eatom{http\_eoh}}}}}\; X
     \\\qquad\mnec{\mret{hPid}{\etuple{\eatom{yaws}, \eatom{do\_recv},\eatom{3},\etuple{\eatom{ok},\etuple{\eatom{http\_header},\_}}}}}\; \mfls)     
\\ \; \Big)}\\
\end{array}\label{eq:altProp4}
\end{equation}
}
\vspace{2mm}

\noindent This property checks that the HTTP headers in every client request follow the correct ordering. The property recurs whenever the following non-violating behaviour is observed:
\begin{enumerate}[label=(\roman*)]\itemsep0em
	\item A data header message (ie, \texttt{http\_header}) is received \emph{after} a request header message (ie, \texttt{http\_req}).
	\item Two data header messages are received \emph{subsequently}.
	\item A data header message is followed by an end-of-header message (ie, \texttt{http\_eoh}), which is then followed by another request header message.
\end{enumerate}
\noindent The property fails when a data header message is received \emph{before} a request header message, and also when a data header message is received \emph{after} an end-of-header notification.

 \section{Detailed Result Tables} \label{sec:app-tab-res}
 The following tables contain the same results plotted in the graphs given in \secref{sec:eval-results} as Figures \ref{fig:unopt_results}, \ref{fig:sopt_results} and \ref{fig:s+dopt_results}. These results are structured in three subsections \ie a subsection for each monitor synthesis algorithm \cite{FraSey14,CasFraSaid15} (\ie unoptimised, statically optimised and dynamically optimised) supported by \detecterGen\ (see \Cref{chp:app-mon-opt} for details).
 In each subsection we present: $(i)$ the average CPU utilization ($\times 10^6$ CPU cycles) per client request; $(ii)$ the average memory consumption (MB) and $(iii)$ the average response time per client request (ms); obtained when monitoring the Yaws webserver for a number properties using different monitoring approaches. 

\subsection{Results for Unoptimised Monitors} \medskip

 	\noindent\begin{tabular*}{\textwidth}{@{\extracolsep\fill}|c||c|c|c|c|c|c|}
 		\hline
 		\multicolumn{7}{|c|}{\textbf{Average CPU utilization per Request ($\mathbf{\times 10^6}$ CPU cycles)}}\\
 		\hline\hline
 		 &\multicolumn{6}{c|}{Number of Requests} \\
 		 \hline		
 			\, Monitoring modes & 50 & 100 & 200 & 500 & 1000 & 2000\\
 		\hline\hline		
 			Baseline & 14.556 & 14.408 & 14.127 & 14.171 & 14.365 & 14.269 \\
 		\hline		
 			Asynchronous& 27.383 & 32.70 & 46.507 & 82.518 & 144.236 & 269.164\\
 		\hline		
 			Hybrid & 28.793 & 34.354 & 48.329 & 84.241 & 148.973 & 277.172\\
 		\hline
 			Synchronous & 31.496 & 39.037 & 53.796 & 97.003 & 175.727 & 308.936\\
 		\hline
 		\multicolumn{7}{c}{}
 		\vspace{-5mm}
 	\end{tabular*}\medskip

 	\noindent\begin{tabular*}{\textwidth}{@{\extracolsep\fill}|c||c|c|c|c|c|c|}
 		\hline
 		\multicolumn{7}{|c|}{\textbf{Average Memory Consumption (MB)}}\\
 		\hline\hline
 		 &\multicolumn{6}{c|}{Number of Requests} \\
 		 \hline		
 			\, Monitoring modes & 50 & 100 & 200 & 500 & 1000 & 2000\\
 		\hline\hline		
 			Baseline & 25.62 & 25.68 & 25.63 & 25.68 & 26.17 & 26.25 \\
 		\hline		
 			Asynchronous  & 30.98 & 32.24 & 32.95 & 35.83 & 39.93 & 45.36 \\
 		\hline
 			Hybrid & 31.41 & 32.39 & 33.38 & 36.26 & 41.02 & 49.02 \\
 		\hline
 			Synchronous & 31.36 & 32.25 & 33.52 & 36.15 & 42.25 & 51.55 \\
 		\hline
 		\multicolumn{7}{c}{}
 		\vspace{-5mm}
 	\end{tabular*}\medskip

 	\noindent\begin{tabular*}{\textwidth}{@{\extracolsep\fill}|c||c|c|c|c|c|c|}
 		\hline
 		\multicolumn{7}{|c|}{\textbf{Average response time per Request (ms)}}\\
 		\hline\hline
 		 &\multicolumn{6}{c|}{Number of Requests} \\
 		 \hline		
 			\, Monitoring modes & 50 & 100 & 200 & 500 & 1000 & 2000\\
 		\hline\hline		
 			Baseline & 11.32 & 11.35 & 11.47 & 11.44 & 11.45 & 11.66\\
 		\hline		 		
 			Asynchronous & 14.54 & 15.82 & 17.75 & 24.74 & 40.45 & 66.02\\
 		\hline
 			Hybrid & 17.3 & 17.331 & 19.75 & 27.3 & 44.82 & 75.62 \\
 		\hline
 			Synchronous & 17.63 & 19.08 & 23 & 36.27 & 60.82 & 112.75\\
 		\hline
 		\multicolumn{7}{c}{}
 		\vspace{-5mm}
 	\end{tabular*}\medskip

\subsection{Results for Statically Optimised Monitors}\medskip

 	\noindent\begin{tabular*}{\textwidth}{@{\extracolsep\fill}|c||c|c|c|c|c|c|}
 		\hline
 		\multicolumn{7}{|c|}{\textbf{Average CPU utilization per Request  ($\mathbf{\times 10^6}$ CPU cycles)}}\\
 		\hline\hline
 		 &\multicolumn{6}{c|}{Number of Requests} \\
 		 \hline		
 			\, Monitoring modes & 50 & 100 & 200 & 500 & 1000 & 2000\\
 		\hline\hline		
 			Baseline & 14.556 & 14.408 & 14.127 & 14.171 & 14.365 & 14.269 \\
 		\hline		
 			Asynchronous & 22.023 & 23.94 & 29.375 & 45.044 & 77.479 & 143.147\\
 		\hline		
 			Hybrid & 22.75 & 24.468 & 30.35 & 45.284 & 78.177 & 144.179 \\
 		\hline
 			Synchronous & 24.581 & 26.11 & 31.721 & 46.239 & 79.098 & 146.011\\
 		\hline
 		\multicolumn{7}{c}{}
 		\vspace{-5mm}
 	\end{tabular*}\medskip

 	\noindent\begin{tabular*}{\textwidth}{@{\extracolsep\fill}|c||c|c|c|c|c|c|}
 		\hline
 		\multicolumn{7}{|c|}{\textbf{Average Memory Consumption (MB)}}\\
 		\hline\hline
 		 &\multicolumn{6}{c|}{Number of Requests} \\
 		 \hline		
 			\, Monitoring modes & 50 & 100 & 200 & 500 & 1000 & 2000\\
 		\hline\hline		
 			Baseline & 25.62 & 25.68 & 25.63 & 25.68 & 26.17 & 26.25 \\
 		\hline		
 			Asynchronous  & 29.4 & 30.5 & 31.87 & 33.79 & 35.78 & 40.09 \\
 		\hline
 			Hybrid & 29.73 & 31.06 & 32.25 & 34.06 & 36.2 & 41.15\\
 		\hline
 			Synchronous & 30.08 & 31.68 & 32.24 & 34.19 & 36.11 & 41.68\\
 		\hline
 		\multicolumn{7}{c}{}
 		\vspace{-5mm}
 	\end{tabular*}\medskip

 	\noindent\begin{tabular*}{\textwidth}{@{\extracolsep\fill}|c||c|c|c|c|c|c|}
 		\hline
 		\multicolumn{7}{|c|}{\textbf{Average response times per Request (ms)}}\\
 		\hline\hline
 		 &\multicolumn{6}{c|}{Number of Requests} \\
 		 \hline		
 			\, Monitoring modes & 50 & 100 & 200 & 500 & 1000 & 2000\\
 		\hline\hline		
 			Baseline & 11.32 & 11.35 & 11.47 & 11.44 & 11.45 & 11.66\\
 		\hline		 		
 			Asynchronous & 13.66 & 14.91 & 16.31 & 20.06 & 27.29 & 48.84 \\
 		\hline
 			Hybrid & 15.66 & 15.94 & 16.85 & 21.41 & 29.95 & 53.08 \\
 		\hline
 			Synchronous & 18.07 & 18.14 & 19.61 & 25.74 & 39.11 & 67.91\\
 		\hline
 		\multicolumn{7}{c}{}
 		\vspace{-5mm}
 	\end{tabular*}\medskip

\subsection{Results for Dynamically Optimised Monitors}\medskip

 	\noindent\begin{tabular*}{\textwidth}{@{\extracolsep\fill}|c||c|c|c|c|c|c|}
 		\hline
 		\multicolumn{7}{|c|}{\textbf{Average CPU utilization per Request ($\mathbf{\times 10^6}$ CPU cycles)}}\\
 		\hline\hline
 		 &\multicolumn{6}{c|}{Number of Requests} \\
 		 \hline		
 			\, Monitoring modes & 50 & 100 & 200 & 500 & 1000 & 2000\\
 		\hline\hline		
 			Baseline & 14.556 & 14.408 & 14.127 & 14.171 & 14.365 & 14.269 \\
 		\hline		
 			Asynchronous & 20.175 & 21.286 & 23.732 & 27.972 & 33.909 & 44.227\\
 		\hline		
 			Hybrid & 20.988 & 21.516 & 24.885 & 29.755 & 37.258 & 50.191\\
 		\hline
 			Synchronous & 23.369 & 22.246 & 26.272 & 34.339 & 46.773 & 69.14\\
 		\hline
 		\multicolumn{7}{c}{}
 		\vspace{-5mm}
 	\end{tabular*}\medskip

 	\noindent\begin{tabular*}{\textwidth}{@{\extracolsep\fill}|c||c|c|c|c|c|c|}
 		\hline
 		\multicolumn{7}{|c|}{\textbf{Average Memory Consumption (MB)}}\\
 		\hline\hline
 		 &\multicolumn{6}{c|}{Number of Requests} \\
 		 \hline		
 			\, Monitoring modes & 50 & 100 & 200 & 500 & 1000 & 2000\\
 		\hline\hline		
 			Baseline & 25.62 & 25.68 & 25.63 & 25.68 & 26.17 & 26.25 \\
 		\hline		
 			Asynchronous  & 28.06 & 28.56 & 29.2 & 30.01 & 30.68 & 31.07\\
 		\hline
 			Hybrid & 28.4 & 28.97 & 29.69 & 30.2 & 31.11 & 31.95 \\
 		\hline
 			Synchronous & 28.88 & 29.19 & 29.62 & 30.23 & 31.33 & 33.01\\
 		\hline
 		\multicolumn{7}{c}{}
 		\vspace{-5mm}
 	\end{tabular*}\medskip

 	\noindent\begin{tabular*}{\textwidth}{@{\extracolsep\fill}|c||c|c|c|c|c|c|}
 		\hline
 		\multicolumn{7}{|c|}{\textbf{Average response times per Request (ms)}}\\
 		\hline\hline
 		 &\multicolumn{6}{c|}{Number of Requests} \\
 		 \hline		
 			\, Monitoring modes & 50 & 100 & 200 & 500 & 1000 & 2000\\
 		\hline\hline		
 			Baseline & 11.32 & 11.35 & 11.47 & 11.44 & 11.45 & 11.66\\
 		\hline		 		
 			Asynchronous & 12.23 & 12.74 & 13.1 & 14.15 & 16.18 & 21.8 \\
 		\hline
 			Hybrid & 13.11 & 13.02 & 13.31 & 15.53 & 18.39 & 23.26 \\
 		\hline
 			Synchronous & 15.05 & 15.014 & 15.14 & 16.74 & 20.91 & 29.68\\
 		\hline
 		\multicolumn{7}{c}{}
 		\vspace{-5mm}
 	\end{tabular*}\medskip

\chapter{Proofs for Auxiliary Lemmas}
\label{chp:aux-lemmas}

\newcommand{\lemmaZ}{\lemmaref{lemmaZ}{}\xspace}
\newcommand{\lemmaH}{\lemmaref{lemmaH}{}\xspace}
\newcommand{\lemmaMA}{\lemmaref{lemmaMA}{}\xspace}
\newcommand{\lemmaMB}{\lemmaref{lemmaMB}{}\xspace}
\newcommand{\lemmaZZZ}{\lemmaref{lemmaZZZ}{}\xspace}
\newcommand{\lemmaZZZA}{\lemmaref{lemmaZZZA}{}\xspace}
\newcommand{\lemmaYYY}{\lemmaref{lemmaYYY}{}\xspace}
\newcommand{\lemmaXXXA}{\lemmaref{lemmaXXXA}{}\xspace}
\newcommand{\lemmaXXXB}{\lemmaref{lemmaXXXB}{}\xspace}
\newcommand{\lemmaXXXC}{\lemmaref{lemmaXXXC}{}\xspace}


We divide this chapter into two main parts. In the first part (\secref{sec:corr-lemma}) we present the proof for \lemmaref{lemmaA}\!\!, an auxiliary lemma used when proving the Theorem of Semantic Correspondence (\ie Thm.~\ref{thm:1} in \secref{sec:derivative-semantics}). 

In the second part (\secref{sec:main-aux-lemmas} and \ref{sec:other-aux-lemmas}) we present the proofs for the auxiliary lemmas required in proving the Theorem of Type Soundness (\ie Thm.~\ref{thm:soundness}). We further divide this part into two sections: in \secref{sec:main-aux-lemmas} we give the proof details about the \emph{primary auxiliary lemmas} \ie lemmas that are used \emph{directly} by the main proofs presented in \secref{sec:soundness}; whereas in \secref{sec:other-aux-lemmas} we prove the \emph{secondary auxiliary lemmas} \ie lemmas that are used by the primary auxiliary lemmas. We now state the auxiliary lemmas that we prove in both parts of this chapter: \medskip

\noindent\underline{\textbf{The Correspondence Auxiliary Lemma:}}\vspace{-2mm}
\begin{description}
	\item[\textnormal{\lemmaref{lemmaA}}] {\small\setstretch{0.5}$\hV\traS{\act}\hV' \; \text{ and } \; \vsat{\actVV}{\sV}{\hV'} \; \text{ and } \; \actV\wtraS{\act}\actVV  \;\imp\; \vsat{\actV}{\act\sV}{\hV}$}
\end{description} \medskip

\noindent\underline{\textbf{The Primary Soundness Auxiliary Lemmas:}}\vspace{-2mm}
\begin{description}
	\item[\textnormal{\lemmaX}] {\small\setstretch{0.5} \begin{math}\\\envVP=\after{\envV,\act}\; \text{ and } \; \wf{\envV}{\stV}\; \text{ and } \; \stV\!\traceEvent{\scriptstyle\act}\!\stV'\,\imp\,\wf{\envVP}{\stV'}\end{math} }
	\item[\textnormal{\lemmaY}] {\small\setstretch{0.5} \begin{math}\\\typeRule{\env{\envVV}{\envV}}{\hV}\; \text{ and } \; \cmon{\envV}{\Used}{\hV}\traceEvent{\scriptstyle\act}\cmon{\envVP}{\UsedP}{\hV'}\, \imp \,\typeRule{\env{\envVV}{\envVP}}{\hV'}\; \text{ and } \; \envVP=\after{\envV,\act}\end{math} }
	\item[\textnormal{\lemmaC}] {\small\setstretch{0.5} \begin{math}\\\typeRule{\env{\envVV}{\envVP}}{\mmax{\hVarX}{\!\hV}}\; \text{ and } \;\typeRule{\env{\envVV}{\envV}}{\hV}\,\imp\,\typeRule{\env{\envVV}{\envV}}{\hV\sub{(\clr{\hVarX}\mmax{\hVarX}{\!\hV})}{\hVarX}}\end{math} }
	\item[\textnormal{\lemmaI}] {\small\setstretch{0.5} \begin{math}\typeRule{\env{\envVVN{\hVarX}{\envVP}}{\envV}}{\hV}\; \text{ and } \; \hVarX\notin\fv(\hV)\, \imp \typeRule{\env{\envVV}{\envV}}{\hV}\end{math} }
	\item[\textnormal{\lemmaD}] {\small\setstretch{0.5} \begin{math}\\\typeRule{\env{\envVV}{(\envVN{\tbnd{\patE}}})}{\hV} \; \text{ and } \;\match{\patE}{\actE}{\sigma} \;\text{ and }\; \dtcheck \\ \qquad \text{ and }\; \dtcheckB \imp\,\typeRule{\env{\envVV}{(\envVN{\tbnd{\patE}}\sigma)}}{\hV\sigma}\end{math} }	
\end{description} 
\medskip

\noindent\underline{\textbf{The Secondary Soundness Auxiliary Lemmas:}}\vspace{-2mm}
\begin{description}
	\item[\textnormal{\lemmaYYY}] {\small\setstretch{0.5} \begin{math}\typeRule{\env{\envVV}{\envVA}}{\hV}\; \text{ and } \; \typeRule{\env{\envVV}{\envVB}}{\hVV}\; \text{ and } \; \envV=\envVA+\envVB\, \imp \,\typeRule{\env{\envVV}{\envV}}{\hV\mand\hVV}\end{math} }
	\item[\textnormal{\lemmaMA}] {\small\setstretch{0.5} \begin{math}\cmon{\envV}{\Used}{\hV}\traceEvent{\scriptstyle\act}\cmon{\envVP}{\UsedP}{\hV'}\; \text{ and } \;\typeRule{\env{\envVV}{\eff{\envV}{\vLstA}}}{\hV}\, \imp\\ \cmon{\eff{\envV}{\vLstA}}{\Used}{\hV}\traceEvent{\scriptstyle\act}\cmon{\eff{\after{\envV,\act}}{\vLstA}}{\UsedP}{\hV'}\; \text{ and } \;\envVP=\after{\envV,\act}\end{math} }
	\item[\textnormal{\lemmaMB}] {\small\setstretch{0.5} \begin{math}\envV=\envVA+\envVB \; \text{ and } \; \cmon{\envV}{\Used}{\hV}\traceEvent{\scriptstyle\act}\cmon{\envVP}{\UsedP}{\hV'}\; \text{ and } \;\typeRule{\env{\envVV}{\envVA}}{\hV}\, \imp \\ \cmon{\envVA}{\Used}{\hV}\traceEvent{\scriptstyle\act}\cmon{\envVPA}{\UsedP}{\hV'} \; \text{ and } \;\envVP=\envVPA+\envVB \end{math} }	
	\item[\textnormal{\lemmaXXXB}] {\small\setstretch{0.5} \begin{math}\excl{\hV,\hVV}=(\vLstA[\hV],\vLstA[\hVV])\; \text{ and } \; \cmon{\envV}{\Used}{\hV}\traceEventTau\cmon{\envV}{\Used}{\hV'}\, \imp \,\excl{\hV',\hVV}=(\vLstA[\hV],\vLstA[\hVV])\end{math} }
	\item[\textnormal{\lemmaXXXA}] {\small\setstretch{0.5} \begin{math}\excl{\hV,\hVV}=(\vLstA[\hV],\vLstA[\hVV])\; \text{ and } \; \cmon{\envV}{\Used}{\hV}\traceEventA\cmon{\envVP}{\UsedP}{\hV'}\; \text{ and } \\ \typeRule{\env{\envVV}{\eff{(\envVP,\envVPP)}{\vLstA[\hVV]}}}{\hV'}\; \text{ and } \; \typeRule{\env{\envVV}{\eff{(\envVP,\envVPP)}{\vLstA[\hV]}}}{\hVV'}\, \imp \,\typeRule{\env{\envVV}{(\envVP,\envVPP)}}{\hV'\mand\hVV'}\end{math} }
	\item[\textnormal{\lemmaH}]  {\small\setstretch{0.5} \begin{math}\excl{\hV,\hVV}=(\vLstA[\hV],\vLstA[\hVV])\; \text{ and } \; \cmon{\envV}{\Used}{\hV}\traceEvent{\scriptstyle\actC}\cmon{\envVP}{\Used}{\hV'}\; \text{ and } \; \typeRule{\env{\envVV}{\eff{\envV}{\vLstA[\hV]}}}{\hVV}\; \text{ and } \\ \typeRule{\env{\envVV}{\eff{\after{\envV,\actC}}{\vLstA[\hVV]}}}{\hV'}\, \imp \,\typeRule{\env{\envVV}{\after{\envV,\actC}}}{\hV'\mand\hVV}\end{math} }
	\item[\textnormal{\lemmaZZZ}] {\small\setstretch{0.5} \begin{math}\excl{\hV,\hVV}=(\vLstA[\hV],\vLstA[\hVV])\; \text{ and } \; \sub{(\clr{\hVarX}\mmax{\hVarX}{\!\hV'})}{\hVarX}\, \imp \\ \excl{\hV\sub{(\clr{\hVarX}\mmax{\hVarX}{\!\hV'})}{\hVarX},\hVV\sub{(\clr{\hVarX}\mmax{\hVarX}{\!\hV'})}{\hVarX}}=(\vLstA[\hV],\vLstA[\hVV])\end{math} }
	\item[\textnormal{\lemmaZZZA}] {\small\setstretch{0.5} \begin{math}\excl{\hV,\hVV}=(\vLstA[\hV],\vLstA[\hVV])\; \text{ and } \; \sub{(\clr{\hVarX}\mmax{\hVarX}{\!\hV'})}{\hVarX}\, \imp \\ \excl{\hV\sub{(\clr{\hVarX}\mmax{\hVarX}{\!\hV'})}{\hVarX},\hVV}=(\vLstA[\hV],\vLstA[\hVV])\end{math} }
\end{description} 

\noindent Note that when proving the Primary and Secondary auxiliary lemmas for Type Soundness, we omit from showing the proofs for cases involving structural equivalence (\ie \rtit{rStr}), idempotency (\rtit{rIdem1} and \rtit{rIdem2}) due to their triviality, as well as for cases that are analogous to other cases, \ie their proof requires minimal effort to be derived from the proof of another case. 

\section{Proving The Correspondence Auxiliary Lemma} \label{sec:corr-lemma}
		
\begin{lemma}\label{lemmaA} $\hV\traS{\act}\hV'\; \text{ and } \;\vsat{\actVV}{\sV}{\hV'} \; \text{ and } \; \actV\wtraS{\act}\actVV  \;\imp\; \vsat{\actV}{\act\sV}{\hV}$ 
\begin{proof} By rule induction on $\hV\traS{\act}\hV'$. The main cases are:

	\begin{Case}[\textsc{rTru}] From our rule premises we know that $\mboolE{b}{\hV}{\hVV}\traS{\tau}\hV$ because
		\begin{gather}
			\mcondEval{b}{true} \label{LA:1:2}
		\end{gather}
		and
		\begin{gather}
			\vsat{\actVV}{\sV}{\hV} \label{LA:1:3}\\
			\actV\wtraS{\tau}\actVV \label{LA:1:4}
		\end{gather}
		By \eqref{LA:1:2}, \eqref{LA:1:3}, \eqref{LA:1:4} and defn of $\vsatL$ we know \;
			$\vsat{\actV}{\tau\sV}{\mboolE{b}{\hV}{\hVV}}$\; as required.
	\end{Case} 
	
	\begin{Case}[\textsc{rFls}] The proof for this case is analogous to that of \textsc{rTru}.
	\end{Case} 
		
	\begin{Case}[\textsc{rCn1}] From our rule premises we know that $\hV\mand\hVV\traS{\actE}\hV'\mand\hVV'$ because
		\begin{gather}
			\hV\traS{\actE}\hV' \label{LA:3:2} \\
			\hVV\traS{\actE}\hVV' \label{LA:3:3} 
		\end{gather}
		and
		\begin{gather}
			\actV\wtraS{\actE}\actVV \label{LA:3:4.5}\\
			\vsat{\actVV}{\sV}{\hV'\mand\hVV'} \label{LA:3:4}
		\end{gather}
		By \eqref{LA:3:4} and defn of $\vsatL$ we know
		\begin{align}
			& \vsat{\actVV}{\sV}{\hV'} \label{LA:3:5} \\
			\text{or } \quad &\vsat{\actVV}{\sV}{\hVV'} \label{LA:3:6}
		\end{align}		
		By \eqref{LA:3:2}, \eqref{LA:3:4.5}, \eqref{LA:3:5} and IH we know		
		\begin{gather}
			\vsat{\actV}{\actE\sV}{\hV} \label{LA:3:7} 
		\end{gather}
		By \eqref{LA:3:3}, \eqref{LA:3:4.5}, \eqref{LA:3:6} and IH we know		
		\begin{gather}
			\vsat{\actV}{\actE\sV}{\hVV} \label{LA:3:8} 
		\end{gather}		
		By \eqref{LA:3:7}, \eqref{LA:3:8} and defn of $\vsatL$ we know $\vsat{\actV}{\actE\sV}{\hV\mand\hVV}$ as required.	
	\end{Case}\vspace{-5mm}
	
	\begin{Case}[\textsc{rCn2}] From our rule premises we know that $\hV\mand\hVV\traS{\tau}\hV'\mand\hVV$ because
		\begin{gather}
			\hV\traS{\tau}\hV' \label{LA:4:2} 
		\end{gather}
		and
		\begin{gather}
			\actV\wtraS{\tau}\actVV \label{LA:4:3.5}\\
			\vsat{\actVV}{\sV}{\hV'\mand\hVV} \label{LA:4:3}
		\end{gather}
		By \eqref{LA:3:3} and defn of $\vsatL$ we know
		\begin{align}
				&\vsat{\actVV}{\sV}{\hV'} \label{LA:4:4} \\
			\text{or } \quad &\vsat{\actVV}{\sV}{\hVV} \label{LA:4:5}
		\end{align}		
		By \eqref{LA:4:2}, \eqref{LA:4:3.5}, \eqref{LA:4:4} and IH we know		
		\begin{gather}
			\vsat{\actV}{\tau\sV}{\hV} \label{LA:4:6} 
		\end{gather}
		By \eqref{LA:4:6} and the defn of $\vsatL$ we know $\vsat{\actV}{\tau\sV}{\hV\mand\hVV}$ as required.		
	\end{Case}
		
	\begin{Case}[\textsc{rMax}] From our rule premises we know 
		\begin{gather}
			\mmax{\hVarX}{\hV}\traS{\tau}\hV\sub{(\mmax{\hVarX}{\hV})}{\hVarX} \\
			\actV\wtraS{\tau}\actVV \label{LA:5:3}\\
			\vsat{\actVV}{\sV}{\hV\sub{(\mmax{\hVarX}{\hV})}{\hVarX}} \label{LA:5:2}
		\end{gather}
		By \eqref{LA:5:2} and defn of $\vsatL$ we know
		\begin{align}
				\vsat{\actVV}{\sV}{\mmax{\hVarX}{\hV}} \label{LA:5:4} 
		\end{align}		
		By \eqref{LA:5:3} and \eqref{LA:5:4} we know $\vsat{\actV}{\tau\sV}{\mmax{\hVarX}{\hV}}$ as required.			
	\end{Case}\medskip
	
	\begin{Case}[\textsc{rNc1}] From our rule premises we know $\mnec{\patE}{\hV}\traS{\actE}\hV\sigma$ because
		\begin{gather}
			\match{\patE}{\actE}{\sigma} \label{LA:6:2}
		\end{gather}
		and
		\begin{gather}
			\actV\wtraS{\actE}\actVV \label{LA:6:4}\\
			\vsat{\actVV}{\sV}{\hV\sigma} \label{LA:6:3}
		\end{gather}		
		By \eqref{LA:6:2}, \eqref{LA:6:4}, \eqref{LA:6:3} and defn of $\vsatL$ we know $\vsat{\actV}{\actE\sV}{\mnec{\patE}{\hV}}$ as required.		
	\end{Case}\vspace{-5mm}
	
	\begin{lastcase}[\textsc{rNc2}] From our rule premises we know $\mnec{\patE}{\hV}\traS{\actE}\mtru$ because
		\begin{gather}
			\match{\patE}{\actE}{\bot} \label{LA:7:2}
		\end{gather}
		and
		\begin{gather}
			\actV\wtraS{\actE}\actVV \label{LA:7:4}\\
			\vsat{\actVV}{\sV}{\mtru} \label{LA:7:3}
		\end{gather}		
		Since by the defn of $\vsatL$ we know that \eqref{LA:7:3} can \emph{never} be true, this means that we have a \emph{contradiction}.
	\end{lastcase} \vspace{-5mm}
	\end{proof}
\end{lemma}

\section{Proving The Primary Soundness Auxiliary Lemmas} \label{sec:main-aux-lemmas}

\begin{lemma} \label{lemmaX} {\bfseries\itshape (System Reduction)} \\[0mm] \indent\indent\indent $\envVP=\after{\envV,\act}\; \text{ and } \; \wf{\envV}{\stV}\; \text{ and } \; s\!\traceEvent{\scriptstyle\act}\!s'\,\imp\,\wf{\envVP}{\stV'}$.
\begin{proof}\emph{By Rule induction on } $s\!\traceEvent{\scriptstyle\act}\!s'$. \medskip 

\begin{Case}[\rtit{sNew}] From our rule premises we know 
\begin{gather}
  s\traceEventTau\state{\mapstate{i}{\unblocked}}{\stV} \label{X:1:1} \\ 
  \wf{\envV}{\stV} \label{X:1:2} \\
  \after{\envV,\tau}=\envVP  \label{X:1:2.5} 
\end{gather}
By \eqref{X:1:2} and \emph{System Well-Formedness} (\defref{def:wf-sys}) we know 
\begin{align}
	\scond{\envV}{\stV} \label{X:1:3}
\end{align}
By \eqref{X:1:2.5} and the defn of \textsf{after} (\defref{def:after}) and since $\act=\tau$ we know that for any $\hV$ 
\begin{align}
	\envVP=\envV \label{X:1:4}
\end{align}
Given that $s\subseteq\state{\mapstate{i}{\unblocked}}{\stV}$ and by \eqref{X:1:3} and \eqref{X:1:4} we know
\begin{align}
	\scond{\envVP}{\state{\mapstate{i}{\unblocked}}{\stV}} \label{X:1:5}
\end{align}
By \eqref{X:1:5} and \emph{System Well-Formedness} (\defref{def:wf-sys}) we know
\begin{align}
	\wf{\envVP}{\state{\mapstate{i}{\unblocked}}{\stV}} \label{X:1:7}
\end{align}
\noindent $\therefore$ Case holds by \eqref{X:1:7}. \end{Case}\vspace{-5mm}

\begin{Case}[\rtit{sAct}] From our rule premises we know 
\begin{align}
  \state{\mapstate{i}{\unblocked}}{\stV} \traceEventA \state{\mapstate{i}{\unblocked}}{\stV} \label{X:2:1}
\end{align}
because
\begin{gather}
  \id{\actE}=i  \label{X:2:2} \\
  \ids{\actE}\subseteq\dom{\state{\mapstate{i}{\unblocked}}{\stV}} \label{X:2:3}
\end{gather}
and
\begin{gather}
  \wf{\envV}{\state{\mapstate{i}{\unblocked}}{\stV}} \label{X:2:4}\\
   \after{\envV,\actE}=\envVP  \label{X:2:4.5} 
\end{gather}
By \eqref{X:2:4} and \emph{System Well-Formedness} (\defref{def:wf-sys}) we know 
\begin{align}
  \scond{\envV}{\state{\mapstate{i}{\unblocked}}{\stV}} \label{X:2:5}
\end{align}
Since $\act=\actE$, by \eqref{X:2:4.5} and by defn of \textsf{after} (\defref{def:after}) we know
\begin{gather}
	\envV\subseteq\envVP \label{X:2:6} 
\end{gather}
Recall that the type system is \emph{only} able to change the type of a linear process id $\typ{i}{\lpid}$ to internal type $\lpidb\,$ whenever a $\mblock{i}$ command is type-checked, so as to signify that the process reference should be \emph{blocked} during runtime. This is because at runtime $\mblock{i}$ generates a $\blk{i}$ event which causes process id $i$ to block. This means that an $\actE$ action can neither directly introduce a process $\typ{j}{\lpidb}$ in the new new type environment $\envVP$, nor block a process in state $\stV$. Hence, from \eqref{X:2:5} and \eqref{X:2:6} we can deduce
\begin{align}
	 \scond{\envVP}{\state{\mapstate{i}{\unblocked}}{\stV}} \label{X:2:8}
\end{align}
By \eqref{X:2:8} and \emph{System Well-Formedness} (\defref{def:wf-sys}) we know
\begin{align}
	\wf{\envVP}{\state{\mapstate{i}{\unblocked}}{\stV}} \label{X:2:10}
\end{align}
\noindent $\therefore$ Case holds by \eqref{X:2:10}. 
\end{Case}

\begin{Case}[\rtit{sBlk}] From our rule premises we know 
\begin{gather}
  \state{\mapstate{\vLstB}{\unblocked}}{\stV}\traceEvent{\scriptstyle\blk{\vLstB}} \state{\mapstate{\vLstB}{\blocked}}{\stV} \label{X:3:1}\\
  \wf{\envV}{\state{\mapstate{\vLstB}{\unblocked}}{\stV}} \label{X:3:2}\\
  \after{\envV,\blk{\vLstB}}=\envVP  \label{X:3:2.5}   
\end{gather}
By \eqref{X:3:2.5} the defn of \textsf{after} (\defref{def:after}) and since $\act=\blk{\vLstB}$ we know 
\begin{gather}
	\envV=(\envVPPN{\typ{\vLstB}{\lpid}}) \label{X:3:3} \\
	\envVP=(\envVPPN{\typ{\vLstB}{\lpidb}}) \label{X:3:4}
\end{gather}
By \eqref{X:3:2}, \eqref{X:3:3} and \emph{System Well-Formedness} (\defref{def:wf-sys}) we know
\begin{align}
	\scond{(\envVPPN{\typ{\vLstB}{\lpid}})}{\state{\mapstate{\vLstB}{\unblocked}}{\stV}} \label{X:3:5}
\end{align}
From \eqref{X:3:4}, \eqref{X:3:5} and since by \eqref{X:3:1} we know that $\state{\mapstate{\vLstB}{\unblocked}}{\stV}$ becomes $\state{\mapstate{\vLstB}{\blocked}}{\stV}$ we can deduce
\begin{align}
	\scond{(\envVPPN{\typ{\vLstB}{\lpidb}})}{\state{\mapstate{\vLstB}{\blocked}}{\stV}} \label{X:3:6}
\end{align}
By \eqref{X:3:4}, \eqref{X:3:6} and \emph{System Well-Formedness} (\defref{def:wf-sys}) we know
\begin{align}
	\wf{\envVP}{\state{\mapstate{\vLstB}{\blocked}}{\stV}} \label{X:3:10}
\end{align}
\noindent $\therefore$ Case holds by \eqref{X:3:10}.
\end{Case}

\begin{Case}[\rtit{sRel}] From our rule premises we know 
\begin{gather}
  \state{\mapstate{\vLstB}{\blocked}}{\stV}\traceEvent{\scriptstyle\rel{\vLstB}} \state{\mapstate{\vLstB}{\unblocked}}{\stV} \label{X:4:1}\\
  \wf{\envV}{\state{\mapstate{\vLstB}{\blocked}}{\stV}} \label{X:4:2}\\
  \after{\envV,\rel{\vLstB}}=\envVP  \label{X:4:2.5}   
\end{gather}
By \eqref{X:4:2.5} the defn of \textsf{after} (\defref{def:after}) and since $\act=\rel{\vLstB}$ we know 
\begin{gather}
	\envV=(\envVPPN{\typ{\vLstB}{\lpidb}}) \label{X:4:3} \\
	\envVP=(\envVPPN{\typ{\vLstB}{\lpid}}) \label{X:4:4}
\end{gather}
By \eqref{X:4:2}, \eqref{X:4:3} and \emph{System Well-Formedness} (\defref{def:wf-sys}) we know
\begin{align}
	\scond{(\envVPPN{\typ{\vLstB}{\lpidb}})}{\state{\mapstate{\vLstB}{\blocked}}{\stV}} \label{X:4:5}
\end{align}
From \eqref{X:4:4}, \eqref{X:4:5} and since by \eqref{X:4:1} we know that $\state{\mapstate{\vLstB}{\blocked}}{\stV}$ becomes $\state{\mapstate{\vLstB}{\unblocked}}{\stV}$ we can deduce
\begin{align}
	\scond{(\envVPPN{\typ{\vLstB}{\lpid}})}{\state{\mapstate{\vLstB}{\unblocked}}{\stV}} \label{X:4:6}
\end{align}
By \eqref{X:4:4}, \eqref{X:4:6} and \emph{System Well-Formedness} (\defref{def:wf-sys}) we know
\begin{align}
	\wf{\envVP}{\state{\mapstate{\vLstB}{\unblocked}}{\stV}} \label{X:4:10}
\end{align}
\noindent $\therefore$ Case holds by \eqref{X:4:10}.
\end{Case}

\begin{Case}[\rtit{sAdS}] From our rule premises we know 
\begin{gather}
  \state{\mapstate{\vLstB}{\blocked}}{\stV}\traceEvent{\scriptstyle\cors{\vLstB}}\state{\mapstate{\vLstB}{\blocked}}{\stV} \label{X:5:1}\\
  \wf{\envV}{\state{\mapstate{\vLstB}{\blocked}}{\stV}} \label{X:5:2}\\
  \after{\envV,\cors{\vLstB}}=\envVP  \label{X:5:2.5}   
\end{gather}
By \eqref{X:5:2.5}, the defn of \textsf{after} (\defref{def:after}) and since $\act=\cors{\vLstB}$
\begin{align}
	\envVP=\envV \label{X:5:3}
\end{align}
By \eqref{X:5:2} and \eqref{X:5:3} we know
\begin{align}
	 \wf{\envVP}{\state{\mapstate{\vLstB}{\blocked}}{\stV}} \label{X:5:4}
\end{align}
\noindent $\therefore$ Case holds by \eqref{X:5:4}. 
\end{Case}

\begin{lastcase}[\rtit{sAdA}] The proof for this case is analogous to that of case \rtit{sAdS} \end{lastcase}
\end{proof}
\end{lemma} \medskip

\begin{lemma} \label{lemmaY} {\bfseries\itshape (Monitor Reduction)} \\[0mm] \indent $\typeRule{\env{\envVV}{\envV}}{\hV}\; \text{ and } \; \cmon{\envV}{\Used}{\hV}\traceEvent{\scriptstyle\act}\cmon{\envVP}{\UsedP}{\hV'}\, \imp \,\typeRule{\env{\envVV}{\envVP}}{\hV'}\; \text{ and } \; \envVP=\after{\envV,\act}$.
\begin{proof}\emph{By Rule induction on} $\cmon{\envV}{\Used}{\hV}\traceEvent{\scriptstyle\act}\cmon{\envVP}{\UsedP}{\hV'}$.\medskip

\noindent\textbf{Note:} This is a continuation of the proof given in \secref{sec:main-mon-red}. \medskip

\begin{Case}[\rtit{rTru}] From our rule premises we know 
\begin{gather}
	\cmon{\envV}{\Used}{\mboolE{c}{\hV}{\hVV}}\traceEvent{\scriptstyle\tau}\cmon{\envV}{\Used}{\hV} \qquad (\text{because  }\mcondEval{c}{true})  \label{Y:2:1} \\ 
 	\typeRule{\env{\envVV}{\envV}}{\mboolE{c}{\hV}{\hVV}}\label{Y:2:2}
\end{gather} 	
By \eqref{Y:2:2} and \rtit{tIf} we know 
\begin{gather}
	\typeRule{\env{\envVV}{\envV}}{\hV} \label{Y:2:3}\\
	\typeRule{\env{\envVV}{\envV}}{\hVV} \label{Y:2:4}
\end{gather}
By the defn of \textsf{after} (\defref{def:after}) and since $\act=\tau$ we know
\begin{align}
	\envV=\after{\envV,\tau} \label{Y:2:5}
\end{align}
$\therefore$ Case holds by \eqref{Y:2:3} and \eqref{Y:2:5}.
\end{Case}

\noindent\textbf{Note:} The proof for cases \rtit{rFls} and \rtit{rClr} are analogous to the proof for case \rtit{rTru}. \medskip

\begin{Case}[\rtit{rCn1}] From our rule premises we know 
\begin{align}
	\cmon{\envV}{\Used}{\hV\mand\hVV}\traceEventA\cmon{(\envVP,\envVPP)}{(\UsedP\cup\UsedPP)}{\hV'\mand\hVV'}  \label{Y:3:1} 
\end{align} 	
because
\begin{gather}
	\cmon{\envV}{\Used}{\hV}\traceEventA\cmon{\envVP}{\UsedP}{\hV'}  \label{Y:3:2} \\
	\cmon{\envV}{\Used}{\hVV}\traceEventA\cmon{\envVPP}{\UsedPP}{\hVV'}  \label{Y:3:3} \\
	\dtAndcheck \label{Y:3:4}
\end{gather} 	
and 
\begin{align}
	\typeRule{\env{\envVV}{\envV}}{\hV\mand\hVV} \label{Y:3:5}
\end{align} 
As $\hV$ and $\hVV$ may either be \emph{exclusive} or not, we must consider the following two subcases: \smallskip

\begin{Subcase}[${\excl{\hV,\hVV}=(\vLstA[\hV],\vLstA[\hVV])}$]
	Since $\excl{\hV,\hVV}=(\vLstA[\hV],\vLstA[\hVV])$, by \eqref{Y:3:5} and \rtit{tCn2} we know
	\begin{gather}
		\typeRule{\env{\envVV}{\eff{\envV}{\vLstA[\hVV]}}}{\hV}  \label{Y:3:6} \\
		\typeRule{\env{\envVV}{\eff{\envV}{\vLstA[\hV]}}}{\hVV}  \label{Y:3:7} 
	\end{gather}
	By \eqref{Y:3:2}, \eqref{Y:3:6} and \lemmaMA we know
	\begin{gather}
		\cmon{\eff{\envV}{\vLstA[\hVV]}}{\Used}{\hV}\traceEventA\cmon{\eff{\after{\envV,\actE}}{\vLstA[\hVV]}}{\UsedP}{\hV'} \label{Y:3:8} \\
		\envVP=\after{\envV,\actE} \label{Y:3:8.5}
	\end{gather} 
	By \eqref{Y:3:3}, \eqref{Y:3:7} and \lemmaMA we know
	\begin{gather}
		\cmon{\eff{\envV}{\vLstA[\hV]}}{\Used}{\hVV}\traceEventA\cmon{\eff{\after{\envV,\actE}}{\vLstA[\hV]}}{\UsedPP}{\hVV'} \label{Y:3:9} \\
		\envVPP=\after{\envV,\actE} \label{Y:3:9.5}
	\end{gather} 
	By \eqref{Y:3:6}, \eqref{Y:3:8} and IH we know
	\begin{gather}
		\typeRule{\env{\envVV}{\eff{\after{\envV,\actE}}{\vLstA[\hVV]}}}{\hV'} \label{Y:3:10}\\
		\after{\eff{\envV}{\vLstA[\hVV]},\actE}=\eff{\after{\envV,\actE}}{\vLstA[\hVV]} \label{Y:3:11}
	\end{gather} 
	By \eqref{Y:3:7}, \eqref{Y:3:9} and IH we know
	\begin{gather}
		\typeRule{\env{\envVV}{\eff{\after{\envV,\actE}}{\vLstA[\hV]}}}{\hVV'} \label{Y:3:12}\\
		\after{\eff{\envV}{\vLstA[\hV]},\actE}=\eff{\after{\envV,\actE}}{\vLstA[\hV]} \label{Y:3:13}
	\end{gather} 
	By \eqref{Y:3:11}, \eqref{Y:3:13} and the defn of \textsf{after} (\defref{def:after}) we know
	\begin{gather}
		\eff{\envV}{\vLstA[\hVV]}\subseteq\eff{\after{\envV,\actE}}{\vLstA[\hVV]} \label{Y:3:14}\\
		\eff{\envV}{\vLstA[\hV]}\subseteq\eff{\after{\envV,\actE}}{\vLstA[\hV]} \label{Y:3:15}
	\end{gather} 
	Hence from \eqref{Y:3:8.5}, \eqref{Y:3:9.5}, \eqref{Y:3:14} and \eqref{Y:3:15} we can deduce
	\begin{gather}
		\envV\subseteq(\envVP,\envVPP) \label{Y:3:18}
	\end{gather} 
	By \eqref{Y:3:18} and the defn of \textsf{after} (\defref{def:after}) we can deduce
	\begin{gather}
		\after{\envV,\actE}=(\envVP,\envVPP) \label{Y:3:19}
	\end{gather} 
	By \eqref{Y:3:10}, \eqref{Y:3:12} and \eqref{Y:3:19} we can deduce
	\begin{gather}
		\typeRule{\env{\envVV}{\eff{(\envVP,\envVPP)}{\vLstA[\hVV]}}}{\hV'} \label{Y:3:20}\\
		\typeRule{\env{\envVV}{\eff{(\envVP,\envVPP)}{\vLstA[\hV]}}}{\hVV'} \label{Y:3:21}
	\end{gather} 	
	By \eqref{Y:3:4} we know that we do not have any type incompatibilities, and since $\excl{\hV,\hVV}=(\vLstA[\hV],\vLstA[\hVV])$, by \eqref{Y:3:2}, \eqref{Y:3:20}, \eqref{Y:3:21} and \lemmaXXXA we know 
	\begin{align}
		\typeRule{\env{\envVV}{(\envVP,\envVPP)}}{\hV'\mand\hVV'} \label{Y:3:22}
	\end{align} 
\noindent$\therefore$ Subcase holds by \eqref{Y:3:19} and \eqref{Y:3:22}.
\end{Subcase}

\begin{lastsubcase}[${\excl{\hV,\hVV}=\undef}$]
Since $\excl{\hV,\hVV}=\undef$, by \eqref{Y:3:5} and \rtit{tCn1} we know 
\begin{gather}
	\envV=(\envVA+\envVB) \label{Y:3:23}\\
	\typeRule{\env{\envVV}{\envVA}}{\hV}  \label{Y:3:24} \\
	\typeRule{\env{\envVV}{\envVB}}{\hVV} \label{Y:3:25}
\end{gather}
By \eqref{Y:3:2}, \eqref{Y:3:23}, \eqref{Y:3:24} and \lemmaMB we know  
\begin{gather}
	\cmon{\envVA}{\Used}{\hV}\traceEventA\cmon{\envVPA}{\UsedP}{\hV'} \label{Y:3:26}\\
	\envVP=(\envVPA+\envVB)	\label{Y:3:26.5}
\end{gather}
By \eqref{Y:3:3}, \eqref{Y:3:23}, \eqref{Y:3:25} and \lemmaMB we know  
\begin{gather}
	\cmon{\envVB}{\Used}{\hV}\traceEventA\cmon{\envVPB}{\UsedP}{\hV'} \label{Y:3:27}\\
	\envVPP=(\envVA+\envVPB)	\label{Y:3:27.5}
\end{gather}
By \eqref{Y:3:24}, \eqref{Y:3:26} and IH we know  
\begin{gather}
	\typeRule{\env{\envVV}{\envVPA}}{\hV'} \label{Y:3:28}\\
	\envVPA=\after{\envVA,\actE}		\label{Y:3:29}
\end{gather}
By \eqref{Y:3:25}, \eqref{Y:3:27} and IH we know  
\begin{gather}
	\typeRule{\env{\envVV}{\envVPB}}{\hVV'} \label{Y:3:30}\\
	\envVPB=\after{\envVB,\actE}		\label{Y:3:31}
\end{gather}
By \eqref{Y:3:29}, \eqref{Y:3:31} and the defn of \textsf{after} (\defref{def:after}) we know
\begin{gather}
	\envVA\subseteq\envVPA  \label{Y:3:32}\\
	\envVB\subseteq\envVPB  \label{Y:3:33}
\end{gather}
Since by \eqref{Y:3:4} we know that we do not have aliasing, then by \eqref{Y:3:26.5}, \eqref{Y:3:27.5}, \eqref{Y:3:32} and \eqref{Y:3:33} we can deduce
\begin{gather}
	(\envVP,\envVPP)=(\envVPA+\envVPB)  \label{Y:3:34}
\end{gather}
By \eqref{Y:3:28}, \eqref{Y:3:30}, \eqref{Y:3:34} and \lemmaYYY we know
\begin{align}
	\typeRule{\env{\envVV}{(\envVP,\envVPP)}}{\hV'\mand\hVV'} \label{Y:3:35}  
\end{align}
By \eqref{Y:3:23}, \eqref{Y:3:26.5}, \eqref{Y:3:27.5}, \eqref{Y:3:32} and \eqref{Y:3:33} we can deduce
\begin{gather}
	\envV\subseteq(\envVP,\envVPP)  \label{Y:3:37}
\end{gather}
By \eqref{Y:3:37} and the defn of \textsf{after} (\defref{def:after}) we know
\begin{gather}
	(\envVP,\envVPP)=\after{\envV,\actE}  \label{Y:3:38}
\end{gather}
\noindent$\therefore$ Subcase holds by \eqref{Y:3:35} and \eqref{Y:3:38}.
\end{lastsubcase}
\end{Case}

\begin{Case}[\rtit{rCn2}] From our rule premises we know 
\begin{align}
	\cmon{\envV}{\Used}{\hV\mand\hVV}\traceEventTau\cmon{\envV}{\UsedP}{\hV'\mand\hVV}  \label{Y:3.5:1} 
\end{align} 	
because
\begin{gather}
	\cmon{\envV}{\Used}{\hV}\traceEventTau\cmon{\envV}{\UsedP}{\hV'}  \label{Y:3.5:2}
\end{gather} 	
and 
\begin{align}
	\typeRule{\env{\envVV}{\envV}}{\hV\mand\hVV} \label{Y:3.5:3}
\end{align} 
As $\hV$ and $\hVV$ may either be \emph{exclusive} or not, we must consider the following two subcases: \medskip

\begin{Subcase}[${\excl{\hV,\hVV}=(\vLstA[\hV],\vLstA[\hVV])}$]
	Since $\excl{\hV,\hVV}=(\vLstA[\hV],\vLstA[\hVV])$, by \eqref{Y:3.5:3} and \rtit{tCn2} we know
	\begin{gather}
		\typeRule{\env{\envVV}{\eff{\envV}{\vLstA[\hVV]}}}{\hV}  \label{Y:3.5:4} \\
		\typeRule{\env{\envVV}{\eff{\envV}{\vLstA[\hV]}}}{\hVV}  \label{Y:3.5:5} 
	\end{gather}
	By \eqref{Y:3.5:2}, \eqref{Y:3.5:4} and \lemmaMA we know	
	\begin{gather}
		\cmon{\eff{\envV}{\vLstA[\hVV]}}{\Used}{\hV}\traceEventTau\cmon{\eff{\envV}{\vLstA[\hVV]}}{\UsedP}{\hV'}  \label{Y:3.5:6}
	\end{gather}	
	By \eqref{Y:3.5:4}, \eqref{Y:3.5:6} and IH we know
	\begin{align}
		\typeRule{\env{\envVV}{\eff{\envV}{\vLstA[\hVV]}}}{\hV'} \label{Y:3.5:7}
	\end{align} 	
	Since $\excl{\hV,\hVV}=(\vLstA[\hV],\vLstA[\hVV])$, by \eqref{Y:3.5:6} and \lemmaXXXB we can deduce 
	\begin{align}
		\excl{\hV',\hVV}=(\vLstA[\hV],\vLstA[\hVV]) \label{Y:3.5:8}
	\end{align} 
	By \eqref{Y:3.5:5}, \eqref{Y:3.5:7}, \eqref{Y:3.5:8} and \rtit{tCn2} we know 
	\begin{align}
		\typeRule{\env{\envVV}{\envV}}{\hV'\mand\hVV} \label{Y:3.5:9}
	\end{align}
	By the defn of \textsf{after} (\defref{def:after}) we know 
	\begin{align}
		\envV=\after{\envV,\tau} \label{Y:3.5:10}
	\end{align}		
\noindent$\therefore$ Subcase holds by \eqref{Y:3.5:9} and \eqref{Y:3.5:10}.
\end{Subcase}

\begin{lastsubcase}[${\excl{\hV,\hVV}=\undef}$]
By \eqref{Y:3.5:3} and \rtit{tCn1} we know 
\begin{gather}
	\envV=(\envVA+\envVB) \label{Y:3.5:11}\\
	\typeRule{\env{\envVV}{\envVA}}{\hV}  \label{Y:3.5:12} \\
	\typeRule{\env{\envVV}{\envVB}}{\hVV} \label{Y:3.5:13}
\end{gather}
By \eqref{Y:3.5:2}, \eqref{Y:3.5:11}, \eqref{Y:3.5:12} and \lemmaMB we know  
\begin{align}
	\cmon{\envVA}{\Used}{\hV}\traceEventTau\cmon{\envVA}{\UsedP}{\hV'} \label{Y:3.5:14} 
\end{align}
By \eqref{Y:3.5:12}, \eqref{Y:3.5:14} and IH we know 
\begin{gather}
	\typeRule{\env{\envVV}{\envVA}}{\hV'} \label{Y:3.5:16}\\ 
	\envVA=\after{\envVA,\tau} \label{Y:3.5:17} 
\end{gather}
By \eqref{Y:3.5:11}, \eqref{Y:3.5:13}, \eqref{Y:3.5:16} and \lemmaYYY we know
\begin{align}
	\typeRule{\env{\envVV}{\envV}}{\hV'\mand\hVV} \label{Y:3.5:18}  
\end{align}
By the defn of \textsf{after} (\defref{def:after}) we know
\begin{align}
	\envV=\after{\envV,\tau} \label{Y:3.5:19}  
\end{align}
\noindent$\therefore$ Subcase holds by \eqref{Y:3.5:18} and \eqref{Y:3.5:19}.
\end{lastsubcase}
\end{Case}

\begin{Case}[\rtit{rCn3}] From our rule premises we know 
\begin{align}
	\cmon{\envV}{\Used}{\hV\mand\hVV}\traceEvent{\scriptstyle\actC}\cmon{\envVP}{\Used}{\hV'\mand\hVV}  \label{Y:4:1} 
\end{align} 	
because
\begin{gather}
	\cmon{\envV}{\Used}{\hV}\traceEvent{\scriptstyle\actC}\cmon{\envVP}{\Used}{\hV'}  \label{Y:4:2}
\end{gather} 	
and 
\begin{align}
	\typeRule{\env{\envVV}{\envV}}{\hV\mand\hVV} \label{Y:4:3}
\end{align} 
As $\hV$ and $\hVV$ may either be \emph{exclusive} or not, we must consider the following two subcases: \smallskip

\begin{Subcase}[${\excl{\hV,\hVV}=(\vLstA[\hV],\vLstA[\hVV])}$]
	Since $\excl{\hV,\hVV}\!=\!(\vLstA[\hV],\vLstA[\hVV])$, by \eqref{Y:4:3} and \rtit{tCn2} we know
	\begin{gather}
		\typeRule{\env{\envVV}{\eff{\envV}{\vLstA[\hVV]}}}{\hV} \label{Y:4:4} \\
		\typeRule{\env{\envVV}{\eff{\envV}{\vLstA[\hV]}}}{\hVV} \label{Y:4:5}
	\end{gather}
	By \eqref{Y:4:2}, \eqref{Y:4:4} and \lemmaMA we know
	\begin{gather}
		\cmon{\eff{\envV}{\vLstA[\hVV]}}{\Used}{\hV}\traceEvent{\scriptstyle\actC}\cmon{\eff{\after{\envV,\actC}}{\vLstA[\hVV]}}{\Used}{\hV'} \label{Y:4:6} \\
		\envVP=\after{\envV,\actC} \label{Y:4:6.5}
	\end{gather}	
	By \eqref{Y:4:4}, \eqref{Y:4:6} and IH we know	
	\begin{gather}
		\typeRule{\env{\envVV}{\eff{\after{\envV,\actC}}{\vLstA[\hVV]}}}{\hV'} \label{Y:4:7} \\
		\after{\eff{\envV}{\vLstA[\hVV]},\actC}=\eff{\after{\envV,\actC}}{\vLstA[\hVV]} \label{Y:4:8}
	\end{gather}	
	Since $\excl{\hV,\hVV}=(\vLstA[\hV],\vLstA[\hVV])$, by \eqref{Y:4:2}, \eqref{Y:4:5}, \eqref{Y:4:7} and \lemmaH we know
	\begin{align}
		\typeRule{\env{\envVV}{\after{\envV,\actC}}}{\hV'\mand\hVV} \label{Y:4:9} 
	\end{align}
	\noindent$\therefore$ Subcase holds by \eqref{Y:4:6.5} and \eqref{Y:4:9}.
\end{Subcase}

\begin{lastsubcase}[${\excl{\hV,\hVV}=\undef}$]
	By \eqref{Y:4:3} and \rtit{tCn1} we know 
	\begin{gather}
		\envV=(\envVA+\envVB) \label{Y:4:10}\\
		\typeRule{\env{\envVV}{\envVA}}{\hV}  \label{Y:4:11} \\
		\typeRule{\env{\envVV}{\envVB}}{\hVV} \label{Y:4:12}
	\end{gather}
	By \eqref{Y:4:2}, \eqref{Y:4:10}, \eqref{Y:4:11} and \lemmaMB we know  
	\begin{gather}
		\cmon{\envVA}{\Used}{\hV}\traceEvent{\scriptstyle\actC}\cmon{\envVPA}{\Used}{\hV'}  \label{Y:4:13}\\
		\envVP=(\envVPA+\envVB)		\label{Y:4:14}
	\end{gather} 
	By \eqref{Y:4:11}, \eqref{Y:4:13} and IH we know
	\begin{gather}
		\typeRule{\env{\envVV}{\envVPA}}{\hV'} \label{Y:4:15}\\
		\envVPA = \after{\envVA,\actC}	\label{Y:4:16}
	\end{gather}
	By \eqref{Y:4:12}, \eqref{Y:4:14}, \eqref{Y:4:15} and \lemmaYYY we know 
	\begin{align}
		\typeRule{\env{\envVV}{\envVP}}{\hV'\mand\hVV} \label{Y:4:17}  
	\end{align}
	By \eqref{Y:4:10}, \eqref{Y:4:14} and \eqref{Y:4:16} we deduce
	\begin{align}
		(\envVPA+\envVB=\after{\envVA,\actC}+\envVB) \; \equiv\; (\envVP=\after{\envV,\actC})	\label{Y:4:18}		
	\end{align}
	\noindent$\therefore$ Subcase holds by \eqref{Y:4:17} and \eqref{Y:4:18}.
\end{lastsubcase}
\end{Case}

\begin{Case}[\rtit{rRel}] From our rule premises we know 
\begin{gather}
	\cmon{(\envVN{\typ{\vLstB}{\lpidb}})}{\Used}{\mrelease{\vLstB}\,\hV}\traceEvent{\scriptstyle\rel{\vLstB}}\cmon{(\envVN{\typ{\vLstB}{\lpid}})}{\Used}{\hV}  \label{Y:10:1} \\
	\typeRule{\env{\envVV}{(\envVN{\typ{\vLstB}{\lpidb}})}}{\mrelease{\vLstB}\,\hV} \label{Y:10:2} 
\end{gather}
By \eqref{Y:10:2} and \rtit{tRel} we know
\begin{gather}	
	\typeRule{\env{\envVV}{(\envVN{\typ{\vLstB}{\lpid}})}}{\hV} \label{Y:10:4} 
\end{gather}
By the defn of \textsf{after} (\defref{def:after}) and since $\act=\rel{\vLstB}$ we know
\begin{align}
	\after{(\envVN{\typ{\vLstB}{\lpidb}}),\rel{\vLstB}}=(\envVN{\typ{\vLstB}{\lpid}}) \label{Y:10:6}
\end{align}
$\therefore$ Case holds by \eqref{Y:10:4} and \eqref{Y:10:6}. 
\end{Case} 

\begin{Case}[\rtit{rBlk}] From our rule premises we know 
\begin{gather}
	\cmon{(\envVN{\typ{\vLstB}{\lpid}})}{\Used}{\mblock{\vLstB}\,\hV}\traceEvent{\scriptstyle\blk{\vLstB}}\cmon{(\envVN{\typ{\vLstB}{\lpidb}})}{\Used}{\hV}  \label{Y:20:1} \\
	\typeRule{\env{\envVV}{(\envVN{\typ{\vLstB}{\lpid}})}}{\mblock{\vLstB}\,\hV} \label{Y:20:2} 
\end{gather}
By \eqref{Y:20:2} and \rtit{tBlk} we know
\begin{gather}	
	\typeRule{\env{\envVV}{(\envVN{\typ{\vLstB}{\lpidb}})}}{\hV} \label{Y:20:4} 
\end{gather}
By the defn of \textsf{after} (\defref{def:after}) and since $\act=\blk{\vLstB}$ we know
\begin{align}
	\after{(\envVN{\typ{\vLstB}{\lpid}}),\blk{\vLstB}}=(\envVN{\typ{\vLstB}{\lpidb}}) \label{Y:20:6}
\end{align}
$\therefore$ Case holds by \eqref{Y:20:4} and \eqref{Y:20:6}. 
\end{Case} 

\begin{Case}[\rtit{rAdA}] From our rule premises we know 
\begin{gather}
	\cmon{\envV}{\Used}{\macor{\vLstB}{\vLstA}\,\hV}\traceEvent{\scriptstyle\cor{\vLstB}}\cmon{\envV}{\Used}{\mrelease{\vLstA}\hV}  \label{Y:5:1} \\
	\typeRule{\env{\envVV}{\envV}}{\macor{\vLstB}{\vLstA}\,\hV} \label{Y:5:2} 
\end{gather}
By \eqref{Y:5:2} and \rtit{tAdA} we know
\begin{gather}	
	\envV=(\envVPN{\typ{\vLstB}{\lpid}}) \label{Y:5:3} \\
	\typeRule{\env{\envVV}{\envV}}{\mrelease{\vLstA}\hV} \label{Y:5:4} 
\end{gather}
By the defn of \textsf{after} (\defref{def:after}) and since $\act=\cor{\vLstB}$ we know
\begin{align}
	\after{\envV,\cor{\vLstB}}=\envV \label{Y:5:5}
\end{align}
$\therefore$ Case holds by \eqref{Y:5:4} and \eqref{Y:5:5}. 
\end{Case} 

\begin{Case}[\rtit{rAdS}] The proof for this case is analogous to the proof of case \rtit{rAdA}. \end{Case}

\begin{lastcase}[\rtit{rNc3}]From our rule premises we know 
\begin{align}
   \cmon{\envV}{\Used}{\mattrDNec{\patE}{\attr}{\stk}{\vLstA}{\hV}}\traceEventA\cmon{\envV}{\Used}{\mrelease{\vLstA}\mtru}  \label{Y:9:1}
\end{align}
because
\begin{align}
   \match{\patE}{\actE}{\undef}  \label{Y:9:2} 
\end{align}
and
\begin{align}   
	\typeRule{\env{\envVV}{\envV}}{\mattrNec{\patE}{\attr}{\vLstA}{\hV}} \label{Y:9:3} 
\end{align}
From \eqref{Y:9:1} we know that $\mattrNec{\patE}{\attr}{\vLstA}{\hV}$ reduces to 
\begin{align}
  \mrelease{\vLstA}\mtru \label{Y:9:6} 
\end{align}
Since $\act=\actE$, by the defn of \textsf{after} (\defref{def:after}) we know
\begin{align}
  \envV\subseteq\after{\envV,\actE} \label{Y:9:4} 
\end{align}
By \eqref{Y:9:1} we know that $\envV$ is not modified by the reduction, and hence from \eqref{Y:9:4} we can deduce
\begin{align}
  \envV=\after{\envV,\actE} \label{Y:9:4.5} 
\end{align}
As $\rho\in\{\agg,\norm\}$ we must consider two subcases: \medskip

\begin{Subcase}[$\rho=`b'$] By \eqref{Y:9:3} and \rtit{tNcB} we know
\begin{gather}
	\typeRule{\env{\envVV}{\envV}}{\mrelease{\vLstA}\mtru} \label{Y:9:5}
\end{gather}
$\therefore$ Subcase holds by \eqref{Y:9:4.5} and \eqref{Y:9:5}.
\end{Subcase}

\begin{lastsubcase}[$\rho=`a'$]
The proof for this subcase is analogous to that of the previous subcase.
\end{lastsubcase}
\end{lastcase}
\end{proof}
\end{lemma}

\begin{lemma} \label{lemma3} {\bfseries\itshape (Formula Substitution)} \\[0mm]  $\typeRule{\env{\envVVN{\hVarX}{\envVP}}{\envV}}{\hVV}\; \text{ and } \; \typeRule{\env{\envVV}{\envVP}}{\mmax{\hVarX}{\!\hV}}\,\imp\,\typeRule{\env{\envVVN{\hVarX}{\envVP}}{\envV}}{\hVV\sub{(\clr{\hVarX}\mmax{\hVarX}{\!\hV})}{\hVarX}}$.
\begin{proof}\emph{ By rule induction on  }$\typeRule{\env{\envVVN{\hVarX}{\envVP}}{\envV}}{\hVV}$.\medskip

\begin{Case}[\rtit{tFls}] From our rule premises we know 
\begin{gather}
	\typeRule{\env{\envVVN{\hVarX}{\envVP}}{\envV}}{\mfls} \label{C:1:1} \\ 
	\typeRule{\env{\envVV}{\envVP}}{\mmax{\hVarX}{\!\hV}} \label{C:1:2} 
\end{gather} 
Since $\mfls\neq X$, by substitution we know 
\begin{align}
 \mfls\sub{(\clr{\hVarX}\mmax{\hVarX}{\!\hV})}{\hVarX}\,\equiv\,\mfls \label{C:1:3} 
\end{align} 
By \eqref{C:1:1} and \eqref{C:1:3} we know
\begin{align}
 \typeRule{\env{\envVVN{\hVarX}{\envVP}}{\envV}}{\mfls\sub{(\clr{\hVarX}\mmax{\hVarX}{\!\hV})}{\hVarX}} \label{C:1:4} 
\end{align} 
\noindent $\therefore$ Case holds by \eqref{C:1:4}. 
\end{Case}

\noindent\textbf{Note:} The proof for case \rtit{tTru} is analogous to the proof for case \rtit{tFls}. \bigskip

\begin{Case}[\rtit{tVar}] We must consider two subcases, one where the formula variable Y is equal to X (\ie, the substitution is applied) and another one where Y is not equal to X.\medskip

\begin{Subcase}[$Y=X$] From our premises we know 
\begin{gather} 
	\typeRule{\env{\envVVN{\hVarX}{\envVP}}{\envV}}{\hVarX} \label{C:2:1} \\ 
	\typeRule{\env{\envVV}{\envVP}}{\mmax{\hVarX}{\!\hV}} \label{C:2:2} 
\end{gather}
Since $Y=X$ and by substitution we know  
\begin{align}
 X\sub{(\clr{\hVarX}\mmax{\hVarX}{\!\hV})}{\hVarX}\,\equiv\,(\clr{\hVarX}\mmax{\hVarX}{\!\hV}) \label{C:2:3} 
\end{align} 
By \eqref{C:2:1} and \rtit{tVar} we know 
\begin{gather}
 \envVV(X)\subseteq\envV\; \equiv\; \envVP\subseteq\envV  \label{C:2:4}
\end{gather} 
By \eqref{C:2:2}, \eqref{C:2:4} and transitivity we know
\begin{align} 
	\typeRule{\env{\envVV}{\envV}}{\mmax{\hVarX}{\!\hV}} \label{C:2:5} 
\end{align}
By \eqref{C:2:5} and \rtit{tClr} we know
\begin{align} 
	\typeRule{\env{\envVV}{\envV}}{\clr{X}\mmax{\hVarX}{\!\hV}} \label{C:2:5.5} 
\end{align}
As we work up-to $\alpha$-equivalence we know that $\envVV$ does not contain an entry for formula variable $X$. In light of this, from \eqref{C:2:5.5} we can deduce
\begin{align} 
	\typeRule{\env{\envVVN{\hVarX}{\envVP}}{\envV}}{\clr{X}\mmax{\hVarX}{\!\hV}} \label{C:2:6} 
\end{align}
By \eqref{C:2:3} and \eqref{C:2:6} we know
\begin{align} 
	\typeRule{\env{\envVVN{\hVarX}{\envVP}}{\envV}}{X\sub{(\clr{\hVarX}\mmax{\hVarX}{\!\hV})}{\hVarX}} \label{C:2:7} 
\end{align}
\noindent $\therefore$ Subcase holds by \eqref{C:2:7}.  
\end{Subcase}

\begin{lastsubcase}[$Y\neq X$] From our premises we know 
\begin{gather} 
	\typeRule{\env{\envVVN{\hVarX}{\envVP}}{\envV}}{Y} \label{C:3:1} \\ 
	\typeRule{\env{\envVV}{\envVP}}{\mmax{\hVarX}{\!\hV}} \label{C:3:2} 
\end{gather}
Since $Y\neq X$, by substitution we know 
\begin{align}
	Y\sub{(\clr{\hVarX}\mmax{\hVarX}{\!\hV})}{\hVarX}\,\equiv\,Y \label{C:3:3} 
\end{align} 
By \eqref{C:3:1} and \eqref{C:3:3} we know
\begin{align}
 \typeRule{\env{\envVVN{\hVarX}{\envVP}}{\envV}}{Y\sub{(\clr{\hVarX}\mmax{\hVarX}{\!\hV})}{\hVarX}} \label{C:3:4} 
\end{align} 
\noindent $\therefore$ Subcase holds by \eqref{C:3:4}. 
\end{lastsubcase}
\end{Case}

\begin{Case}[\rtit{tMax}] From our rule premises we know 
\begin{align}
	\typeRule{\env{\envVVN{\hVarX}{\envVP}}{\envV}}{\mmax{Y}{\hVV}} \label{C:4:1} 
\end{align} 
because
\begin{align} 
	\typeRule{\env{\envVVNC{\Sigma}{\{\hVarX\mapsto\envVP,\hVarY\mapsto\envV\}}}{\envV}}{\hVV} \label{C:4:2} 
\end{align} 
and
\begin{align} 
	\typeRule{\env{\envVV}{\envVP}}{\mmax{\hVarX}{\!\hV}} \label{C:4:3} 
\end{align}
By substitution we know 
\begin{align}
	(\mmax{Y}{\hVV})\sub{(\clr{\hVarX}\mmax{\hVarX}{\!\hV})}{\hVarX}\,\equiv\,\mmax{Y}{(\hVV\sub{(\clr{\hVarX}\mmax{\hVarX}{\!\hV})}{\hVarX})}  \label{C:4:4} 
\end{align}
By \eqref{C:4:2}, \eqref{C:4:3} and IH we know 
\begin{align} 
	\typeRule{\env{\envVVNC{\Sigma}{\{\hVarX\mapsto\envVP,\hVarY\mapsto\envV\}}}{\envV}}{\hVV\sub{(\clr{\hVarX}\mmax{\hVarX}{\!\hV})}{\hVarX}} \label{C:4:5} 
\end{align} 
By \eqref{C:4:5} and \rtit{tMax} we know 
\begin{align} 
	\typeRule{\env{\envVVN{\hVarX}{\envVP}}{\envV}}{\mmax{Y}{(\hVV\sub{(\clr{\hVarX}\mmax{\hVarX}{\!\hV})}{\hVarX})}} \label{C:4:6} 
\end{align} 
By \eqref{C:4:4} and \eqref{C:4:6} we know 
\begin{align} 
	\typeRule{\env{\envVVN{\hVarX}{\envVP}}{\envV}}{(\mmax{Y}{\hVV})\sub{(\clr{\hVarX}\mmax{\hVarX}{\!\hV})}{\hVarX}} \label{C:4:7} 
\end{align} 
\noindent $\therefore$ Case holds by \eqref{C:4:7}.  
\end{Case}

\begin{Case}[\rtit{tClr}] The proof for this case is analogous to that of case \rtit{tMax} \end{Case}

\begin{Case}[\rtit{tAdS}] From our rule premises we know 
\begin{align} 
	\typeRule{\env{\envVVN{\hVarX}{\envVP}}{\envV}}{\mscor{\vLstB}{\vLstA}{\hVV}} \label{C:5:1} 
\end{align}
because 
\begin{gather} 
	\envV=\envVPP,\typ{\vLstB}{\lpidb} \label{C:5:1.5}\\
	\typeRule{\env{\envVVN{\hVarX}{\envVP}}{\envV}}{\mrelease{\vLstA}\hVV} \label{C:5:2}
\end{gather}
and
\begin{gather} 	
	\typeRule{\env{\envVV}{\envVP}}{\mmax{\hVarX}{\!\hV}} \label{C:5:3}  
\end{gather}
By substitution we know 
\begin{align}
	\mscor{\vLstB}{\vLstA}{(\hVV\sub{(\clr{\hVarX}\mmax{\hVarX}{\!\hV})}{\hVarX})}\,\equiv\,(\mscor{\vLstB}{\vLstA}{\hVV})\sub{(\clr{\hVarX}\mmax{\hVarX}{\!\hV})}{\hVarX}  \label{C:5:4} 
\end{align}
By \eqref{C:5:2}, \eqref{C:5:3} and IH we know 
\begin{align} 
	\typeRule{\env{\envVVN{\hVarX}{\envVP}}{\envV}}{\mrelease{\vLstA}\hVV\sub{(\clr{\hVarX}\mmax{\hVarX}{\!\hV})}{\hVarX}} \label{C:5:5} 
\end{align} 
By \eqref{C:5:1.5}, \eqref{C:5:5} and \rtit{tAdS} we know 
\begin{align} 
	\typeRule{\env{\envVVN{\hVarX}{\envVP}}{\envV}}{\mscor{\vLstB}{\vLstA}{(\hVV\sub{(\clr{\hVarX}\mmax{\hVarX}{\!\hV})}{\hVarX}})} \label{C:5:6} 
\end{align} 
By \eqref{C:5:4} and \eqref{C:5:6} we know 
\begin{align} 
	\typeRule{\env{\envVVN{\hVarX}{\envVP}}{\envV}}{(\mscor{\vLstB}{\vLstA}{\hVV})\sub{(\clr{\hVarX}\mmax{\hVarX}{\!\hV})}{\hVarX}} \label{C:5:7} 
\end{align} 
\noindent $\therefore$ Case holds by \eqref{C:5:7}.
\end{Case}

\begin{Case}[\rtit{tAdA}] The proof for this case is analogous to the proof of cases \rtit{tAdS}. \end{Case}

\begin{Case}[\rtit{tIf}] From our rule premises we know 
\begin{gather}
	\typeRule{\env{\envVVN{\hVarX}{\envVP}}{\envV}}{\mboolE{c}{\hVV_{1}}{\hVV_{2}}} \label{C:7:1}  
\end{gather}
because
\begin{gather} 
	\typeRule{\env{\envVVN{\hVarX}{\envVP}}{\envV}}{\hVV_{1}} \label{C:7:2} \\
	\typeRule{\env{\envVVN{\hVarX}{\envVP}}{\envV}}{\hVV_{2}} \label{C:7:3}
\end{gather} 
and
\begin{align}
	\typeRule{\env{\envVV}{\envVP}}{\mmax{\hVarX}{\!\hV}} \label{C:7:4} 	
\end{align}
By substitution we know 
\begin{align}
	& (\mboolE{c}{\hVV_{1}}{\hVV_{2}})\sub{(\clr{\hVarX}\mmax{\hVarX}{\!\hV})}{\hVarX} \nonumber\\ \equiv \qquad &\qquad\mboolE{c}{(\hVV_{1}\sub{(\clr{\hVarX}\mmax{\hVarX}{\!\hV})}{\hVarX})}{(\hVV_{2}\sub{(\clr{\hVarX}\mmax{\hVarX}{\!\hV})}{\hVarX})}  \label{C:7:5} 
\end{align}
By \eqref{C:7:2}, \eqref{C:7:4} and IH we know 
\begin{align} 
	\typeRule{\env{\envVVN{\hVarX}{\envVP}}{\envV}}{\hVV_{1}\sub{(\clr{\hVarX}\mmax{\hVarX}{\!\hV})}{\hVarX}} \label{C:7:6} 
\end{align} 
By \eqref{C:7:3}, \eqref{C:7:4} and IH we know 
\begin{align} 
	\typeRule{\env{\envVVN{\hVarX}{\envVP}}{\envV}}{\hVV_{2}\sub{(\clr{\hVarX}\mmax{\hVarX}{\!\hV})}{\hVarX}} \label{C:7:7} 
\end{align} 
By \eqref{C:7:6}, \eqref{C:7:7} and \rtit{tIf} we know 
\begin{align} 
	\typeRule{\env{\envVVN{\hVarX}{\envVP}}{\envV}}{\mboolE{c}{(\hVV_{1}\sub{(\clr{\hVarX}\mmax{\hVarX}{\!\hV})}{\hVarX})}{(\hVV_{2}\sub{(\clr{\hVarX}\mmax{\hVarX}{\!\hV})}{\hVarX}})} \label{C:7:8} 
\end{align} 
By \eqref{C:7:5} and \eqref{C:7:8} we know 
\begin{align} 
	\typeRule{\env{\envVVN{\hVarX}{\envVP}}{\envV}}{(\mboolE{c}{\hVV_{1}}{\hVV_{2}})\sub{(\clr{\hVarX}\mmax{\hVarX}{\!\hV})}{\hVarX}} \label{C:7:9} 
\end{align} 
\noindent $\therefore$ Case holds by \eqref{C:7:9}.  
\end{Case}

\begin{Case}[\rtit{tCn1}] From our rule premises we know 
\begin{gather}
	 \typeRule{\env{\envVVN{\hVarX}{\envVP}}{\envV}}{\hVV_{1}\mand\hVV_{2}} \label{C:8:1}  
\end{gather}
because
\begin{gather} 
	\envV=(\envVA+\envVB) \label{C:8:2} \\
	\typeRule{\env{\envVVN{\hVarX}{\envVP}}{\envVA}}{\hVV_{1}} \label{C:8:3} \\
	\typeRule{\env{\envVVN{\hVarX}{\envVP}}{\envVB}}{\hVV_{2}} \label{C:8:4} \\
	\excl{\hV,\hVV}=\undef  \label{C:8:5}
\end{gather} 
and
\begin{align} 
	\typeRule{\env{\envVV}{\envVP}}{\mmax{\hVarX}{\!\hV}} \label{C:8:7}	
\end{align}
By substitution we know 
\begin{align}
	(\hVV_{1}\mand\hVV_{2})\sub{(\clr{\hVarX}\mmax{\hVarX}{\!\hV})}{\hVarX}\,\equiv\,(\hVV_{1}\sub{(\clr{\hVarX}\mmax{\hVarX}{\!\hV})}{\hVarX})\mand(\hVV_{2}\sub{(\clr{\hVarX}\mmax{\hVarX}{\!\hV})}{\hVarX})  \label{C:8:8} 
\end{align}
By \eqref{C:8:3}, \eqref{C:8:7} and IH we know 
\begin{align} 
	\typeRule{\env{\envVVN{\hVarX}{\envVP}}{\envVA}}{\hVV_{1}\sub{(\clr{\hVarX}\mmax{\hVarX}{\!\hV})}{\hVarX}} \label{C:8:9} 
\end{align} 
By \eqref{C:8:4}, \eqref{C:8:7} and IH we know 
\begin{align} 
	\typeRule{\env{\envVVN{\hVarX}{\envVP}}{\envVB}}{\hVV_{2}\sub{(\clr{\hVarX}\mmax{\hVarX}{\!\hV})}{\hVarX}} \label{C:8:10} 
\end{align} 
By \eqref{C:8:2}, \eqref{C:8:9}, \eqref{C:8:10} and \lemmaYYY we know 
\begin{align} 
	\typeRule{\env{\envVVN{\hVarX}{\envVP}}{(\envVA+\envVB)}}{(\hVV_{1}\sub{(\clr{\hVarX}\mmax{\hVarX}{\!\hV})}{\hVarX})\mand(\hVV_{2}\sub{(\clr{\hVarX}\mmax{\hVarX}{\!\hV})}{\hVarX})} \label{C:8:11} 
\end{align} 
By \eqref{C:8:2}, \eqref{C:8:8} and \eqref{C:8:11} we know 
\begin{align} 
	\typeRule{\env{\envVVN{\hVarX}{\envVP}}{\envV}}{(\hVV_{1}\mand\hVV_{2})\sub{(\clr{\hVarX}\mmax{\hVarX}{\!\hV})}{\hVarX}} \label{C:8:12} 
\end{align} 
\noindent $\therefore$ Case holds by \eqref{C:8:12}.  
\end{Case}

\begin{Case}[\rtit{tCn2}] From our rule premises we know 
\begin{gather}
	 \typeRule{\env{\envVVN{\hVarX}{\envVP}}{\envV}}{\hVV_{1}\mand\hVV_{2}} \label{C:8.5:1}  
\end{gather}
because
\begin{gather} 
	\excl{\hVV_{1},\hVV_{2}}=(\vLstA[1],\vLstA[2]) \label{C:8.5:2} \\
	\typeRule{\env{\envVVN{\hVarX}{\envVP}}{\eff{\envV}{\vLstA[2]}}}{\hVV_{1}} \label{C:8.5:3} \\
	\typeRule{\env{\envVVN{\hVarX}{\envVP}}{\eff{\envV}{\vLstA[1]}}}{\hVV_{2}} \label{C:8.5:4} 
\end{gather} 
and
\begin{align} 
	\typeRule{\env{\envVV}{\envVP}}{\mmax{\hVarX}{\!\hV}} \label{C:8.5:5}	
\end{align}
By substitution we know 
\begin{align}
	(\hVV_{1}\mand\hVV_{2})\sub{(\clr{\hVarX}\mmax{\hVarX}{\!\hV})}{\hVarX}\,\equiv\,(\hVV_{1}\sub{(\clr{\hVarX}\mmax{\hVarX}{\!\hV})}{\hVarX})\mand(\hVV_{2}\sub{(\clr{\hVarX}\mmax{\hVarX}{\!\hV})}{\hVarX})  \label{C:8.5:6} 
\end{align}
By \eqref{C:8.5:3}, \eqref{C:8.5:5} and IH we know 
\begin{align} 
	\typeRule{\env{\envVVN{\hVarX}{\envVP}}{\eff{\envV}{\vLstA[2]}}}{\hVV_{1}\sub{(\clr{\hVarX}\mmax{\hVarX}{\!\hV})}{\hVarX}} \label{C:8.5:7} 
\end{align} 
By \eqref{C:8.5:4}, \eqref{C:8.5:5} and IH we know 
\begin{align} 
	\typeRule{\env{\envVVN{\hVarX}{\envVP}}{\eff{\envV}{\vLstA[1]}}}{\hVV_{2}\sub{(\clr{\hVarX}\mmax{\hVarX}{\!\hV})}{\hVarX}} \label{C:8.5:8}
\end{align} 
Since $\sub{(\clr{\hVarX}\mmax{\hVarX}{\!\hV})}{\hVarX}$, by \eqref{C:8.5:2} and \lemmaZZZ we know 
\begin{align} 
	\excl{\hVV_{1}\sub{(\clr{\hVarX}\mmax{\hVarX}{\!\hV})}{\hVarX},\hVV_{2}\sub{(\clr{\hVarX}\mmax{\hVarX}{\!\hV})}{\hVarX}}=(\vLstA[1],\vLstA[2]) \label{C:8.5:9}
\end{align} 
By \eqref{C:8.5:7}, \eqref{C:8.5:8}, \eqref{C:8.5:9} and \rtit{tCn2} we know 
\begin{align} 
	\typeRule{\env{\envVVN{\hVarX}{\envVP}}{\envV}}{(\hVV_{1}\sub{(\clr{\hVarX}\mmax{\hVarX}{\!\hV})}{\hVarX})\mand(\hVV_{2}\sub{(\clr{\hVarX}\mmax{\hVarX}{\!\hV})}{\hVarX})} \label{C:8.5:10} 
\end{align} 
From \eqref{C:8.5:6} and \eqref{C:8.5:10} we can deduce 
\begin{align} 
	\typeRule{\env{\envVVN{\hVarX}{\envVP}}{\envV}}{(\hVV_{1}\mand\hVV_{2})\sub{(\clr{\hVarX}\mmax{\hVarX}{\!\hV})}{\hVarX}} \label{C:8.5:11} 
\end{align} 
\noindent $\therefore$ Case holds by \eqref{C:8.5:11}.  
\end{Case}

\begin{Case}[\rtit{tNcB}] From our rule premises we know 
\begin{align}
	\typeRule{\env{\envVVN{\hVarX}{\envVP}}{\envV}}{\mBNec{\patE}{\vLstA}{\hVV}} \label{C:9:1}
\end{align}
because
\begin{gather} 	
	\id{\patE}=\varid \label{C:9:2} \\
	\typeRule{\env{\envVVN{\hVarX}{\envVP}}{((\envVN{\tbnd{\patE}}))}}{\mblock{\varid}\hVV} \label{C:9:4} \\
	\typeRule{\env{\envVV}{\envV}}{\mrelease{\vLstA}\mtru}			\label{C:9:5}
\end{gather} 
and
\begin{align} 
	\typeRule{\env{\envVV}{\envVP}}{\mmax{\hVarX}{\!\hV}} \label{C:9:6} 
\end{align}
By substitution we know 
\begin{align}
	(\mBNec{\patE}{\vLstA}{\hVV})\sub{(\clr{\hVarX}\mmax{\hVarX}{\!\hV})}{\hVarX}\,\equiv\,\mBNec{\patE}{\vLstA}{(\hVV\sub{(\clr{\hVarX}\mmax{\hVarX}{\!\hV})}{\hVarX})}  \label{C:9:7} 
\end{align}
By \eqref{C:9:4}, \eqref{C:9:6} and IH we know 
\begin{align} 
	\typeRule{\env{\envVVN{\hVarX}{\envVP}}{(\envVN{\tbnd{\patE}})}}{(\mblock{\varid}\hVV)\sub{(\clr{\hVarX}\mmax{\hVarX}{\!\hV})}{\hVarX}} \label{C:9:8} 
\end{align} 
By substitution we also know 
\begin{align}
	(\mblock{\vLstB}\hVV)\sub{(\clr{\hVarX}\mmax{\hVarX}{\!\hV})}{\hVarX}\,\equiv\,\mblock{\varid}(\hVV\sub{(\clr{\hVarX}\mmax{\hVarX}{\!\hV})}{\hVarX})  \label{C:9:9} 
\end{align}
By \eqref{C:9:8} and \eqref{C:9:9} we know 
\begin{align} 
	\typeRule{\env{\envVVN{\hVarX}{\envVP}}{(\envVN{\tbnd{\patE}})}}{\mblock{\varid}(\hVV\sub{(\clr{\hVarX}\mmax{\hVarX}{\!\hV})}{\hVarX})} \label{C:9:10} 
\end{align} 
By \eqref{C:9:2}, \eqref{C:9:5}, \eqref{C:9:10} \rtit{tNcB} we know 
\begin{align} 
	\typeRule{\env{\envVVN{\hVarX}{\envVP}}{(\envVN{\tbnd{\patE}})}}{\mBNec{\patE}{\vLstA}{(\hVV\sub{(\clr{\hVarX}\mmax{\hVarX}{\!\hV})}{\hVarX})}} \label{C:9:11} 
\end{align} 
By \eqref{C:9:7} and \eqref{C:9:11} we know 
\begin{align} 
	\typeRule{\env{\envVVN{\hVarX}{\envVP}}{(\envVN{\tbnd{\patE}})}}{(\mBNec{\patE}{\vLstA}{\hVV})\sub{(\clr{\hVarX}\mmax{\hVarX}{\!\hV})}{\hVarX}} \label{C:9:12} 
\end{align} 
\noindent $\therefore$ Case holds by \eqref{C:9:12}.  \end{Case}

\begin{Case}[\rtit{tNcA}] The proof for this case is analogous to that of case \rtit{tNcB}. \end{Case}

\begin{Case}[\rtit{tRel}] From our rule premises we know 
\begin{align} 
	\typeRule{\env{\envVVN{\hVarX}{\envVP}}{\envV}}{\mrelease{\vLstB}{\hVV}} \label{C:6:1} 
\end{align}
because 
\begin{gather} 
	\envV=(\envVPPN{\typ{\vLstB}{\lpidb}}) \label{C:6:1.5}\\
	\typeRule{\env{\envVVN{\hVarX}{\envVP}}{\envVPPN{\typ{\vLstB}{\lpid}}}}{\hVV} \label{C:6:2}
\end{gather}
and
\begin{align} 	
	\typeRule{\env{\envVV}{\envVP}}{\mmax{\hVarX}{\!\hV}} \label{C:6:3} 
\end{align}
By substitution we know 
\begin{align}
	\mrelease{\vLstB}{(\hVV\sub{(\clr{\hVarX}\mmax{\hVarX}{\!\hV})}{\hVarX})}\,\equiv\,(\mrelease{\vLstB}{\hVV})\sub{(\clr{\hVarX}\mmax{\hVarX}{\!\hV})}{\hVarX}  \label{C:6:4} 
\end{align}
By \eqref{C:6:2}, \eqref{C:6:3} and IH we know 
\begin{align} 
	\typeRule{\env{\envVVN{\hVarX}{\envVP}}{\envVPPN{\typ{\vLstB}{\lpid}}}}{\hVV\sub{(\clr{\hVarX}\mmax{\hVarX}{\!\hV})}{\hVarX}} \label{C:6:5} 
\end{align} 
By \eqref{C:6:1.5}, \eqref{C:6:5} and \rtit{tRel} we know 
\begin{align} 
	\typeRule{\env{\envVVN{\hVarX}{\envVP}}{\envV}}{\mrelease{\vLstB}{(\hVV\sub{(\clr{\hVarX}\mmax{\hVarX}{\!\hV})}{\hVarX}})} \label{C:6:6} 
\end{align} 
By \eqref{C:6:4} and \eqref{C:6:6} we know 
\begin{align} 
	\typeRule{\env{\envVVN{\hVarX}{\envVP}}{\envV}}{(\mrelease{\vLstB}{\hVV})\sub{(\clr{\hVarX}\mmax{\hVarX}{\!\hV})}{\hVarX}} \label{C:6:7} 
\end{align} 
\noindent $\therefore$ Case holds by \eqref{C:6:7}.
\end{Case}

\begin{lastcase}[\rtit{tBlk}] From our rule premises we know 
\begin{align}
	\typeRule{\env{\envVVN{\hVarX}{\envVP}}{\envV}}{\mblock{\vLstB}\hVV} \label{C:10:1}
\end{align}
because
\begin{gather} 	
	\envV=(\envVPPN{\typ{\vLstB}{\lpid}}) \label{C:10:1.5} \\
	\typeRule{\env{\envVVN{\hVarX}{\envVP}}{\envVPPN{\typ{\vLstB}{\lpidb}}}}{\hVV} \label{C:10:2} 
\end{gather} 
and
\begin{align} 
	\typeRule{\env{\envVV}{\envVP}}{\mmax{\hVarX}{\!\hV}} \label{C:10:3}	
\end{align}
By substitution we know 
\begin{align}
	(\mblock{\vLstB}\hVV)\sub{(\clr{\hVarX}\mmax{\hVarX}{\!\hV})}{\hVarX}\,\equiv\,\mblock{\vLstB}(\hVV\sub{(\clr{\hVarX}\mmax{\hVarX}{\!\hV})}{\hVarX})  \label{C:10:4} 
\end{align}
By \eqref{C:10:2}, \eqref{C:10:3} and IH we know 
\begin{align} 
	\typeRule{\env{\envVVN{\hVarX}{\envVP}}{\envVPPN{\typ{\vLstB}{\lpidb}}}}{\hVV\sub{(\clr{\hVarX}\mmax{\hVarX}{\!\hV})}{\hVarX}} \label{C:10:5} 
\end{align} 
By \eqref{C:10:1.5}, \eqref{C:10:5} and \rtit{tBlk} we know 
\begin{align} 
	\typeRule{\env{\envVVN{\hVarX}{\envVP}}{\envV}}{\mblock{\vLstB}(\hVV\sub{(\clr{\hVarX}\mmax{\hVarX}{\!\hV})}{\hVarX})} \label{C:10:6} 
\end{align} 
By \eqref{C:10:4} and \eqref{C:10:6} we know 
\begin{align} 
	\typeRule{\env{\envVVN{\hVarX}{\envVP}}{\envV}}{(\mblock{\vLstB}\hVV)\sub{(\clr{\hVarX}\mmax{\hVarX}{\!\hV})}{\hVarX}} \label{C:10:7} 
\end{align} 
\noindent $\therefore$ Case holds by \eqref{C:10:7}. 
\end{lastcase}
\end{proof}

\end{lemma} \medskip
 
 \begin{lemma}\label{lemmaI} {\bfseries\itshape (Weakening)} $\\$\indent\indent\indent\indent$\typeRule{\env{\envVVN{\hVarX}{\envVP}}{\envV}}{\hV}\; \text{ and } \; \hVarX\notin\fv(\hV)\, \imp \typeRule{\env{\envVV}{\envV}}{\hV}$.
\begin{proof} By rule induction on $\typeRule{\env{\envVVN{\hVarX}{\envVP}}{\envV}}{\hV}$.

	\begin{Case}[\rtit{tNcA}]	From our rule premises we know 
		\begin{gather}
			\typeRule{\env{\envVVN{\hVarX}{\envVP}}{\envV}}{\mANec{\patE}{\vLstA}\hV} \label{I:1:1} 
		\end{gather}
		because
		\begin{gather}
			\typeRule{\env{\envVVN{\hVarX}{\envVP}}{(\envVN{\tbnd{\patE}})}}{\hV} \label{I:1:2} \\
			\typeRule{\env{\envVVN{\hVarX}{\envVP}}{\envV}}{\mrelease{\vLstA}\mtru} \label{I:1:3} 
		\end{gather}
		and
		\begin{gather}
			\hVarX\notin\fv(\mANec{\patE}{\vLstA}\hV) \label{I:1:4} 
		\end{gather}
		From \eqref{I:1:4} we can deduce 
		\begin{gather}
			\hVarX\notin\fv(\hV) \label{I:1:5} 
		\end{gather}
		Since formula variable $\hVarX$ does not feature in formula $\mrelease{\vLstA}\mtru$ we can therefore deduce
		\begin{gather}
			\hVarX\notin\fv(\mrelease{\vLstA}\mtru) \label{I:1:7} 
		\end{gather}
		By \eqref{I:1:2}, \eqref{I:1:5} and IH we know 
		\begin{gather}
			\typeRule{\env{\envVV}{(\envVN{\tbnd{\patE}})}}{\hV} \label{I:1:6} 
		\end{gather}
		By \eqref{I:1:3}, \eqref{I:1:7} and IH we know 
		\begin{gather}
			\typeRule{\env{\envVV}{\envV}}{\mrelease{\vLstA}\mtru} \label{I:1:8} 
		\end{gather}
		By \eqref{I:1:6}, \eqref{I:1:8} and \rtit{tNcA} we know
		\begin{gather}
			\typeRule{\env{\envVV}{\envV}}{\mANec{\patE}{\vLstA}\hV} \label{I:1:9} 
		\end{gather}
		\noindent$\therefore$ Case holds by \eqref{I:1:9}.
	\end{Case}
	
	\begin{Case}[\rtit{tNcB}] The proof for this case is analogous to that of case \rtit{tNcA}. \end{Case} 
	
	\begin{Case}[\rtit{tFls}]	From our rule premises we know 
		\begin{gather}
			\typeRule{\env{\envVVN{\hVarX}{\envVP}}{\envV}}{\mfls} \label{I:2:1} \\
			\hVarX\notin\fv(\mfls) \label{I:2:2} 
		\end{gather}
		Since \eqref{I:2:1} holds (typechecks) regardless of the contents of the formula environment $\envVV$ we can conclude $\typeRule{\env{\envVV}{\envV}}{\mfls}$ as required.
	\end{Case}
		
	\begin{Case}[\rtit{tTru}] The proof for this case is analogous to that of case \rtit{tFls}. \end{Case}
	
	\begin{Case}[\rtit{tIf}]	From our rule premises we know 
		\begin{gather}
			\typeRule{\env{\envVVN{\hVarX}{\envVP}}{\envV}}{\mboolE{b}{\hV}{\hVV}} \label{I:3:1} 
		\end{gather}
		because
		\begin{gather}
			\typeRule{\env{\envVVN{\hVarX}{\envVP}}{\envV}}{\hV} \label{I:3:2} \\
			\typeRule{\env{\envVVN{\hVarX}{\envVP}}{\envV}}{\hVV} \label{I:3:3} 
		\end{gather}
		and
		\begin{gather}
			\hVarX\notin\fv(\mboolE{b}{\hV}{\hVV}) \label{I:3:4} 
		\end{gather}
		From \eqref{I:3:4} we can deduce 
		\begin{gather}
			\hVarX\notin\fv(\hV) \label{I:3:5} \\
			\hVarX\notin\fv(\hVV) \label{I:3:6} 
		\end{gather}		
		By \eqref{I:3:2}, \eqref{I:3:5} and IH we know 
		\begin{gather}
			\typeRule{\env{\envVV}{\envV}}{\hV} \label{I:3:7} 
		\end{gather}
		By \eqref{I:3:3}, \eqref{I:3:6} and IH we know 
		\begin{gather}
			\typeRule{\env{\envVV}{\envV}}{\hVV} \label{I:3:8} 
		\end{gather}
		By \eqref{I:3:7}, \eqref{I:3:8} and \rtit{tIf} we know
		\begin{gather}
			\typeRule{\env{\envVV}{\envV}}{\mboolE{b}{\hV}{\hVV}} \label{I:3:9} 
		\end{gather}
		\noindent$\therefore$ Case holds by \eqref{I:3:9}.
	\end{Case}
	
	\begin{Case}[\rtit{tCn1} and \rtit{tCn2}] The proofs for these cases are analogous to that of case \rtit{tIf}. \end{Case}
	
	\begin{Case}[\rtit{tBlk}]	From our rule premises we know 
		\begin{gather}
			\typeRule{\env{\envVVN{\hVarX}{\envVP}}{\envV}}{\mblock{\vLstB}{\hV}} \label{I:4:1} 
		\end{gather}
		because
		\begin{gather}
			\typeRule{\env{\envVVN{\hVarX}{\envVP}}{(\envVPPN{\typ{\vLstB}{\lpidb}})}}{\hV} \label{I:4:2} \\
			\envV=\envVPPN{\typ{\vLstB}{\lpid}} \label{I:4:3} 
		\end{gather}
		and
		\begin{gather}
			\hVarX\notin\fv(\mblock{\vLstB}{\hV}) \label{I:4:4} 
		\end{gather}
		From \eqref{I:4:4} we can deduce 
		\begin{gather}
			\hVarX\notin\fv(\hV) \label{I:4:5} 
		\end{gather}		
		By \eqref{I:4:2}, \eqref{I:4:5} and IH we know 
		\begin{gather}
			\typeRule{\env{\envVV}{(\envVPPN{\typ{\vLstB}{\lpidb}})}}{\hV} \label{I:4:6} 
		\end{gather}
		By \eqref{I:4:3}, \eqref{I:4:6} and \rtit{tBlk} we know 
		\begin{gather}
			\typeRule{\env{\envVV}{\envV}}{\mblock{\vLstB}{\hV}} \label{I:4:7} 
		\end{gather}		
		\noindent$\therefore$ Case holds by \eqref{I:4:7}.
	\end{Case}
	
	\begin{Case}[\rtit{tRel}, \rtit{tAdA} and \rtit{tAdS}] The proofs for these cases are analogous to that of case \rtit{tBlk}. \end{Case}

	\begin{Case}[\rtit{tMax}]	From our rule premises we know 
		\begin{gather}
			\typeRule{\env{\envVVN{\hVarX}{\envVP}}{\envV}}{\mmax{\hVarY}{\hV}} \label{I:5:1} 
		\end{gather}
		because
		\begin{gather}
			\typeRule{\env{(\envVVNC{\envVV}{\{\hVarX\mapsto\envVP,\hVarY\mapsto\envV\}})}{\envVP}}{\hV} \label{I:5:2} 
		\end{gather}
		and
		\begin{gather}
			\hVarX\notin\fv(\mmax{\hVarY}{\hV}) \label{I:5:3} 
		\end{gather}
		From \eqref{I:5:3} we can deduce 
		\begin{gather}
			\hVarX\notin\fv(\hV) \label{I:5:4} 
		\end{gather}		
		By \eqref{I:5:2}, \eqref{I:5:4} and IH we know 
		\begin{gather}
			\typeRule{\env{\envVVN{\hVarY}{\envV}}{\envV}}{\hV} \label{I:5:5} 
		\end{gather}
		By \eqref{I:5:5} and \rtit{tMax} we know 
		\begin{gather}
			\typeRule{\env{\envVV}{\envV}}{\mmax{\hVarY}{\hV}} \label{I:5:6} 
		\end{gather}		
		\noindent$\therefore$ Case holds by \eqref{I:5:6}.
	\end{Case}
	
	\begin{Case}[\rtit{tClr}] The proofs for this case is analogous to that of case \rtit{tMax}. \end{Case}

	\begin{lastcase}[\rtit{tVar}]	From our rule premises we know 
		\begin{gather}
			\typeRule{\env{\envVVN{\hVarX}{\envVP}}{\envV}}{\hVarY} \label{I:6:1} 
		\end{gather}
		because
		\begin{gather}
			\envVVN{\hVarX}{\envVP}(\hVarY)\,\subseteq\,\envV \label{I:6:2} 
		\end{gather}
		and
		\begin{gather}
			\hVarX\notin\fv(\hVarY) \label{I:6:3} 
		\end{gather}
		We now consider two subcases:\\
		
		\begin{Subcase}[$\hVarY\neq\hVarX$]
			Since $\hVarY\neq\hVarX$ we know that \eqref{I:6:2} still holds if we remove entry $\hVarX\mapsto\envVP$ from the formula environment, and hence we deduce
			\begin{gather}
				\envVV(\hVarY)\,\subseteq\,\envV \label{I:6:4} 
			\end{gather}
			Hence by \eqref{I:6:4} and \rtit{tVar} we can conclude $\typeRule{\env{\envVV}{\envV}}{\hVarY}$ as required.
		\end{Subcase}
		
		\begin{lastsubcase}[$\hVarY\eq\hVarX$]
			Since $\hVarY\eq\hVarX$ we know that \eqref{I:6:3} cannot ever be true. Hence this case holds by contradiction.
		\end{lastsubcase}		
	\end{lastcase}
\end{proof}
\end{lemma}
 
\begin{lemma}\label{lemma4} {\bfseries\itshape (Term Substitution)}  $\\$\indent$\typeRule{\env{\envVV}{(\envVN{\tbnd{\patE}}})}{\hV} \text{ and } \match{\patE}{\actE}{\sigma} \text{ and } \dtcheck\\$\indent$\qquad  \text{ and } \dtcheckB \imp\,\typeRule{\env{\envVV}{(\envVN{\tbnd{\patE}}\sigma)}}{\hV\sigma}$
\begin{proof}\emph{ By rule induction on  }$\typeRule{\env{\envVV}{(\envVN{\tbnd{\patE}}\sigma)}}{\hV}$.

\begin{Case}[\rtit{tFls}] From our rule premises we know 
\begin{gather}
	\typeRule{\env{\envVV}{(\envVN{\tbnd{\patE}})}}{\mfls} \label{D:1:1} \\ 
	\match{\patE}{\actE}{\sigma} \label{D:1:2} \\
	\dtcheck  \label{D:1:3} \\
	\dtcheckB  \label{D:1:3.5}
\end{gather} 
Since $\mfls$ has no term variables that can be substituted, we know
\begin{align}
 \mfls\sigma\,\equiv\,\mfls \label{D:1:4} 
\end{align} 
By \eqref{D:1:1}, \eqref{D:1:4} and given that $\mfls$ typechecks with any type environment $\envV$ we can conclude
\begin{align}
 \typeRule{\env{\envVV}{(\envVN{\tbnd{\patE}\sigma})}}{\mfls\sigma} \label{D:1:5} 
\end{align} 
\noindent $\therefore$ Case holds by \eqref{D:1:5}. 
\end{Case}

\noindent\textbf{Note:} The proof for cases \rtit{tTru} and \rtit{tVar} are analogous to the proof for case \rtit{tFls}. \medskip

\begin{Case}[\rtit{tMax}] From our rule premises we know 
\begin{align}
	\typeRule{\env{\envVV}{(\envVN{\tbnd{\patE}})}}{\mmax{\hVarX}{\!\hV}} \label{D:4:1} 
\end{align} 
because
\begin{align} 
	\typeRule{\env{\envVV}{(\envVN{\tbnd{\patE}})}}{\hV} \label{D:4:2} 
\end{align} 
and
\begin{gather} 
	\match{\patE}{\actE}{\sigma} \label{D:4:3} \\
	\dtcheck  \label{D:4:4} \\
	\dtcheckB  \label{D:4:4.5}
\end{gather}
By \eqref{D:4:2}, \eqref{D:4:3}, \eqref{D:4:4}, \eqref{D:4:4.5} and IH we know 
\begin{align} 
	\typeRule{\env{\envVV}{(\envVN{\tbnd{\patE}\sigma})}}{(\hV\sigma)} \label{D:4:5} 
\end{align} 
By \eqref{D:4:5} and \rtit{tMax} we know 
\begin{align} 
	\typeRule{\env{\envVV}{(\envVN{\tbnd{\patE}\sigma})}}{\mmax{\hVarX}{(\hV\sigma)}} \label{D:4:6} 
\end{align} 
By \eqref{D:4:6} and data substitution we know 
\begin{align} 
	\typeRule{\env{\envVV}{(\envVN{\tbnd{\patE}\sigma})}}{(\mmax{\hVarX}{\!\hV})\sigma} \label{D:4:7} 
\end{align} 
\noindent $\therefore$ Case holds by \eqref{D:4:7}.  
\end{Case}

\begin{Case}[\rtit{tClr}] The proof for this case is analogous to that of case \rtit{tMax} \end{Case}

\begin{Case}[\rtit{tAdS}] From our rule premises we know 
\begin{align} 
	\typeRule{\env{\envVV}{(\envVN{\tbnd{\patE}})}}{\mscor{\vLstB}{\vLstA}{\hV}} \label{D:6:1} 
\end{align}
because 
\begin{gather} 
	\envVN{\tbnd{\patE}} = \envVPN{\typ{\vLstB}{\lpidb}} \label{D:6:2}\\
	\typeRule{\env{\envVV}{(\envVN{\tbnd{\patE}})}}{\mrelease{\vLstA}\hV} \label{D:6:3}
\end{gather}
and
\begin{gather} 	
	\match{\patE}{\actE}{\sigma} \label{D:6:4} \\ 
	\dtcheck  \label{D:6:5} \\
	\dtcheckB  \label{D:6:5.5}
\end{gather}
By \eqref{D:6:3}, \eqref{D:6:4}, \eqref{D:6:5}, \eqref{D:6:5.5} and IH we know 
\begin{align} 
	\typeRule{\env{\envVV}{(\envVN{\tbnd{\patE}\sigma})}}{(\mrelease{\vLstA}\hV)\sigma} \label{D:6:6} 
\end{align} 
By \eqref{D:6:5} and \eqref{D:6:5.5} we know that we do not have any aliasing, hence we can safely apply substitution $\sigma$ on both sides of equation \eqref{D:6:2}, such that we know
\begin{gather} 
	\envVN{\tbnd{\patE}}\sigma=(\envVPN{\typ{\vLstB}{\lpidb}})\sigma  \label{D:6:7}
\end{gather} 
By \eqref{D:6:6}, \eqref{D:6:7} and \rtit{tAdS} we know 
\begin{align} 
	\typeRule{\env{\envVV}{(\envVN{\tbnd{\patE}\sigma})}}{\mscor{\vLstB}{\vLstA}{(\hV\sigma)}} \label{D:6:8} 
\end{align} 
By \eqref{D:6:8} and data substitution we know 
\begin{align} 
	\typeRule{\env{\envVV}{(\envVN{\tbnd{\patE}\sigma})}}{(\mscor{\vLstB}{\vLstA}{\hV})\sigma} \label{D:6:9} 
\end{align} 
\noindent $\therefore$ Case holds by \eqref{D:6:9}. 
\end{Case}

\begin{Case}[\rtit{tAdA}] The proof for this case is very similar to the proof of cases \rtit{tAdS}.\end{Case}

\begin{Case}[\rtit{tIf}] From our rule premises we know 
\begin{gather}
	\typeRule{\env{\envVV}{(\envVN{\tbnd{\patE}})}}{\mboolE{c}{\hV_{1}}{\hV_{2}}} \label{D:7:1}  
\end{gather}
because
\begin{gather} 
	\typeRule{\env{\envVV}{(\envVN{\tbnd{\patE}})}}{\hV_{1}} \label{D:7:2} \\
	\typeRule{\env{\envVV}{(\envVN{\tbnd{\patE}})}}{\hV_{2}} \label{D:7:3}
\end{gather} 
and
\begin{gather}
	\match{\patE}{\actE}{\sigma}		 \label{D:7:4}\\
	\dtcheck	\label{D:7:5} \\
	\dtcheckB  	\label{D:7:5.5}
\end{gather}
By \eqref{D:7:2}, \eqref{D:7:4}, \eqref{D:7:5}, \eqref{D:7:5.5} and IH we know 
\begin{align} 
	\typeRule{\env{\envVV}{(\envVN{\tbnd{\patE}\sigma})}}{\hV_{1}\sigma} \label{D:7:6} 
\end{align} 
By \eqref{D:7:3}, \eqref{D:7:4}, \eqref{D:7:5}, \eqref{D:7:5.5} and IH we know 
\begin{align} 
	\typeRule{\env{\envVV}{(\envVN{\tbnd{\patE}\sigma})}}{\hV_{2}\sigma} \label{D:7:7} 
\end{align} 
By \eqref{D:7:6}, \eqref{D:7:7} and \rtit{tIf} we know 
\begin{align} 
	\typeRule{\env{\envVV}{(\envVN{\tbnd{\patE}\sigma})}}{\mboolE{c\sigma}{(\hV_{1}\sigma)}{(\hV_{2}\sigma)}} \label{D:7:8} 
\end{align} 
By \eqref{D:7:8} and data substitution we know 
\begin{align} 
	\typeRule{\env{\envVV}{(\envVN{\tbnd{\patE}\sigma})}}{(\mboolE{c}{\hV_{1}}{\hV_{2}})\sigma} \label{D:7:9} 
\end{align} 
\noindent $\therefore$ Case holds by \eqref{D:7:9}. 
\end{Case}

\begin{Case}[\rtit{tCn1}] From our rule premises we know 
\begin{gather}
	 \typeRule{\env{\envVV}{(\envVN{\tbnd{\patE}})}}{\hV_{1}\mand\hV_{2}} \label{D:8:1}  
\end{gather}
because
\begin{gather} 
	\envVN{\tbnd{\patE}}=(\envVA+\envVB) \label{D:8:2} \\
	\typeRule{\env{\envVV}{\envVA}}{\hV_{1}} \label{D:8:3} \\
	\typeRule{\env{\envVV}{\envVB}}{\hV_{2}} \label{D:8:4} \\
	\excl{\hV,\hVV}=\undef  \label{D:8:5} 
\end{gather} 
and
\begin{gather} 
	\match{\patE}{\actE}{\sigma}  \label{D:8:7}	\\
	\dtcheck  \label{D:8:8} \\
	\dtcheckB  \label{D:8:8.5}	
\end{gather}
By \eqref{D:8:3}, \eqref{D:8:7}, \eqref{D:8:8}, \eqref{D:8:8.5} and IH we know 
\begin{align} 
	\typeRule{\env{\envVV}{(\envVA\sigma)}}{\hV_{1}\sigma} \label{D:8:9} 
\end{align} 
By \eqref{D:8:4}, \eqref{D:8:7}, \eqref{D:8:8}, \eqref{D:8:8.5} and IH we know 
\begin{align} 
	\typeRule{\env{\envVV}{(\envVB\sigma)}}{\hV_{2}\sigma} \label{D:8:10} 
\end{align} 
By \eqref{D:8:8} and \eqref{D:8:8.5} we know that we do not have any aliasing, hence we can safely apply substitution $\sigma$ on both sides of equation \eqref{D:8:2}, such that we know
\begin{gather} 
	\envVN{\tbnd{\patE}}\sigma=(\envVA+\envVB)\sigma  \label{D:8:12}
\end{gather} 
By \eqref{D:8:9}, \eqref{D:8:10}, \eqref{D:8:12} and \lemmaYYY we know 
\begin{align} 
	\typeRule{\env{\envVV}{(\envVA+\envVB)\sigma}}{(\hV_{1}\sigma)\mand(\hV_{2}\sigma)} \label{D:8:13} 
\end{align} 
By \eqref{D:8:2}, \eqref{D:8:13} and data substitution we know
\begin{align} 
	\typeRule{\env{\envVV}{(\envVN{\tbnd{\patE}})\sigma}}{(\hV_{1}\mand\hV_{2})\sigma} \label{D:8:14} 
\end{align} 
\noindent $\therefore$ Case holds by \eqref{D:8:14}.
\end{Case}

\begin{Case}[\rtit{tCn2}] From our rule premises we know 
\begin{gather}
	 \typeRule{\env{\envVV}{(\envVN{\tbnd{\patE}})}}{\hV_{1}\mand\hV_{2}} \label{D:8.5:1}  
\end{gather}
because
\begin{gather} 
	\excl{\hVV_{1},\hVV_{2}}=(\vLstA[1],\vLstA[2]) \label{D:8.5:2} \\
	\typeRule{\env{\envVV}{\eff{(\envVN{\tbnd{\patE}})}{\vLstA[2]}}}{\hV_{1}} \label{D:8.5:3} \\
	\typeRule{\env{\envVV}{\eff{(\envVN{\tbnd{\patE}})}{\vLstA[1]}}}{\hV_{2}} \label{D:8.5:4}
\end{gather} 
and
\begin{gather} 
	\match{\patE}{\actE}{\sigma}  \label{D:8.5:5}	\\
	\dtcheck  \label{D:8.5:6} \\
	\dtcheckB  \label{D:8.5:6.5}	
\end{gather}
As by \eqref{D:8.5:2} we know that patterns $\hVV_{1}$ and $\hVV_{2}$ are exclusive, if we add more information by applying substitution \eqref{D:8.5:5}, we \emph{do not affect} their mutual exclusion. Moreover, the substitution can safely be applied as by \eqref{D:8.5:6} and \eqref{D:8.5:6.5} we know that we do not have aliasing. Hence we can deduce 
\begin{align} 
	\excl{\hV_{1}\sigma,\hV_{2}\sigma}=(\vLstA[1]\sigma,\vLstA[2]\sigma) \label{D:8.5:7}
\end{align} 
By \eqref{D:8.5:3}, \eqref{D:8.5:5}, \eqref{D:8.5:6}, \eqref{D:8.5:6.5} and IH we know 
\begin{align} 
	\typeRule{\env{\envVV}{\eff{\envVN{\tbnd{\patE}}}{\vLstA[2]}\sigma}}{\hV_{1}} \label{D:8.5:8} 
\end{align} 
By \eqref{D:8.5:4}, \eqref{D:8.5:5}, \eqref{D:8.5:6}, \eqref{D:8.5:6.5} and IH we know 
\begin{align} 
	\typeRule{\env{\envVV}{\eff{\envVN{\tbnd{\patE}}}{\vLstA[1]}\sigma}}{\hV_{2}} \label{D:8.5:9} 
\end{align} 
By \eqref{D:8.5:7}, \eqref{D:8.5:8}, \eqref{D:8.5:9} and \rtit{tCn2} we know 
\begin{align} 
	\typeRule{\env{\envVV}{(\envVN{\tbnd{\patE}\sigma})}}{(\hV_{1}\sigma)\mand(\hV_{2}\sigma)} \label{D:8.5:10} 
\end{align} 
By \eqref{D:8.5:10} and substitution we know 
\begin{align} 
	\typeRule{\env{\envVV}{(\envVN{\tbnd{\patE}\sigma}})}{(\hV_{1}\mand\hV_{2})\sigma} \label{D:8.5:11} 
\end{align} 
\noindent $\therefore$ Case holds by \eqref{D:8.5:11}.
\end{Case}

\begin{Case}[\rtit{tNcB}] From our rule premises we know 
\begin{align}
	\typeRule{\env{\envVV}{(\envVN{\tbnd{\patE}})}}{\mBNec{\patEE}{\vLstA}{\hV}} \label{D:9:1}
\end{align}
because
\begin{gather} 	
	\id{\patEE}=\varid \label{D:9:2} \\
	\typeRule{\env{\envVV}{((\envVN{\tbnd{\patE}}),\tbnd{\patEE})}}{\mblock{\varid}\hV} \label{D:9:4} \\
	\typeRule{\env{\envVV}{(\envVN{\tbnd{\patE}})}}{\mrelease{\vLstA}\mtru}			\label{D:9:5}
\end{gather} 
and
\begin{gather} 
	\match{\patE}{\actE}{\sigma} \label{D:9:6} \\
	\dtcheck  \label{D:9:7} \\
	\dtcheckB  \label{D:9:7.5}		
\end{gather}
By \eqref{D:9:7} and \eqref{D:9:7.5} we know that we do not have any aliasing, hence we can safely apply substitution $\sigma$ on both sides of equation \eqref{D:9:2}, such that we know
\begin{align}
	\id{\patEE\sigma}=\varid\sigma \label{D:9:8} 
\end{align}
By \eqref{D:9:4}, \eqref{D:9:6}, \eqref{D:9:7}, \eqref{D:9:7.5} and IH we know 
\begin{align} 
	&\typeRule{\env{\envVV}{((\envVN{\tbnd{\patE}}),\tbnd{\patEE})\sigma}}{(\mblock{\varid}\hV)\sigma}  \nonumber \\ 
	\equiv  \qquad &\typeRule{\env{\envVV}{((\envVN{\tbnd{\patE}\sigma}),\tbnd{\patEE}\sigma)}}{\mblock{\varid\sigma}(\hV\sigma)} \label{D:9:10} 
\end{align} 
By \eqref{D:9:5}, \eqref{D:9:6}, \eqref{D:9:7}, \eqref{D:9:7.5} and IH we know 
\begin{align} 
	\typeRule{\env{\envVV}{(\envVN{\tbnd{\patE}\sigma})}}{\mrelease{\vLstA\sigma}\mtru} \label{D:9:11}
\end{align}
By \eqref{D:9:8}, \eqref{D:9:10}, \eqref{D:9:11} and \rtit{tNcB} we know 
\begin{align} 
	\typeRule{\env{\envVV}{(\envVN{\tbnd{\patE}\sigma})}}{\mBNec{\patEE}{\vLstA}{\!\sigma\,}{(\hV\sigma)}} \label{D:9:12} 
\end{align} 
By \eqref{D:9:12} and data substitution we know 
\begin{align} 
	\typeRule{\env{\envVV}{(\envVN{\tbnd{\patE}\sigma})}}{(\mBNec{\patEE}{\vLstA}{\hV})\sigma} \label{D:9:13} 
\end{align} 
\noindent $\therefore$ Case holds by \eqref{D:9:13}. 
\end{Case}

\begin{Case}[\rtit{tNcA}] The proof for this case is analogous to that of case \rtit{tNcB}. \end{Case}

\begin{Case}[\rtit{tBlk}] From our rule premises we know 
\begin{align}
	\typeRule{\env{\envVV}{(\envVN{\tbnd{\patE}})}}{\mblock{\vLstB}\hV} \label{D:10:1}
\end{align}
because
\begin{gather} 	
	\envVN{\tbnd{\patE}}=\envVPN{\typ{\vLstB}{\lpid}} \label{D:10:2} \\
	\typeRule{\env{\envVV}{(\envVPN{\typ{\vLstB}{\lpidb}})}}{\hV} \label{D:10:3}
\end{gather} 
and
\begin{gather} 
	\match{\patE}{\actE}{\sigma} \label{D:10:4} \\
	\dtcheck  \label{D:10:5} \\
	\dtcheckB  \label{D:10:5.5}	
\end{gather}
By \eqref{D:10:3}, \eqref{D:10:4}, \eqref{D:10:5}, \eqref{D:10:5.5} and IH we know 
\begin{align} 
	\typeRule{\env{\envVV}{(\envVPN{\typ{\vLstB}{\lpidb}})\sigma}}{\hV\sigma} \label{D:10:6} 
\end{align} 
By \eqref{D:10:5} and \eqref{D:10:5.5} we know that we do not have any aliasing, hence we can safely apply substitution $\sigma$ on both sides of equation \eqref{D:10:2}, such that we know
\begin{align}
	\envVN{\tbnd{\patE}\sigma}=(\envVPN{\typ{\vLstB}{\lpid}})\sigma \label{D:10:7} 
\end{align}
By \eqref{D:10:6}, \eqref{D:10:7} and \rtit{tBlk} we know 
\begin{align} 
	\typeRule{\env{\envVV}{(\envVN{\tbnd{\patE}\sigma})}}{\mblock{\vLstB}(\hV\sigma)} \label{D:10:8} 
\end{align} 
By \eqref{D:10:8} and data substitution we know
\begin{align} 
	\typeRule{\env{\envVV}{(\envVN{\tbnd{\patE}\sigma})}}{(\mblock{\vLstB}\hV)\sigma} \label{D:10:9} 
\end{align} 
\noindent $\therefore$ Case holds by \eqref{D:10:9}. 
\end{Case}

\begin{lastcase}[\rtit{tRel}] The proof for this case is analogous to that of case \rtit{tBlk}. \end{lastcase}
\end{proof}
\end{lemma} \medskip

\section[Proving The Secondary Soundness Auxiliary Lemmas]{\setstretch{1.2}Proving The Secondary Soundness Auxiliary \\ Lemmas} \label{sec:other-aux-lemmas}

\begin{lemma}\label{lemmaYYY} $\typeRule{\env{\envVV}{\envVA}}{\hV}\; \text{ and } \; \typeRule{\env{\envVV}{\envVB}}{\hVV}\; \text{ and } \; \envV=\envVA+\envVB\, \imp \,\typeRule{\env{\envVV}{\envV}}{\hV\mand\hVV}$.
\begin{proof} \setstretch{1}
	From our premises we know 
	\begin{gather}
		\typeRule{\env{\envVV}{\envVA}}{\hV} \label{YYY:1} \\
		\typeRule{\env{\envVV}{\envVB}}{\hVV} \label{YYY:2} \\
		\envV=\envVA+\envVB \label{YYY:3} 
	\end{gather}
	We consider two cases:\\
	
	\begin{Case}[$\excl{\hV,\hVV}=\undef$]
		By \eqref{YYY:1}, \eqref{YYY:2}, \eqref{YYY:3} and \rtit{tCn1} we know 
		\begin{align}
			\typeRule{\env{\envVV}{\envV}}{\hV\mand\hVV} \label{YYY:7} 
		\end{align}	
	\noindent$\therefore$ Case holds by \eqref{YYY:7}.		
	\end{Case} \medskip
	
	\begin{lastcase}[${\excl{\hV,\hVV}=(\vLstA[\hV],\vLstA[\hVV])}$]
		By \eqref{YYY:1} and since $\excl{\hV,\hVV}=(\vLstA[\hV],\vLstA[\hVV])$ we can deduce that since $\vLstA[\hV]$ represents a \emph{set of processes} that \emph{may be released by }$\hV$ in case this is \emph{trivially satisfied}, then we know that these processes are \emph{linear typed}. Hence since by \eqref{YYY:3} we know that $\envVA$ and $\envVB$ do not share linear typed references we can conclude
		\begin{align}
			\forall\varid\in\vLstA[\hV]\cdot\typ{\varid}{\Lin}\in\envVA\,\land\,\typ{\varid}{\Lin}\notin\envVB \;\; (\text{where }\Lin=\lpid \text{ or } \lpidb)\label{YYY:8} 
		\end{align}
		Same argument applies for \eqref{YYY:2}, such that we know
		\begin{align}
			\forall\varid\in\vLstA[\hVV]\cdot\typ{\varid}{\Lin}\in\envVB\,\land\,\typ{\varid}{\Lin}\notin\envVA \label{YYY:9} 
		\end{align}		
		Hence by \eqref{YYY:1}, \eqref{YYY:3}, \eqref{YYY:9} and defn of \textsf{eff} (\defref{def:eff}) we know that the side-effects of $\hVV$ (\ie when the processes in $\vLstA[\hVV]$ are released) do not interfere with $\hV$, as these branches do not share linear variables, such that we can deduce 
		\begin{align}
			\typeRule{\env{\envVV}{\eff{(\envVA+\envVB)}{\vLstA[\hVV]}}}{\hV} \, \equiv\, \typeRule{\env{\envVV}{\eff{\envV}{\vLstA[\hVV]}}}{\hV} \label{YYY:10}
		\end{align}	
		Similarly by \eqref{YYY:2}, \eqref{YYY:3}, \eqref{YYY:8} and defn of \textsf{eff} (\defref{def:eff}) we can deduce 
		\begin{align}
			\typeRule{\env{\envVV}{\eff{(\envVA+\envVB)}{\vLstA[\hV]}}}{\hVV} \, \equiv\, \typeRule{\env{\envVV}{\eff{\envV}{\vLstA[\hV]}}}{\hVV} \label{YYY:11}
		\end{align}	
		Since $\excl{\hV,\hVV}=(\vLstA[\hV],\vLstA[\hVV])$, by \eqref{YYY:10}, \eqref{YYY:11} and \rtit{tCn2} we know 
		\begin{align}
			\typeRule{\env{\envVV}{\envV}}{\hV\mand\hVV} \label{YYY:12}
		\end{align}
	\noindent$\therefore$ Case holds by \eqref{YYY:12}.				
	\end{lastcase}
\end{proof}
\end{lemma}\smallskip

\begin{lemma}\label{lemmaMA} $\cmon{\envV}{\Used}{\hV}\traceEvent{\scriptstyle\act}\cmon{\envVP}{\UsedP}{\hV'}\; \text{ and } \;\typeRule{\env{\envVV}{\eff{\envV}{\vLstA}}}{\hV}\, \imp \\$\indent\indent$\cmon{\eff{\envV}{\vLstA}}{\Used}{\hV}\traceEvent{\scriptstyle\act}\cmon{\eff{\after{\envV,\act}}{\vLstA}}{\UsedP}{\hV'}\; \text{ and } \;\envVP=\after{\envV,\act}$
\begin{proof}By rule induction on $\cmon{\envV}{\Used}{\hV}\traceEvent{\scriptstyle\act}\cmon{\envVP}{\UsedP}{\hV'}$.
	
	\begin{Case}[\rtit{rTru}] From our rule premises we know
		\begin{gather}
			\cmon{\envV}{\Used}{\mboolE{b}{\hV}{\hVV}}\traceEvent{\scriptstyle\tau}\cmon{\envV}{\Used}{\hV} \label{M1:1:1} \\
			\typeRule{\env{\envVV}{\eff{\envV}{\vLstA}}}{\mboolE{b}{\hV}{\hVV}}	\label{M1:1:2}
		\end{gather}
		Since $\act=\tau$ and by defn of \textsf{after} (\defref{def:after}) we know
		\begin{gather}
			\envVP=\after{\envV,\tau}	\label{M1:1:3}
		\end{gather}
		By \eqref{M1:1:2} and \rtit{tIf} we know
		\begin{gather}
			\typeRule{\env{\envVV}{\eff{\envV}{\vLstA}}}{\hV}	\label{M1:1:4}
		\end{gather}
		Since reduction \eqref{M1:1:1} can be performed regardless of the contents of $\envV$ and by \eqref{M1:1:2}, \eqref{M1:1:3} and \eqref{M1:1:4} we can deduce
		\begin{gather}
			\cmon{\eff{\envV}{\vLstA}}{\Used}{\mboolE{b}{\hV}{\hVV}}\traceEvent{\scriptstyle\tau}\cmon{\eff{\after{\envV,\tau}}{\vLstA}}{\Used}{\hV} \label{M1:1:5}
		\end{gather}
		\noindent$\therefore$ Case holds by \eqref{M1:1:3} and \eqref{M1:1:5}.
	\end{Case}	
	
	\noindent\textbf{Note:} The proofs for cases \rtit{rFls}, \rtit{rClr} and \rtit{rMax} are analogous to the proof of case \rtit{rTru}.\medskip
	
	\begin{Case}[\rtit{rRel}] From our rule premises we know
		\begin{gather}
			\cmon{(\envVN{\typ{\vLstB}{\lpidb}})}{\Used}{\mrelease{\vLstB}{\hV}}\traceEvent{\scriptstyle\rel{\vLstB}}\cmon{(\envVN{\typ{\vLstB}{\lpid}})}{\Used}{\hV} \label{M1:2:1} \\
			\typeRule{\env{\envVV}{\eff{(\envVN{\typ{\vLstB}{\lpidb}})}{\vLstA}}}{\mrelease{\vLstB}{\hV}}	\label{M1:2:2}
		\end{gather}
		Since $\act=\rel{\vLstB}$ and by defn of \textsf{after} (\defref{def:after}) we know
		\begin{gather}
			(\envVN{\typ{\vLstB}{\lpid}})=\after{(\envVN{\typ{\vLstB}{\lpidb}}),\rel{\vLstB}}	\label{M1:2:3}
		\end{gather}
		By \eqref{M1:2:2} and \rtit{tRel} we know
		\begin{gather}
			\typeRule{\env{\envVV}{\eff{(\envVN{\typ{\vLstB}{\lpid}})}{\vLstA}}}{\hV}	\label{M1:2:4}
		\end{gather}
		By \eqref{M1:2:2} we know that even if we apply the \textsf{eff} function on environment $(\envVN{\typ{\vLstB}{\lpidb}})$, reduction \eqref{M1:2:1} can still be performed. Hence by \eqref{M1:2:4} and the defn of \textsf{after} (\defref{def:after}) we can deduce
		\begin{gather}
			\cmon{\eff{(\envVN{\typ{\vLstB}{\lpidb}})}{\vLstA}}{\Used}{\mrelease{\vLstB}{\hV}}\traceEvent{\scriptstyle\rel{\vLstB}}\cmon{\eff{(\after{\envVN{\typ{\vLstB}{\lpidb}}),\rel{\vLstB}}}{\vLstA}}{\Used}{\hV} \label{M1:2:5}
		\end{gather}
		\noindent$\therefore$ Case holds by \eqref{M1:2:3} and \eqref{M1:2:5}.
	\end{Case}	
	
	\noindent\textbf{Note:} The proofs for cases \rtit{rBlk}, \rtit{rAdA} and \rtit{rAdS} are analogous to the proof of case \rtit{rRel}.\medskip
	
	\begin{Case}[\rtit{rNc1}] From our rule premises we know
		\begin{gather}
			\cmon{\envV}{\Used}{\mBDNec{\patE}{\stk}{\vLstB}\hV}\traceEventA\cmon{(\envVN{\envVP})}{\UsedP}{\mblock{i}\hV\sigma} \label{M1:3:1} 
		\end{gather}
		because
		\begin{gather}
			\match{\patE}{\actE}{\sigma}	\label{M1:3:2}\\
			\id{\actE}=i		\label{M1:3:3}\\
			\envVP=\tbnd{\patE}\sigma \label{M1:3:4}\\
			\dtcheck	\label{M1:3:5} \\
			\dtcheckB  \label{M1:3:5.5} 
		\end{gather}
		and
		\begin{gather}
			\typeRule{\env{\envVV}{\eff{\envV}{\vLstA}}}{\mBNec{\patE}{\vLstB}\hV}	\label{M1:3:6}
		\end{gather}		
		Since $\act=\actE$,\, $\envV\subseteq(\envVN{\envVP})$ and by defn of \textsf{after} (\defref{def:after}) we know
		\begin{gather}
			(\envVN{\envVP})=\after{\envV,\actE} 	\label{M1:3:7}
		\end{gather}
		By \eqref{M1:3:6} and \rtit{tNcB} we know
		\begin{gather}
			\id{\patE}=\varid \label{M1:3:7.5} \\
			\typeRule{\env{\envVV}{(\eff{\envV}{\vLstA},\tbnd{\patE})}}{\mblock{\varid}\hV}	\label{M1:3:8} 
		\end{gather}
		Since newly bound variables cannot be of the type \lpidb, then by the defn of \textsf{eff} (\defref{def:eff}) we know
		\begin{gather}
			\eff{\tbnd{\patE}}{\vLstA}\equiv\tbnd{\patE}	\label{M1:3:9}
		\end{gather}
		Hence by \eqref{M1:3:8} and \eqref{M1:3:9} we know 
		\begin{gather}
			\typeRule{\env{\envVV}{(\eff{(\envVN{\tbnd{\patE}})}{\vLstA})}}{\mblock{\varid}\hV}	\label{M1:3:10}
		\end{gather}		
		By \eqref{M1:3:3} and \eqref{M1:3:7.5} we know that $\varid\sigma=i$, hence by \eqref{M1:3:2}, \eqref{M1:3:5}, \eqref{M1:3:5.5}, \eqref{M1:3:10} and \lemmaD we know 
		\begin{gather}
			\typeRule{\env{\envVV}{(\eff{(\envVN{\tbnd{\patE}\sigma})}{\vLstA})}}{\mblock{i}\hV\sigma}	\label{M1:3:11}
		\end{gather}		
		By \eqref{M1:3:6} and \eqref{M1:3:10} we know that even if we apply the \textsf{eff} function on environment $\envV$, reduction \eqref{M1:3:1} can still be performed, as by \eqref{M1:3:11} we know that it reduces into a form which also typechecks. Hence by \eqref{M1:3:11} and the defn of \textsf{after} (\defref{def:after}) we can deduce
		\begin{gather}
			\cmon{\eff{\envV}{\vLstA}}{\Used}{\mBDNec{\patE}{\stk}{\vLstB}\hV}\traceEventA\cmon{\eff{\after{\envV,\actE}}{\vLstA}}{\Used}{\mblock{i}\hV\sigma} \label{M1:3:12}
		\end{gather}
		\noindent$\therefore$ Case holds by \eqref{M1:3:7} and \eqref{M1:3:12}.
	\end{Case}
	
	\begin{Case}[\rtit{rNc2}] The proof for this case is analogous to the proof for case \rtit{rNc1}. \end{Case}
	
	\begin{Case}[\rtit{rNc3}] From our rule premises we know
		\begin{gather}
			\cmon{\envV}{\Used}{\mattrDNec{\patE}{\attr}{\stk}{\vLstB}\hV}\traceEventA\cmon{\envV}{\Used}{\mrelease{\vLstB}\mtru} \label{M1:4:1} 
		\end{gather}
		because
		\begin{gather}
			\match{\patE}{\actE}{\undef}	\label{M1:4:2}
		\end{gather}
		and
		\begin{gather}
			\typeRule{\env{\envVV}{\eff{\envV}{\vLstA}}}{\mattrNec{\patE}{\attr}{\vLstB}\hV}	\label{M1:4:3}
		\end{gather}		
		Since $\act=\actE$, \, $\envV\subseteq\envV$ and by defn of \textsf{after} (\defref{def:after}) we know
		\begin{gather}
			\envV=\after{\envV,\actE} 	\label{M1:4:3.5}
		\end{gather}
		Since $\attr\in\sset{\agg,\norm}$ we must consider two subcases: \medskip
		
		\begin{Subcase}[$\attr=\agg$]				
			By \eqref{M1:4:3} and \rtit{tNcB} we know
			\begin{gather}
				\typeRule{\env{\envVV}{\eff{\envV}{\vLstA}}}{\mrelease{\vLstB}\mtru}	\label{M1:4:4}
			\end{gather}			
			By \eqref{M1:4:3} and \eqref{M1:4:4} we know that reduction \eqref{M1:4:1} can still be performed regardless of the contents of $\envV$, hence by \eqref{M1:4:3.5} we can deduce
			\begin{gather}
				\cmon{\eff{\envV}{\vLstA}}{\Used}{\mattrDNec{\patE}{\attr}{\stk}{\vLstB}\hV}\traceEventA\cmon{\eff{\envV}{\vLstA}}{\Used}{\mrelease{\vLstB}\mtru} \label{M1:4:5} 
			\end{gather}
			\noindent$\therefore$ Subcase holds by \eqref{M1:4:3.5} and \eqref{M1:4:5}.		
		\end{Subcase}
		
		\begin{lastsubcase}[$\attr=\norm$] The proof for this subcase is analogous to that of the previous subcase (\ie	$\attr=\agg$).	\end{lastsubcase}
	\end{Case}
	
	\begin{Case}[\rtit{rCn1}] From our rule premises we know
		\begin{gather}
			\cmon{\envV}{\Used}{\hV\mand\hVV}\traceEventA\cmon{(\envVPN{\envVPP})}{\UsedP\cup\UsedPP}{\hV'\mand\hVV'} \label{M1:5:1} 
		\end{gather}
		because
		\begin{gather}
			\cmon{\envV}{\Used}{\hV}\traceEventA\cmon{\envVP}{\UsedP}{\hV'}	\label{M1:5:2}\\
			\cmon{\envV}{\Used}{\hVV}\traceEventA\cmon{\envVPP}{\UsedPP}{\hVV'}	\label{M1:5:3}\\
			\dtAndcheck \label{M1:5:4}
		\end{gather}
		and
		\begin{gather}
			\typeRule{\env{\envVV}{\eff{\envV}{\vLstA}}}{\hV\mand\hVV}	\label{M1:5:5}
		\end{gather}		
		
		\noindent We must consider two subcases: \medskip
		
		\begin{Subcase}[{$\excl{\hV,\hVV}=(\vLstA[\hV],\vLstA[\hVV])$}]		
			Since $\excl{\hV,\hVV}=(\vLstA[\hV],\vLstA[\hVV])$, by \eqref{M1:5:5} and \rtit{tCn2} we know
			\begin{gather}
				\typeRule{\env{\envVV}{\eff{\envV}{\vLstA\vLstA[\hVV]}}}{\hV}	\label{M1:5:6}\\
				\typeRule{\env{\envVV}{\eff{\envV}{\vLstA\vLstA[\hV]}}}{\hVV}	\label{M1:5:7}
			\end{gather}
			By \eqref{M1:5:2}, \eqref{M1:5:6} and IH we know 
			\begin{gather}
				\cmon{\eff{\envV}{\vLstA\vLstA[\hVV]}}{\Used}{\hV}\traceEventA\cmon{\eff{\after{\envV,\actE}}{\vLstA\vLstA[\hVV]}}{\UsedP}{\hV'} \label{M1:5:8} \\
				\envVP=\after{\envV,\actE}	\label{M1:5:9}
			\end{gather}
			By \eqref{M1:5:3}, \eqref{M1:5:7} and IH we know 
			\begin{gather}
				\cmon{\eff{\envV}{\vLstA\vLstA[\hV]}}{\Used}{\hVV}\traceEventA\cmon{\eff{\after{\envV,\actE}}{\vLstA\vLstA[\hV]}}{\UsedPP}{\hVV'} \label{M1:5:10} \\
				\envVPP=\after{\envV,\actE}	\label{M1:5:11}
			\end{gather}
			By \eqref{M1:5:2} and \eqref{M1:5:8} we know that formula $\hV$ can reduce into $\hV'$ regardless of whether we apply the side-effects or not. Hence we can deduce
			\begin{gather}
				\cmon{\eff{\envV}{\vLstA}}{\Used}{\hV}\traceEventA\cmon{\eff{\after{\envV,\actE}}{\vLstA}}{\UsedP}{\hV'} \label{M1:5:12}
			\end{gather}
			The same argument applies by \eqref{M1:5:3} and \eqref{M1:5:10}, such that we know we know
			\begin{gather}
				\cmon{\eff{\envV}{\vLstA}}{\Used}{\hVV}\traceEventA\cmon{\eff{\after{\envV,\actE}}{\vLstA}}{\UsedPP}{\hVV'} \label{M1:5:13}
			\end{gather}
			By \eqref{M1:5:9}, \eqref{M1:5:11} and the defn of \textsf{after} (\defref{def:after}) we know
			\begin{gather}
				\envV\subseteq\envVP \label{M1:5:13.1}\\
				\envV\subseteq\envVPP \label{M1:5:13.2}
			\end{gather}
			By \eqref{M1:5:13.1} and \eqref{M1:5:13.2} we know
			\begin{gather}
				\envV\subseteq(\envVP,\envVPP) \label{M1:5:13.3}
			\end{gather}
			By \eqref{M1:5:13.3} and and the defn of \textsf{after} (\defref{def:after}) we know
			\begin{gather}				
				(\envVP,\envVPP)=\after{\envV,\actE}	\label{M1:5:14}
			\end{gather}			
			By \eqref{M1:5:4}, \eqref{M1:5:12}, \eqref{M1:5:13}, \eqref{M1:5:14} and \rtit{rCn1} we know
			\begin{gather}
				\cmon{\eff{\envV}{\vLstA}}{\Used}{\hV\mand\hVV}\traceEventA\cmon{\eff{\after{\envV,\actE}}{\vLstA}}{\UsedP\cup\UsedPP}{\hV'\mand\hVV'} \label{M1:5:15}
			\end{gather}
			\noindent$\therefore$ Subcase holds by \eqref{M1:5:14} and \eqref{M1:5:15}.		
		\end{Subcase}
		
		\begin{lastsubcase}[$\excl{\hV,\hVV}=\undef$]		
			Since $\excl{\hV,\hVV}=\undef$, by \eqref{M1:5:5} and \rtit{tCn1} we know
			\begin{gather}
				\typeRule{\env{\envVV}{\eff{\envV_{1}}{\vLstA}}}{\hV}	\label{M1:5:16}\\
				\typeRule{\env{\envVV}{\eff{\envV_{2}}{\vLstA}}}{\hVV}	\label{M1:5:17}\\
				\eff{\envV}{\vLstA}=\eff{\envV_{1}}{\vLstA}+\eff{\envV_{2}}{\vLstA} \label{M1:5:18}
			\end{gather}
			Since by \eqref{M1:5:18} we know $\eff{\envV_{1}}{\vLstA}\subseteq\eff{\envV}{\vLstA}$ and $\eff{\envV_{2}}{\vLstA}\subseteq\eff{\envV}{\vLstA}$, then by \eqref{M1:5:16} and \eqref{M1:5:17} we can deduce
			\begin{gather}
				\typeRule{\env{\envVV}{\eff{\envV}{\vLstA}}}{\hV}	\label{M1:5:19}\\
				\typeRule{\env{\envVV}{\eff{\envV}{\vLstA}}}{\hVV}	\label{M1:5:20}			
			\end{gather}
			By \eqref{M1:5:2}, \eqref{M1:5:19} and IH we know 
			\begin{gather}
				\cmon{\eff{\envV}{\vLstA}}{\Used}{\hV}\traceEventA\cmon{\eff{\after{\envV,\actE}}{\vLstA}}{\UsedP}{\hV'} \label{M1:5:21} \\
				\envVP=\after{\envV,\actE}	\label{M1:5:22}
			\end{gather}
			By \eqref{M1:5:3}, \eqref{M1:5:20} and IH we know 
			\begin{gather}
				\cmon{\eff{\envV}{\vLstA}}{\Used}{\hV}\traceEventA\cmon{\eff{\after{\envV,\actE}}{\vLstA}}{\UsedPP}{\hVV'} \label{M1:5:23} \\
				\envVPP=\after{\envV,\actE}	\label{M1:5:24}
			\end{gather}
			By \eqref{M1:5:22}, \eqref{M1:5:24} and the defn of \textsf{after} (\defref{def:after}) we know
			\begin{gather}
				(\envVP,\envVPP)=\after{\envV,\actE}	\label{M1:5:25}
			\end{gather}
			By \eqref{M1:5:4}, \eqref{M1:5:21}, \eqref{M1:5:23}, \eqref{M1:5:25} and \rtit{rCn1} we know
			\begin{gather}
				\cmon{\eff{\envV}{\vLstA}}{\Used}{\hV\mand\hVV}\traceEventA\cmon{\eff{\after{\envV,\actE}}{\vLstA}}{\UsedP\cup\UsedPP}{\hV'\mand\hVV'} \label{M1:5:26}
			\end{gather}
			\noindent$\therefore$ Subcase holds by \eqref{M1:5:25} and \eqref{M1:5:26}.	
		\end{lastsubcase}		
	\end{Case}

	\begin{Case}[\rtit{rCn2}] From our rule premises we know
		\begin{gather}
			\cmon{\envV}{\Used}{\hV\mand\hVV}\traceEventTau\cmon{\envV}{\UsedP}{\hV'\mand\hVV} \label{M1:6:1} 
		\end{gather}
		because
		\begin{gather}
			\cmon{\envV}{\Used}{\hV}\traceEventTau\cmon{\envVP}{\UsedP}{\hV'}	\label{M1:6:2}
		\end{gather}
		and
		\begin{gather}
			\typeRule{\env{\envVV}{\eff{\envV}{\vLstA}}}{\hV\mand\hVV}	\label{M1:6:3}
		\end{gather}		
		
		\noindent We must consider two subcases: \medskip
		
		\begin{Subcase}[{$\excl{\hV,\hVV}=(\vLstA[\hV],\vLstA[\hVV])$}]		
			Since $\excl{\hV,\hVV}=(\vLstA[\hV],\vLstA[\hVV])$, by \eqref{M1:6:3} and \rtit{tCn2} we know
			\begin{gather}
				\typeRule{\env{\envVV}{\eff{\envV}{\vLstA\vLstA[\hVV]}}}{\hV}	\label{M1:6:4}\\
				\typeRule{\env{\envVV}{\eff{\envV}{\vLstA\vLstA[\hV]}}}{\hVV}	\label{M1:6:5}
			\end{gather}
			By \eqref{M1:6:2}, \eqref{M1:6:4} and IH we know 
			\begin{gather}
				\cmon{\eff{\envV}{\vLstA\vLstA[\hVV]}}{\Used}{\hV}\traceEventTau\cmon{\eff{\after{\envV,\tau}}{\vLstA\vLstA[\hVV]}}{\UsedP}{\hV'} \label{M1:6:6} \\
				\envV=\after{\envV,\tau}	\label{M1:6:7}
			\end{gather}			
			By \eqref{M1:6:2} and \eqref{M1:6:6} we know that formula $\hV$ can reduce into $\hV'$ regardless of whether we apply the side-effects or not. Hence we can deduce
			\begin{gather}
				\cmon{\eff{\envV}{\vLstA}}{\Used}{\hV}\traceEventTau\cmon{\eff{\after{\envV,\tau}}{\vLstA}}{\UsedP}{\hV'} \label{M1:6:8}
			\end{gather}
			By \eqref{M1:6:8} and \rtit{rCn2} we know
			\begin{gather}
				\cmon{\eff{\envV}{\vLstA}}{\Used}{\hV\mand\hVV}\traceEventTau\cmon{\eff{\after{\envV,\tau}}{\vLstA}}{\UsedP}{\hV'\mand\hVV} \label{M1:6:9}
			\end{gather}
			\noindent$\therefore$ Subcase holds by \eqref{M1:6:7} and \eqref{M1:6:9}.		
		\end{Subcase}
		
		\begin{lastsubcase}[$\excl{\hV,\hVV}=\undef$]		
			Since $\excl{\hV,\hVV}=\undef$, by \eqref{M1:6:3} and \rtit{tCn1} we know
			\begin{gather}
				\typeRule{\env{\envVV}{\eff{\envV_{1}}{\vLstA}}}{\hV}	\label{M1:6:10}\\
				\typeRule{\env{\envVV}{\eff{\envV_{2}}{\vLstA}}}{\hVV}	\label{M1:6:11}\\
				\eff{\envV}{\vLstA}=\eff{\envV_{1}}{\vLstA}+\eff{\envV_{2}}{\vLstA} \label{M1:6:12}
			\end{gather}
			Since by \eqref{M1:6:12} we know $\eff{\envV_{1}}{\vLstA}\subseteq\eff{\envV}{\vLstA}$ and $\eff{\envV_{2}}{\vLstA}\subseteq\eff{\envV}{\vLstA}$, then by \eqref{M1:6:10} we can deduce
			\begin{gather}
				\typeRule{\env{\envVV}{\eff{\envV}{\vLstA}}}{\hV}	\label{M1:6:13}			
			\end{gather}
			By \eqref{M1:6:2}, \eqref{M1:6:13} and IH we know 
			\begin{gather}
				\cmon{\eff{\envV}{\vLstA}}{\Used}{\hV}\traceEventTau\cmon{\eff{\after{\envV,\tau}}{\vLstA}}{\UsedP}{\hV'} \label{M1:6:14} \\
				\envV=\after{\envV,\tau}	\label{M1:6:15}
			\end{gather}						
			By \eqref{M1:6:14} and \rtit{rCn2} we know
			\begin{gather}
				\cmon{\eff{\envV}{\vLstA}}{\Used}{\hV\mand\hVV}\traceEventTau\cmon{\eff{\after{\envV,\tau}}{\vLstA}}{\UsedP}{\hV'\mand\hVV} \label{M1:6:16}
			\end{gather}
			\noindent$\therefore$ Subcase holds by \eqref{M1:6:15} and \eqref{M1:6:16}.	
		\end{lastsubcase}		
	\end{Case}
	
	\begin{lastcase}[\rtit{rCn3}] The proof for this case is analogous to the proof of case \rtit{rCn2}.	\end{lastcase} 
\end{proof} 
\end{lemma}

\begin{lemma}\label{lemmaMB} $\envV=\envVA+\envVB \; \text{ and } \; \cmon{\envV}{\Used}{\hV}\traceEvent{\scriptstyle\act}\cmon{\envVP}{\UsedP}{\hV'}\; \text{ and } \;\typeRule{\env{\envVV}{\envVA}}{\hV}\\$ \indent\indent $\imp \,\cmon{\envVA}{\Used}{\hV}\traceEvent{\scriptstyle\act}\cmon{\envVPA}{\UsedP}{\hV'}\; \text{ and } \;\envVP=\envVPA+\envVB$
\begin{proof}By rule induction on $\cmon{\envV}{\Used}{\hV}\traceEvent{\scriptstyle\act}\cmon{\envVP}{\UsedP}{\hV'}$.
	
	\begin{Case}[\rtit{rTru}] From our rule premises we know
		\begin{gather}
			\cmon{\envV}{\Used}{\mboolE{b}{\hV}{\hVV}}\traceEvent{\scriptstyle\tau}\cmon{\envV}{\Used}{\hV} \label{M2:1:1} \\
			\typeRule{\env{\envVV}{\envVA}}{\mboolE{b}{\hV}{\hVV}}	\label{M2:1:2}\\
			\envV=\envVA+\envVB	\label{M2:1:3}
		\end{gather}
		By \eqref{M2:1:2} and \rtit{tIf} we know
		\begin{gather}
			\typeRule{\env{\envVV}{\envVA}}{\hV}	\label{M2:1:4}
		\end{gather}
		Since reduction \eqref{M2:1:1} can be performed regardless of the contents of $\envV$ and by \eqref{M2:1:2} and \eqref{M2:1:4} we can deduce
		\begin{gather}
			\cmon{\envVA}{\Used}{\mboolE{b}{\hV}{\hVV}}\traceEvent{\scriptstyle\tau}\cmon{\envVA}{\Used}{\hV} \label{M2:1:5}
		\end{gather}
		\noindent$\therefore$ Case holds by \eqref{M2:1:3} and \eqref{M2:1:5}.
	\end{Case}
	
	\noindent\textbf{Note:} The proofs for cases \rtit{rFls}, \rtit{rClr} and \rtit{rMax} are analogous to the proof of case \rtit{rTru}.\medskip
	
	\begin{Case}[\rtit{rRel}] From our rule premises we know
		\begin{gather}
			\cmon{(\envVN{\typ{\vLstB}{\lpidb}})}{\Used}{\mrelease{\vLstB}{\hV}}\traceEvent{\scriptstyle\rel{\vLstB}}\cmon{(\envVN{\typ{\vLstB}{\lpid}})}{\Used}{\hV} \label{M2:2:1} \\
			\typeRule{\env{\envVV}{\envVA}}{\mrelease{\vLstB}{\hV}}	\label{M2:2:2}\\
			(\envVN{\typ{\vLstB}{\lpid}})=\envVA+\envVB	\label{M2:2:3}
		\end{gather}
		By \eqref{M2:2:2} and \rtit{tRel} we know
		\begin{gather}
			\envVA=(\envVPPA,\typ{\vLstB}{\lpidb}) \label{M2:2:4}\\
			\envVPA=(\envVPPA,\typ{\vLstB}{\lpid}) \label{M2:2:5}\\
			\typeRule{\env{\envVV}{\envVPA}}{\hV}	\label{M2:2:6}
		\end{gather}		
		By \eqref{M2:2:4} and \eqref{M2:2:5} we know that reduction \eqref{M2:2:1} can still be performed using subset \envVA instead of \envV. Hence we can deduce
		\begin{gather}
			\cmon{\envVA}{\Used}{\mrelease{\vLstB}{\hV}}\traceEvent{\scriptstyle\rel{\vLstB}}\cmon{\envVPA}{\Used}{\hV} \label{M2:2:7}
		\end{gather}	
		By \eqref{M2:2:3} and \eqref{M2:2:5} we can conclude 
		\begin{gather}
			(\envVN{\typ{\vLstB}{\lpid}})=\envVPA+\envVB \label{M2:2:8}
		\end{gather}	
		\noindent$\therefore$ Case holds by \eqref{M2:2:7} and \eqref{M2:2:8}.
	\end{Case}\\
	
	\noindent\textbf{Note:} The proofs for cases \rtit{rBlk}, \rtit{rAdA} and \rtit{rAdS} are analogous to the proof of case \rtit{rRel}.\medskip
	
	\begin{Case}[\rtit{rNc1}] From our rule premises we know
		\begin{gather}
			\cmon{\envV}{\Used}{\mBDNec{\patE}{\stk}{\vLstA}\hV}\traceEventA\cmon{(\envVN{\envVP})}{\UsedP}{\mblock{i}\hV\sigma} \label{M2:3:1} 
		\end{gather}
		because
		\begin{gather}
			\match{\patE}{\actE}{\sigma}	\label{M2:3:2}\\
			\id{\actE}=i		\label{M2:3:3}\\
			\envVP=\tbnd{\patE}\sigma \label{M2:3:4}\\
			\dtcheck	\label{M2:3:5} \\
			\dtcheckB	\label{M2:3:5.5}
		\end{gather}
		and
		\begin{gather}
			\typeRule{\env{\envVV}{\envVA}}{\mBNec{\patE}{\vLstA}\hV}	\label{M2:3:6}	\\
			 \envV=\envVA+\envVB	\label{M2:3:7}
		\end{gather}
		By \eqref{M2:3:6} and \rtit{tNcB} we know
		\begin{gather}
			\typeRule{\env{\envVV}{(\envVA,\tbnd{\patE})}}{\mblock{\varid}\hV}	\label{M2:3:8}
		\end{gather}
		By \eqref{M2:3:2}, \eqref{M2:3:5}, \eqref{M2:3:5.5}, \eqref{M2:3:8} and \lemmaD we know 
		\begin{gather}
			\typeRule{\env{\envVV}{(\envVA,\tbnd{\patE}\sigma)}}{\mblock{i}\hV\sigma}	\label{M2:3:9}
		\end{gather}
		Since by \eqref{M2:3:4} and \eqref{M2:3:9} we know that $\mblock{i}\hV\sigma$ typechecks with $(\envVA,\envVP)$, this means that reduction \eqref{M2:3:1} can still be performed using subset $\envVA$ instead of $\envV$, such that we know
		\begin{gather}
			\cmon{\envVA}{\Used}{\mBDNec{\patE}{\stk}{\vLstA}\hV}\traceEvent{\scriptstyle\actE}\cmon{(\envVA,\envVP)}{\UsedP}{\mblock{i}\hV\sigma} \label{M2:3:10}
		\end{gather}
		By \eqref{M2:3:7} and \eqref{M2:3:4} we can conclude
		\begin{gather}
			 (\envV,\envVP)=(\envVA,\envVP)+\envVB	\label{M2:3:11}
		\end{gather}
		\noindent$\therefore$ Case holds by \eqref{M2:3:10} and \eqref{M2:3:11}.
	\end{Case}
	
	\begin{Case}[\rtit{rNc2}] The proof for this case is analogous to the proof for case \rtit{rNc1}. \end{Case}
	
	\begin{Case}[\rtit{rNc3}] From our rule premises we know
		\begin{gather}
			\cmon{\envV}{\Used}{\mattrDNec{\patE}{\attr}{\stk}{\vLstA}\hV}\traceEventA\cmon{\envV}{\Used}{\mrelease{\vLstA}\mtru} \label{M2:4:1} 
		\end{gather}
		because
		\begin{gather}
			\match{\patE}{\actE}{\undef}	\label{M2:4:2}
		\end{gather}
		and
		\begin{gather}
			\typeRule{\env{\envVV}{\envVA}}{\mattrNec{\patE}{\attr}{\vLstA}\hV}	\label{M2:4:3}		\\
			\envV=\envVA+\envVB 	\label{M2:4:3.5}
		\end{gather}
		Since $\attr\in\sset{\agg,\norm}$ we must consider two subcases: \medskip
		
		\begin{Subcase}[$\attr=\agg$]				
			By \eqref{M2:4:3} and \rtit{tNcB} we know
			\begin{gather}
				\typeRule{\env{\envVV}{\envVA}}{\mrelease{\vLstA}\mtru}	\label{M2:4:4}
			\end{gather}			
			By \eqref{M2:4:3} and \eqref{M2:4:4} we know that reduction \eqref{M2:4:1} can still be performed regardless of the contents of $\envV$, hence by \eqref{M2:4:3.5} we can deduce
			\begin{gather}
				\cmon{\envVA}{\Used}{\mattrDNec{\patE}{\attr}{\stk}{\vLstA}\hV}\traceEventA\cmon{\envVA}{\Used}{\mrelease{\vLstA}\mtru} \label{M2:4:5} 
			\end{gather}
			\noindent$\therefore$ Subcase holds by \eqref{M2:4:3.5} and \eqref{M2:4:5}.		
		\end{Subcase}\vspace{-5mm}
		
		\begin{lastsubcase}[$\attr=\norm$] The proof for this subcase is analogous to that of the previous subcase (\ie	$\attr=\agg$).	\end{lastsubcase}
	\end{Case}
	
	\begin{Case}[\rtit{rCn1}] From our rule premises we know
		\begin{gather}
			\cmon{\envV}{\Used}{\hV\mand\hVV}\traceEventA\cmon{(\envVPN{\envVPP})}{\UsedP\cup\UsedPP}{\hV'\mand\hVV'} \label{M2:5:1} 
		\end{gather}
		because
		\begin{gather}
			\cmon{\envV}{\Used}{\hV}\traceEventA\cmon{\envVP}{\UsedP}{\hV'}	\label{M2:5:2}\\
			\cmon{\envV}{\Used}{\hVV}\traceEventA\cmon{\envVPP}{\UsedPP}{\hVV'}	\label{M2:5:3}\\
			\dtAndcheck \label{M2:5:4}
		\end{gather}
		and
		\begin{gather}
			\typeRule{\env{\envVV}{\envVA}}{\hV\mand\hVV}	\label{M2:5:5} \\
			\envV=\envVA+\envVB	\label{M2:5:6}
		\end{gather}		
		
		\noindent We must consider two subcases: \medskip
		
		\begin{Subcase}[{$\excl{\hV,\hVV}=(\vLstA[\hV],\vLstA[\hVV])$}]		
			Since $\excl{\hV,\hVV}=(\vLstA[\hV],\vLstA[\hVV])$, by \eqref{M2:5:5} and \rtit{tCn2} we know
			\begin{gather}
				\typeRule{\env{\envVV}{\eff{\envVA}{\vLstA[\hVV]}}}{\hV}	\label{M2:5:7}\\
				\typeRule{\env{\envVV}{\eff{\envVA}{\vLstA[\hV]}}}{\hVV}	\label{M2:5:8}
			\end{gather}
			Since by \eqref{M2:5:6} we know that $\envVA\subseteq\envV$, then from \eqref{M2:5:7} and \eqref{M2:5:8} we can deduce
			\begin{gather}
				\typeRule{\env{\envVV}{\eff{\envV}{\vLstA[\hVV]}}}{\hV}	\label{M2:5:9}\\
				\typeRule{\env{\envVV}{\eff{\envV}{\vLstA[\hV]}}}{\hVV}	\label{M2:5:10}
			\end{gather}
			By \eqref{M2:5:2}, \eqref{M2:5:9} and \lemmaMA we know
			\begin{gather}
				\cmon{\eff{\envV}{\vLstA[\hVV]}}{\Used}{\hV}\traceEventA\cmon{\eff{\envVP}{\vLstA[\hVV]}}{\UsedP}{\hV'} \label{M2:5:11}
			\end{gather}
			By \eqref{M2:5:3}, \eqref{M2:5:10} and \lemmaMA we know
			\begin{gather}
				\cmon{\eff{\envV}{\vLstA[\hV]}}{\Used}{\hVV}\traceEventA\cmon{\eff{\envVPP}{\vLstA[\hV]}}{\UsedPP}{\hVV'} \label{M2:5:12}
			\end{gather}
			By \eqref{M2:5:6}, \eqref{M2:5:7}, \eqref{M2:5:8} and defn of \textsf{eff} (\defref{def:eff}) we know
			\begin{gather}
				\eff{\envV}{\vLstA[\hVV]}=\eff{\envVA}{\vLstA[\hVV]}+\envVB	\label{M2:5:13}\\
				\eff{\envV}{\vLstA[\hV]}=\eff{\envVA}{\vLstA[\hV]}+\envVB	\label{M2:5:14}
			\end{gather}			
			By \eqref{M2:5:7}, \eqref{M2:5:11}, \eqref{M2:5:13} and IH we know 
			\begin{gather}
				\cmon{\eff{\envVA}{\vLstA[\hVV]}}{\Used}{\hV}\traceEventA\cmon{\eff{\envVPA}{\vLstA[\hVV]}}{\UsedP}{\hV'} \label{M2:5:15} \\
				\eff{\envVP}{\vLstA[\hVV]}=\eff{\envVPA}{\vLstA[\hVV]}+\envVB \; \implies \; \envVP=\envVPA+\envVB	\label{M2:5:16}
			\end{gather}
			By \eqref{M2:5:8}, \eqref{M2:5:12}, \eqref{M2:5:14} and IH we know 
			\begin{gather}
				\cmon{\eff{\envVA}{\vLstA[\hV]}}{\Used}{\hVV}\traceEventA\cmon{\eff{\envVPPA}{\vLstA[\hV]}}{\UsedPP}{\hVV'} \label{M2:5:17} \\
				\eff{\envVPP}{\vLstA[\hV]}=\eff{\envVPPA}{\vLstA[\hV]}+\envVB \; \implies \; \envVPP=\envVPPA+\envVB	\label{M2:5:18}
			\end{gather}	
			From \eqref{M2:5:2}, \eqref{M2:5:15} and \eqref{M2:5:16} we know that formula $\hV$ can reduce into $\hV'$ regardless of whether we apply the side-effects or not. Hence we can deduce
			\begin{gather}
				\cmon{\envVA}{\Used}{\hV}\traceEventA\cmon{\envVPA}{\UsedP}{\hV'} \label{M2:5:19}
			\end{gather}			
			The same argument applies by \eqref{M2:5:3}, \eqref{M2:5:17} and \eqref{M2:5:18}, such that we know we know
			\begin{gather}
				\cmon{\envVA}{\Used}{\hVV}\traceEventA\cmon{\envVPPA}{\UsedPP}{\hVV'} \label{M2:5:20}
			\end{gather}			
			By \eqref{M2:5:4}, \eqref{M2:5:19}, \eqref{M2:5:20} and \rtit{rCn1} we know
			\begin{gather}
				\cmon{\envVA}{\Used}{\hV\mand\hVV}\traceEventA\cmon{(\envVPA,\envVPPA)}{\UsedP\cup\UsedPP}{\hV'\mand\hVV'} \label{M2:5:21}
			\end{gather}
			From \eqref{M2:5:16} and \eqref{M2:5:18} we can conclude
			\begin{gather}
				(\envVP,\envVPP)=(\envVPA,\envVPPA)+\envVPB	\label{M2:5:22}
			\end{gather}
			\noindent$\therefore$ Subcase holds by \eqref{M2:5:21} and \eqref{M2:5:22}.		
		\end{Subcase}
		
		\begin{lastsubcase}[$\excl{\hV,\hVV}=\undef$]		
			Since $\excl{\hV,\hVV}=\undef$, by \eqref{M2:5:5} and \rtit{tCn1} we know
			\begin{gather}
				\typeRule{\env{\envVV}{\envV_{3}}}{\hV}	\label{M2:5:23}\\
				\typeRule{\env{\envVV}{\envV_{4}}}{\hVV}	\label{M2:5:24}\\
				\envVA=\envV_{3}+\envV_{4} \label{M2:5:25}
			\end{gather}
			Since by \eqref{M2:5:25} we know $\envV_{3}\subseteq\envVA$ and $\envV_{4}\subseteq\envVA$, then by \eqref{M2:5:23} and \eqref{M2:5:24} we can deduce
			\begin{gather}
				\typeRule{\env{\envVV}{\envVA}}{\hV}	\label{M2:5:26}\\
				\typeRule{\env{\envVV}{\envVA}}{\hVV}	\label{M2:5:27}			
			\end{gather}
			By \eqref{M2:5:2}, \eqref{M2:5:6}, \eqref{M2:5:26} and IH we know 
			\begin{gather}
				\cmon{\envVA}{\Used}{\hV}\traceEventA\cmon{\envVPA}{\UsedP}{\hV'} \label{M2:5:28} \\
				\envVP=\envVPA+\envVB	\label{M2:5:29}
			\end{gather}
			By \eqref{M2:5:3}, \eqref{M2:5:6}, \eqref{M2:5:27} and IH we know 
			\begin{gather}
				\cmon{\envVA}{\Used}{\hV}\traceEventA\cmon{\envVPPA}{\UsedPP}{\hVV'} \label{M2:5:30} \\
				\envVPP=\envVPPA+\envVB	\label{M2:5:31}
			\end{gather}
			By \eqref{M2:5:4}, \eqref{M2:5:28}, \eqref{M2:5:30} and \rtit{rCn1} we know
			\begin{gather}
				\cmon{\envVA}{\Used}{\hV\mand\hVV}\traceEventA\cmon{(\envVPA,\envVPPA)}{\UsedP\cup\UsedPP}{\hV'\mand\hVV'} \label{M2:5:32}
			\end{gather}
			From \eqref{M2:5:29} and \eqref{M2:5:31} we can deduce
			\begin{gather}
				(\envVP,\envVPP)=(\envVPA,\envVPPA)+\envVB	\label{M2:5:33} 
			\end{gather}
			\noindent$\therefore$ Subcase holds by \eqref{M2:5:32} and \eqref{M2:5:33}.	
		\end{lastsubcase}		
	\end{Case}

	\begin{Case}[\rtit{rCn2}] From our rule premises we know
		\begin{gather}
			\cmon{\envV}{\Used}{\hV\mand\hVV}\traceEventTau\cmon{\envV}{\UsedP}{\hV'\mand\hVV} \label{M2:6:1} 
		\end{gather}
		because
		\begin{gather}
			\cmon{\envV}{\Used}{\hV}\traceEventTau\cmon{\envV}{\UsedP}{\hV'}	\label{M2:6:2}
		\end{gather}
		and
		\begin{gather}
			\typeRule{\env{\envVV}{\envVA}}{\hV\mand\hVV}	\label{M2:6:3} \\
			\envV=\envVA+\envVB	\label{M2:6:4}
		\end{gather}		
		
		\noindent We must consider two subcases: \medskip
		
		\begin{Subcase}[{$\excl{\hV,\hVV}=(\vLstA[\hV],\vLstA[\hVV])$}]		
			Since $\excl{\hV,\hVV}=(\vLstA[\hV],\vLstA[\hVV])$, by \eqref{M2:6:3} and \rtit{tCn2} we know
			\begin{gather}
				\typeRule{\env{\envVV}{\eff{\envVA}{\vLstA[\hVV]}}}{\hV}	\label{M2:6:5}\\
				\typeRule{\env{\envVV}{\eff{\envVA}{\vLstA[\hV]}}}{\hVV}	\label{M2:6:6}
			\end{gather}
			Since by \eqref{M2:6:4} we know that $\envVA\subseteq\envV$, then from \eqref{M2:6:5} we can deduce
			\begin{gather}
				\typeRule{\env{\envVV}{\eff{\envV}{\vLstA[\hVV]}}}{\hV}	\label{M2:6:7}
			\end{gather}
			By \eqref{M2:6:2}, \eqref{M2:6:7} and \lemmaMA we know
			\begin{gather}
				\cmon{\eff{\envV}{\vLstA[\hVV]}}{\Used}{\hV}\traceEventTau\cmon{\eff{\envV}{\vLstA[\hVV]}}{\UsedP}{\hV'} \label{M2:6:8}
			\end{gather}	
			By \eqref{M2:6:4}, \eqref{M2:6:5} and the defn of \textsf{eff} (\defref{def:eff}) we know
			\begin{gather}
				\eff{\envV}{\vLstA[\hVV]}=\eff{\envVA}{\vLstA[\hVV]}+\envVB	\label{M2:6:9}
			\end{gather}							
			By \eqref{M2:6:5}, \eqref{M2:6:8}, \eqref{M2:6:9} and IH we know 
			\begin{gather}
				\cmon{\eff{\envVA}{\vLstA[\hVV]}}{\Used}{\hV}\traceEventTau\cmon{\eff{\envVA}{\vLstA[\hVV]}}{\UsedP}{\hV'} \label{M2:6:10} \\
				\eff{\envV}{\vLstA[\hVV]}=\eff{\envVA}{\vLstA[\hVV]}+\envVB \; \implies \; \envV=\envVA+\envVB	\label{M2:6:11}
			\end{gather}			
			From \eqref{M2:6:2}, \eqref{M2:6:10} and \eqref{M2:6:11} we know that formula $\hV$ can reduce into $\hV'$ using subset $\envVA$, as the reduction can still be applied regardless of whether we apply the side-effects or not. Hence we can deduce
			\begin{gather}
				\cmon{\envVA}{\Used}{\hV}\traceEventTau\cmon{\envVA}{\UsedP}{\hV'} \label{M2:6:12}
			\end{gather}	
			By \eqref{M2:6:12} and \rtit{rCn2} we know
			\begin{gather}
				\cmon{\envVA}{\Used}{\hV\mand\hVV}\traceEventTau\cmon{\envVA}{\UsedP}{\hV'\mand\hVV} \label{M2:6:13}
			\end{gather}
			\noindent$\therefore$ Subcase holds by \eqref{M2:6:11} and \eqref{M2:6:13}.		
		\end{Subcase}
		
		\begin{lastsubcase}[$\excl{\hV,\hVV}=\undef$]		
			Since $\excl{\hV,\hVV}=\undef$, by \eqref{M2:6:3} and \rtit{tCn1} we know
			\begin{gather}
				\typeRule{\env{\envVV}{\envV_{3}}}{\hV}	\label{M2:6:15}\\
				\typeRule{\env{\envVV}{\envV_{4}}}{\hVV}	\label{M2:6:16}\\
				\envVA=\envV_{3}+\envV_{4} \label{M2:6:17}
			\end{gather}
			Since by \eqref{M2:6:17} we know $\envV_{3}\subseteq\envVA$, then from \eqref{M2:6:15} we can deduce
			\begin{gather}
				\typeRule{\env{\envVV}{\envVA}}{\hV}	\label{M2:6:18}		
			\end{gather}			
			By \eqref{M2:6:2}, \eqref{M2:6:4}, \eqref{M2:6:18} and IH we know 
			\begin{gather}
				\cmon{\envVA}{\Used}{\hV}\traceEventTau\cmon{\envVA}{\UsedP}{\hV'} \label{M2:6:19} \\
				\envV=\envVA+\envVB	\label{M2:6:20}
			\end{gather}		
			By \eqref{M2:6:19} and \rtit{rCn2} we know
			\begin{gather}
				\cmon{\envVA}{\Used}{\hV\mand\hVV}\traceEventTau\cmon{\envVA}{\UsedP}{\hV'\mand\hVV} \label{M2:6:21}
			\end{gather}
			\noindent$\therefore$ Subcase holds by \eqref{M2:6:20} and \eqref{M2:6:21}.	
		\end{lastsubcase}		
	\end{Case}

	\begin{lastcase}[\rtit{rCn3}] The proof for this case is analogous to the proof of case \rtit{rCn2}.	\end{lastcase} 
\end{proof} 
\end{lemma}

\begin{lemma}\label{lemmaXXXB} $\\$\indent\indent$\excl{\hV,\hVV}=(\vLstA[\hV],\vLstA[\hVV])\; \text{ and } \; \cmon{\envV}{\Used}{\hV}\traceEventTau\cmon{\envV}{\UsedP}{\hV'}\, \imp \,\excl{\hV',\hVV}=(\vLstA[\hV],\vLstA[\hVV])$.
\begin{proof}By rule induction on $\cmon{\envV}{\Used}{\hV}\traceEventTau\cmon{\envV}{\UsedP}{\hV'}$.\\

	\begin{Case}[\rtit{rTru}] From our rule premises we know
		\begin{gather}
			\cmon{\envV}{\Used}{\mboolE{c}{\hV_{1}}{\hV_{2}}}\traceEventTau\cmon{\envV}{\Used}{\hV_{1}} \label{XXXB:1:1} \\
			\excl{\mboolE{c}{\hV_{1}}{\hV_{2}},\hVV}=(\vLstA[\hV],\vLstA[\hVV]) \label{XXXB:1:2} 
		\end{gather}
		By \eqref{XXXB:1:2} and defn of \textsf{excl} (\defref{def:excl}) we know 
		\begin{align}
			\excl{\hV_{1},\hVV}=\excl{\hV_{2},\hVV}=(\vLstA[\hV],\vLstA[\hVV]) \label{XXXB:1:3}			
		\end{align}
		\noindent$\therefore$ Case holds by \eqref{XXXB:1:3}.
	\end{Case}
	
	\begin{Case}[\rtit{rFls}] The proof for this case is analogous to the proof for \rtit{rTru}. \end{Case}
	
	\begin{Case}[\rtit{rMax}] From our rule premises we know
		\begin{gather}
			\cmon{\envV}{\Used}{\mmax{\hVarX}{\!\hV}}\traceEventTau\cmon{\envV}{\Used}{\hV\sub{(\clr{\hVarX}\mmax{\hVarX}{\!\hV})}{\hVarX}} \label{XXXB:2:1} \\
			\excl{\mmax{\hVarX}{\!\hV},\hVV}=(\vLstA[\hV],\vLstA[\hVV]) \label{XXXB:2:2} 
		\end{gather}
		By \eqref{XXXB:2:2} and defn of \textsf{excl} (\defref{def:excl}) we know 
		\begin{align}
			\excl{\hV,\hVV}=(\vLstA[\hV],\vLstA[\hVV]) \label{XXXB:2:3}			
		\end{align}
		By $\sub{(\clr{\hVarX}\mmax{\hVarX}{\!\hV})}{\hVarX}$, \eqref{XXXB:2:3} and \lemmaZZZA we know 
		\begin{align}
			\excl{\hV\sub{(\clr{\hVarX}\mmax{\hVarX}{\!\hV})}{\hVarX},\hVV}=(\vLstA[\hV],\vLstA[\hVV]) \label{XXXB:2:4}			
		\end{align}
		\noindent$\therefore$ Case holds by \eqref{XXXB:2:4}.		
	\end{Case}
	
	\begin{Case}[\rtit{rClr}] The proof for this case is analogous to the proof for \rtit{rMax}. \end{Case} \vspace{-5mm}
		
	\begin{Case}[\rtit{rCn2}] From our rule premises we know
		\begin{align}
			\cmon{\envV}{\Used}{\hV_{1}\mand\hV_{2}}\traceEventTau\cmon{\envV}{\UsedP}{\hV'_{1}\mand\hV_{2}} \label{XXXB:3:1} 
		\end{align}
		because
		\begin{align}
			\cmon{\envV}{\Used}{\hV_{1}}\traceEventTau\cmon{\envV}{\UsedP}{\hV'_{1}} \label{XXXB:3:2}
		\end{align}
		and
		\begin{align}
			\excl{(\hV_{1}\mand\hV_{2}),\hVV}=(\vLstA[\scriptstyle(\hV_{1}\mand\hV_{2})],\vLstA[\hVV]) \label{XXXB:3:3}		
		\end{align}
		By \eqref{XXXB:3:3} and defn of \textsf{excl} (\defref{def:excl}) we know 
		\begin{gather}
			\excl{\hV_{1},\hVV}=(\vLstA[\hV_{1}],\vLstA[\hVV]) \label{XXXB:3:4}	\\	
			\excl{\hV_{2},\hVV}=(\vLstA[\hV_{2}],\vLstA[\hVV]) \label{XXXB:3:5}	\\	
			\vLstA[\scriptstyle(\hV_{1}\mand\hV_{2})] = \vLstA[\hV_{1}]\cup\vLstA[\hV_{2}] \label{XXXB:3:6}		
		\end{gather} 
		By \eqref{XXXB:3:2}, \eqref{XXXB:3:4} and IH we know
		\begin{align}
			\excl{\hV'_{1},\hVV}=(\vLstA[\hV_{1}],\vLstA[\hVV]) \label{XXXB:3:7}		
		\end{align}
		By \eqref{XXXB:3:5}, \eqref{XXXB:3:6}, \eqref{XXXB:3:7} and defn of \textsf{excl} (\defref{def:excl}) we know 
		\begin{align}
			\excl{(\hV'_{1}\mand\hV_{2}),\hVV}=(\vLstA[\scriptstyle(\hV_{1}\mand\hV_{2})],\vLstA[\hVV]) \label{XXXB:3:8}		
		\end{align}
		\noindent$\therefore$ Case holds by \eqref{XXXB:3:8}.		
	\end{Case}
	
	\noindent \textbf{Note:} The remaining cases \emph{do not apply} as we only consider $\tau$ actions.
\end{proof}
\end{lemma}

\begin{lemma}\label{lemmaXXXA} $\excl{\hV,\hVV}=(\vLstA[\hV],\vLstA[\hVV])\; \text{ and } \; \cmon{\envV}{\Used}{\hV}\traceEventA\cmon{\envVP}{\UsedP}{\hV'}\; \text{ and } \\$\indent\indent$ \typeRule{\env{\envVV}{\eff{(\envVP,\envVPP)}{\vLstA[\hVV]}}}{\hV'}\; \text{ and } \; \typeRule{\env{\envVV}{\eff{(\envVP,\envVPP)}{\vLstA[\hV]}}}{\hVV'}\, \imp \,\typeRule{\env{\envVV}{(\envVP,\envVPP)}}{\hV'\mand\hVV'}$.
\begin{proof}By rule induction on $\cmon{\envV}{\Used}{\hV}\traceEventA\cmon{\envVP}{\UsedP}{\hV'}$.
	
	\begin{Case}[\rtit{rNc1}] From our rule premises we know
		\begin{align}
			\cmon{\envV}{\Used}{\mBDNec{\patE}{\stk}{\vLstA}{\hV}}\traceEventA\cmon{\envVP}{\UsedP}{\mblock{i}\hV\sigma} \label{XXXA:1:1} 
		\end{align}
		because
		\begin{gather}
			\match{\patE}{\actE}{\sigma} \label{XXXA:1:2} \\
			\text{ and others \ldots} \nonumber
		\end{gather}
		and
		\begin{gather}
			 \excl{\mBNec{\patE}{\vLstA}{\hV},\hVV}=(\vLstA,\vLstA[\hVV]) \label{XXXA:1:3} \\ 
			 \typeRule{\env{\envVV}{\eff{(\envVP,\envVPP)}{\vLstA[\hVV]}}}{\mblock{i}\hV\sigma} \label{XXXA:1:4} \\ 
			 \typeRule{\env{\envVV}{\eff{(\envVP,\envVPP)}{\vLstA}}}{\hVV'} \label{XXXA:1:5}  
		\end{gather}
		By \eqref{XXXA:1:3} and defn of \textsf{excl} (\defref{def:excl}) we know 
		\begin{align}
			\forall\patE_{1}\in\fps{\hVV}\cdot\match{\patE}{\patE_{1}}{\undef} \label{XXXA:1:6} 
		\end{align}
		By \eqref{XXXA:1:2} we know that system event $\actE$ \emph{matches} pattern $\patE$, but by \eqref{XXXA:1:6} we also know that $\actE$ \emph{does not match} any of the concurrent necessities defined in $\hVV$. This means that all the other necessities defined in $\hVV$ are \emph{trivially satisfied}. Hence we can deduce
		\begin{align}
			\hVV'=\trivSat \label{XXXA:1:7} 
		\end{align}
		Since $\mBNec{\patE}{\vLstA}{\hV}$ reduces into $\mblock{i}\hV\sigma$, by \eqref{XXXA:1:7} and defn of \textsf{excl} (\defref{def:excl}) we know 
		\begin{align}
			\excl{\mblock{i}\hV\sigma,\hVV'}=(\varepsilon,\vLstA[\hVV]) \label{XXXA:1:8} 
		\end{align}
		Since $\varepsilon\subseteq\vLstA$, and by defn of \textsf{eff} (\defref{def:eff}), we know that \eqref{XXXA:1:5} still typechecks using $\varepsilon$ instead of $\vLstA$ given that we know \eqref{XXXA:1:7}. (Remember that the \textsf{eff} function ``releases'' a number of blocked ids from a type environment $\envV$, so if we don't release them $\hVV'=\trivSat$ still typechecks). Hence we can deduce
		\begin{align}
			\typeRule{\env{\envVV}{\eff{(\envVP,\envVPP)}{\varepsilon}}}{\hVV'} \label{XXXA:1:9} 
		\end{align}
		By \eqref{XXXA:1:4}, \eqref{XXXA:1:8}, \eqref{XXXA:1:9} and \rtit{tCn2} we know
		\begin{align}
			\typeRule{\env{\envVV}{(\envVP,\envVPP)}}{\mblock{i}\hV\sigma\mand\hVV'} \label{XXXA:1:10} 
		\end{align}
		\noindent$\therefore$ Case holds by \eqref{XXXA:1:10}.
	\end{Case}
	
	\begin{Case}[\rtit{rNc2}] The proof for this case is analogous to that of case \rtit{rNc1}.\end{Case}
	
	\begin{Case}[\rtit{rNc3}] From our rule premises we know
		\begin{align}
			\cmon{\envV}{\Used}{\mattrDNec{\patE}{\attr}{\stk}{\vLstA}{\hV}}\traceEventA\cmon{\envV}{\Used}{\mrelease{\vLstA}\mtru} \label{XXXA:2:1} 
		\end{align}
		because
		\begin{align}
			\match{\patE}{\actE}{\undef} \label{XXXA:2:2} 
		\end{align}
		and
		\begin{gather}
			 \excl{\mattrNec{\patE}{\attr}{\vLstA}{\hV},\hVV}=(\vLstA,\vLstA[\hVV]) \label{XXXA:2:4} \\ 
			 \typeRule{\env{\envVV}{\eff{(\envVP,\envVPP)}{\vLstA[\hVV]}}}{\mrelease{\vLstA}\mtru} \label{XXXA:2:3} \\ 
			 \typeRule{\env{\envVV}{\eff{(\envVP,\envVPP)}{\vLstA}}}{\hVV'} \label{XXXA:2:4.5}  
		\end{gather}
		By \eqref{XXXA:2:4} and defn of \textsf{excl} (\defref{def:excl}) we know 
		\begin{align}
			\forall\patE_{1}\in\fps{\hVV}\cdot\match{\patE}{\patE_{1}}{\undef} \label{XXXA:2:5} 
		\end{align}
		By \eqref{XXXA:2:2} we know that system event $\actE$ \emph{does not match} pattern $\patE$, and by \eqref{XXXA:2:5} we know that the opposing branch $\hVV$ may have lived on. We must therefore consider these two cases:
		
		\begin{Subcase}[$\hVV$ also died $-$ $\hVV'=\trivSat$]
			Since $\mattrDNec{\patE}{\attr}{\stk}{\vLstA}{\hV}$ reduces to $\mrelease{\vLstA}\mtru$,\, $\hVV=\trivSat$, and by defn of \textsf{excl} (\defref{def:excl}) we know 
			\begin{align}
				\excl{\mrelease{\vLstA}\mtru,\hVV'}=(\vLstA,\vLstA[\hVV]) \label{XXXA:2:5.1} 
			\end{align}
			By \eqref{XXXA:2:3}, \eqref{XXXA:2:4.5}, \eqref{XXXA:2:5.1} and \rtit{tCn2} we know
			\begin{align}
				\typeRule{\env{\envVV}{(\envVP,\envVPP)}}{\mrelease{\vLstA}\mtru\mand\hVV'} \label{XXXA:2:5.2} 
			\end{align}
			\noindent $\therefore$ Subcase holds by \eqref{XXXA:2:5.2}.
		\end{Subcase}
		
		\begin{lastsubcase}[$\hVV$ lived on $-$ $\hVV\neq\trivSat$]
			Since $\mattrDNec{\patE}{\attr}{\stk}{\vLstA}{\hV}$ reduces to $\mrelease{\vLstA}\mtru$\, $\hVV\neq\trivSat$, and by defn of \textsf{excl} (\defref{def:excl}) we know 
			\begin{align}
				\excl{\mrelease{\vLstA}\mtru,\hVV'}=(\vLstA,\varepsilon) \label{XXXA:2:6} 
			\end{align}
			Since $\varepsilon\subseteq\vLstA[\hVV]$, and by defn of \textsf{eff} (\defref{def:eff}), we know that \eqref{XXXA:2:3} still typechecks using $\varepsilon$ instead of $\vLstA[\hVV]$ given that we know \eqref{XXXA:2:6}. Hence we can deduce
			\begin{align}
				\typeRule{\env{\envVV}{\eff{(\envVP,\envVPP)}{\varepsilon}}}{\mrelease{\vLstA}\mtru} \label{XXXA:2:7} 
			\end{align}
			By \eqref{XXXA:2:4.5}, \eqref{XXXA:2:6}, \eqref{XXXA:2:7} and \rtit{tCn2} we know
			\begin{align}
				\typeRule{\env{\envVV}{(\envVP,\envVPP)}}{\mrelease{\vLstA}\mtru\mand\hVV'} \label{XXXA:2:8} 
			\end{align}
		\noindent$\therefore$ Subcase holds by \eqref{XXXA:2:8}.
		\end{lastsubcase}
	\end{Case}
	
	\begin{Case}[\rtit{rCn1}] From our rule premises we know
		\begin{gather}
			\cmon{\envV}{\Used}{\hV_{1}\mand\hV_{2}}\traceEventA\cmon{\envVP}{\UsedP\cup\UsedPP}{\hV'_{1}\mand\hV'_{2}} \label{XXXA:3:1} \\
			 \excl{\hV_{1}\mand\hV_{2},\hVV}=(\vLstA[\hV],\vLstA[\hVV]) \label{XXXA:3:4} \\ 
			 \typeRule{\env{\envVV}{\eff{(\envVP,\envVPP)}{\vLstA[\hVV]}}}{\hV'_{1}\mand\hV'_{2}} \label{XXXA:3:2} \\ 
			 \typeRule{\env{\envVV}{\eff{(\envVP,\envVPP)}{\vLstA[\hV]}}}{\hVV'} \label{XXXA:3:3}  
		\end{gather}
		As by \eqref{XXXA:3:4} we know that branch $\hV_{1}\mand\hV_{2}$ is \emph{mutually exclusive} to $\hVV$, then we have to consider the following three subcases:
		
		\begin{Subcase}[$\hV'_{1}\mand\hV'_{2}=\trivSat$ and $\hVV'\neq\trivSat$]
			As we assume that $\hV_{1}\mand\hV_{2}$ reduces into $\trivSat$, by \eqref{XXXA:3:4} and defn of \textsf{excl} (\defref{def:excl}) we know 
			\begin{align}
				\excl{\trivSat,\hVV'}=(\vLstA[\hV],\varepsilon) \label{XXXA:3:5} 
			\end{align}
			Since $\hV'_{1}\mand\hV'_{2}=\trivSat$, and defn of \textsf{eff} (\defref{def:eff}) we know that \eqref{XXXA:3:2} still holds if $\vLstA[\hV]=\varepsilon$. (Remember that the \textsf{eff} function ``releases'' a number of blocked ids from a type environment $\envV$, so if we don't release them $\hVV'=\trivSat$ still typechecks). Hence we can deduce
			\begin{align}
				\typeRule{\env{\envVV}{\eff{(\envVP,\envVPP)}{\varepsilon}}}{\hV'_{1}\mand\hV'_{2}} \label{XXXA:3:6} 
			\end{align}
			By \eqref{XXXA:3:3}, \eqref{XXXA:3:5}, \eqref{XXXA:3:6} and \rtit{tCn2} we know
			\begin{align}
				\typeRule{\env{\envVV}{(\envVP,\envVPP)}}{(\hV'_{1}\mand\hV'_{2})\mand\hVV'} \label{XXXA:3:7} 
			\end{align}
			\noindent$\therefore$ Subcase holds by \eqref{XXXA:3:7}.
		\end{Subcase}
		
		\begin{Subcase}[$\hVV'=\trivSat$ and $\hV'_{1}\mand\hV'_{2}\neq\trivSat$] The proof for this subcase is analogous to that of the previous subcase.	\end{Subcase}
		
		\begin{lastsubcase}[$\hV'_{1}\mand\hV'_{2}=\trivSat$ and $\hVV'=\trivSat$]
			As we assume that \emph{both branches} $\hV_{1}\mand\hV_{2}$ and $\hVV$ are trivially satisfied, \ie reduce into the form $\trivSat$, by \eqref{XXXA:3:4} and defn of \textsf{excl} (\defref{def:excl}) we know 
			\begin{align}			
				\excl{\hV'_{1}\mand\hV'_{2},\hVV'}=(\vLstA[\hV],\vLstA[\hVV]) \label{XXXA:3:8} 
			\end{align}
			By \eqref{XXXA:3:2}, \eqref{XXXA:3:3}, \eqref{XXXA:3:8} and \rtit{tCn2} we know
			\begin{align}
				\typeRule{\env{\envVV}{(\envVP,\envVPP)}}{(\hV'_{1}\mand\hV'_{2})\mand\hVV'} \label{XXXA:3:9} 
			\end{align}
			\noindent$\therefore$ Subcase holds by \eqref{XXXA:3:9}.
		\end{lastsubcase}		
	\end{Case}	
	
	\noindent \textbf{Note:} The remaining cases \emph{do not apply} as we only consider $\actE$ actions.	
\end{proof}
\end{lemma}

\begin{lemma}\label{lemmaH} $\excl{\hV,\hVV}\!=\!(\vLstA[\hV],\vLstA[\hVV])\; \text{ and } \; \cmon{\envV}{\Used}{\hV}\traceEvent{\scriptstyle\actC}\cmon{\envVP}{\Used}{\hV'}\; \text{ and } \\$\indent\indent$ \typeRule{\env{\envVV}{\eff{\envV}{\vLstA[\hV]}}}{\hVV} \; \text{ and }\; \typeRule{\env{\envVV}{\eff{\after{\envV,\actC}}{\vLstA[\hVV]}}}{\hV'}\, \imp \,\typeRule{\env{\envVV}{\after{\envV,\actC}}}{\hV'\mand\hVV}$
\begin{proof}By rule induction on $\cmon{\envV}{\Used}{\hV}\traceEvent{\scriptstyle\actC}\cmon{\envVP}{\Used}{\hV'}$.

	\begin{Case}[\rtit{rBlk}] From our rule premises we know
		\begin{gather}
			\cmon{(\envVN{\typ{\vLstB}{\lpid}})}{\Used}{\mblock{\vLstB}\hV}\traceEvent{\scriptstyle\blk{\vLstB}}\cmon{(\envVN{\typ{\vLstB}{\lpidb}})}{\Used}{\hV} \label{H:1:1} \\
			\excl{\mblock{\vLstB}\hV,\hVV}=(\varepsilon,\vLstA[\hVV]) \label{H:1:2} 		\\
			\typeRule{\env{\envVV}{\eff{\after{(\envVN{\typ{\vLstB}{\lpid}}),\blk{\vLstB}}}{\vLstA[\hVV]}}}{\hV} \label{H:1:4} \\	
			\typeRule{\env{\envVV}{\eff{(\envVN{\typ{\vLstB}{\lpid}})}{\varepsilon}}}{\hVV} \label{H:1:5} 	
		\end{gather}
		By \eqref{H:1:4} and the defn of \textsf{after} (\defref{def:after}) we know
		\begin{gather}
			\typeRule{\env{\envVV}{\eff{(\envVN{\typ{\vLstB}{\lpidb}})}{\vLstA[\hVV]}}}{\hV} \label{H:1:4.5} 	
		\end{gather}		
		We know that by the defn of \textsf{excl} (\defref{def:excl}), \eqref{H:1:2} can only hold if $\hVV$ represents a collection of trivially satisfied branches, hence we can deduce
		\begin{gather}
			\excl{\hV,\hVV}=(\varepsilon,\vLstA[\hVV]) \label{H:1:6} 		\\
			\hVV=\trivSat \label{H:1:7}
		\end{gather}
		By \eqref{H:1:7} we know that for \eqref{H:1:5} to hold, then every identifier in the release set $(\vLstA[\hVV])$ of $\hVV$ must be of the type $\lpidb$. Since we know that $\typ{\vLstB}{\lpid}$, then we know for sure that release-set $\vLstA[\hVV]$ \emph{does not} intersect with blocking set $\vLstB$ (otherwise \eqref{H:1:5} would not hold). Hence we know that $\hVV$ can typecheck regardless of the type of $\vLstB$ (\ie it typechecks both if we have $\typ{\vLstB}{\lpid}$ or $\typ{\vLstB}{\lpidb}$), such that we can conclude
		\begin{gather}
			\typeRule{\env{\envVV}{\eff{(\envVN{\typ{\vLstB}{\lpidb}})}{\varepsilon}}}{\hVV} \label{H:1:8}
		\end{gather}
		By \eqref{H:1:4.5}, \eqref{H:1:6}, \eqref{H:1:8}, \rtit{tCn2} and the defn of \textsf{after} (\defref{def:after}) we know
		\begin{gather}
			\typeRule{\env{\envVV}{\after{(\envVN{\typ{\vLstB}{\lpid}}),\blk{\vLstB}}}}{\hV\mand\hVV} \label{H:1:9} 	
		\end{gather}
		\noindent$\therefore$ Case holds by \eqref{H:1:9}.
	\end{Case}
	
	\begin{Case}[\rtit{rRel}] From our rule premises we know
		\begin{gather}
			\cmon{(\envVN{\typ{\vLstA}{\lpidb}})}{\Used}{\mrelease{\vLstA}\hV}\traceEvent{\scriptstyle\rel{\vLstA}}\cmon{(\envVN{\typ{\vLstA}{\lpid}})}{\Used}{\hV} \label{H:2:1} \\
			\excl{\mrelease{\vLstA}\hV,\hVV}=(\vLstA[\hV],\vLstA[\hVV]) \label{H:2:2} 		\\
			\typeRule{\env{\envVV}{\eff{\after{(\envVN{\typ{\vLstA}{\lpidb}}),\rel{\vLstA}}}{\vLstA[\hVV]}}}{\hV} \label{H:2:3} 	\\
			\typeRule{\env{\envVV}{\eff{(\envVN{\typ{\vLstA}{\lpidb}})}{\vLstA}}}{\hVV} \label{H:2:3.5} 	
		\end{gather}
		By \eqref{H:2:2} and the defn of \textsf{excl} (\defref{def:excl}) we must consider 2 subcases:\\
		
		\begin{Subcase}[$\hV\neq\mtru$\, and \,$\hVV=\trivSat$]
			The proof for this subcase is analogous to the proof of case \rtit{rBlk}.
		\end{Subcase}
		
		\begin{lastsubcase}[$\hV=\mtru$]
			Since $\hV=\mtru$, by \eqref{H:2:2}, \eqref{H:2:3} and \eqref{H:2:3.5} we know
			\begin{gather}
				\excl{\mrelease{\vLstA}\mtru,\hVV}=(\vLstA,\vLstA[\hVV])	\label{H:2:4}\\ 	
				\typeRule{\env{\envVV}{\eff{\after{(\envVN{\typ{\vLstA}{\lpidb}}),\rel{\vLstA}}}{\vLstA[\hVV]}}}{\mtru} \label{H:2:6}\\ 	
				\typeRule{\env{\envVV}{\eff{(\envVN{\typ{\vLstA}{\lpidb}})}{\vLstA}}}{\hVV} \label{H:2:7} 	
			\end{gather}
			By \eqref{H:2:4} and the defn of \textsf{excl} (\defref{def:excl}) we know
			\begin{gather}
				\excl{\mtru,\hVV}=(\varepsilon,\vLstA[\hVV]) \label{H:2:8} 		
			\end{gather}			
			By \eqref{H:2:6} and the defn of \textsf{after} (\defref{def:after}) we know
			\begin{gather}
				\typeRule{\env{\envVV}{\eff{(\envVN{\typ{\vLstA}{\lpid}})}{\vLstA[\hVV]}}}{\mtru} \label{H:2:9}
			\end{gather}			
			By \eqref{H:2:7} and the defn of \textsf{eff} (\defref{def:eff}) we know
			\begin{gather}
				\typeRule{\env{\envVV}{\eff{(\envVN{\typ{\vLstA}{\lpid}})}{\varepsilon}}}{\hVV} \label{H:2:10} 	
			\end{gather}			
			By \eqref{H:2:8}, \eqref{H:2:9}, \eqref{H:2:10}, \rtit{tCn2} and the defn of \textsf{after} (\defref{def:after}) we know
			\begin{gather}
				\typeRule{\env{\envVV}{\after{(\envVN{\typ{\vLstA}{\lpidb}}),\rel{\vLstA}}}}{\mtru\mand\hVV} \label{H:2:11} 	
			\end{gather}
			\noindent$\therefore$ Case holds by \eqref{H:2:11}.
		\end{lastsubcase}
	\end{Case}
	
	\begin{Case}[\rtit{rAdS}] From our rule premises we know
		\begin{gather}
			\cmon{(\envVN{\typ{\vLstB}{\lpidb}})}{\Used}{\mscor{\vLstB}{\vLstA}\hV}\traceEvent{\scriptstyle\cors{\vLstB}}\cmon{(\envVN{\typ{\vLstB}{\lpid}})}{\Used}{\mrelease{\vLstA}\hV} \label{H:3:1} \\
			\excl{\mscor{\vLstB}{\vLstA}\hV,\hVV}=(\varepsilon,\vLstA[\hVV]) \label{H:3:2} 		\\
			\typeRule{\env{\envVV}{\eff{\after{(\envVN{\typ{\vLstB}{\lpidb}}),\cors{\vLstB}}}{\vLstA[\hVV]}}}{\mrelease{\vLstA}\hV} \label{H:3:6}\\ 	
			\typeRule{\env{\envVV}{\eff{(\envVN{\typ{\vLstB}{\lpidb}})}{\varepsilon}}}{\hVV} \label{H:3:7} 	 	
		\end{gather}
		By \eqref{H:3:2} and the defn of \textsf{excl} (\defref{def:excl}) we know
		\begin{gather}
			\excl{\mrelease{\vLstA}\hV,\hVV}=(\varepsilon,\vLstA[\hVV])	\label{H:3:4}\\ 	
			\hVV=\trivSat  \label{H:3:5} 	
		\end{gather}
		By \eqref{H:3:6} and the defn of \textsf{after} (\defref{def:after}) we know
		\begin{gather}
			\typeRule{\env{\envVV}{\eff{(\envVN{\typ{\vLstB}{\lpidb}})}{\vLstA[\hVV]}}}{\mrelease{\vLstA}\hV} \label{H:3:8} 		
		\end{gather}			
		By \eqref{H:3:4}, \eqref{H:3:7}, \eqref{H:3:8}, \rtit{tCn2} and the defn of \textsf{after} (\defref{def:after}) we know
		\begin{gather}
			\typeRule{\env{\envVV}{\after{(\envVN{\typ{\vLstB}{\lpidb}}),\cors{\vLstB}}}}{\mrelease{\vLstA}\hV\mand\hVV} \label{H:3:9} 	
		\end{gather}
		\noindent$\therefore$ Case holds by \eqref{H:3:9}.		
	\end{Case}
	
	\begin{Case}[\rtit{rAdA}] The proof for this case is analogous to the proof for case \rtit{rAdS}. \end{Case}
	
	\begin{Case}[\rtit{rCn3}] From our rule premises we know
		\begin{gather}
			\cmon{\envV}{\Used}{\hV_{1}\mand\hV_{2}}\traceEvent{\actC}\cmon{\envVP}{\Used}{\hV'_{1}\mand\hV_{2}} \label{H:4:1} \\
			 \excl{\hV_{1}\mand\hV_{2},\hVV}=(\vLstA[\hV],\vLstA[\hVV]) \label{H:4:3} \\ 
			 \typeRule{\env{\envVV}{\eff{\after{\envV,\actC}}{\vLstA[\hVV]}}}{\hV'_{1}\mand\hV_{2}} \label{H:4:5} \\
			 \typeRule{\env{\envVV}{\eff{\envV}{\vLstA[\hV]}}}{\hVV} \label{H:4:6}
		\end{gather}		

		\noindent As by \eqref{H:4:3} we know that branch $\hV_{1}\mand\hV_{2}$ is \emph{mutually exclusive} to $\hVV$, based on the defn of \textsf{excl} (\defref{def:excl}) we have to consider the following three subcases:\\
		
		\begin{Subcase}[$\hV_{1}\mand\hV_{2}=\trivSat$ and $\hVV\neq\trivSat$] By our subcase assumptions and \eqref{H:4:1} we can deduce that the reduction can only be made to issue a \emph{release of one} of the releases sets declared in the collection of trivially satisfied branches $\hV_{1}\mand\hV_{2}$. Hence we can conclude
			\begin{gather}
				\envV=\envVPPN{\typ{\vLstA[\hV]}{\lpidb}}  \label{H:4:6.5}  \\
				 \actC = \vLstA[\hV_{1}] \label{H:4:7}  \\
				 \vLstA[\hV]=\vLstA[\hV_{1}],\vLstA[\hV_{n-1}] \label{H:4:8} \\ 
				\hV_{1}\mand\hV_{2}=\Big(\mrelease{\vLstA[\hV_{1}]}\mand\bigAnd\!{\scriptsize(\mrelease{\vLstA}\!\mtru)_{n-1}}\Big) \label{H:4:9}\\ 
				\hV'_{1}\mand\hV_{2}=\bigAnd\!{\scriptsize(\mrelease{\vLstA}\!\mtru)_{n-1}}  \label{H:4:10} \\
				\text{ where }\vLstA[\hV_{1}] \text{ is the release set of an \emph{arbitrary} trivially satisfied branch }\mrelease{\vLstA[\hV_{1}]} \nonumber
			\end{gather}				
			By \eqref{H:4:3}, \eqref{H:4:8}, \eqref{H:4:9}, \eqref{H:4:10} and the defn of \textsf{excl} (\defref{def:excl}) we know 
			\begin{align}
				\excl{\hV'_{1}\mand\hV_{2},\hVV}\equiv\excl{\bigAnd\!{\scriptsize(\mrelease{\vLstA}\!\mtru)_{n-1}},\hVV}\,=\,(\vLstA[\hV_{n-1}],\varepsilon) \label{H:4:11} 
			\end{align}
			Since $\hVV\neq\trivSat$, by \eqref{H:4:5}, \eqref{H:4:6}, \eqref{H:4:6.5} and \eqref{H:4:7} we can deduce
			\begin{gather}
				\typeRule{\env{\envVV}{\eff{\after{(\envVPPN{\typ{\vLstA[\hV]}{\lpidb}}),\rel{\hV_{1}}}}{\varepsilon}}}{\hV'_{1}\mand\hV_{2}} \label{H:4:12} \\
				\typeRule{\env{\envVV}{\eff{(\envVPPN{\typ{\vLstA[\hV]}{\lpidb}})}{\vLstA[\hV]}}}{\hVV} \label{H:4:13} 				
			\end{gather}
			By \eqref{H:4:12}, \eqref{H:4:8} and the defn of \textsf{after} (\defref{def:after}) we know
			\begin{align}
				\typeRule{\env{\envVV}{\eff{(\envVPPN{\typ{\vLstA[\hV_{n-1}]}{\lpidb},\typ{\vLstA[\hV_{1}]}{\lpid}})}{\varepsilon}}}{\hV'_{1}\mand\hV_{2}} \label{H:4:14} 
			\end{align}
			By \eqref{H:4:13}, \eqref{H:4:8} and the defn of \textsf{eff} (\defref{def:eff}) we know
			\begin{align}
				\typeRule{\env{\envVV}{\eff{(\envVPPN{\typ{\vLstA[\hV_{n-1}]}{\lpidb},\typ{\vLstA[\hV{1}]}{\lpid}})}{\vLstA[\hV_{n-1}]}}}{\hVV} \label{H:4:15} 
			\end{align}			
			By \eqref{H:4:11}, \eqref{H:4:14}, \eqref{H:4:15}, \rtit{tCn2} and the defn of \textsf{after} (\defref{def:after}) we know
			\begin{align}
				&\typeRule{\env{\envVV}{(\envVPPN{\typ{\vLstA[\hV_{n-1}]}{\lpidb},\typ{\vLstA[\hV{1}]}{\lpid}})}}{(\hV'_{1}\mand\hV_{2})\mand\hVV} \nonumber \\
				\equiv \; &\typeRule{\env{\envVV}{(\after{(\envVPPN{\typ{\vLstA[\hV_{n-1}]}{\lpidb}}),\rel{\vLstA[\hV{1}]}})}}{(\hV'_{1}\mand\hV_{2})\mand\hVV} \label{H:4:16} 
			\end{align}
			\noindent$\therefore$ Subcase holds by \eqref{H:4:17}.
		\end{Subcase}
		
		\begin{Subcase}[$\hV_{1}\mand\hV_{2}=\trivSat$ and $\hVV=\trivSat$] The proof for this subcase follows a similar argument as per the previous subcase.	\end{Subcase}
		
		\begin{lastsubcase}[$\hV_{1}\mand\hV_{2}\neq\trivSat$ and $\hVV=\trivSat$]
			As we assume that $\hVV=\trivSat$, if $\hV_{1}$ commits an action $\actC\in\sset{\blk{\vLstA[\hV]},\rel{\vLstA[\hV]},\ldots}$ and becomes $\hV'_{1}$, then we are certain that $\vLstA$ does not conflict with any release set in pertaining to $\hVV$, as otherwise \eqref{H:4:6} would not hold. Therefore we can deduce that after committing an action $\actC$, $\hVV=\trivSat$ would not be effected, such that we can conclude 
 			\begin{gather}
				\typeRule{\env{\envVV}{\eff{\after{\envV,\actC}}{\varepsilon}}}{\hVV} \label{H:4:17}\\ 
				\excl{\hV'_{1}\mand\hV_{2},\hVV}=(\varepsilon,\vLstA[\hVV]) \label{H:4:17.5} 
			\end{gather}
			By \eqref{H:4:5}, \eqref{H:4:17}, \eqref{H:4:17.5} and \rtit{tCn2} we conclude
			\begin{align}
				\typeRule{\env{\envVV}{(\after{\envV,\actC})}}{(\hV'_{1}\mand\hV_{2})\mand\hVV} \label{H:4:18} 
			\end{align}
			\noindent$\therefore$ Subcase holds by \eqref{H:4:18}.
		\end{lastsubcase}		
	\end{Case}	
	
	\noindent \textbf{Note:} The remaining cases \emph{do not apply} as we only consider $\actC$ actions.	
\end{proof}
\end{lemma}

\begin{lemma}\label{lemmaZZZ} $\excl{\hV,\hVV}=(\vLstA[\hV],\vLstA[\hVV])\; \text{ and } \; \sub{(\clr{\hVarX}\mmax{\hVarX}{\!\hV'})}{\hVarX}\, \imp \\$\indent\indent$\excl{\hV\sub{(\clr{\hVarX}\mmax{\hVarX}{\!\hV'})}{\hVarX},\hVV\sub{(\clr{\hVarX}\mmax{\hVarX}{\!\hV'})}{\hVarX}}=(\vLstA[\hV],\vLstA[\hVV])$.
\begin{proof} \setstretch{0.5} From our premises we know 
	\begin{gather}
		\excl{\hV,\hVV}=(\vLstA[\hV],\vLstA[\hVV]) \label{ZZZ:1} \\
		\sub{(\clr{\hVarX}\mmax{\hVarX}{\!\hV'})}{\hVarX} \label{ZZZ:2}
	\end{gather}
	By \eqref{ZZZ:1}, \eqref{ZZZ:2} and \lemmaZZZA we know 
	\begin{align}
		\excl{\hV\sub{(\clr{\hVarX}\mmax{\hVarX}{\!\hV'})}{\hVarX},\hVV}=(\vLstA[\hV],\vLstA[\hVV]) \label{ZZZ:3}
	\end{align}
	By \eqref{ZZZ:3} and symmetry we know 
	\begin{align}
		\excl{\hVV,\hV\sub{(\clr{\hVarX}\mmax{\hVarX}{\!\hV'})}{\hVarX}}=(\vLstA[\hVV],\vLstA[\hV]) \label{ZZZ:4}
	\end{align}
	By \eqref{ZZZ:2}, \eqref{ZZZ:4} and \lemmaZZZA we know 
	\begin{align}
		\excl{\hVV\sub{(\clr{\hVarX}\mmax{\hVarX}{\!\hV'})}{\hVarX},\hV\sub{(\clr{\hVarX}\mmax{\hVarX}{\!\hV'})}{\hVarX}}=(\vLstA[\hVV],\vLstA[\hV]) \label{ZZZ:5}
	\end{align}
	By \eqref{ZZZ:5} and symmetry we know 
	\begin{align}
		\excl{\hV\sub{(\clr{\hVarX}\mmax{\hVarX}{\!\hV'})}{\hVarX},\hVV\sub{(\clr{\hVarX}\mmax{\hVarX}{\!\hV'})}{\hVarX}}=(\vLstA[\hV],\vLstA[\hVV]) \label{ZZZ:6}
	\end{align}
	\noindent$\therefore$ Lemma holds by \eqref{ZZZ:6}.
\end{proof}
\end{lemma}

\begin{lemma}\label{lemmaZZZA} $\excl{\hV,\hVV}=(\vLstA[\hV],\vLstA[\hVV])\; \text{ and } \; \sub{(\clr{\hVarX}\mmax{\hVarX}{\!\hV'})}{\hVarX}\, \imp \\$\indent\indent$ \excl{\hV\sub{(\clr{\hVarX}\mmax{\hVarX}{\!\hV'})}{\hVarX},\hVV}=(\vLstA[\hV],\vLstA[\hVV])$.
\begin{proof} By structural induction on $\hV$.

	\begin{Case}[$\hV=\mmax{Y}{\hV}$]	From our rule premises we know 
		\begin{gather}
			\excl{\mmax{Y}{\hV},\hVV}=(\vLstA[\hV],\vLstA[\hVV]) \label{ZZZA:1:1} \\
			\sub{(\clr{\hVarX}\mmax{\hVarX}{\!\hV'})}{\hVarX} \label{ZZZA:1:2}
		\end{gather}
		We must consider two subcases.\bigskip
		
		\begin{Subcase}[$\hVV\neq\trivSat$]
			By \eqref{ZZZA:1:1} and defn of \textsf{excl} (\defref{def:excl}) we know 
			\begin{align}
				\excl{\hV,\hVV}=(\vLstA[\hV],\vLstA[\hVV]) \label{ZZZA:1:3} 
			\end{align}
			By \eqref{ZZZA:1:2}, \eqref{ZZZA:1:3} and IH we know 
			\begin{align}
				\excl{\hV\sub{(\clr{\hVarX}\mmax{\hVarX}{\!\hV'})}{\hVarX},\hVV}=(\vLstA[\hV],\vLstA[\hVV]) \label{ZZZA:1:4} 
			\end{align}
			By \eqref{ZZZA:1:4} and defn of \textsf{excl} (\defref{def:excl}) we can deduce 		
			\begin{align}
				\excl{(\mmax{Y}{\!\hV\sub{(\clr{\hVarX}\mmax{\hVarX}{\!\hV'})}{\hVarX}}),\hVV}=(\vLstA[\hV],\vLstA[\hVV]) \label{ZZZA:1:5} 
			\end{align}
			By \eqref{ZZZA:1:5} and substitution we know
			\begin{align}
				\excl{(\mmax{Y}{\!\hV})\sub{(\clr{\hVarX}\mmax{\hVarX}{\!\hV'})}{\hVarX},\hVV}=(\vLstA[\hV],\vLstA[\hVV]) \label{ZZZA:1:6} 
			\end{align}
			\noindent$\therefore$ Subcase holds by 	\eqref{ZZZA:1:6}.						
		\end{Subcase}		

		\begin{lastsubcase}[$\hVV=\trivSat$]
			By defn of \textsf{excl} (\defref{def:excl}) and since $\hVV=\trivSat$ we know that no matter what $\hV$ is, the \textsf{excl} function always returns $\varepsilon$ as the release set of $\hV$. Hence by \eqref{ZZZA:1:1} and \eqref{ZZZA:1:2} we know  
			\begin{align}
				\excl{(\mmax{Y}{\!\hV}),\;\trivSat} &= \nonumber 		\\
				\excl{(\mmax{Y}{\!\hV})\sub{(\clr{\hVarX}\mmax{\hVarX}{\!\hV'})}{\hVarX},\;\trivSat}&= \;(\varepsilon,\vLstA[\hVV]) \label{ZZZA:1:8} 
			\end{align}
			\noindent$\therefore$ Subcase holds by 	\eqref{ZZZA:1:8}.
		\end{lastsubcase}
	\end{Case}
	
	\begin{Case}[$\hV=\clr{Y}\hV$] The proof for this case is analogous to that of case $\hV=\mmax{Y}{\hV}$ \end{Case}
	
	\begin{Case}[$\hV=\hV_{1}\mand\hV_{2}$]
		From our rule premises we know 
		\begin{gather}
			\excl{(\hV_{1}\mand\hV_{2}),\hVV}=(\vLstA[\hV],\vLstA[\hVV]) \label{ZZZA:2:1} \\
			\sub{(\clr{\hVarX}\mmax{\hVarX}{\!\hV'})}{\hVarX} \label{ZZZA:2:2}
		\end{gather}
		By \eqref{ZZZA:2:1} and defn of \textsf{excl} (\defref{def:excl}) we know 
		\begin{gather}
			\excl{\hV_{1},\hVV}=(\vLstA[\hV_{1}],\vLstA[\hVV]) \label{ZZZA:2:3} \\ 
			\excl{\hV_{2},\hVV}=(\vLstA[\hV_{2}],\vLstA[\hVV]) \label{ZZZA:2:4} \\ 
			\vLstA[\hV] = (\vLstA[\hV_{1}]\cup\vLstA[\hV_{2}]) \label{ZZZA:2:5}  
		\end{gather}
		By \eqref{ZZZA:2:2}, \eqref{ZZZA:2:3} and IH we know 
		\begin{align}
			\excl{\hV_{1}\sub{(\clr{\hVarX}\mmax{\hVarX}{\!\hV'})}{\hVarX},\hVV}=(\vLstA[\hV_{1}],\vLstA[\hVV]) \label{ZZZA:2:6} 
		\end{align}
		By \eqref{ZZZA:2:2}, \eqref{ZZZA:2:4} and IH we know 
		\begin{align}
			\excl{\hV_{2}\sub{(\clr{\hVarX}\mmax{\hVarX}{\!\hV'})}{\hVarX},\hVV}=(\vLstA[\hV_{2}],\vLstA[\hVV]) \label{ZZZA:2:7} 
		\end{align}
		By \eqref{ZZZA:2:5}, \eqref{ZZZA:2:6}, \eqref{ZZZA:2:7} and defn of \textsf{excl} (\defref{def:excl}) we know
		\begin{align}
			\excl{(\hV_{1}\sub{(\clr{\hVarX}\mmax{\hVarX}{\!\hV'})})\mand(\hV_{2}\sub{(\clr{\hVarX}\mmax{\hVarX}{\!\hV'})}{\hVarX}),\hVV}=(\vLstA[\hV],\vLstA[\hVV]) \label{ZZZA:2:8} 
		\end{align}
		By \eqref{ZZZA:2:8} and substitution we know
		\begin{align}
			\excl{(\hV_{1}\mand\hV_{2})\sub{(\clr{\hVarX}\mmax{\hVarX}{\!\hV'})}{\hVarX},\hVV}=(\vLstA[\hV],\vLstA[\hVV]) \label{ZZZA:2:9} 
		\end{align}		
		\noindent$\therefore$ Case holds by 	\eqref{ZZZA:2:9}.						
	\end{Case}
	
	\begin{Case}[$\hV=\mattrNec{\patE}{\attr}{\vLstA}{\hV}$]
		From our rule premises we know 
		\begin{gather}
			\excl{\mattrNec{\patE}{\attr}{\vLstA}{\hV},\hVV}=(\vLstA[\hV],\vLstA[\hVV]) \label{ZZZA:3:1} \\
			\sub{(\clr{\hVarX}\mmax{\hVarX}{\!\hV'})}{\hVarX} \label{ZZZA:3:2}
		\end{gather}
		We must consider two subcases.\medskip
		
		\begin{Subcase}[$\hVV\neq\trivSat$]
			By \eqref{ZZZA:3:1} and defn of \textsf{excl} (\defref{def:excl}) we know 
			\begin{gather}
				\excl{\mrelease{\vLstA}\mtru,\hVV}=(\vLstA,\vLstA[\hVV]) \label{ZZZA:3:3} \\
				\forall\patE_{1}\in\fps{\hVV}\cdot\match{\patE}{\patE_{1}}{\undef} \label{ZZZA:3:4} 
			\end{gather}
			Since $\mrelease{\vLstA}\mtru\equiv(\mrelease{\vLstA}\mtru)\sub{(\clr{\hVarX}\mmax{\hVarX}{\!\hV'})}{\hVarX}$, by \eqref{ZZZA:3:3} we know 
			\begin{align}
				\excl{(\mrelease{\vLstA}\mtru)\sub{(\clr{\hVarX}\mmax{\hVarX}{\!\hV'})}{\hVarX},\hVV}=(\vLstA,\vLstA[\hVV]) \label{ZZZA:3:5} 
			\end{align}			
			By \eqref{ZZZA:3:4}, \eqref{ZZZA:3:5} and defn of \textsf{excl} (\defref{def:excl}) we can deduce 		
			\begin{align}
				\excl{(\mattrNec{\patE}{\attr}{\vLstA}{\hV})\sub{(\clr{\hVarX}\mmax{\hVarX}{\!\hV'})}{\hVarX},\hVV}=(\vLstA,\vLstA[\hVV]) \label{ZZZA:3:6}				
			\end{align}
			\noindent$\therefore$ Subcase holds by \eqref{ZZZA:3:6}.						
		\end{Subcase}
		
		\begin{lastsubcase}[$\hVV=\trivSat$]
			By defn of \textsf{excl} (\defref{def:excl}) and since $\hVV=\trivSat$ we know that no matter what $\hV$ is, the \textsf{excl} function always returns $\varepsilon$ as the release set of $\hV$. Hence by \eqref{ZZZA:3:1} and \eqref{ZZZA:3:2} we know  
			\begin{align}
				\excl{\mattrNec{\patE}{\attr}{\vLstA}{\hV},\;\trivSat} &= \nonumber 		\\
				\excl{(\mattrNec{\patE}{\attr}{\vLstA}{\hV})\sub{(\clr{\hVarX}\mmax{\hVarX}{\!\hV'})}{\hVarX},\;\trivSat}&= \;(\varepsilon,\vLstA[\hVV]) \label{ZZZA:3:8} 
			\end{align}
			\noindent$\therefore$ Subcase holds by 	\eqref{ZZZA:3:8}.
		\end{lastsubcase}
	\end{Case}
	
	\begin{Case}[$\hV=\mscor{\vLstA}{\vLstB}\hV$]
		From our rule premises we know 
		\begin{gather}
			\excl{\mscor{\vLstA}{\vLstB}\hV,\hVV}=(\vLstA[\hV],\vLstA[\hVV]) \label{ZZZA:4:1} \\
			\sub{(\clr{\hVarX}\mmax{\hVarX}{\!\hV'})}{\hVarX} \label{ZZZA:4:2}
		\end{gather}
		By defn of \textsf{excl} (\defref{def:excl}) we know that \eqref{ZZZA:3:1} can only be true if $\hVV\eq\trivSat$ (because otherwise, \textsf{excl} would have returned \undef). Hence we know
		\begin{align}
			 \hVV\eq\trivSat \label{ZZZA:4:3} 
		\end{align}
		Since we know \eqref{ZZZA:4:3}, then we also know that no matter what $\hV$ is, the \textsf{excl} function always returns $\varepsilon$ as the release set of $\hV$. Hence by \eqref{ZZZA:4:1}, \eqref{ZZZA:4:2} and defn of \textsf{excl} (\defref{def:excl}) we can deduce 
		\begin{align}
			\excl{\mscor{\vLstA}{\vLstB}\hV,\;\trivSat} &= \nonumber 		\\
			\excl{(\mscor{\vLstA}{\vLstB}\hV)\sub{(\clr{\hVarX}\mmax{\hVarX}{\!\hV'})}{\hVarX},\trivSat}&=(\varepsilon,\vLstA[\hVV]) \label{ZZZA:4:4} 
		\end{align}
		\noindent$\therefore$ Case holds by \eqref{ZZZA:4:4}.		
	\end{Case}
	
	\noindent\textbf{Note:} The proofs for cases $\hV\!=\!\macor{\vLstA}{\vLstB}\hV$,\, $\hV\!=\!\mblock{i}\hV$ and $\hV\!=\!\mfls$ are analogous to the proof for case $\hV=\mscor{\vLstA}{\vLstB}\hV$.\bigskip
	
	\begin{Case}[$\hV=\mrelease{\vLstA}{\hV}$]
		From our rule premises we know 
		\begin{gather}
			\excl{\mrelease{\vLstA}{\hV},\hVV}=(\vLstA[\hV],\vLstA[\hVV]) \label{ZZZA:5:1} \\
			\sub{(\clr{\hVarX}\mmax{\hVarX}{\!\hV'})}{\hVarX} \label{ZZZA:5:2}
		\end{gather}
		By defn of \textsf{excl} (\defref{def:excl}) we know that \eqref{ZZZA:5:1} can only be true in the following two subcases (because otherwise, \textsf{excl} would have returned \undef).\bigskip
		
		\begin{Subcase}[$\hV=\mtru$]
			Since $\hV=\mtru$, by \eqref{ZZZA:5:1} and defn of \textsf{excl} (\defref{def:excl}) we know 
			\begin{align}
				\excl{\mrelease{\vLstA}\mtru,\hVV}=(\vLstA,\vLstA[\hVV]) \label{ZZZA:5:3} 
			\end{align}
			By defn of substitution we know 
			\begin{align}
				(\mrelease{\vLstA}\mtru)\sub{(\clr{\hVarX}\mmax{\hVarX}{\!\hV'})}{\hVarX},\hVV \equiv \mrelease{\vLstA}\mtru \label{ZZZA:5:4} 
			\end{align}			
			By \eqref{ZZZA:5:3} and \eqref{ZZZA:5:4} we deduce
			\begin{align}
				\excl{(\mrelease{\vLstA}\mtru)\sub{(\clr{\hVarX}\mmax{\hVarX}{\!\hV'})}{\hVarX},\hVV}=(\vLstA,\vLstA[\hVV]) \label{ZZZA:5:5} 
			\end{align}	
			\noindent$\therefore$ Subcase holds by \eqref{ZZZA:5:5}.						
		\end{Subcase}
		
		\begin{lastsubcase}[$\hVV=\trivSat$]
			By defn of \textsf{excl} (\defref{def:excl}) and since $\hVV=\trivSat$ we know that no matter what $\hV$ is, the \textsf{excl} function always returns $\varepsilon$ as the release set of $\hV$. Hence by \eqref{ZZZA:5:1} and \eqref{ZZZA:5:2} we know  
			\begin{align}
				\excl{\mrelease{\vLstA}\hV,\;\trivSat} &= \nonumber 		\\
				\excl{(\mrelease{\vLstA}\hV)\sub{(\clr{\hVarX}\mmax{\hVarX}{\!\hV'})}{\hVarX},\;\trivSat}&= \;(\varepsilon,\vLstA[\hVV]) \label{ZZZA:5:8} 
			\end{align}
			\noindent$\therefore$ Subcase holds by 	\eqref{ZZZA:5:8}.
		\end{lastsubcase}
	\end{Case}
	
	\begin{Case}[$\hV=\mboolE{c}{\hV}{\hVV}$]
		From our rule premises we know 
		\begin{gather}
			\excl{\mboolE{c}{\hV}{\hVV},\hVV}=(\vLstA[\hV],\vLstA[\hVV]) \label{ZZZA:6:1} \\
			\sub{(\clr{\hVarX}\mmax{\hVarX}{\!\hV'})}{\hVarX} \label{ZZZA:6:2}
		\end{gather}
		We must consider the following two subcases: \medskip
		
		\begin{Subcase}[$\hVV\neq\trivSat$]
			By \eqref{ZZZA:6:1} and defn of \textsf{excl} (\defref{def:excl}) we know 
			\begin{gather}
				\excl{\hV_{1},\hVV}=(\vLstA[\hV],\vLstA[\hVV]) \label{ZZZA:6:3} \\
				\excl{\hV_{2},\hVV}=(\vLstA[\hV],\vLstA[\hVV]) \label{ZZZA:6:4} 
			\end{gather}
			By \eqref{ZZZA:6:2}, \eqref{ZZZA:6:3} and IH we know 
			\begin{align}
				\excl{\hV_{1}\sub{(\clr{\hVarX}\mmax{\hVarX}{\!\hV'})}{\hVarX},\hVV}=(\vLstA[\hV],\vLstA[\hVV]) \label{ZZZA:6:5}
			\end{align}
			By \eqref{ZZZA:6:2}, \eqref{ZZZA:6:4} and IH we know 
			\begin{align}
				\excl{\hV_{2}\sub{(\clr{\hVarX}\mmax{\hVarX}{\!\hV'})}{\hVarX},\hVV}=(\vLstA[\hV],\vLstA[\hVV]) \label{ZZZA:6:6}
			\end{align}
			By \eqref{ZZZA:6:2}, \eqref{ZZZA:6:4} and defn of \textsf{excl} (\defref{def:excl}) we can conclude 
			\begin{align}
				\excl{\mboolE{c}{(\hV_{1}\sub{(\clr{\hVarX}\mmax{\hVarX}{\!\hV'})}{\hVarX})}{(\hV_{2}\sub{(\clr{\hVarX}\mmax{\hVarX}{\!\hV'})}{\hVarX})},\hVV}=(\vLstA[\hV],\vLstA[\hVV]) \label{ZZZA:6:7}
			\end{align}
			By \eqref{ZZZA:6:7} and substitution we know 
			\begin{align}
				\excl{(\mboolE{c}{\hV_{1}}{\hV_{2})\sub{(\clr{\hVarX}\mmax{\hVarX}{\!\hV'})}{\hVarX}},\hVV}=(\vLstA[\hV],\vLstA[\hVV]) \label{ZZZA:6:8}				
			\end{align}			
			\noindent$\therefore$ Subcase holds by \eqref{ZZZA:6:8}.						
		\end{Subcase}
		
		\begin{lastsubcase}[$\hVV=\trivSat$]
			By defn of \textsf{excl} (\defref{def:excl}) and since $\hVV=\trivSat$ we know that no matter what $\hV$ is, the \textsf{excl} function always returns $\varepsilon$ as the release set of $\hV$. Hence by \eqref{ZZZA:6:1} and \eqref{ZZZA:6:2} we know  
			\begin{align}
				\excl{\mboolE{c}{\hV_{1}}{\hV_{2}},\;\trivSat} &= \nonumber 		\\
				\excl{(\mboolE{c}{\hV_{1}}{\hV_{2}})\sub{(\clr{\hVarX}\mmax{\hVarX}{\!\hV'})}{\hVarX},\;\trivSat}&= \;(\varepsilon,\vLstA[\hVV]) \label{ZZZA:6:9} 
			\end{align}
			\noindent$\therefore$ Subcase holds by 	\eqref{ZZZA:6:9}.
		\end{lastsubcase}
	\end{Case}
	
	\begin{Case}[$\hV=\mtru$]
		From our rule premises we know 
		\begin{gather}
			\excl{\mtru,\hVV}=(\vLstA[\hV],\vLstA[\hVV]) \label{ZZZA:7:1} \\
			\sub{(\clr{\hVarX}\mmax{\hVarX}{\!\hV'})}{\hVarX} \label{ZZZA:7:2}
		\end{gather}
		From \eqref{ZZZA:7:1} and since $\mtru\equiv\mrelease{\varepsilon}\mtru$ we can deduce 
		\begin{align}
			\excl{\mrelease{\varepsilon}\mtru,\hVV}=(\vLstA[\hV],\vLstA[\hVV]) \label{ZZZA:7:3}
		\end{align}
		By \eqref{ZZZA:7:2}, \eqref{ZZZA:7:3} and IH we know 
		\begin{align}
			\excl{(\mrelease{\varepsilon}\mtru)\sub{(\clr{\hVarX}\mmax{\hVarX}{\!\hV'})}{\hVarX},\hVV}=(\vLstA[\hV],\vLstA[\hVV]) \nonumber \\\equiv\quad \excl{\mtru\sub{(\clr{\hVarX}\mmax{\hVarX}{\!\hV'})}{\hVarX},\hVV}=(\vLstA[\hV],\vLstA[\hVV]) \label{ZZZA:7:4}
		\end{align}
		\noindent$\therefore$ Case holds by \eqref{ZZZA:7:4}.
	\end{Case}
	
	\begin{lastcase}[$\hV=Y$]
		From our rule premises we know 
		\begin{gather}
			\excl{Y,\hVV}=(\vLstA[\hV],\vLstA[\hVV]) \label{ZZZA:8:1} \\
			\sub{(\clr{\hVarX}\mmax{\hVarX}{\!\hV'})}{\hVarX} \label{ZZZA:8:2}
		\end{gather}
		By defn of \textsf{excl} (\defref{def:excl}) we know that \eqref{ZZZA:8:1} can only be true if $\hVV\eq\trivSat$ (because otherwise, \textsf{excl} would have returned \undef). Hence we know
		\begin{align}
			 \hVV\eq\trivSat \label{ZZZA:8:3} 
		\end{align}
		Since we know \eqref{ZZZA:8:3}, then we also know that no matter what $\hV$ is (\ie irrespective of whether $Y=X$ or not), the \textsf{excl} function always returns $\varepsilon$ as the release set of $\hV$. Hence by \eqref{ZZZA:8:1}, \eqref{ZZZA:8:2} and defn of \textsf{excl} (\defref{def:excl}) we can deduce 
		\begin{align}
			\excl{Y,\;\trivSat} &= \nonumber 		\\
			\excl{Y\sub{(\clr{\hVarX}\mmax{\hVarX}{\!\hV'})}{\hVarX},\trivSat}&=(\varepsilon,\vLstA[\hVV]) \label{ZZZA:8:4} 
		\end{align}
		\noindent$\therefore$ Case holds by \eqref{ZZZA:8:4}.		
	\end{lastcase}		
\end{proof}
\end{lemma}

\bibliomatter
\bibliographystyle{abbrv}
\bibliography{references}
 
\end{document}